\documentclass[11pt, a4paper]{report}
\usepackage[T1]{fontenc} 
\pdfoutput=1
\usepackage[utf8x]{inputenc} 
\usepackage[spanish]{babel}

\usepackage{amsfonts}
\usepackage{amssymb}

\usepackage{setspace}

\usepackage[T1]{fontenc} 
\usepackage{slashed}
\usepackage{color}

\usepackage{epsfig,graphicx,bm,color} 
\usepackage{dcolumn}
\usepackage{bm}
\usepackage{hyperref}
\usepackage{color}
\usepackage{breakurl}
\usepackage{soul} 

\usepackage{subfigure}
\usepackage{multirow}
\usepackage{amsmath}
\usepackage{float}
\usepackage{mathtools}

\usepackage{babelbib}

\newcommand{\newc}{\newcommand}

\newc{\Fermi}{\textit{Fermi}-}
\def    \beq            {\begin{equation}}
\def    \eeq            {\end{equation}}
\def    \bea           {\begin{eqnarray}}
\def    \eea           {\end{eqnarray}}

\begin{document}

\thispagestyle{empty} 

\begin{center}

\begin{figure}
\centering
\includegraphics[width=4cm]{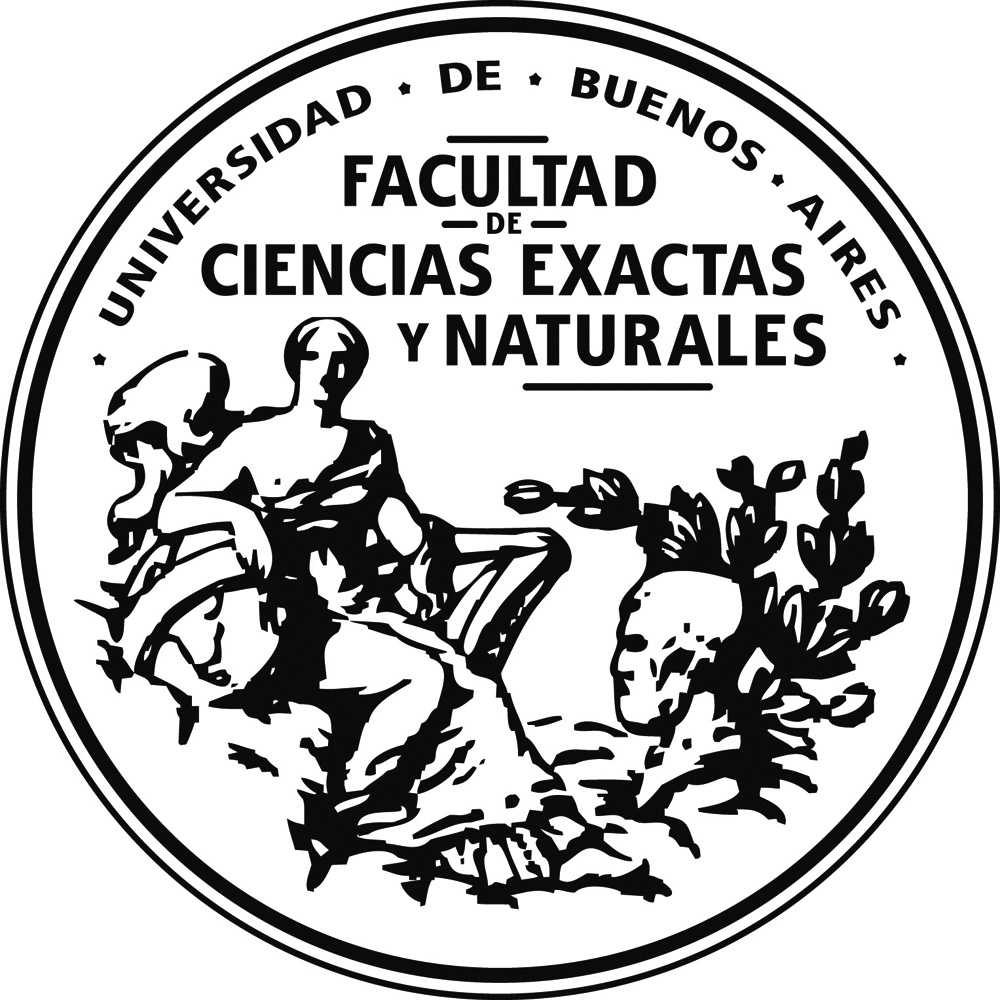}
\end{figure}

\begin{normalsize}
\textbf{UNIVERSIDAD DE BUENOS AIRES}\\
Facultad de Ciencias Exactas y Naturales\\
Departamento de Física
\end{normalsize}

\vspace{25mm}

\huge{\textbf{Fenomenología de modelos supersimétricos: partículas y materia oscura}}

\vspace{35mm}

\begin{normalsize}
Tesis presentada para optar al título de\\
\textit{Doctor de la Universidad de Buenos Aires en el área Ciencias Físicas}\\
por \large \textbf{Lic. Andres Daniel PEREZ}
\end{normalsize}

\vspace{30mm}
\end{center}
\begin{normalsize}
Director de Tesis: \textbf{Dr. Daniel Elbio LOPEZ-Fogliani}\\
Consejero de Estudios: \textbf{Dr. Claudio M. SIMEONE}

\vspace{4mm}
\begin{flushleft}
Lugar de Trabajo: Instituto de Física de Buenos Aires, UBA-CONICET, Departamento de Física, Facultad de Ciencias Exactas y Naturales, Universidad de Buenos Aires.\\
Buenos Aires, 2020
\end{flushleft}

\end{normalsize}

\newpage

\thispagestyle{empty} 
\null

\newpage

\newpage
\pagestyle{plain}
\pagenumbering{roman}
\setcounter{page}{1}

\spacing{1.36}
\abstract{

Dilucidar la composición de la materia oscura es uno de los temas abiertos más importantes en la fenomenología de partículas y astropartículas. Para ello, es necesario contar con un marco teórico de trabajo, y poder contrastar los resultados con experimentos capaces de detectar una señal proveniente de la materia oscura.

Con respecto al marco teórico, los modelos supersimétricos se encuentran muy bien motivados y ofrecen candidatos viables a materia oscura. Por consiguiente, el compañero supersimétrico del gravitón, conocido como gravitino, surge como un candidato natural ya que solo es necesario asumir la existencia de gravedad en un contexto supersimétrico.
	
En el marco de las extensiones supersimétricas mínimas del modelo estándar de partículas fundamentales, el `mu-from-nu supersymmetric standard model', $\mu\nu$SSM, es capaz de reproducir toda la fenomenología conocida, incluyendo la masa y ángulos de mezcla de los neutrinos, dando además la posibilidad de encontrar señales de nueva física en un futuro cercano. 

Trabajando en el modelo mencionado, se estudió la detección en experimentos como \Fermi-LAT de señales de rayos gamma provenientes del decaimiento del gravitino como materia oscura. Para ello se tuvo en cuenta todos los decaimientos del gravitino, complementando trabajos anteriores.

La posibilidad de que se construya una nueva generación de detectores de rayos gamma en un futuro próximo, con la presentación a nivel internacional de propuestas concretas, motivó la inclusión del axino como candidato a materia oscura. El axino es el compañero supersimétrico del axión, y este último surge a partir de la conocida solución al problema denominado `strong CP problem' del modelo estándar de partículas fundamentales. En esta tesis se ha analizado al axino como candidato a materia oscura el contexto del $\mu\nu$SSM por primera vez. 

Finalmente, se han estudiado escenarios con componente múltiple de materia oscura, en particular gravitinos y axinos ambos coexistiendo como constituyentes de la misma. Este escenario novedoso, incluye dos casos muy diferentes, cada uno con características distintivas: el axino más ligero que el gravitino y vice versa. Teniendo en cuenta la sensibilidad de futuros telescopios de rayos gamma en el rango energético MeV-GeV, se analizó la detectabilidad del axino y gravitino en los escenarios mixtos previamente mencionados. Se concluye que en casos especiales ambos pueden contribuir con una linea de rayos gamma como señal clara que indicaría la composición de la materia oscura.}

\bigskip

\noindent Palabras clave: {\it Fenomenología, Materia Oscura, Supersimetría, Gravitino, Axino, Rayos gamma}.

\spacing{1}

\newpage
\pagestyle{plain}
\pagenumbering{gobble}
\spacing{1.36}

\begin{center}
    \textbf{Abstract:}
\end{center}

To elucidate the composition of dark matter is one of the most important open questions in particle and astroparticle phenomenology. To answer it, is necessary to work in a theoretical framework and to be able to experimentally test the theoretical results through dark matter signals.

	With respect to the theoretical framework, supersymmetric models are very appealing since are well motivated offering viable dark matter candidates. In this context, the supersymmetric partner of the graviton, known as gravitino, becomes a natural candidate. The existence of gravity in a supersymmetric context guarantees its existence.
	
	Within the framework of minimal supersymmetric extensions of the standard model of fundamental particles, the `mu-from-nu supersymmetric standard model', $\mu\nu$SSM,  reproduces all known particle phenomenology including neutrino masses and mixing angles, giving also the possibility of detectable signals in the near future.
	
	Working in the mentioned model, in this thesis we analyze the detection of gamma-ray signals arising from decaying gravitino dark matter considering experiments such as \Fermi LAT. For this purpose all gravitino decays have been taken into account, complementing previous works.
	
	The possibility of an upcoming new generation of gamma-ray detectors, with concrete proposals at the international level submitted, has motivated the inclusion of axino as a dark matter candidate. The axino is the supersymmetric partner of the axion, and the later   arise as a well-known solution to the  `strong CP problem' of the standard model of fundamental particles. In this thesis, the axino as a dark matter candidate has been analyzed in the context of the $\mu\nu$SSM for the first time.
	
	Finally, scenarios with multicomponent dark matter have been studied, focusing on gravitinos and axinos  coexisting as dark matter constituents. This novel scenario includes two very different cases, each one with distinctive features: the axino lighter than the gravitino and vice versa. Considering the sensitivity of future MeV-GeV gamma-ray telescopes, axino and gravitino signals were analyzed in the mixed scenarios mentioned previously. In special cases both dark matter candidates can produce a gamma-ray line as a smoking gun that would uncover the dark matter composition.

\bigskip


\spacing{1}



\newpage
\pagenumbering{roman}
\setcounter{page}{3}
 
\chapter*{Agradecimientos} 
\spacing{1.5}
Al universo por ser bastante interesante.

A mi familia, en especial a mis viejos que hicieron siempre lo imposible y más.

A Iris, que se rehúsa a rimar, por el tremendo apoyo a lo largo de todos estos años.

A Zoster, que hizo todo relativo.

A los muchachos de la oficina y mis amigos, que siempre han sido indispensables.

Al Rotary Club, en particular al ingeniero Mignone por el apoyo durante la secundaria y la licenciatura.

A Daniel L-F, por su buena onda y por compartir su trabajo. Al igual que a mis colaboradores Carlos M, Germán G-V y Roberto RdA, por mostrarme los procesos de la ciencia, haciendo.

A la UBA, en particular a FCEyN y a todos los que la componen, por ofrecerme mucho más que educación. No imagino un mejor ámbito para estudiar.

A toda Argentina, que funciona por razones cósmicas.
\bigskip

El resultado de mi tesis es más que la consecuencia de mi esfuerzo individual.

\bigskip
\bigskip
\bigskip
\bigskip
\bigskip
\bigskip
\bigskip
\bigskip
\bigskip
\bigskip
\bigskip
\bigskip
\bigskip
\bigskip
\bigskip


\spacing{1}
\newpage
\thispagestyle{empty} 
\null

\newpage

\tableofcontents

\newpage

\pagenumbering{arabic}

\chapter{Introducción}\label{introduccion}

\spacing{1.5}

El modelo estándar de partículas fundamentales, o SM por sus siglas en inglés, provee una descripción notablemente exitosa y sin precedentes de los fenómenos explorados hasta el momento. La correspondencia entre las predicciones del SM y los experimentos es asombrosa.

La base del SM se encuentra en los trabajos de Glashow~\cite{Glashow:1961}, Weinberg~\cite{Weinberg:1967} y Salam~\cite{Salam:1964}, que introdujeron por primera vez la teoría unificada de las interacciones electrodébiles con grupos de recalibración o de `\textit{gauge}' $SU(2)_L \otimes  U(1)_Y$. Dicha simetría prohíbe la inclusión de términos de masa para bosones de gauge y fermiones, necesarios para describir las observaciones experimentales. 
La solución a este problema se da a través del mecanismo de Higgs~\cite{Englert:1964,Higgs:1964,Guralnik:1964}, el cual implica la ruptura espontánea de la simetría electrodébil (EWSB). Un doblete escalar de $SU(2)$, $\phi$, es introducido para tal fin de manera que el Lagrangiano del SM respeta la simetría de gauge, pero el estado de más baja energía para el escalar $\phi$ es distinto de cero, es decir, posee un valor de expectación de vacío (VEV) que notamos $v$. Luego de aplicar el teorema de Nambu-Goldstone~\cite{Goldstone:1961,Nambu:1960}, se introducen términos de masa para los bosones de gauge electrodébiles a partir de los términos cinéticos del potencial escalar. Al mismo tiempo, el subgrupo $U(1)_{\text{em}}$ permanece intacto, por lo cual el fotón no adquiere masa. Finalmente, la ruptura espontánea de simetría gauge se puede representar como $SU(2)_L \otimes U(1)_Y \xrightarrow[]{v \neq 0} U(1)_\text{{em}}$, donde el subíndice $Y$ refiere al número cuántico de hipercarga, y el subíndice $L$ especifica las componentes levógiras, izquierdas, o `\textit{left-handed}' de las tres generaciones de leptones y quarks organizadas en dobletes de $SU(2)$, mientras que las componentes dextrógiras, derechas, o `\textit{right-handed}' son singletes.

La teoría permite en el Lagrangiano términos invariantes de gauge del tipo de Yukawa, $Y_{ij} \, \psi_{i,L} \cdot \phi \, \psi_{j,R}$, que mezclan las componentes izquierdas y derechas de los fermiones $\psi$ con el doblete escalar. El punto indica el producto antisimétrico de $SU(2)$. De este modo, se genera un término de masa para los fermiones dependiente de $v$, $m_{\psi}=Yv/\sqrt{2}$, que estaría prohibido antes de EWSB, es decir, antes que $\phi$ adquiera un VEV.

Con la inclusión de la teoría cuántica de campos que describe las interacciones fuertes, `\textit{Quantum chromodynamics}' o QCD (teoría de gauge no abeliana $SU(3)$), la estructura de grupo del SM antes de EWSB resulta:
\begin{equation}
    SU(3)_{C} \otimes SU(2)_L \otimes U(1)_Y,
\end{equation}
donde el subíndice $C$ se refiere a la carga de color de los quarks y gluones. Los quarks pueden adquirir masa mediante términos de Yukawa, mientras que los gluones resultan no masivos dado que $SU(3)_C$ no está involucrado en EWSB.

Una de las predicciones más importantes del SM ha sido la existencia de un escalar neutro, necesario como hemos dicho para generar la ruptura espontánea de simetría. En 2012, el Gran Colicionador de Hadrones (Large Hadron Collider, LHC) descubrió una nueva resonancia~\cite{ATLAS:2012,CMS:2012}, con un valor esperado de masa $m_h\simeq 125$ GeV. Luego, con más mediciones sobre sus acoples y espín (`\textit{spin}') se pudo confirmar que la partícula producida es compatible con el bosón de Higgs, la última partícula predicha por el SM sin detectar hasta ese entonces, y notablemente la única partícula escalar de la teoría, dando comienzo a una nueva era en la física de partículas.

Los 19 parámetros libres del SM (3 acoples de gauge, 9 masas para los fermiones, 3 ángulos de mezcla del sector de quarks, 1 fase de violación de CP, 1 VEV del Higgs, 1 masa del Higgs, y 1 ángulo del vacío de QCD), explican con gran precisión los fenómenos explorados a bajas energías. A pesar de su rotundo éxito y del espectacular hito histórico del descubrimiento del bosón de Higgs, existen motivos para creer que el SM debe ser extendido. En primer lugar, el modelo no incluye a la gravedad. Un nuevo marco teórico deberá ser formulado, al menos, a la escala de la energía de Planck $M_{Pl} c^2=(\hbar c^5 /8\pi G_{\text{N}})^{1/2}=2.4 \times 10^{18}$ $\text{GeV}$ con $c$ la velocidad de la luz, $\hbar$ la constante de Planck reducida, y $G_{\text{N}}$ la constante de gravitación universal de Newton, donde los efectos cuántico-gravitacionales no incluidos en el SM se tornan relevantes. Pero esa escala de energía está muy alejada de los 14 TeV que puede alcanzar el Gran Colisionador de Hadrones, y su exploración está fuera del alcance de experimentos realizables en un futuro próximo.

Si pudiésemos interpretar el SM como válido hasta la escala de Planck, podríamos trabajar desde el punto de vista fenomenológico sin inconvenientes con el SM, dada la imposibilidad de verificar o refutar los fenómenos más allá de dicha escala. Esto nos lleva al segundo inconveniente, el problema de las jerarquías (aunque técnicamente no es una dificultad del SM en sí mismo, sino de la interpretación del modelo como válido hasta la escala donde los efectos cuántico-gravitacionales son importantes). La dependencia cuadrática con la escala y acoplamientos de nueva física del cuadrado de la masa del bosón de Higgs, hace que dicha escala sea la natural para la masa del Higgs. Por ejemplo, para una teoría conteniendo gravedad debido a correcciones cuadráticas, sería de esperar que la masa del Higgs fuese del orden de la masa de Planck, anulando de esta manera la ruptura electrodébil tan exitosa del SM.

Por otra parte, a escalas órdenes de magnitud más grandes que la longitud de Planck $l_{Pl}=(\hbar G_{\text{N}} / c^3)^{1/2}=1,6 \times 10^{-35}$ m, la teoría general de la relatividad describe exitosamente todos los fenómenos gravitatorios\footnote{También es necesario que $l_{Pl}^2$ sea más grande que la curvatura del espacio-tiempo.}. En particular, los desarrollos en cosmología de los últimos 50 años también han sido asombrosos. Gracias a una cantidad enorme de datos y observaciones que describen la composición del universo con precisiones incluso inferiores al por ciento, tenemos un modelo estándar cosmológico denominado $\Lambda$CDM (`\textit{Lambda Cold Dark Matter}'), donde $\Lambda$ se refiere a la constante cosmológica (energía oscura) y CDM a materia oscura fría.

En este contexto, la identidad de la materia oscura o DM es uno de los grandes enigmas de la física contemporánea. Durante los últimos cien años la comunidad científica ha acumulado una enorme cantidad de evidencias mediante distintas observaciones sobre su existencia. Describiendo las observaciones con simples argumentos gravitacionales, se ha revelado que es necesaria la inclusión de un nuevo tipo de materia no bariónica presente desde escalas galácticas (por ejemplo a través de curvas de rotación de galaxias, lentes gravitacionales, colisiones de clusters de galaxias) hasta escalas cosmológicas (en la estructura a gran escala del universo o en el espectro de la radiación cósmica de fondo, entre otros). Para todas las observaciones la hipótesis de materia oscura provee la explicación más satisfactoria y simple. Sin embargo, el SM no nos brinda ningún candidato viable ni descripción alguna sobre la composición de la materia faltante.

Otra importante evidencia experimental involucra al sector de los neutrinos. En el SM dichas partículas no poseen masa, pero durante los últimos veinte años distintas colaboraciones han demostrado que son masivos. Como consecuencia, todos los modelos teóricos deben incluir este resultado, al menos en alguna extensión. Se ha comprobado que el sector posee una estructura compleja con oscilaciones y ángulos de mezcla entre los estados, permitiendo una singular y notable fenomenología que puede brindar pistas sobre física más allá del SM.

Finalmente, el SM tiene otro inconveniente que involucra violación de carga-paridad (CP) en el sector de QCD, llamado `\textit{strong CP-problem}'. La teoría permite incluir un término en el Lagrangiano que respeta todas las simetrías del SM: $\bar{\theta} \; F_a^{\mu\nu} \; \widetilde{F}_{\mu\nu a}$, donde $F_a^{\mu\nu}$ y $\widetilde{F}_{\mu\nu a}$
son el tensor de fuerza de los gluones y su dual, y $\bar{\theta}$ está relacionado con la topología del vacío de la teoría. Dicho término es fuente de violación de CP y debería generar un momento dipolar eléctrico del neutrón medible. Sin embargo, experimentalmente se ve que $\bar{\theta}<10^{-10}$~\cite{Afach:2015sja}, lo que constituye un problema de naturalidad. La manera más simple de solucionar el inconveniente es mediante física más allá del SM con la introducción de un campo pseudo-escalar llamado axión, y una simetría global $U(1)_{PQ}$ extra que se encuentra rota~\cite{Peccei:1977hh,Peccei:1977ur,Weinberg:1977ma,Wilczek:1977pj}.

Como explicaremos más adelante, los modelos supersimétricos o SUSY cumplen un rol muy importante en la física teórica, al proveer de soluciones a los problemas que el SM no puede resolver. Mediante una transformación supersimétrica los campos bosónicos rotan a fermiónicos y viceversa. En un modelo SUSY los campos que rotan entre sí a través de esta transformación se dicen que pertenecen a un mismo supermultiplete. Como consecuencia importante de SUSY el número de grados de libertad bosónicos y fermiónicos de la teoría son necesariamente iguales.
Entonces, la cancelación sistemática de contribuciones del cuadrado de la masa del bosón de Higgs puede ser explicada por medio de esta simetría, ya que las contribuciones de loops de campos contenidos en el mismo supermultiplete se cancelan, resolviendo el problema de las jerarquías. Cualidad que torna muy interesante a los modelos SUSY.

Pero como es bien sabido tal simetría no se observa en la naturaleza, por lo tanto, si existe debe estar rota. Una idea de cómo llevar a cabo una ruptura espontánea de supersimetría consiste en postular un sector oculto de partículas donde SUSY se rompe espontáneamente, para ser mediado al sector visible. El resultado a baja escala es una teoría SUSY más términos denominados suaves o `\textit{soft}', cuyos coeficientes tienen dimensión positiva de masa y rompen SUSY explícitamente sin introducir divergencias cuadráticas. Trabajaremos con este tipo de modelos, pues nos interesa la descripción de fenómenos físicos por debajo de la escala de Planck.

Otro punto a favor de las teorías SUSY que las convierte en candidatos serios, es que poseen la característica de poder ser verificadas o refutadas con experimentos realizables a corto plazo. Ya que en sus versiones más puras o naturales la escala de ruptura de SUSY es del orden del TeV, cercana a la escala de ruptura electrodébil.


Las teorías SUSY además nos pueden brindar candidatos para explicar la composición de la materia oscura. Cuando se impone una simetría discreta llamada paridad-R, existe un número cuántico que distingue entre partículas supersimétricas y partículas del SM. Además, en una interacción la cantidad de partículas supersimétricas debe ser par. Entonces, la partícula supersimétrica más ligera o LSP, por sus siglas en inglés, es estable y serviría como materia oscura si cumple con otros requisitos experimentales, como por ejemplo que se genere la cantidad de reliquia correcta.

La mínima extensión del SM, llamado `\textit{Minimal Supersymmetric Standard Model}' o MSSM ha sido profundamente analizada en la literatura. En este modelo, el LSP y candidato a materia oscura es típicamente un neutralino, es decir una combinación de los compañeros fermiónicos neutros de los bosones de gauge B, W y los bosones de Higgs.

En el superpotencial del MSSM se encuentra presente un término de masa entre los supermultipletes de Higgs, $\mu \, \hat{H}_{1} \, \hat{H}_{2}$. El parámetro $\mu$ debe ser del orden de la ruptura electrodébil para producir la ruptura espontánea correcta de dicha simetría. Sin embargo, los únicos valores naturales para $\mu$ son cero o la masa de Planck (o bien la escala `\textit{Grand Unification Theory}', GUT). El primero introduce un bosón de Goldstone experimentalmente excluido, y el segundo introduce un problema de ajuste fino de los parámetros, debido a que $\mu$ entra en las ecuaciones de minimización. Por lo tanto, existe un problema de naturalidad en el MSSM denominado `problema-$\mu$'. Además, al igual que para el SM, el MSSM no puede explicar el patrón de masas de neutrinos observado.


Para lograr describir la física de neutrinos dentro del marco de las teorías SUSY, podemos extender los modelos y agregar neutrinos `\textit{right-handed}' guiados por la existencia de partes dextrógiras de los demás fermiones. De esta forma, mediante un mecanismo denominado de balancín o `\textit{seesaw}', los neutrinos activos pueden adquirir masa. Más aún, las componentes escalares de los neutrinos derechos pueden ser candidatos viables a materia oscura.

Por otro lado, es posible utilizar dicho campo no solo para obtener la física de neutrinos sino que al mismo tiempo se puede solucionar el problema-$\mu$ acoplándolo con los Higgses, mediante un término $\hat{\nu}_i^c \, \hat{H}_{1} \, \hat{H}_{2}$, donde $\hat{\nu}_i^c$ representa al supercampo de neutrinos derechos. Entonces, el término $\mu$ es generado dinámicamente cuando la partícula escalar compañera supersimétrica del neutrino adquiere VEV. 
Dicho modelo mínimo es denominado `\textit{$\mu$-from-$\nu$ Supersymmetric Standard Model}' o $\mu \nu$SSM, y la única escala presente es la escala de los términos de ruptura \textit{soft}, la cual se espera cercana a la escala electrodébil.


En el marco del $\mu \nu$SSM la simetría discreta paridad-R, que nos otorgaba candidatos a materia oscura estables en el MSSM, no se conserva. En dicho contexto, el LSP decae a partículas del SM, por lo tanto el candidato a materia oscura a considerar tiene que tener un tiempo de vida mucho más grande que la edad del universo. El gravitino, compañero supersimétrico del gravitón en modelos de supergravedad, como así también el axión y su compañero supersimétrico el axino, cumplen con todos los requisitos para ser buenos candidatos. En principio tanto para modelos con paridad-R rota como conservada, aunque su validez debe ser demostrada.

Asimismo, en el $\mu \nu$SSM el decaimiento del gravitino y del axino están íntimamente relacionados con la física de los neutrinos. Entonces, si queremos evaluar la viabilidad de dichos candidatos a materia oscura, tendremos que tener en cuenta el sector de los neutrinos.

El contenido de esta tesis es el siguiente. En el Capítulo~\ref{cosmo} presentamos un breve resumen de la cosmología estándar, algunas etapas en la historia térmica del universo relevantes para nuestro trabajo, y finalmente describimos las principales evidencias experimentales que implican la necesidad de incluir un nuevo tipo de materia en el contenido energético del universo.

En el Capítulo~\ref{del lado de alla} describimos la relación entre la física de partículas y las evidencias más importantes de física más allá del SM: las oscilaciones de neutrinos masivos y la existencia de materia oscura. Además presentamos detalles del `\textit{strong CP-problem}', su evidencia experimental, y la inclusión de un nuevo campo, el axión, como consecuencia de una posible la solución al problema CP.

El problema de las jerarquías del SM es introducido brevemente en el Capítulo~\ref{SUSY} así como también la solución que brindan las teorías supersimétricas. A continuación, describimos cómo construir modelos SUSY en cuatro dimensiones a bajas energías, y discutimos los modelos más simples, el MSSM y el $\mu\nu$SSM, estudiando la fenomenología de distintos sectores y las importantes diferencias entre modelos debido a la conservación o no de la paridad-R.

En los capítulos siguientes nos concentramos especialmente en la viabilidad y perspectivas de detección o exclusión de distintos candidatos a materia oscura en el marco de modelos supersimétricos que no conservan paridad-R, trabajando en el $\mu \nu$SSM. En el Capítulo~\ref{gravitinoresults1} consideramos al gravitino como único candidato a materia oscura. En primer lugar estudiamos el espacio de parámetros buscando reproducir la física de los neutrinos. Para ello, utilizamos las cotas experimentales actuales de sus masas y ángulos de mezcla. Luego, se enfatizan las cotas experimentales extraídas del estudio de rayos gamma, producto del decaimiento de gravitinos tanto provenientes de nuestra galaxia como de origen extragaláctico, principalmente obtenidos por el satélite Fermi. Se tienen en cuenta señales generadas por el decaimiento del gravitino a dos y tres cuerpos, para imponer los límites actuales y utilizar las perspectivas futuras a partir de un análisis completo de la morfología de la señal.

En el Capítulo~\ref{axinoDMchapter} consideramos al axino como único candidato a materia oscura para los dos modelos más populares de axiones: KSVZ y DFSZ, en el marco de teorías supersimétricas, y el $\mu\nu$SSM. Calculamos el flujo de rayos gamma producto del decaimiento del axino y comparamos las predicciones con las búsquedas de líneas espectrales. De este modo, imponemos límites sobre la masa y el tiempo de vida media del axino, para obtener las perspectivas de detección consistentes con una nueva generación de detectores de rayos gamma. Con este fin se tomó como referencia las especificaciones del proyecto e-ASTROGAM.

Una vez tratados en los dos capítulos anteriores escenarios con un único candidato a materia oscura, pasamos al estudio de la viabilidad de escenarios con múltiples candidatos a materia oscura en el Capítulo~\ref{axinogravitinoDMchapter} donde se analizó en rigor axinos y gravitinos coexistiendo.

El gravitino (axino) Next-to-LSP puede decaer a axino (gravitino) LSP, junto con un axión. Como consecuencia produce especies ultrarelativistas no interactuantes, las cuales poseen importantes cotas experimentales que aplicamos a nuestro análisis. Al igual que en los casos antes estudiados, tanto el gravitino como el axino pueden decaer a fotones y neutrinos, dando posibles señales en rayos gamma. Mostramos las señales que caracterizan y discriminan este escenario múltiple, el espacio de parámetros permitido por las observaciones actuales, y presentamos las perspectivas de detección que tendrá la próxima generación de telescopios de rayos gamma.

Finalmente, se presenta un resumen y conclusión del análisis del trabajo de esta tesis en el Capítulo~\ref{conclusiones}.

A continuación utilizamos unidades naturales donde $\hbar=c=1$.

\spacing{1}

\chapter{
Gravitación, cosmología, y el problema de la materia faltante
}\label{cosmo}

\spacing{1.5}

En este capítulo presentamos un breve resumen de la cosmología estándar. Discutimos la dinámica del universo en expansión y algunas etapas importantes en la historia térmica del universo, las cuales son relevantes para el análisis de gravitinos y axinos en los Capítulos~\ref{gravitinoresults1} y \ref{axinoDMchapter}, y para los límites impuestos por el decaimiento de un candidato a otro en el Capítulo~\ref{axinogravitinoDMchapter}. Finalmente, describimos las evidencias observacionales que llevan a incluir un nuevo tipo de materia no bariónica en el contenido energético del universo: la materia oscura.

\section{Universo en expansión}\label{cosmostandar}

Para estudiar la dinámica del universo es necesario utilizar la teoría general de la relatividad. Las ecuaciones de Einstein que relacionan la geometría espacio-temporal con el contenido de energía del universo son:
\begin{equation}
R_{\mu\nu}-\frac{1}{2}g_{\mu\nu}R+ \Lambda g_{\mu\nu} \; = \; 8 \pi G_N T_{\mu\nu},
\end{equation}
donde $R_{\mu\nu}$ y $R$ representan el tensor y el escalar de Ricci respectivamente, $g_{\mu\nu}$ la métrica del espacio-tiempo, $\Lambda$ la constante cosmológica, $G_N$ la constante gravitacional de Newton, y $T_{\mu\nu}$ el tensor de energía-momento el cual describe el contenido de energía del universo.

Podemos asumir una serie de simetrías para simplificar las ecuaciones acopladas de Einstein. Las mediciones del fondo cósmico de microondas (CMB) junto con los catálogos de distribución de galaxias muestran que el universo es isotrópico y homogéneo en grandes escalas, $O(100\text{Mpc})$. Entonces, la métrica más general bajo dichas hipótesis es la métrica de Friedmann-Lema$\hat{\text{i}}$tre–Robertson–Walker, cuyo intervalo está dado por
\begin{equation}
ds^2 \, = \, g_{\mu\nu} \, dx^{\mu}dx^{\nu} \, = \, dt^2 - a^2(t) \left[ \frac{dr^2}{1-k \, r^2} + r^2 \left( d\theta^2 + \sin^2 \theta d\phi^2 \right)  \right],
\end{equation}
donde $a(t)$ es el factor de escala, $r$, $\theta$ y $\phi$ son las coordenadas espaciales comóviles, y la constante $k$ caracteriza la curvatura espacial del universo: $k=-1, 0, +1$ corresponden a un universo abierto, plano y cerrado, respectivamente.

En un universo isótropo y homogéneo el tensor de energía-momento tiene que ser diagonal. Para un fluido perfecto tenemos
\begin{equation}
T_{\mu\nu}=\begin{pmatrix}
\rho & 0 & 0 & 0 \\
0 & p & 0 & 0 \\
0 & 0 & p & 0 \\
0 & 0 & 0 & p
\end{pmatrix},
\end{equation}
donde $\rho$ y $p$ son la densidad de energía y la presión del fluido.

Entonces, las ecuaciones de Einstein dan como resultado las ecuaciones de Friedmann que determinan la evolución del factor de escala y por lo tanto la dinámica del universo,
\begin{equation}
    H^2 \, = \, \left( \frac{\dot{a}}{a} \right)^2 = \frac{8 \pi G_N}{3} \sum_i \rho_i - \frac{k}{a^2}+\frac{\Lambda}{3},
    \label{friedmann1}
\end{equation}
\begin{equation}
    \dot{H} + H^2 \, = \, \frac{\ddot{a}}{a} \, = \, - \frac{4 \pi G_N}{3} \sum_i (\rho_i + 3 p_i) + \frac{\Lambda}{3}, 
    \label{friedmann2}
\end{equation}
donde los puntos denotan la derivada temporal y definimos el parámetro de Hubble $H \, = \, \dot{a}/a$ que caracteriza la expansión del universo. Podemos definir la densidad crítica, la cual corresponde a la densidad de energía para un universo plano, como
\begin{equation}
    \rho_c \, = \, \frac{3 H^2}{8 \pi G_N} \simeq 1.05 \times 10^{-5} h^2 \text{GeV cm}^{-3},
\end{equation}
con
\begin{equation}
    H= 100 \, h \text{ km s}^{-1} \text{ Mpc}^{-1}.
\end{equation}
A partir de la densidad crítica podemos definir el parámetro de densidad de energía
\begin{equation}
    \Omega_i=\frac{\rho_i}{\rho_c}.
\end{equation}
Reemplazando el parámetro de densidad en Ec.~(\ref{friedmann1}) tenemos
\begin{equation}
    1\, = \, \sum_i \Omega_i - \frac{k}{a^2 H^2} \, = \, \Omega_{tot} - \frac{k}{a^2 H^2},
    \label{friedmann3}
\end{equation}
donde consideramos que la constante cosmológica es parte de la densidad de energía total. En este caso los valores $\Omega_{tot}<1$, $\Omega_{tot}=1$ y $\Omega_{tot}>1$ corresponden a universos abiertos, planos y cerrados.

Además de las ecuaciones de Friedmann, Ec.~(\ref{friedmann1}) y (\ref{friedmann2}), es necesaria una relación entre la densidad de energía y la presión para cada uno de los fluidos que componen el universo. Dicha relación es la denominada ecuación de estado, la cual puede ser generalizada como
\begin{equation}
    p_i \, = \, w_i \, \rho_i.
\end{equation}
Para radiación y partículas relativistas tenemos $w_r=1/3$; las partículas no relativistas tienen presión despreciable, por lo tanto $w_m=0$; para representar la energía que describe la componente con constante cosmológica tenemos $w_{\Lambda}=-1$.

La ecuación de continuidad puede ser derivada a partir de Ec.~(\ref{friedmann1}) y (\ref{friedmann2}),
\begin{equation}
    \dot{\rho}_i + 3 H (\rho_i + p_i) \, = \, \dot{\rho}_i + 3 \, \frac{\dot{a}}{a} \, \rho_i \, (1+w_i) \, = \, 0,
    \label{eccontinuidad}
\end{equation}
donde hemos utilizado la ecuación de estado en la segunda igualdad. Integrando Ec.~(\ref{eccontinuidad}) podemos encontrar la dependencia de la densidad de energía de cada componente en función del factor de escala,
\begin{equation}
    \rho_i \, \propto \, a^{-3(1+w_i)} = (1+z)^{3(1+w_i)}, 
    \label{densidadenergia}
\end{equation}
con $z$ el parámetro del corrimiento al rojo definido como
\begin{equation}
    1+z \, = \, \frac{\lambda(t_0)}{\lambda(t_e)} \, = \, \frac{a(t_0)}{a(t_e)},
    \label{redshift}
\end{equation}
suponiendo una fuente de fotones y un observador con coordenadas comóviles fijas, entonces $\lambda(t_0)$ y $a(t_0)$ son la longitud de onda y el factor de escala medidos por el observador, mientras que $\lambda(t_e)$ y $a(t_e)$ son los valores en el lugar de emisión. Tomamos por convención $a(t_0)=1$. Como consecuencia, sin tener en cuenta los movimientos relativos peculiares, los espectros de los objetos astrofísicos se encuentran corridos al rojo.

A partir de Ec.~(\ref{densidadenergia}) notamos que la densidad de energía de la materia no relativista decrece con un factor $(1+z)^3$ debido a la dilución de la densidad numérica de partículas por la expansión del universo. La energía de los fotones y otras partículas relativistas decrece con un factor adicional $(1+z)$ debido al corrimiento al rojo. Finalmente, la densidad energética correspondiente a la constante cosmológica es independiente de la dinámica del universo. Estos distintos comportamientos hace que la fracción respecto del total de las densidades energéticas de cada uno de los fluidos cambie con el tiempo. En consecuencia, el universo temprano está dominado por radiación, luego debido a la expansión encontramos una etapa dominada por materia, y finalmente una era dominada por constante cosmológica.

\section{Historia térmica del universo} \label{historiatermica}

Durante la era dominada por la radiación, el universo se encuentra compuesto por partículas relativistas en equilibrio térmico. La densidad total de energía está dada por
\begin{equation}
    \rho \, = \, \frac{\pi^2}{30} g_* T^4,
\end{equation}
donde $T$ es la temperatura y $g_*$ es el número efectivo de grados de libertad relativistas, definido como
\begin{equation}
    g_* \, = \, \sum_{\text{bosones}} g_i \left( \frac{T_i}{T} \right)^4 \, + \, \frac{7}{8} \sum_{\text{fermiones}} g_i \left( \frac{T_i}{T} \right)^4 ,
    \label{gradosdelibertad1}
\end{equation}
donde $T_i$ y los factores $g_i$ son la temperatura y los números de libertad de cada una de las especies relativistas bosónicas y fermiónicas. Si consideramos que la expansión del universo es adiabática, la densidad comóvil de entropía se conserva, es decir,
\begin{equation}
    \frac{d(s \, a^3)}{dt} = 0,
\end{equation}
Según el primer principio de la termodinámica, la entropía para partículas relativistas está dada por
\begin{equation}
    s \, = \, \frac{\rho + p}{ T} \,  = \,  \frac{4}{3}\frac{\rho}{T} \, = \, \frac{2 \pi^2}{45} g_{*S} T^3,
\end{equation}
y $g_{*S}$ es el número efectivo de grados de libertad de partículas relativistas con respecto a la entropía (igual a Ec.~(\ref{gradosdelibertad1}) reemplazando la potencia 4 por 3). Se puede ver que la temperatura es inversamente proporcional al factor de escala mientras $g_{*S}$ se mantenga constante, es decir,
\begin{equation}
    T \, \propto \, g_{*S}^{1/3} \, a^{-1} \, = \, g_{*S}^{-1/3} \, (1+z).
\end{equation}
Cuando un tipo de partículas se desacopla del plasma, ya sea por aniquilación o decaimiento, su entropía es transferida al resto de las partículas del baño térmico y su temperatura aumenta levemente. Sin embargo, este efecto es típicamente una corrección de $O(1)$, entonces podemos describir la temperatura del baño térmico como
\begin{equation}
    T \, = \, T_0 \, (1+z),
\end{equation}
donde $T_0$ es la temperatura actual del fondo cósmico de microondas. Por lo tanto, la temperatura de los fotones y las especies relativistas decrece con la expansión del universo con el corrimiento al rojo.

A continuación discutiremos brevemente algunas etapas del universo temprano relevantes para esta tesis.

\subsection{Inflación} \label{inflacion}

Las observaciones indican que nuestro universo es espacialmente plano y homogéneo, lo cual implica para un universo en expansión una cosmología con condiciones iniciales muy específicas. El valor inicial de la densidad de energía total debería tener un ajuste fino para obtener el valor actual observado $\Omega_{tot}\simeq 1$. Por ejemplo, si en Ec.~(\ref{friedmann3}) inicialmente tenemos $k=0$, entonces $\sum_i \Omega_i(t) = \Omega_{tot}(t) =1$. Pero si la curvatura no es exactamente cero, entonces $\Omega_{tot}(t)$ se aleja de 1 a medida que el universo se expande. Para un universo dominado por materia
\begin{equation}
    |\Omega_{tot}(t) -1 | \, \propto \, t,
\end{equation}
como actualmente observamos $k\simeq 0$, en la época del desacople de los fotones del CMB deberíamos haber tenido
\begin{equation}
    |\Omega_{tot}(t_{CMB} ) -1 | \, \leq 10^{-16},
\end{equation}
lo cual constituye un problema de naturalidad denominado \textit{el problema de la planitud}.

Por otra parte, los experimentos que han medido el fondo cósmico de microondas indican que el universo era isotrópico antes de la formación de estructuras. La homogeneidad de la temperatura del CMB no pudo ser causada por interacciones físicas, pues la totalidad del cielo es homogéneo, incluso las zonas que se encuentran fuera de sus respectivos horizontes causales por varios órdenes de magnitud, en el momento del desacople de los fotones que conforman el CMB. Nuevamente, una de las soluciones posibles es comenzar con condiciones iniciales ajustadas finamente, lo cual constituye otro problema de naturalidad llamado \textit{el problema del horizonte}.

Una propuesta para solucionar ambos problemas sin tener que recurrir a condiciones iniciales específicas consiste en introducir una fase de rápida expansión en el universo temprano. Dicha expansión resulta exponencial considerando un fluido con $w_{eff} \simeq -1$~\cite{Guth:1981,Albrecht:1982,Linde:1983}. Debido a la expansión exponencial, se obtiene $\Omega_{tot} \simeq 1$ y un universo espacialmente plano, sin importar la cantidad de materia-energía inicial. La isotropía del CMB también puede ser explicada, ya que todo el universo observable inicialmente estaba contenido en una pequeña región causalmente conectada que luego fue desconectada durante la fase de expansión exponencial.

Una etapa inflacionaria al comienzo del universo puede ser generada mediante distintos mecanismos, como por ejemplo, por un campo escalar llamado inflatón que pasa por un proceso denominado de `\textit{slow-roll}'. Las fluctuaciones de dicho campo producen anisotropías en el CMB, las cuales son experimentalmente buscadas, y que eventualmente generaron las semillas para la formación de estructuras a gran escala del universo.

Por último, al finalizar la etapa de inflación, las partículas que inicialmente estaban presentes en el universo son diluidas. Sin embargo, el inflatón decae y produce el baño térmico de partículas elementales, en un proceso denominado de `\textit{reheating}' o \textit{recalentamiento}. La temperatura de equilibrio del plasma luego de inflación es llamada \textit{temperatura de recalentamiento} y luego de dicha fase la evolución del universo y del baño térmico es descrita por la cosmología explicada en las secciones anteriores.

\subsection{Nucleosíntesis} \label{nucleosintesis}

Se denomina Big Bang nucleosíntesis (BBN) a la etapa del universo durante la cual se generan los elementos ligeros del universo. Provee una de las pruebas más confiables del universo temprano debido a que ocurre a temperaturas $T \simeq 1- 0.1$ MeV, y por lo tanto se basa en física de partículas del modelo estándar~\cite{Alpher:1948,Wagoner:1967}. A temperaturas mayores a 1 MeV, los neutrones y protones se encuentran en equilibrio térmico debido a las interacciones débiles. Al descender la temperatura, los neutrones se apartan del equilibrio y la relación entre las densidades numéricas de neutrones, $n_n$, y protones, $n_p$, queda congelada por el siguiente factor de Boltzmann
\begin{equation}
    \frac{n_n}{n_p} \, = \, \exp \left( -\frac{m_n - m_p}{T_{fr}} \right) \, \simeq \, \frac{1}{6},
\end{equation}
donde $T_{fr}$ es la temperatura de congelamiento o `\textit{freeze-out}' y depende de los grados de libertad de las especies relativistas. Luego de dicho proceso, la gran cantidad de fotones del plasma, mediante procesos de fotodisociación, evitan la formación eficiente de núcleos de deuterio de manera eficiente hasta que la temperatura disminuye por debajo de 0,1 MeV. A este efecto se lo denomina cuello de botella del deuterio. Debido al decaimiento de neutrones, la relación entre las densidades numéricas de neutrones y protones para ese momento es de $1/7$. Si consideramos que todos los neutrones disponibles se combinan con protones para terminar formando $^4$He, la abundancia puede ser estimada fácilmente como
\begin{equation}
    Y_{^4\text{He}} \, \simeq \, \frac{2 n_n}{n_p + n_n} \, = \, \frac{2 n_n/n_p}{1+n_n/n_p} \, \simeq \, 0.25,
    \label{helioBBN}
\end{equation}
donde en el numerador de la primera igualdad tenemos que todo neutrón se encuentra en núcleos de deuterio y por lo tanto una cantidad igual de protones también se encuentra en núcleos de deuterio, y el denominador representa la cantidad total de nucleones en el universo. Con esta simple estimación tendríamos un universo formado por un 75\% de Hidrógeno y 25\% de Helio. 

El cálculo completo de BBN tiene en cuenta los detalles de las interacciones nucleares y en particular el valor de la relación entre bariones y fotones, el cual determina la temperatura a la cual comienza el proceso de nucleosíntesis. La estimación (\ref{helioBBN}) es suficiente para el caso de $^4$He, pero otros elementos ligeros son más sensibles a dichas consideraciones.

La predicción de elementos ligeros se encuentra en buen acuerdo con las observaciones astrofísicas (salvo el caso del litio), y corresponden a una densidad de bariones de
\begin{equation}
    0.019 \, \leq \, \Omega_b h^2 \, \leq \, 0.024,
\end{equation}
compatible con el valor determinado por las observaciones del fondo cósmico de microondas.

\subsection{Fondo cósmico de microondas} \label{CMB}

La radiación de fondo cósmico de microondas (CMB) es una herramienta muy poderosa que en la actualidad brinda la mayor precisión para poner a prueba el origen de las fluctuaciones primordiales del universo y los parámetros cosmológicos en una época temprana del cosmos.

El CMB consiste en la radiación termal reliquia del momento de recombinación. El universo temprano, con temperaturas mayores a $T \simeq 0.25$ eV, era opaco para la radiación electromagnética debido al scattering de Thomson con los electrones libres. Sin embargo, eventualmente la temperatura disminuyó por debajo de dicho valor, el cual corresponde a un corrimiento al rojo de $z\simeq 1100$, y la formación de hidrógeno eléctricamente neutro se tornó favorable, causando una disminución de electrones libres. Entonces, el camino libre medio de los fotones aumentó hasta alcanzar el tamaño del universo y se desacoplaron del baño térmico. Dichos fotones viajan libremente y constituyen el fondo cósmico de microondas presente en la actualidad, cuya energía disminuye por efectos del corrimiento al rojo.

Como los fotones se encontraban en equilibrio térmico antes de su último scattering, el CMB tiene un espectro de cuerpo negro, con temperatura $\sim2,72$
K uniforme en la actualidad. Sin embargo se han detectado pequeñas diferencias entre direcciones distintas
del cielo, del orden $\Delta T/T \simeq 10^{−5}$. Dichas variaciones son esperadas, pues tienen como origen las fluctuaciones de los pozos de potencial gravitatorio necesarios para la evolución y crecimiento de las estructuras de clusters y galaxias. Los procesos de atracción gravitatoria y repulsión por presión de radiación, tienen como resultado oscilaciones acústicas que dejan una huella característica en el CMB.

\section{Evidencia observacional de materia oscura}\label{Moscura}

La existencia de materia oscura en el universo es uno de los grandes enigmas de la física contemporánea. Desde hace casi un siglo, por argumentos gravitacionales, observaciones en galaxias y cúmulos de galaxias han revelado que es necesaria la presencia de un nuevo tipo de materia. Su presencia tiene consecuencias cosmológicas, al ser uno de los componentes predominantes del universo, y consecuencias astrofísicas al ser un elemento indispensable para la formación de estructuras como galaxias. Por ello, el análisis de los distintos tipos de evidencia observacional de materia oscura y sus características es un tema activo en la física moderna.

\begin{figure}
\centering
\includegraphics[scale=0.3]{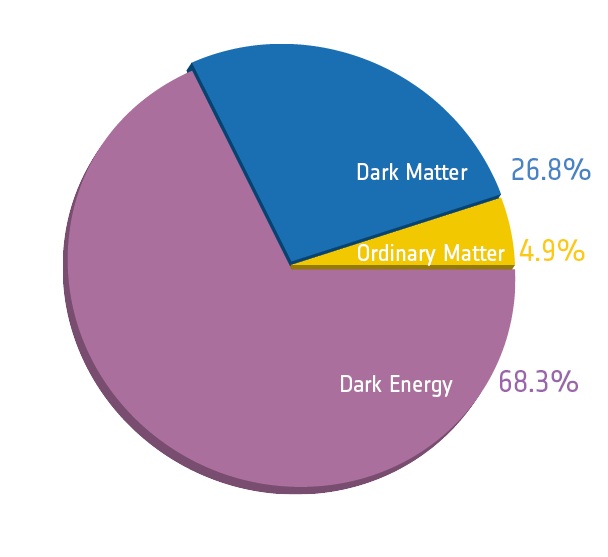}%
\caption{Gráfico de la composición de la densidad de energía del universo obtenido a partir de
mediciones de las fluctuaciones del fondo de radiación de microondas con Planck~\cite{Aghanim:2018eyx}}%
\label{planckpie}%
\end{figure}

En la Figura~\ref{planckpie} se muestra un gráfico de la composición actual de la densidad de energía
del universo, a partir de mediciones de las fluctuaciones del CMB con Planck~\cite{Aghanim:2018eyx}, notamos que
\begin{itemize}
\item 4,9 \% está compuesta por átomos o materia bariónica (4 \% de hidrógeno y helio libre, 0,5 \% en forma de estrellas, 0,3 \% de neutrinos, 0,03\% de elementos pesados, y 0,01 \% de fotones en forma de radiación de fondo),
esto significa que más del 95 \% de la densidad de energía del universo se encuentra en una
forma que nunca ha sido detectada directamente en un laboratorio.
\item 26,8 \% está compuesta por materia oscura.
\item 68,3 \% está compuesta por energía oscura, responsable por la expansión acelerada del universo. Las
primeras evidencias experimentales de energía oscura datan de la década de 1980’.
\end{itemize}

La mayoría de la materia del universo es oscura, es decir, la sección eficaz entre partículas de materia oscura y fotones no debe ser significativa. A partir de los datos del CMB se puede ver que $\sigma_{DM-\gamma} \leq 2,25 \times 10^{6} \sigma_{\text{Th}} (m_{DM}/\text{GeV})$~\cite{Stadler:2018}, donde $\sigma_{\text{Th}} \sim 10^3$ GeV$^{-2}$ es la sección eficaz de Thomson. En cambio, su presencia es inferida indirectamente por su influencia gravitacional en el movimiento de galaxias y cúmulos de galaxias, o por las fluctuaciones
del CMB y nuestro entendimiento sobre la formaciones de estructuras a grandes escalas.

Sabemos que la materia oscura no está compuesta por estrellas o planetas.
Las observaciones indican que no hay suficiente materia visible en el universo para completar
el 31,7\% requerido (cantidad de materia total). Tampoco se encuentra en forma de nubes
de gas de materia bariónica, las cuales pueden ser detectadas mediante la absorción de radiación
que las atraviesa, ni antimateria, ya que se distinguiría un exceso
significativo de rayos gamma generados en procesos de aniquilación de materia-antimateria.

De todas maneras, aún es viable que materia bariónica pueda completar parte de la densidad de energía faltante en forma de enanas marrones u otros objetos masivos compactos,
conocidos como `MACHOs' (en inglés, MAssive Compact Halo Objects), pero no la totalidad. La hipótesis más popular es que la materia oscura no sea bariónica, sino partículas
fundamentales exóticas, como axiones o `WIMPs' (Weakly Interacting Massive Particles). En el marco de modelos supersimétricos, dada la existencia de gravedad se obtiene la presencia del gravitino (compañero supersimétrico del gravitón) y por lo tanto un candidato que surge de forma natural para las construcciones más generales de este tipo de teorías. Tanto el gravitino como el axino (compañero supersimétrico del axión), debido a la escala de sus interacciones son ejemplos de `superWIMPs'. Discutiremos con mayor detalle los candidatos a materia oscura en el Capitulo~\ref{del lado de alla}.


En las próximas secciones describiremos algunas de las evidencias recogidas durante casi un siglo
que llevan a la necesidad de incluir materia oscura en el modelo cosmológico. Los detalles y el desarrollo histórico de las evidencias descritas a continuación no pretenden ser un estudio completo, para más información sobre este tema se sugieren los siguientes reviews~\cite{Bergstrom:2000,Bertone:2005,Bertone:2018}.

\subsection{Escala de clusters de galaxias}
\label{DMclusters}

\subsubsection{Materia faltante en los clusters de galaxias}

Un cluster o cúmulo de galaxias es una estructura astrofísica que consiste de cientos a miles
de galaxias unidas gravitatoriamente, con una masa típica entre $10^{14}-10^{15} \, M_{\odot}$, donde $M_{\odot}$ representa la masa solar, y un diámetro entre $2-10$ Mpc. El medio intracluster no está vacío, sino que consiste de gas a alta temperatura, entre $2-15$ keV independientemente de la masa total del cluster.

Los primeros en mencionar la contribución dinámica de una componente invisible en el
cosmos fueron Kapteyn~\cite{Kapteyn:1922} y Oort~\cite{Oort:1927} en la década de 1920. En 1933 el astrónomo Fritz Zwicky,
uno de los pioneros más famosos en el campo de la materia oscura, estudió el corrimiento al
rojo de varios clusters de galaxias y observó una amplia dispersión en la velocidad aparente
de varias galaxias dentro del cluster Coma, con diferencias excediendo los $2000$km/s~\cite{Zwicky:1933,Zwicky:1937}.
Este hecho había sido observado por Hubble y Humason~\cite{Hubble:1931}, pero Zwicky fue el primero
en aplicar el teorema del virial para estimar la masa total de un cluster. No fue el primero
en utilizar dicho teorema en astronomía, pues Poincare lo había hecho más de 20 años antes.

El teorema del virial aplicado a sistemas astrofísicos, determina que en un
sistema en equilibrio aislado y autogravitante el promedio temporal de la energía cinética
total, $T$, y el promedio temporal de la energía potencial, $U$, se encuentran relacionados:
\begin{equation}
2\langle T \rangle = -\sum^{N}_{k=1} \langle r_k F_k \rangle,
\end{equation}
donde $k = 1, ..., N$ es el número de partículas del sistema, y $r_k , F_k$ la posición y fuerza de la partícula $k$ respectivamente. Para fuerzas conservativas $F_k=-\frac{\partial U}{\partial r_k}$. Si consideramos que
\begin{equation}
U(r)=Ar^{p} \hspace{1cm} \Rightarrow \hspace{1cm} 2\langle T \rangle = -\sum^{N}_{k=1} \langle r_k F_k \rangle = p \langle U_{total} \rangle.
\end{equation}
Entonces, para un potencial gravitatorio, $p=-1$, obtenemos:
\begin{equation}
2\langle T \rangle + \langle U_{total} \rangle = 0.
\end{equation}

Una de las hipótesis del teorema del virial es que el sistema se encuentra en equilibrio,
veamos si se cumple para el caso de cluster de galaxias. Consideremos un cluster de radio
$R_c \sim 1 \text{Mpc} = 3,08 \times 10^{19}$ km y la velocidad de una galaxia que lo compone $v_g \sim 10^3$ km/s. El tiempo que tarda una galaxia en recorrer el cluster es
\begin{equation}
\tau_{cross}\sim \frac{R_c}{v_g} \sim 3,08 \times 10^{16} \text{s}.
\end{equation}
Como los cúmulos de galaxias se formaron hace $\tau_{cluster}\sim 10^{10}$ años$=3,15\times 10^{17}$ s, el número de veces que cada galaxia componente ha recorrido el cluster es
\begin{equation}
\frac{\tau_{cluster}}{\tau_{cross}} \sim 10,
\end{equation}
cantidad que podemos considerar suficiente para suponer que los clusters de galaxias son sistemas autogravitantes en equilibrio.

Empleando el teorema del virial se puede estimar la masa de un cluster de galaxias, a partir de sus distancias características y la velocidad de sus constituyentes. Tenemos que,
\begin{equation}
\langle T \rangle = \frac{1}{2} \sum^N_{k=1} m_k v_k^2 = \frac{1}{2} N m \langle v^2 \rangle = \frac{1}{2} M \langle v^2 \rangle ,
\end{equation}
\begin{equation}
\langle U \rangle = - \frac{1}{2} \sum^N_{k=1,k\neq j} \frac{G \, m_k \, m_j}{r_k-r_j} \simeq -G \frac{N^2 m^2}{2d} = -G \frac{M^2}{2d},
\end{equation}
donde $m_k$ y $v_k$ representan la masa y velocidad de la partícula $k$ respectivamente, $d$ la distancia típica entre galaxias, $m$
la masa típica de una galaxia, $M \simeq N \,m$ la masa total del sistema y $G$ la constante de gravitación universal, mientras que $\frac{N^2}{2} \sim$ cantidad de pares de galaxias. Entonces
\begin{equation}
2\langle T \rangle + \langle U_{total} \rangle = 0 \hspace{1cm} \Rightarrow \hspace{1cm} M = \frac{2 \, \langle v^2 \rangle \, d}{G}.
\end{equation}

Para un cluster formado por $N$ galaxias en una esfera homogénea de radio $R$
\begin{equation}
d=\left( \frac{V}{N} \right)^{1/3}=\left( \frac{4 \pi}{3N} \right)^{1/3}R.
\end{equation}

Recordando que un cluster está formado por $N \sim $ cientos a miles de galaxias, obtenemos
\begin{equation}
M=O(1) \frac{\langle v^2 \rangle R}{G}.
\end{equation}

Podemos ver que la estimación de la masa del cluster depende de:

i. $R$, inferido por la distancia y el tamaño angular del sistema,

ii. $v^2$ , inferido por el corrimiento Doppler en el espectro,

iii. débilmente de la geometría/distribución de galaxias del cluster ($O(1)$).

El mismo resultado puede obtenerse si estimamos el potencial gravitatorio de una esfera
homogénea de radio R autogravitante
\begin{equation}
\langle U \rangle \simeq - \frac{3}{5} \frac{G \, M^2}{R}
\end{equation}

Consideremos el sistema estudiado por Zwicky en 1933~\cite{Zwicky:1933}, el cluster Coma de $\sim$ 800
galaxias, un radio aproximado de un millón de años luz ($10^{22}$m) y luminosidad típica de
$10^9 M_{\odot}$ por galaxia. La masa total del sistema es:
\begin{equation}
M \sim 800 \times 10^9 M_{\odot} = 1,6 \times 10^{42} \text{kg},
\end{equation}
con el teorema del virial obtenemos
\begin{equation}
\langle v^2 \rangle = - \frac{\langle U \rangle}{M} \simeq \frac{3}{5} \frac{G \, M}{R} \sim 6,4 \times 10^9 \text{m}^2/\text{s}^2 \hspace{1cm} \Rightarrow \hspace{1cm} \langle v^2 \rangle ^{1/2} \simeq 80 \text{km/s}.
\end{equation}

Sin embargo mediante el efecto Doppler se mide que las velocidades aparentes son del
orden de $1000$ km/s. Para obtener tales velocidades la masa necesaria de cada galaxia tiene
que ser por lo menos $400$ veces mayor a la cantidad de materia luminosa observada. Este
resultado de principio de siglo XX no solo nos indica que existe un tipo de materia no
luminosa en los clusters, sino que sería 2 o 3 ordenes de magnitud más abundante.

En 1937 Zwicky extendió su análisis~\cite{Zwicky:1937} a partir de la velocidad de dispersión observada (excluyendo galaxias con $v \sim 5000$ km/s) y obtuvo un límite inferior de $4,5\times 10^{13} M_{\odot}$ para el cluster, es decir $4,5\times 10^{10} M_{\odot}$ por galaxia. Si la luminosidad promedio por galaxia es $8,5\times 10^{7} M_{\odot}$ la relación luminosidad-masa resulta $\sim$500 y no 3 como era esperado. El parámetro de Hubble usado por Zwicky fue $H_0 = 558$ km/s/Mpc, y está involucrado en la determinación de la distancia con el corrimiento al rojo. Corregido por el valor actual, $H_0 = 67,27 \pm 0,66$ km/s/Mpc, la
relación luminosidad-masa fue sobre estimada por un factor $558/67,27=8,3$. Igualmente se
obtiene una discrepancia en la masa de un orden de magnitud entre el valor esperado y el calculado.

\subsubsection{Gas intergaláctico}

En la actualidad, se conoce que la mayoría de la masa bariónica de
un cluster está presente en forma de gas caliente intergaláctico.
El satélite Uhuru en 1970~\cite{Giacconi:1971}, estudió el gas intergaláctico mediante rayos X (6,3 keV)
brindando información sobre el perfil de densidades y la cantidad de materia. Según las observaciones, aún es necesaria $\sim$ 7 veces más masa que la inferida en forma de gas para explicar el movimiento de los clusters.

Una manera complementaria de medir las propiedades del gas caliente del cluster emplea
el efecto Sunyaev-Zeldovich~\cite{Sunyaev:1970}. Consiste en la distorsión del espectro de fondo cósmico de
microondas (CMB) en la dirección del cluster. Los fotones del CMB que atraviesan el cluster
interactúan con los electrones de alta energía del gas caliente por medio de scattering de
Compton inverso. Entonces los fotones adquieren en promedio un aumento de energía al
cruzar el cluster. Dichas distorsiones se utilizan para detectar perturbaciones en la densidad
de materia y están en buen acuerdo con las mediciones por emisión de rayos X de los clusters.

\subsubsection{Lentes gravitacionales}

Estudiando el efecto de lentes gravitacionales se puede estimar la distribución y cantidad de materia total de un sistema. El efecto fue observado por primera vez en 1979~\cite{Walsh:1979}, y consiste en detectar la deflexión por efectos gravitatorios de la
luz proveniente de fuentes ubicadas detrás de un objeto masivo que actúa como lente. A partir de la deflexión observada utilizando clusters de galaxias como lentes, se encontró en 1997 una proporción masa-luz de alrededor de 200 para estos sistemas~\cite{Fischer:1997}, solidificando progresivamente la necesidad de grandes cantidades de materia faltante en clusters, distribuida de manera suave entre las galaxias.

\begin{figure}
\centering
\includegraphics[scale=0.4]{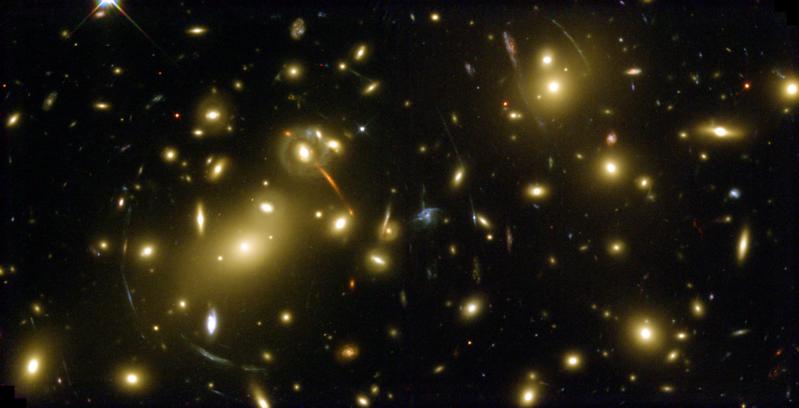}%
\caption{Cluster de galaxias Abell 2218 y los arcos formados por el efecto de lente gravitacional por las fuentes de luz detrás del cluster. Imagen tomada por el telescopio Hubble~\cite{hubbleimage}.}%
\label{abell2218}%
\end{figure}

En la Figura~\ref{abell2218} se muestra una imagen del cluster de galaxias Abell 2218 del telescopio
Hubble~\cite{hubbleimage}. Es uno de los cúmulos más lejanos observados, con redshift $z\simeq 0,17$, y utilizando este cluster como lente se ha podido detectar una de las galaxias más lejanas, con
redshift $z\sim 6$. En la imagen se pueden observar los arcos formados por efecto de lente gravitacional provenientes de fuentes de luz detrás del cluster, por ejemplo, el arco naranja es
una galaxia elíptica con redshift $z\simeq 0,7$, los arcos azules son galaxias con redshift $z\simeq 1−2,5$.

\subsubsection{Microlentes gravitacionales}

Podríamos considerar que la materia faltante está constituida por
objetos compactos mucho menos luminosos que las estrellas ordinarias, como planetas, enanas
marrones, enanas blancas, estrellas de neutrones y agujeros negros. Dichos candidatos a
materia oscura son conocidos como `MACHOs' (en inglés, MAssive Compact Halo Objects).

En la actualidad, experimentalmente los MACHOs no pueden constituir parte importante
de la materia oscura. En particular el uso de microlentes gravitacionales en la búsqueda de
MACHOs y la determinación de la densidad bariónica del universo basada en la abundancia de
elementos ligeros primordiales y en el fondo cósmico de microondas, han sido muy importantes
para descartar a los MACHOs como principal componente de la materia faltante.

En la década de 1980 se propuso monitorear variaciones en el brillo de un gran número de estrellas
en una galaxia cercana (como la Nube de Magallanes), si el halo de materia faltante de la
Vía Láctea estuviese comprendido enteramente de MACHOs, estos actuarían como lentes y aproximadamente una de 2 millones de estrellas debería experimentar un aumento en su brillo aparente en cada momento. La duración de un evento de microlente sería $\sim$130 días $\times \left(\frac{M}{M_{\odot}} \right)^{0.5}$ es decir que se podrían detectar objetos con masas en el rango de $\sim 10^{−7} \, M_{\odot}$ hasta $\sim 10^{2} \, M_{\odot}$ con escalas temporales de variación en el brillo de horas a un año.

Las colaboraciones MACHO, EROS y OGLE comenzaron en la década de 1990. En
1993 reportaron las primeras detecciones de eventos de microlentes gravitacionales, consistentes
con un halo dominado por objetos compactos. Sin embargo, luego de 5.7 años monitoreando
40 millones de estrellas, MACHO identificó entre 14 y 17 candidatos, concluyendo que entre
8 y 50\% del halo galáctico está compuesto por objetos compactos de entre 0.15 a 0.9 $M_{\odot}$~\cite{MACHO:2000}. En cambio, la colaboración EROS durante 6.7 años solo identificó un evento candidato,
lo que implica un límite superior del 8\%~\cite{EROS:2007}. Dentro del rango de masas estudiado, los
MACHOs no constituyen una fracción dominante de la materia faltante de la Vía Láctea.

\subsubsection{Fusión de clusters}

Un tipo de experimento que permite combinar las distintas técnicas discutidas previamente son las observaciones de clusters fusionándose o que recientemente, en el sentido
cosmológico, han chocado. Estos sistemas son de suma importancia pues los bariones observados y la materia oscura inferida se encuentran espacialmente segregados, permitiendo
contrastar la hipótesis de materia oscura contra teorías de gravedad modificada con solo materia visible.

Dado el tiempo suficiente, las galaxias (1-2\% de la masa de un cluster), el gas (5-15\%)
y la materia oscura adquieren una distribución similar, espacialmente simétrica y
centrada haciendo que los tres componentes sigan al potencial gravitacional total.
Considerando las distancias típicas, la probabilidad de que dos galaxias colisionen en la
fusión de clusters es extremadamente baja y por lo tanto se comportan como partículas sin colisiones. Por otra parte el gas intergaláctico, la mayoría de la materia visible de un cluster, tiene interacciones significativas.

Como resultado, luego de la colisión entre dos clusters, la distribución de galaxias
(determinada en el espectro óptico), que no interactúan, se encuentra desacoplada de la distribución de
gas caliente (medida por rayos X) debido a que el gas queda `rezagado' en el choque pues interactúan las nubes de gas de ambos clusters. Por otra parte, se pueden establecer superficies
de contorno que determinen el potencial gravitatorio utilizando lentes gravitacionales.

Si la materia visible actuase como fuente de la totalidad del potencial gravitatorio del
sistema, se espera que las superficies de contorno del potencial coincidan con la distribución
del gas caliente inferida por rayos X. Sin embargo esto no es lo observado.

\begin{figure}
\centering
\includegraphics[scale=0.55]{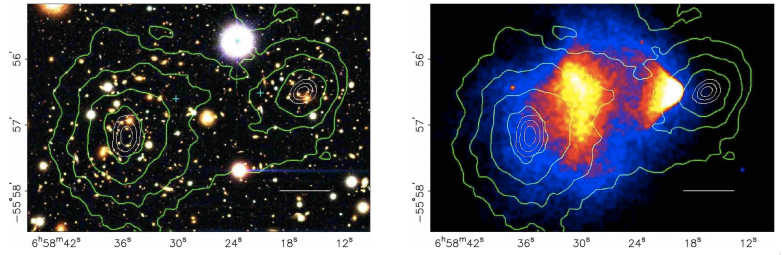}%
\caption{El \textit{bullet cluster} (o 1E 0657-558)~\cite{Clowe:2006}. Izquierda: imagen en el espectro visible tomada por Magallanes. Derecha: imagen en rayos X tomada por Chandra usada para medir la cantidad de gas.
En verde se muestran los mapas de contorno del potencial gravitatorio proporcional a la masa total
del sistema, determinado mediante lentes gravitacionales. La barra blanca indica 200~kpc.}%
\label{bulletcluster}%
\end{figure}

En la Figura~\ref{bulletcluster} se muestra el bullet cluster, las primeras observaciones de precisión de
un par de clusters fusionándose, por Clowe et al en 2006~\cite{Clowe:2006}. El panel de la izquierda corresponde a
la imagen en el espectro visible tomada por Magallanes. El panel de la derecha corresponde
a la imagen en rayos X tomada por Chandra, la escala de colores es usada para medir la
cantidad de gas: su concentración es mayor en la región blanca/amarilla. La forma de la
nube derecha, que le da su nombre al cluster, indica que se alejan a $\sim$ 4700 km/s, en cambio
la velocidad en la linea de visión es de $\sim$ 600 km/s por lo que el choque ocurre casi en el
plano del cielo y los núcleos de los clusters se atravesaron hace 100 millones de años. Se muestra
en verde los mapas de contorno del potencial gravitatorio proporcional a la masa total del
sistema. Vemos que los mapas de potencial no siguen las distribuciones de gas.

El potencial gravitatorio no queda `rezagado' junto con el gas, sino que es independiente
de distribución del mismo (nuevamente remarcamos que el gas caliente aporta la mayoría de
la masa bariónica del sistema) y se ubica aproximadamente donde están las galaxias de cada
uno de los clusters. A partir de estas observaciones se puede inferir que es necesario otro tipo
de fuente para el potencial gravitatorio, como materia oscura con baja autointeracción (casi
sin colisiones, $\sigma /m \leq O(1)\text{cm}^2$/g), y que debe ser la principal contribución a la masa del sistema.

En Ref.~\cite{Clowe:2006} se determinó que el pico
del centro de masa total y el pico del centro de la masa bariónica se encuentran separados
por 8$\sigma$ de significancia. La evidencia observacional de dicha separación nos permite inferir
la necesidad de materia oscura asumiendo las condiciones más generales posibles respecto
al comportamiento de la gravedad, comprendiendo también un serio y grave problema para propuestas alternativas de gravedad modificada.

Existen varios ejemplos de fusiones de clusters como 1E 0657-558 o bullet cluster, MACS
J00254.4-1222, Abell 520, para los cuales son compatibles tanto la determinación de la masa faltante no bariónica, como la cota para su autointeracción.

\subsection{Escala de galaxias}
\label{DMgalaxias}

\subsubsection{Curvas de rotación}

Las curvas de rotación relacionan el perfil de velocidades de las estrellas y el gas de una
galaxia en función de la distancia al centro galáctico, constituyendo una de las evidencias experimentales más importantes en el problema de la materia oscura. Como veremos, es posible inferir
la distribución de masa de una galaxia a partir de su curva de rotación bajo simples hipótesis.

Una galaxia puede ser modelada por un sistema sin colisiones y en estado
estacionario. Tomemos por ejemplo el caso de la Vía Láctea. Posee $N_∗ \simeq 10^{11}$ estrellas y un
disco estelar con radio $R_d \simeq 10$ kpc y altura $h_d \simeq 0.5$ kpc. Para una estrella típica, similar al Sol su período orbital, $t_{orb}$ , es
\begin{equation}
t_{orb}= \frac{2\pi R_o}{v_{rot}} \sim 246 \text{Myr} \ll t_{MW}\sim 10\text{Gyr},
\end{equation}
donde $R_o\simeq 8$ kpc es la distancia entre el Sol y el centro galáctico, es decir, el radio de la
órbita solar, $v_{rot}\simeq 200$ km/s es la velocidad orbital promedio y $t_{MW}$ es el tiempo transcurrido
desde al formación de la Vía Láctea. Por lo tanto, una estrella típica ha tenido tiempo de
completar $\frac{t_{MW}}{t_{orb}}\sim 41$ órbitas, justificando la hipótesis de estado estacionario o de equilibrio.

En cuanto a la hipótesis de constituir un sistema sin colisiones, calculemos el camino
libre medio de una estrella. Para ello es necesario estimar la sección eficaz y la densidad
numérica de las mismas. Si consideramos a las estrellas como esferas su sección eficaz, $\sigma_∗$,
es:
\begin{equation}
\sigma_* \sim \pi \, R^2_{\odot}=\pi \, 10^{20} \, \text{cm}^2,
\end{equation}
donde $R_{\odot}$ es el radio del Sol.

La densidad numérica de estrellas, $n_∗$ , en la galaxia la podemos estimar como:
\begin{equation}
n_* \sim \frac{N_*}{Volumen}= \frac{N_*}{\pi \, R^2_d \, h_d} \sim \frac{10^{-52}}{5 \pi} \, \text{cm}^{-3}.
\end{equation}
Por ende el camino libre medio de una estrella típica en la Vía Láctea es
\begin{equation}
\lambda_* \sim \frac{1}{\sigma_* \, n_*} \sim 5\times 10^{32} \, \text{cm} \simeq 5 \times 10^{12} \, \text{kpc},
\end{equation}
es decir, varios órdenes de magnitud más grande que las distancias típicas de una galaxia.

Por otra parte, no es necesario hacer uso de la relatividad general, pues las velocidades
típicas de las galaxias y los elementos que la conforman son mucho menores que la velocidad
de la luz.

\begin{figure}
\centering
\includegraphics[scale=0.45]{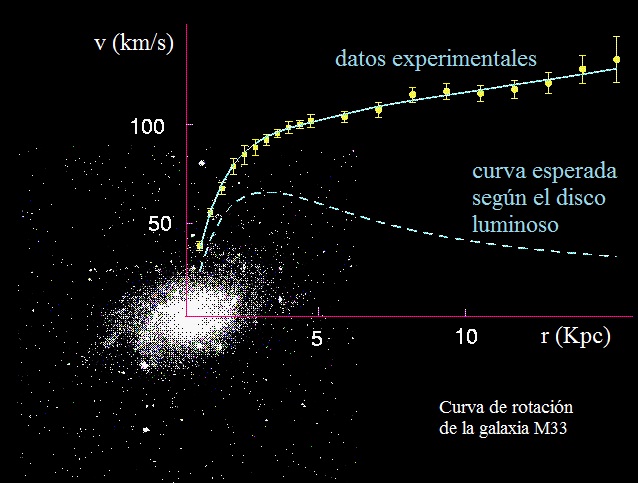}%
\caption{Curva de rotación de la galaxia M33 (del Triángulo) observada junto con la curva de rotación calculada teniendo en cuenta solo la cantidad de materia luminosa~\cite{rotacion}. Ambas curvas se encuentras superpuestas sobre la imagen óptica de la galaxia.}%
\label{m33}%
\end{figure}

Si consideramos una distribución de masa aproximadamente esférica, podemos conocer la
velocidad orbital promedio de un objeto que orbita alrededor de una galaxia. Para un sistema
autogravitante, reemplazando en la ley de Newton la aceleración centrípeta y la fuerza de
gravedad tenemos que:
\begin{equation}
	\frac{v^{2}(r)}{r}=\frac{GM(r)}{r^{2}}
\end{equation}
 donde $\textit{v(r)}$ es la velocidad orbital promedio, $\textit{r}$ el radio de la órbita y $\textit{M(r)}$ la cantidad total de masa dentro de la órbita:
\begin{equation}
	M(r)=\int_0^r 4 \, \pi \, \hat{r}^2 \, \rho (\hat{r}) \, d\hat{r},
\end{equation}
con $\rho (r)$ la densidad de materia en el radio $r$. Suponiendo que la masa de la galaxia está concentrada en su parte visible esperamos:
\begin{equation}
v(r) \, \alpha \,
\begin{dcases}
	r \qquad \qquad  \quad\text{si } r<R_d \\
	\frac{1}{\sqrt{r}} \qquad \qquad \text{si } r>R_d \\
    \end{dcases} \, ,
\end{equation}
donde $R_d$ es el radio visible de la galaxia. Por lo tanto, se espera que los objetos por fuera del radio visible tengan velocidades orbitales que disminuyan con la distancia. En cambio, experimentalmente que esto no ocurre. Mediante efecto Doppler se observa que la velocidad orbital aumenta hasta un valor constante ($v_c$ $\approx$ 100-200 km/s). En la Figura~\ref{m33} (de Ref.~\cite{rotacion}) se muestra la curva de rotación de la galaxia M33 junto con la curva de velocidad calculada teniendo en cuenta solo la cantidad de materia luminosa para comparación. Este fenómeno se ha observado en cientos de galaxias~\cite{rotacion3,rotacion4}, incluso la Vía Láctea.

Una explicación posible es asumir que los discos galácticos se encuentran inmersos en halos de materia oscura que se extienden más allá de la región compuesta por la materia visible. Para distancias grandes la cantidad $M(r)/r$ es constante si la masa aumenta linealmente con $r$, es decir $M(r)=\frac{v_{c}^{2}r}{G}$ o la densidad de masa $\rho(r)\propto 1/r^{2}$. Un gas ideal autogravitante con temperatura uniforme tendría dicho perfil de masas~\cite{galactdinamic}.

Tomando en cuenta el análisis de las curvas de rotación, obtenemos que el 90$\%$ de la masa que conforma una galaxia es materia oscura.

Históricamente, en 1939 Babcock presentó la primera la curva de rotación de la galaxia
M31, mostrando velocidades orbitales hasta distancias de $\sim$20 kpc del centro galáctico~\cite{Babcock:1939}.
En 1970 Rubin y Ford publicaron la curva de rotación de M31 extendiendo el radio de la
órbita de objetos medidos en un trabajo de alto impacto~\cite{Rubin:1970}. Se comenzó a argumentar
firmemente que era necesaria masa adicional no solo en clusters de galaxias sino incluso en
las galaxias mismas para explicar la dinámica de cada sistema.

En 1978 Rubin, Ford y Thonnard, en uno de los trabajos más citados del área, publicaron
curvas de rotación en el espectro óptico de 21 galaxias de distinta masa y luminosidad~\cite{Rubin:1978}. Concluye que tan solo a unos kpc del núcleo, la velocidad de rotación deja de estar correlacionada con el radio para mantenerse constante en las regiones externas, alcanzando velocidades similares para todas las galaxias independientemente del tamaño. Esto indica que todas las galaxias se ubican en pozos de potencial similares y que materia no luminosa debe existir por fuera de la región visible.

\subsubsection{Vía Láctea}

Datos cinemáticos brindan información sobre el potencial gravitacional total, mientras que
datos de fotometría dan el contenido de materia bariónica. Existen dos clases de métodos
para inferir la distribución de materia oscura en la Vía Láctea: locales y globales.

En el primer método se utiliza que la densidad y distribución de velocidades de una
población de estrellas en función de su altura sobre el plano galáctico están relacionadas con
el potencial gravitatorio total del disco galáctico.

Los métodos globales se basan en el análisis de la curva de rotación de la Vía Láctea.
Requieren 2 tipos de datos: la curva de rotación y la distribución espacial de estrellas y gas.

Una manera de validar las técnicas es medir la densidad de materia en la vecindad del sistema solar, $\rho (R_o)$, para ambos métodos. Un review completo puede hallarse en Ref~\cite{Read:2014}. Ambos métodos son consistentes con $\rho (R_o) \sim 0,3$-$0,4$ GeV/cm$^2$.

Si bien esperamos una cantidad significativa de materia oscura en nuestra galaxia, la
cantidad esperada en la vecindad del sistema solar es pequeña. Esto se debe a que el gas
es disipativo y, a diferencia de la materia oscura, puede condensar para formar un disco
rotacional. Estimemos la masa encerrada en la posición del Sol, $R_o \simeq 8$ kpc, asumiendo
simetría esférica:
\begin{equation}
M_{dm}(R_o) \sim \frac{v_c^2 \, R_o}{G} - M_d,
\end{equation}
donde $v_c(R_o)\simeq 220$ km/s es la velocidad circular local y $M_d \sim 6\times 10^{10} M_{\odot}$ es la masa visible encerrada $\Rightarrow M_{dm}(R_o) \sim 3 \times 10^{10} M_{\odot}$. Considerando que la mayoría de la materia visible está en un disco de $\sim$500 pc de ancho, la materia oscura es aproximadamente
10 veces menos abundante en la vecindad del sistema solar que la materia bariónica.

\subsection{Escala cosmológica}
\label{DMcosmologica}

\subsubsection{Abundancia bariónica del universo}

Desde principios del siglo XX se argumentó que la fusión de hidrógeno en núcleos de helio podrían ser la principal fuente de energía de las estrellas, y que elementos más pesados también podrían generarse en los interiores de las estrellas. En 1939, Bethe describió los procesos y ciclos que dominan la producción de energía de las estrellas de la secuencia principal~\cite{Bethe:1939}. Pero a fines de 1950, se estableció que el mecanismo de nucleosíntesis estelar es insuficiente para generar la cantidad observada de helio~\cite{Burbidge:1957}.

El descubrimiento del fondo cósmico de microondas en 1965 refinó las predicciones de las
abundancias de elementos ligeros del modelo de nucleosíntesis del Big Bang (ver la Sección~\ref{nucleosintesis}), en particular se
calculó una abundancia de helio consistente con las observaciones~\cite{Peebles:1966}. En 1973 con la medición
de abundancias de elementos ligeros se derivó un límite superior para la densidad de materia
bariónica del universo~\cite{Reeves:1973}: $\Omega_b \lesssim 0.1$.

A finales de la década de 1990, mediciones de precisión en la abundancia primordial
del deuterio determinaron la densidad bariónica con 10\% de precisión $\Omega_b \, h^2 = 0.020 \pm 0.002$
(95\% C.L.)~\cite{Burles:2001}, dejando poco lugar para que la materia bariónica pueda ser parte significativa de la materia faltante en el universo.

\subsubsection{Anisotropías del fondo cósmico de microondas}

\begin{figure}[t!]
\begin{center}
 \begin{tabular}{cc}
 \hspace*{-4mm}
 \epsfig{file=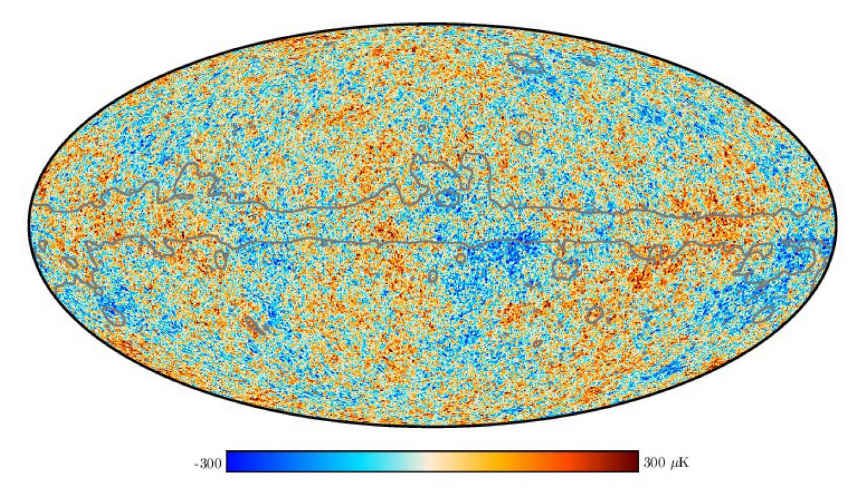,height=4.5cm} 
       \epsfig{file=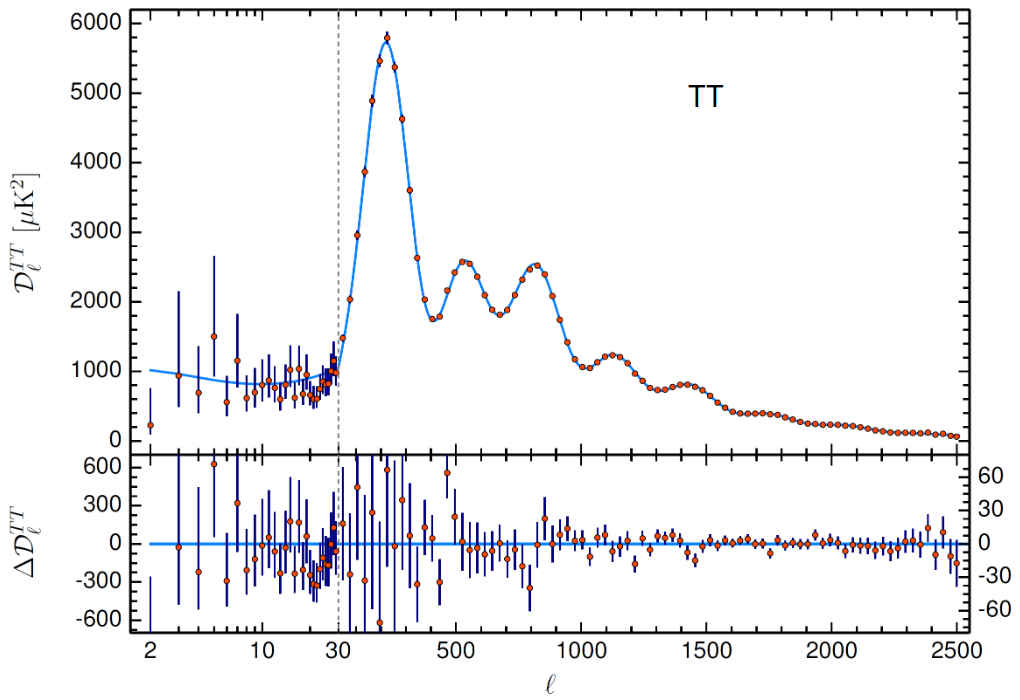,height=5cm}   
    \end{tabular}
    \caption{El panel de la izquierda muestra el mapa del cielo de las diferencias de temperatura con
respecto al promedio del CMB a partir de un análisis conjunto de Planck y WMAP. El panel de la derecha presenta el espectro de potencias de la temperatura obtenido
a partir de datos de Planck 2018; la curva celeste representa el mejor ajuste con respecto al modelo
cosmológico estándar $\Lambda$CDM. Las barras de error representan 1$\sigma$ de incerteza, y en el panel inferior
se muestran los residuos de los datos con respecto al modelo~\cite{Aghanim:2018eyx}.}
    \label{cmbPLANCK}
\end{center}
\end{figure}

Como hemos visto en la Sección~\ref{CMB} las fluctuaciones de temperatura en el CMB corresponden a las oscilaciones acústicas del baño térmico de fotones y bariones. Las anisotropías fueron detectadas por primera vez por el satélite COBE~\cite{COBE:1992} en 1992, y los satélites WMAP~\cite{WMAP} a partir del 2001 y Planck~\cite{Aghanim:2018eyx} desde 2009 incrementaron la precisión de las observaciones. Dichas anisotropías
otorgan información sobre las inhomogeneidades de la materia que luego se amplificaron y
formaron las estructuras del universo. En la Figura~\ref{cmbPLANCK} se muestra una imagen del cielo de las
diferencias de temperaturas para el CMB a partir de un análisis conjunto de observaciones. El
panel de la derecha presenta el espectro de potencias de la temperatura, relacionado con la función de correlación de dos puntos de las fluctuaciones de temperatura, a partir de
datos de Planck 2018~\cite{Aghanim:2018eyx} (se incluyen los errores en la figura), el cual es muy sensible a los parámetros cosmológicos del modelo.
El primer pico brinda información sobre la curvatura del universo, el segundo pico y la
relación entre los picos impares con los pares dan información sobre la densidad de bariones,
mientras que el tercer pico es más sensible a la densidad de materia oscura.

La curva celeste de la Figura~\ref{cmbPLANCK} representa el mejor ajuste al modelo cosmológico estándar
$\Lambda$CDM. Pequeños cambios en los parámetros cosmológicos, por ejemplo en
la densidad de materia oscura del universo, producen una variación notable en esta curva.

Las observaciones de materia luminosa en galaxias determinan $\Omega_{lum} \lesssim 0.01$, mientras que el análisis de las curvas de rotación implican que la densidad total de materia es $\Omega_{m} \gtrsim 0.1$ como límite inferior pues las curvas de rotación se mantienen constantes hasta el máximo valor de $r$ para el cual se pueden encontrar objetos orbitando la galaxia. Las observaciones sobre lentes gravitacionales son consistentes con los resultados anteriores teniendo en cuenta el halo de materia oscura de las galaxias. En estos casos encontramos $\Omega_m \approx0.2-0.3$. Entonces, $0.1 \lesssim \Omega_{\text{DM}} \lesssim 0.3$, donde $\Omega_{\text{DM}}$ hace referencia a la densidad de materia oscura actual sobre la densidad crítica.

El análisis sobre la anisotropía de la radiación cósmica de fondo a partir de los datos brindados por la colaboración Planck~\cite{Aghanim:2018eyx} determinó que la densidad de materia total es
\begin{equation}
\Omega_m \, h^2 = 0.1428 \pm 0.0011 \hspace{1cm}(68 \%\text{ C.L.}),
\end{equation}
donde $h\sim 0.7$. En cambio, la densidad de materia bariónica $\Omega_b$ y de materia oscura $\Omega_{\text{DM}}$ resultan
\begin{equation}
	\Omega_b \, h^2 = 0.02233 \pm 0.00015, \hspace{2cm} \Omega_{\text{DM}}h^{2}= 0.1198\pm 0.0012,
	\label{barionicandDMdensities}
\end{equation}
es decir $\sim 85\%$ de la materia total no es bariónica.

El cálculo de la densidad reliquia de materia oscura para distintos candidatos y mecanismos de producción se detallará en la Sección~\ref{candiDM} donde se abordará el tema en el marco de la física de partículas.

\subsubsection{Estructuras a gran escala}\label{formacionestructuras}

Aunque el universo es isótropo y homogéneo a escalas cosmológicas, no lo es a escalas astrofísicas, donde galaxias y cúmulos de galaxias pueden ser observadas. Estas estructuras se originan a partir de las fluctuaciones amplificadas durante inflación, las cuales podrían comenzar a crecer eficientemente al comienzo de la era dominada por materia, $z \simeq 2700$. Sin embargo, como los bariones se encuentran acoplados al plasma de fotones cuya presión de radiación se opone al colapso gravitatorio, las perturbaciones de densidad pueden comenzar a crecer luego del desacople de los fotones, es decir luego de la emisión del CMB, $z \simeq 1100$.

Para explicar las estructuras observadas en el presente, las densidades de las fluctuaciones debería ser significativas en el momento del desacople de los fotones, correspondiendo a fluctuaciones de temperatura en el CMB del orden $\Delta T/T \simeq 10^{−3}$, es decir, dos ordenes de magnitud superior al valor observado. Sin embargo, como las partículas de materia oscura se encuentran desacopladas del baño térmico antes del desacople de los fotones, sus fluctuaciones de densidad pudieron comenzar a crecer desde el comienzo de la era dominada por materia.

\begin{figure}
\centering
\includegraphics[scale=0.2]{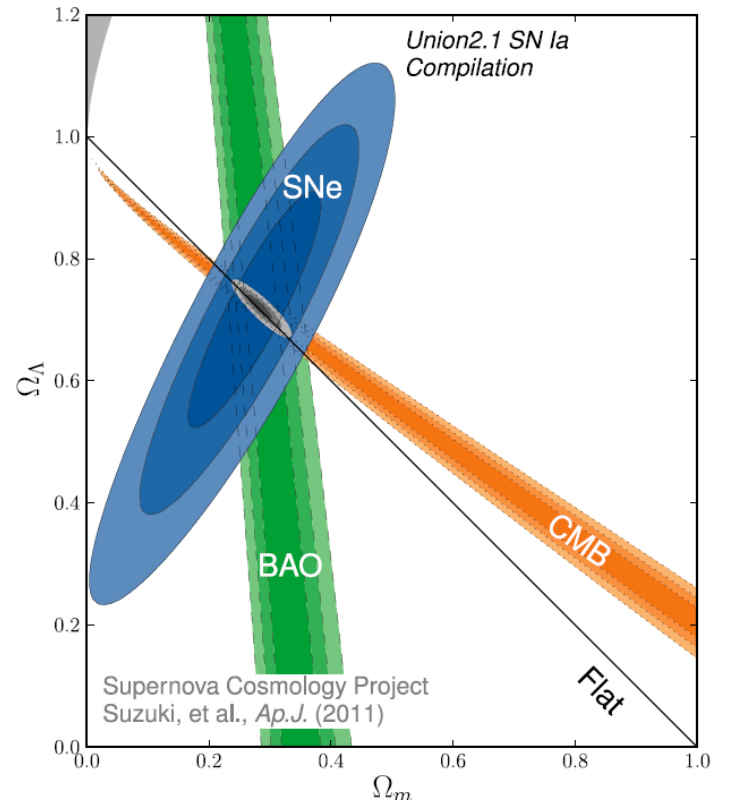}%
\caption{Límites con 68\%, 95\%, y 99.7\% C.L. de $\Omega_m$ y $\Omega_{\Lambda}$ correspondientes a CMB (naranja), BAO (verde), Supernovas (azul). En gris se muestran los límites conjuntos~\cite{SNproject:2012}.}%
\label{sn}%
\end{figure}

Otros tipos de observaciones, como por ejemplo las Oscilaciones Acústicas de Bariones,
BAO por sus siglas en inglés detectadas por primera vez en 2005 en la distribución a gran escala de galaxias, o el análisis de explosiones de Supernovas tipo IA de alto corrimiento al rojo por Supernova Cosmology Project en 1998, también son sensibles a los valores de los parámetros
cosmológicos. Son consistentes con una densidad de materia total $\Omega_m \sim 0.3$ y de constante
cosmológica $\Omega_{\Lambda} \sim 0.7$. En la Figura~\ref{sn} se muestran los límites con 68\%, 95\%, y 99.7\% C.L. de $\Omega_m$ y $\Omega_{\Lambda}$ correspondientes a CMB (naranja), BAO (verde), Supernovas (azul)~\cite{SNproject:2012}.

\subsubsection{Simulaciones numéricas}

Si bien las simulaciones numéricas de N-cuerpos no son evidencia observacional constituyen una importante
herramienta para reconstruir la formación de estructuras y sus características principales a
partir de la evolución de una cantidad de partículas y principios básicos que rigen su dinámica. De esta manera se puede estudiar la distribución de materia a diferentes escalas espaciales y temporales,
el perfil de densidad de DM en las estructuras y el número de halos.

En general las simulaciones solo incluyen materia oscura fría (no relativista) al tener en cuenta que es mucho más abundante que la materia bariónica, y
sin otra interacción además de la gravitatoria (no colosional). Sin embargo, a
medida que aumenta la resolución, es indispensable incluir efectos bariones pues dominan la evolución
a escalas astrofísicas, como en galaxias.

Las simulaciones muestran gran diferencia al considerar materia oscura fría o caliente (relativista). En el paradigma de materia oscura fría las
estructuras crecen jerárquicamente, de `abajo hacia arriba': objetos más pequeños colapsan
bajo su propia gravedad para fusionarse y dar lugar a estructuras más grandes y a objetos
más masivos. En cambio, en el paradigma de materia oscura caliente las estructuras se
forman por medio de fragmentación de `arriba hacia abajo': los superclusters se formarían
en primer lugar y darían paso a objetos más pequeños. A partir de la década de 1980’ las
observaciones indicaron progresivamente que la evolución de estructuras del universo se daba
jerárquicamente, descartando una contribución dominante de especies relativistas.

\begin{figure}[t!]
\begin{center}
 \begin{tabular}{cc}
 \hspace*{-4mm}
 \epsfig{file=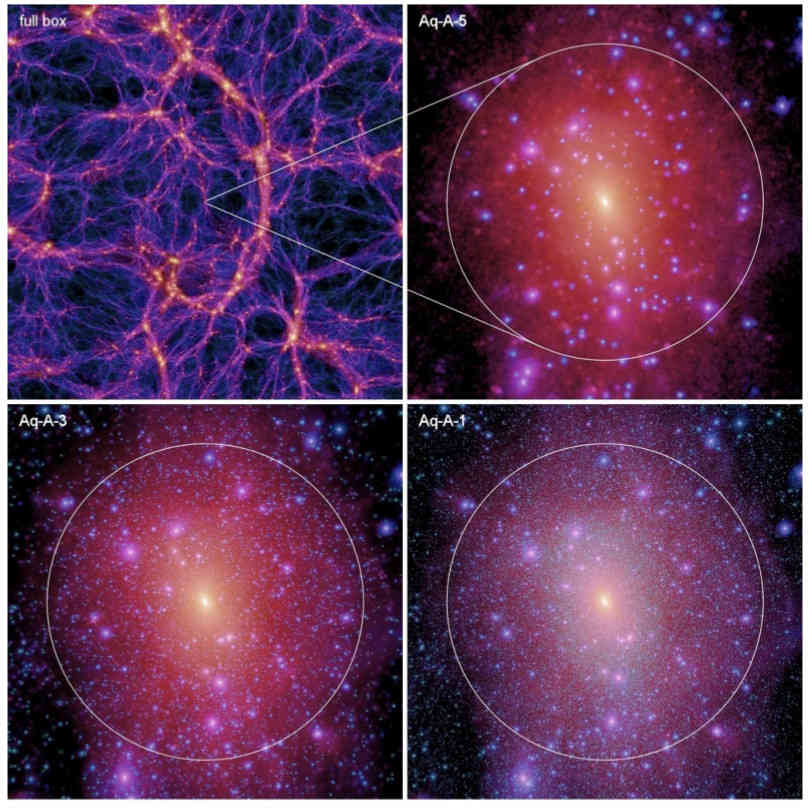,height=5.5cm} \hspace{2cm}
       \epsfig{file=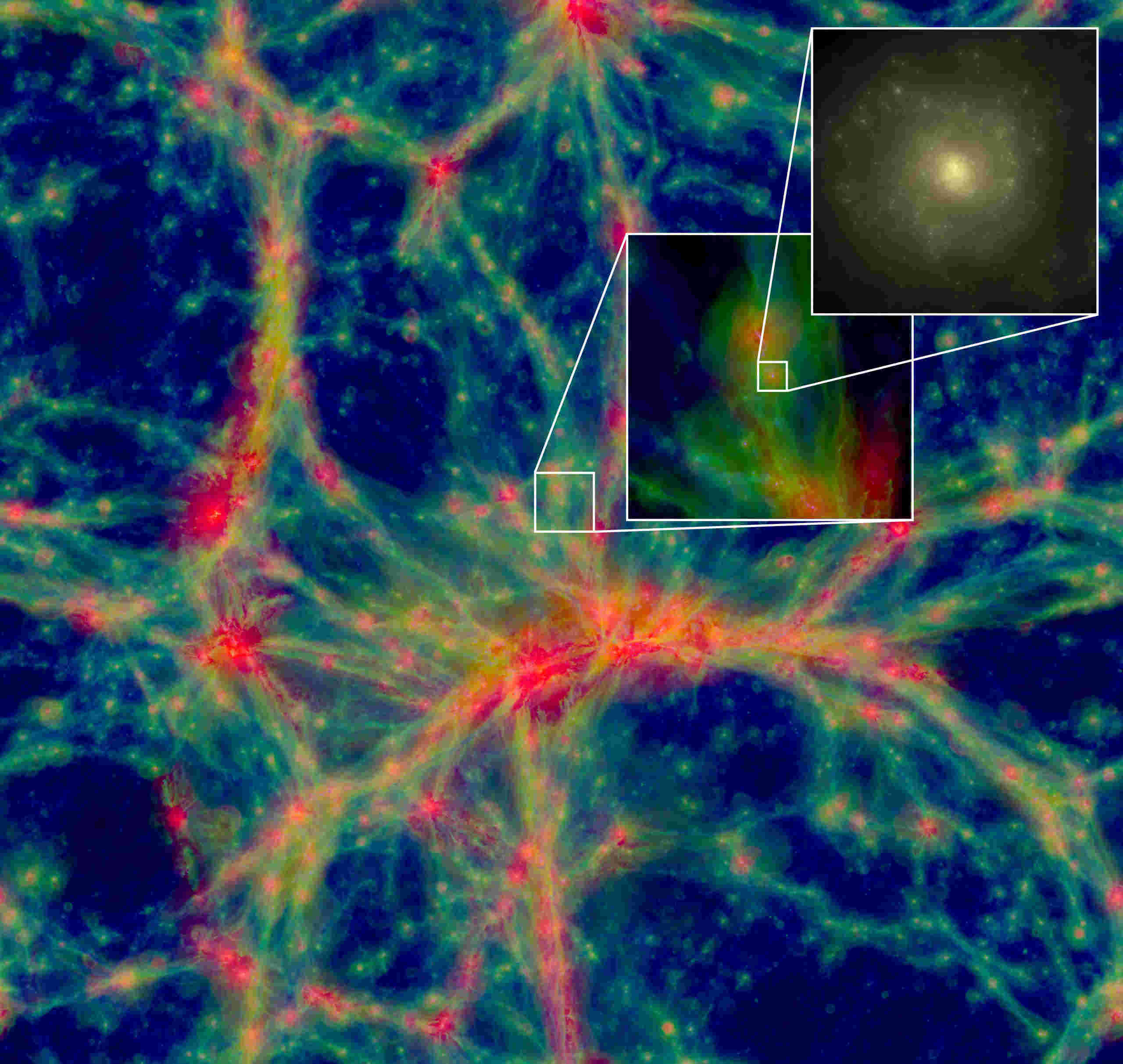,height=5.4cm}   
    \end{tabular}
    \caption{Panel izquierdo: simulación del Proyecto Aquarius~\cite{Aquarius} de un halo de masa similar a la
Vía Láctea. Un solo halo se simula con distinta cantidad de partículas: desde 800 millones (\textit{Aq-A-5})
hasta 1500 millones (\textit{Aq-A-1}). Panel derecho: simulación del Proyecto Eagle~\cite{Virgo} del patrón de filamentos
de materia oscura incluyendo efectos bariónicos; las galaxias y el gas se ubican en las regiones más densas formando clusters de
galaxias (sectores brillantes). El detalle muestra una galaxia individual, similar a la Vía Láctea}
    \label{simulations}
\end{center}
\end{figure}

En la Figura~\ref{simulations}  izquierda se muestra una imagen de una simulación del Proyecto
Aquarius~\cite{Aquarius}. El panel \textit{Aq-A-1}, 1500 millones de partículas, simula un solo halo de
materia oscura de una galaxia espiral típica en el centro, de masa similar a la Vía Láctea.
La gran cantidad de puntos brillantes representan los subhalos donde galaxias enanas satélites podrían formarse.

En la Figura~\ref{simulations} derecha se muestra una imagen de una simulación del Proyecto
Eagle~\cite{Virgo}. Consistió de una caja de 100 Mpc de lado y $\sim$ 10000 galaxias de tamaño mayor o igual a
la Vía Láctea, con más de 7 mil millones de partículas. 
La simulación Eagle contempla efectos causados por materia bariónica que brinda
energía al sistema. Las galaxias y el gas se ubican en las regiones más densas de los filamentos
de materia oscura, por lo tanto los sectores brillantes constituyen los clusters de galaxias en
las intersecciones de dichos filamentos. Se pueden estudiar y reproducir las propiedades y
morfología de galaxias individuales, como se ve en el detalle.

Actualmente las simulaciones han sido exitosas al describir la evolución espacial y temporal de las estructuras al compararlas con las observaciones de distribuciones 3D de galaxias~\cite{Springel:2006}, pero no están libres de inconvenientes, en particular a escalas galácticas, donde los efectos de la física bariónica es relevante.

\textbf{Problema núcleo/cúspide:} (`\textit{core/cust problem}') refiere a la discrepancia entre los perfiles
de densidades de materia oscura predichos por simulaciones y el perfil de masa observado
de galaxias, en particular de baja masa. Solo con materia oscura fría, la densidad de los
halos aumenta cada vez más para radios pequeños, `cúspide' (como los perfiles NFW y
Einasto, ver Sección~\ref{perfilesDM}). Empero, las observaciones en galaxias enanas indican que los perfiles
de densidades de materia oscura poseen una zona central achatada, `núcleo'.

\textbf{Problema de los satélites faltantes:} (`\textit{missing satellite problem}'), refiere a la superabundancia de subhalos predichos por simulaciones de materia oscura fría en comparación
con galaxias satélites conocidas. A pesar que parece haber suficientes galaxias de tamaño
normal, el número de galaxias enanas (de subhalos) observado es órdenes de magnitud inferior al predicho. En el Grupo Local se observan $\sim$40-50 galaxias enanas de las cuales $\sim$11
orbitan la Vía Láctea, mientras que algunas simulaciones predicen $O(500)$. Sin embargo, en la última década se han observado un número creciente de galaxias enanas con brillo ultra tenue (`\textit{ultra faint dwarf galaxies}') donde más del 99.9\% de su masa está compuesta por materia oscura, incluso algunas orbitando la Vía Láctea.

\textbf{Ploblema demasiado grande para fallar:} (`\textit{too big to fail problem}') se puede considerar que los halos de materia oscura pequeños existen pero son extremadamente ineficientes para
formar estrellas y galaxias enanas satélites o que han perdidos sus estrellas durante interacciones con la galaxia principal, en efectos de mareas. Sin embargo, se argumenta que muchos de los satélites predichos por $\Lambda$CDM serían tan masivos que no habría manera que tengan una formación de estrellas ineficientes.

Si se incluyen efectos causados por bariones, mediante supernovas o núcleos de galaxias activos, se puede
brindar energía al medio generando un potencial gravitacional variable que achate los perfiles
de densidades de las simulaciones. Estos efectos modifican la distribución de materia oscura al igual que la evolución de la galaxia, lo que a su vez también disminuyen el número de subhalos predichos. Otras posibles soluciones consisten en considerar materia
oscura tibia (`\textit{warm dark matter}'), o materia oscura con autointeracción.

\subsection{Perfiles de densidades de materia oscura}
\label{perfilesDM}

La distribución de materia oscura en la Vía Láctea no es conocida exactamente, por lo tanto presenta una fuente de incertidumbre en el análisis de las posibles señales. Sin embargo, existen diferentes perfiles comúnmente utilizados en la literatura. La forma funcional de los más importantes es:
\begin{equation}
\begin{array}{r r l}
\text{NFW:} & \rho_{\text{NFW}}(r) = & \frac{\rho_{h}}{\frac{r}{r_{c}}\left(1+\frac{r}{r_{c}}\right)^{2}}, \\
\text{NFW generalizado:} & \rho_{\text{NFWg}}(r) = & \frac{\rho_{h}}{\left(\frac{r}{r_{c}}\right)^{\gamma}\left(1+\frac{r}{r_{c}}\right)^{3-\gamma}}, \\
\text{Einasto:} & \rho_{\text{Ein}}(r) = & \rho_{h}\exp\left\{-\frac{2}{\alpha}\left[\left(\frac{r}{r_{c}}\right)^{\alpha}-1\right]\right\}, \\
\text{Isotermal:} & \rho_{\text{Iso}}(r) = & \frac{\rho_{h}}{1+\left(\frac{r}{r_{c}}\right)^{2}}, \\
\text{Burkert:} & \rho_{\text{Bur}}(r) = & \frac{\rho_{h}}{\left( 1+ \frac{r}{r_c} \right) \left(1+\left(\frac{r}{r_{c}}\right)^{2}\right)},
\end{array}
\label{}
\end{equation}

El perfil denominado Navarro-French-White (NFW) obtenido a partir de simulaciones de N-cuerpos de halos de galaxias y clusters~\cite{nfw}, tiende como $r^{-1}$ hacia el centro galáctico, es uno de los perfiles más populares y sirve como punto de referencia en la literatura. La generalización del perfil NFW permite mayor flexibilidad para poder ajustar distintas simulaciones más recientes, si $\gamma$=1.3, se lo denomina NFW contraído o NFWc, mientras que si $\gamma$=1.16 se lo llama perfil de Moore~\cite{Diemand:2004} y generan perfiles más empinados. El perfil Einasto describe una ley de potencias logarítmica y no converge a una ley de potencias para el centro de la galaxia~\cite{Graham:2006,Navarro:2010}. El parámetro $\alpha$ varía entre distintas simulaciones, pero $\alpha$=0.17 tiende a ser el valor central de dicha variación. Re-simulaciones recientes~\cite{Tissera:2010} de Refs.~\cite{Graham:2006,Navarro:2010} incluyendo bariones, obtienen un parámetro más chico, $\alpha=0.11$, al cual denominaremos EinastoB o EinB. El perfil Isotermal~\cite{Begeman:1991} y el perfil Burkert~\cite{Burkert:2004} poseen un mayor núcleo que los anteriores y se encuentran motivados principalmente por las curvas de rotación de las galaxias en vez de solo simulaciones.

\begin{figure}
\centering
\includegraphics[scale=0.8]{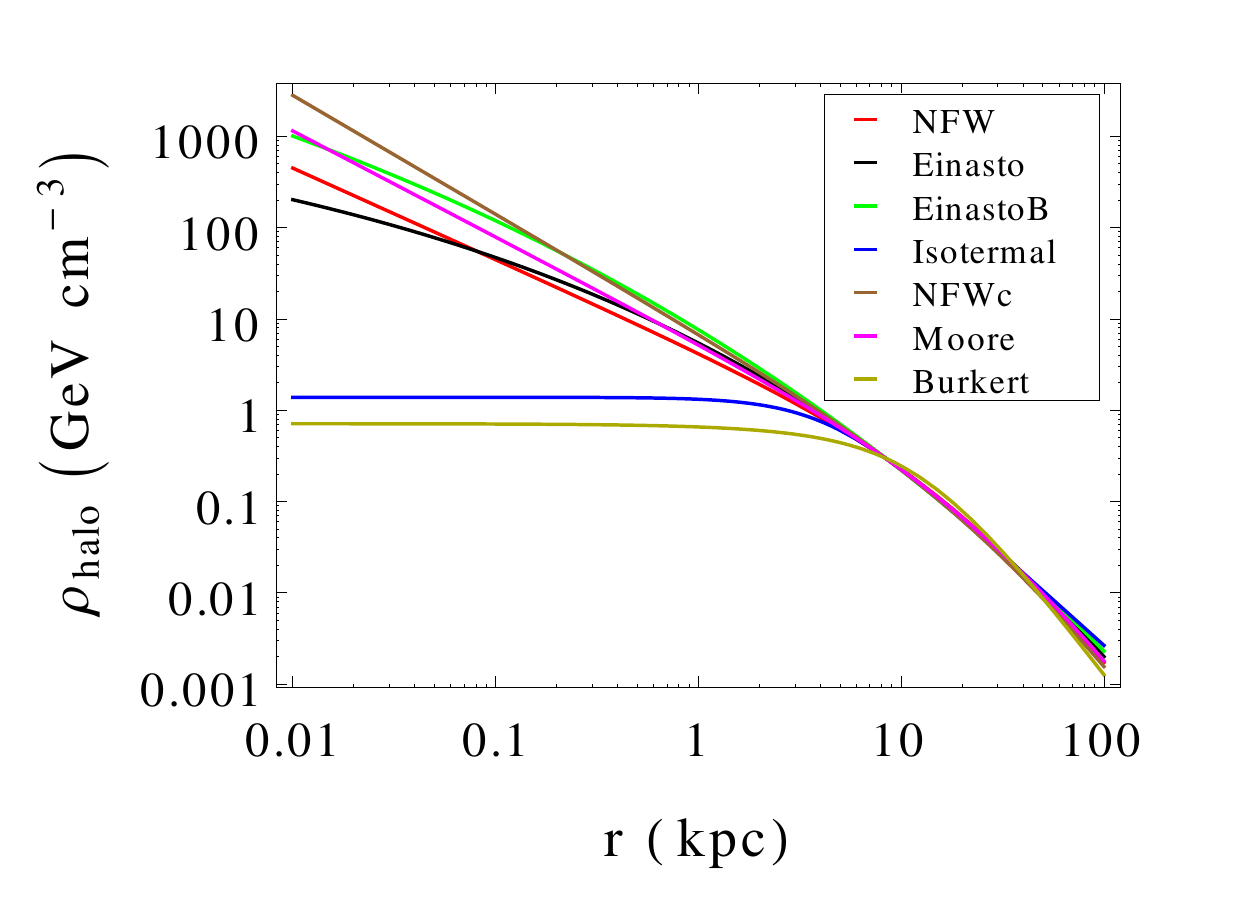}%
\caption{Densidad de materia oscura en función de la distancia al centro galáctico para la Vía
Láctea. Parámetros y normalizaciones para los distintos perfiles tomados de Ref.~\cite{cirelli}.}
\label{profiles}%
\end{figure}

En la Figura~\ref{profiles} se muestra la comparación entre los diferentes perfiles de densidades
de materia oscura en función de la distancia al centro galáctico.
Para todos los perfiles se han tomado los parámetros y la normalización de la densidad, $\rho_{h}$, de Ref.~\cite{cirelli}, considerando el caso de la Vía Láctea de manera que la densidad de materia oscura en la posición del sistema solar sea $\rho(r=R_{\odot}\simeq8.5 \, \text{kpc})\simeq0.3 \, \text{GeV} \, \text{cm}^{-3}$.
En general los perfiles se comportan de manera similar en la región externa de la galaxia,
pero difieren significativamente en la vecindad del centro galáctico.

\chapter{Física más allá del modelo estándar}\label{del lado de alla}

\spacing{1.5}

En esta sección describimos desde el punto de vista de la física de partículas dos evidencias confirmadas de física más allá del SM: la existencia de masas para los neutrinos, y la presencia de materia oscura, para la cual analizamos distintos candidatos y métodos de detección. Al final del capítulo, también tratamos un problema de naturalidad en el SM, el problema CP de QCD, que se manifiesta en la no detección del momento dipolar eléctrico del neutrón, y una posible solución incluyendo un nuevo campo, los axiones.

\section{Física de neutrinos} \label{Fneutrinos}

Un problema que posee el SM con respecto a los neutrinos es que dichas partículas se consideran sin masa, pero a finales de la década de 1990 se ha demostrado experimentalmente lo contrario~\cite{superkamio1,sno1,kam}. En la actualidad existen distintos tipos de experimentos que exploran el área, brindando información sobre la oscilación de los neutrinos, tanto de los ángulos de mezcla del sector como de sus diferencias de masas.

La oscilación de los neutrinos consiste en que un neutrino creado con un sabor leptónico específico (electrónico, muónico, o tauónico) posee probabilidad de ser medido con un sabor leptónico distinto dada una distancia $L$ de propagación. La oscilación no solo depende de la distancia, sino de la energía de los neutrinos $E$, los ángulos de mezcla y la diferencia de masas entre los neutrinos de distintos sabores, por lo tanto la medición de oscilaciones implica que los neutrinos necesariamente tienen que tener masa distinta de cero.

Si definimos $| \nu_{\alpha} \rangle$ la base de autoestados de neutrinos de sabor con $\alpha=e, \mu,\tau$, y $| \nu_{i} \rangle$ la base de autoestados de masa con $i=1,2,3$,
\begin{equation}
| \nu_{i} \rangle = \sum_{\alpha} U_{\alpha i} \, | \nu_{\alpha} \rangle,
\end{equation}
donde $U_{\alpha i}$ está representa la matriz unitaria de Pontecorvo–Maki–Nakagawa–Sakata (PMNS) dada por
\begin{equation}
\begin{split}
U&=
\begin{pmatrix}
1 & 0 & 0\\
0 & c_{23} & s_{23} \\
0 & -s_{23} & c_{23}
\end{pmatrix}
\begin{pmatrix}
c_{13} & 0 & s_{13}e^{-i\delta}\\
0 & 1 & 0 \\
-s_{13}e^{i\delta} & 0 & c_{13}
\end{pmatrix}
\begin{pmatrix}
c_{12} & s_{12} & 0\\
-s_{12} & c_{12} & 0 \\
0 & 0 & 1
\end{pmatrix}
\begin{pmatrix}
e^{i\alpha_1 /2} & 0 & 0\\
0 & e^{i\alpha_2 /2} & 0 \\
0 & 0 & 1
\end{pmatrix}
\\
	&=
\begin{pmatrix}
	c_{12}c_{13} & s_{12}c_{13} & s_{13}e^{-i\delta} \\
	-s_{12}c_{23} - c_{12}s_{23}s_{13}e^{-i\delta} & c_{12}c_{23} - s_{12}s_{23}s_{13}e^{-i\delta} & s_{23}c_{13} \\
	s_{12}s_{23} - c_{12}c_{23}s_{13}e^{-i\delta} & -c_{12}s_{23} - s_{12}c_{23}s_{13}e^{-i\delta} & c_{23}c_{13} 
\end{pmatrix}
\begin{pmatrix}
e^{i\alpha_1 /2} & 0 & 0\\
0 & e^{i\alpha_2 /2} & 0 \\
0 & 0 & 1
\end{pmatrix},
\end{split}
\label{PMNSmatrix}
\end{equation}
donde $c_{ij}$=$\cos\theta_{ij}$, $s_{ij}$=$\sin\theta_{ij}$ corresponden al coseno y seno de los ángulos de mezcla de los neutrinos; $\delta$ es la fase de Dirac, la cual viola CP, y $\alpha_1$ y $\alpha_2$ son fases que tienen significado físico solo si los neutrinos son partículas de Majorana (es decir, si los neutrinos son idénticos a sus antipartículas). Notamos que las fases de Majorana no están involucradas en la oscilación de neutrinos.

La probabilidad de que un neutrino originalmente de sabor leptónico $\alpha$ sea observado con sabor $\beta$ luego de que se haya propagado durante un tiempo $t$, está dada por
\begin{equation}
P_{\alpha \rightarrow  \beta} = | \langle \nu_{\beta}(t)| \nu_{\alpha} \rangle |^2 = \left| \sum_i U^*_{\alpha i} U_{\beta i} e^{-i \frac{m_i^2 t}{2E}} \right|^2,
\end{equation}
pues $| \nu_{\beta}(t) \rangle = \sum_i U^*_{\beta i} e^{-i E_i t} | \nu_i \rangle$, donde $E_i$ es la energía del autoestado de masa. Además hemos utilizado que en el límite ultrarelativista $|\overline{p}_i| \gg m_i$, con $\overline{p}_i$ y $m_i$ el momento y masa del neutrino $i$, por lo tanto,
\begin{equation}
E_i=\sqrt{p_i^2 + m_i^2} \simeq E + \frac{m_i^2}{2 E},
\end{equation}
donde $E$ es la energía total de la partícula.

Si consideramos un modelo con solo dos neutrinos,
\begin{equation}
U=\begin{pmatrix}
 \cos \theta & \sin \theta \\
 -\sin \theta & \cos \theta
\end{pmatrix},
\end{equation}
\begin{equation}
P_{\alpha \rightarrow  \beta, \alpha \neq \beta} = \sin^2(2\theta ) \sin^2 \left( \frac{\Delta m_{12}^2 \, L}{4 \, E \, c} \right) \simeq \sin^2(2\theta ) \sin^2 \left(1.27 \, \frac{\Delta m_{12}^2}{1 \text{ eV}^2} \,  \frac{L}{1 \text{ km}} \, \frac{1 \text{ GeV}}{E} \right),
\label{2neutrinoprob}
\end{equation}
donde $L=c \, t$ y $\Delta m_{12}^2= |m_1^2 - m_2^2|$. Debido a que $\theta_{13} \sim 0$, el modelo de dos neutrinos describe apropiadamente varias transiciones.

Como se puede ver en Ec.~(\ref{2neutrinoprob}), colocando detectores a distintas distancias para fuentes de neutrinos de distintas energías, se puede estudiar distintos parámetros del sector de neutrinos. Si el cociente $\frac{\Delta m^2 \, L}{E}\sim 1$, se puede inferir $\theta$ a partir de la frecuencia de las oscilaciones, y $\Delta m^2$ de la amplitud, con $L$ y $E$ como datos del experimento. Por ejemplo, gracias a las diferencias de masas, empleando los neutrinos producidos en reactores con energías del orden $O(100)$ MeV, si $L\sim O(10)$ km se puede probar $\theta_{13}$, y si $L\sim O(1000)$ km se puede medir $\theta_{12}$.

Debido a los valores afortunados de los ángulos de mezcla y masas del sector de neutrinos, junto con la distancia tierra-sol y las energías de los neutrinos generados en el sol, se pueden estudiar las transiciones $\nu_{\mu}\leftrightarrow\nu_{\tau}$ con el sol como fuente. De igual manera los neutrinos generados en la atmósfera por el choque de rayos cósmicos, permiten extraer información de la transición $\nu_{e}\leftrightarrow\nu_{x}$, donde $\nu_{x}$ es una superposición de los neutrinos muónicos y tauónicos.

Experimentos con reactores nucleares, como Double Chooz en Francia~\cite{chooz}, Daya Bay en China~\cite{daya}, y RENO en Corea del Sur~\cite{reno}, buscan la desaparición de antineutrinos provenientes de un reactor en longitudes del orden del kilómetro.

También existen experimentos de oscilación en aceleradores como T2K en Japón~\cite{t2k}, y NOvA en Estados Unidos~\cite{NOvA:2018}, donde se observa la aparición de neutrinos electrónicos en un rayo de neutrinos muónicos.

Como hemos mencionado, se realizan experimentos radioquímicos de neutrinos solares y atmosféricos, como Super-Kamiokande en Japón~\cite{superkamio1,superkamio2} y Sudbury Neutrino Experiment en Canadá~\cite{sno1,sno2}. Los detectores generalmente poseen grandes depósitos de algún material, por ejemplo agua pesada o hielo ultrapuro, y la interacción de los neutrinos con los núcleos produce partículas cargadas de alta energía que luego son detectadas al producir radiación de Cherenkov.

En el futuro cercano, la próxima generación de detectores incluirá a JUNO~\cite{JUNO:2016}, DUNE~\cite{DUNE:2015} e Hyper-K~\cite{Hyper-k:2018} que reemplazará a Super-Kamiokande en 2025. Por otra parte, Super-Kamiokande y experimentos como IceCube~\cite{IceCube:2014,IceCube:2018} estudian también neutrinos ultra-energéticos provenientes de procesos astrofísicos.

\begin{table}
\begin{center}
   \begin{tabular}{| c | c | c |}
     \hline
     Parámetro             & Mejor ajuste           & Rango 3$\sigma$     \\ \hline
     & & \\
		$\Delta m^{2}_{\text{solar}}/10^{-5} \, \text{eV}^{2}$    & $7.39^{+0.21}_{-0.20}$    &  6.79 - 8.01                    \\ 
		 & & \\
     $\Delta m^{2}_{\text{atm}}/10^{-3} \, \text{eV}^{2}$   & $2.525^{+0.033}_{-0.031}$     & 2.431 - 2.622     \\ 
		  & $2.512^{+0.031}_{-0.034}$ & 2.413 - 2.606 \\ 
		 & &	\\
		 $\sin^{2}\theta_{12}$      & $0.310^{+0.013}_{-0.012}$ & 0.275 - 0.350     \\
		 & & \\
		$\sin^{2}\theta_{23}$      & $0.582^{+0.015}_{-0.019}$  &  0.428 - 0.624    \\
		& $0.582^{+0.015}_{-0.018}$ & 0.433 - 0.623 \\ 
		 &  & \\
		$\sin^{2}\theta_{13}$      & $0.02240^{+0.00065}_{-0.00066}$  & 0.02044 - 0.02437     \\
		& $0.02263^{+0.00065}_{-0.00066}$ & 0.02067 - 0.02461 \\  
		&  &  \\  \hline
		& \multicolumn{2}{c|}{Cota experimental}	 \\ \hline
		$\sum m_{\nu}$ & \multicolumn{2}{c|}{$<$0.12 eV} 	\\
     \hline
   \end{tabular}
	\caption{Masas y ángulos de mezcla experimentales para los neutrinos. La fila de arriba (abajo) corresponde a jerarquía normal (invertida)~\cite{neutrinoSalas:2018,neutrinoEsteban:2019,Aghanim:2018eyx}.}
\label{tablaneutrinos} 
\end{center}
\end{table}

En la Tabla~\ref{tablaneutrinos} se muestra el resumen de los parámetros medidos experimentalmente para el sector de neutrinos~\cite{neutrinoSalas:2018,neutrinoEsteban:2019}. La cota sobre la suma total de las masas de los neutrinos proviene de mediciones realizadas por el satélite Planck~\cite{Aghanim:2018eyx}, y se infiere al determinar cómo los neutrinos masivos modifican el espectro del fondo cósmico de microondas. Para explicar dichos resultados necesitamos una teoría más allá del SM. Notamos que $\theta_{13}$=0 se encuentra excluido por más de 10$\sigma$. En la literatura pueden encontrarse muchos trabajos antiguos donde se consideraba un escenario maximal, es decir $\theta_{13}$=0, en concordancia con los datos experimentales presentes al momento de elaboración de cada estudio.

\begin{figure}
\centering
\includegraphics[scale=0.7]{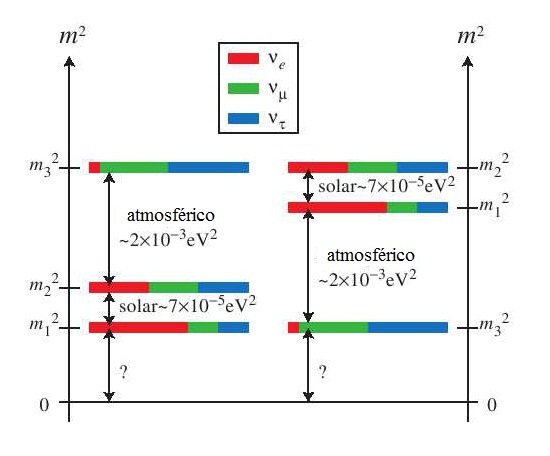}%
\caption{Las dos posibles jerarquías de las masas de los neutrinos. El patrón de la izquierda corresponde a jerarquía normal, el estado más pesado tiene poca componente de neutrino electrónico y los dos estados más livianos son casi degenerados. El patrón de la derecha corresponde a jerarquía invertida, el estado más liviano tiene poca componente de neutrino electrónico y los dos estados más pesados son casi degenerados. En cada caso, los datos solares corresponden a la diferencia de masas entre los estados casi degenerados, y los datos atmosféricos corresponden a la diferencia de masas entre el estado más liviano y el pesado~\cite{esquemaneutrino}.}%
\label{figuraneutrinos}%
\end{figure}

En la Figura~\ref{figuraneutrinos} se muestra esquemáticamente la composición de los autoestados de masa para los casos de jerarquía normal e invertida~\cite{esquemaneutrino}. Dado que se conocen las diferencias de masas al cuadrado, ambos ordenamientos son actualmente permitidos por los experimentos.

\newpage

\section{Candidatos a materia oscura}
\label{candiDM}

Cualquier candidato a materia oscura debe cumplir las siguientes condiciones:
\begin{itemize}
	\item Debe ser estable para escalas temporales cosmológicas. Podría ser completamente estable o su tiempo de vida debe ser al menos un orden de magnitud superior a la edad actual del universo ($\sim 13.7$ Gyr $=4.3 \times 10^{17}$s~\cite{Amigo:2009}).
	\item La interacción con la radiación electromagnética y la materia bariónica debe ser del orden de la fuerza débil o inferior. Si la materia oscura tuviese interacciones electromagnéticas se mezclaría con materia ordinaria y existiría en abundancia en la Tierra formando estados ligados, como por ejemplo átomos de hidrógeno pesados~\cite{Ellis:1983ew}. De manera similar, materia oscura con interacciones de color formaría estados ligados de carga de color neutra con la materia ordinaria, causando por ejemplo isótopos pesados~\cite{Ellis:1983ew}. Para todos estos casos existen cotas experimentales~\cite{Wolfram:1979,Javorsek:2001}.
	\item Debe dar la correcta densidad de reliquia. El dato actualmente más preciso está dado en Ec.~(\ref{barionicandDMdensities}) por la colaboración Planck~\cite{Aghanim:2018eyx}.
	\item Debe ser no relativista, para estar de acuerdo con la	formación de estructuras a gran escala al permitir el crecimiento de las perturbaciones de densidad como ha sido discutido en la Sección~\ref{formacionestructuras}.
\end{itemize}

Existen además distintos límites para muchos modelos de materia oscura al imponer que la física de partículas del modelo sea consistente con todas las observaciones experimentales. Por ejemplo, Big Bang nucleosíntesis es muy sensible a desviaciones del modelo estándar en el universo temprano. Considerando la presencia de partículas cargadas durante BBN o la inyección de energía debido al decaimiento de partículas metaestables durante o luego de BBN se puede alterar significativamente la abundancia de elementos ligeros predicha, en conflicto con las observaciones~\cite{Jedamzik:2009}.

Asimismo, debemos notar que no hay razones para asumir que la materia oscura está compuesta por un único candidato.

Dentro del marco de la física de partículas encontramos como principales candidatos a axiones, neutrinos, WIMPs (partículas masivas débilmente interactuantes por sus siglas en inglés) como neutralinos, y superWIMPs como axinos y gravitinos (con interacciones órdenes de magnitud más chicas que las interacciones electrodébiles), cuyas masas son del orden de $10^{-5}$ eV, 1 eV, 1-1000 GeV y 1 KeV - 1 TeV respectivamente.

Existen más candidatos propuestos en la literatura como por ejemplo SIMPs, CHAMPs, SIDM, neutrinos estériles y partículas de Kaluza-Klein (ver~\cite{darkmattermunoz,darkmatter,minidarkmatter}), entre otros pero su discusión se encuentra fuera del interés de esta tesis.

Como veremos en la Sección~\ref{axiones}, los axiones son partículas de spin 0 sin carga, introducidos para solucionar problemas de CP en QCD~\cite{Peccei:1977hh,Peccei:1977ur,axion6}. Serían producidos en grandes cantidades en el Big-Bang, siempre han sido no relativistas, y poseen tiempos de vida varios órdenes de magnitud superiores a la vida del universo.

Los neutrinos, el único candidato experimentalmente comprobado, pertenece al grupo de materia oscura caliente, ya que su velocidad al comenzar el proceso de formación de galaxias era relativista. Por otro lado, la densidad reliquia de los neutrinos está determinada por
\begin{equation}
\Omega_{\nu}h^2=\frac{\sum_i m_i}{93.14 \, \text{eV}},
\end{equation}
donde $m_i$ es la masa del neutrino $i=e, \mu, \tau$. El límite superior para la sumatoria de las masas es de aproximadamente $< 0.11$ eV, por lo tanto $\Omega_{\nu}h^2 < 2,6\times 10^{-3}$. Como conclusión ya no se lo considera como candidato principal a materia oscura.

Los gravitinos son los compañeros supersimétricos de los gravitones, por lo tanto su interacción con el resto de las partículas es del orden de la fuerza gravitatoria. Para modelos supersimétricos donde la paridad-R no se conserva, el gravitino conforma un buen candidato a materia oscura~\cite{gravitinonoR,gravitinook}. De manera similar, los axinos son los compañeros supersimétricos de los axiones, con interacciones suprimidas por la escala característica de los modelos axiónicos $\sim 10^{11}$-$10^{13}$ GeV. Ambos candidatos serán el foco de esta tesis en los Capítulos~\ref{gravitinoresults1}, \ref{axinoDMchapter} y \ref{axinogravitinoDMchapter}.

\subsection{Densidad reliquia por `\textit{freeze-out}'}

Como mencionamos, los WIMPs son candidatos con sección eficaz comparable con la fuerza débil. Cuando la temperatura $T$ del universo era mayor a la masa de los WIMPs, estas partículas se aniquilaban con sus antipartículas dando como resultado partículas más ligeras, pero al mismo tiempo ocurría el proceso inverso manteniendo el equilibrio. Sin embargo, la expansión del universo causó un continuo decrecimiento de la temperatura, y cuando esta se redujo por debajo de la masa de los WIMPs, su densidad fue suprimida exponencialmente debido a que la aniquilación de WIMPs no podía ser compensada por la pequeña fracción de las partículas más livianas con la energía cinética suficiente para crear WIMPs.

Pero eventualmente también la aniquilación de WIMPs resulta ineficiente, cuando la densidad disminuye tal que la tasa de aniquilación $\Gamma$ es menor que la tasa de expansión del universo, el parámetro de Hubble $H$. Como consecuencia, a partir de ese momento la densidad de WIMPs queda estable y se congela (proceso denominado `\textit{freeze-out}'~\cite{Arcadi:2018}).

 En el marco de la cosmología estándar, la densidad de reliquia de los WIMPs puede ser calculada si en el universo temprano se encontraban en equilibrio térmico y químico con las partículas del modelo estándar. Utilizando la ecuación de Boltzmann, se determina la densidad numérica de materia oscura, $n_{\chi}$, en un universo en expansión
\begin{equation}
\frac{d n_{\chi}}{dt} + 3 H(T) n_{\chi} = -\langle \sigma v \rangle (n_{\chi}^2 - (n_{\chi}^{eq})^2),
\label{DMboltzmann1}
\end{equation}
donde $\sigma$ es la sección eficaz de aniquilación, $v$ la velocidad relativa y $n_{\chi}^{eq}$ el valor de equilibrio. El segundo término de la izquierda denota la dilución de la densidad numérica debido a la expansión del universo, mientras que el término de la derecha representa las fuentes, en este caso la creación y aniquilación de las partículas de materia oscura.

\begin{figure}
\centering
\includegraphics[scale=0.5]{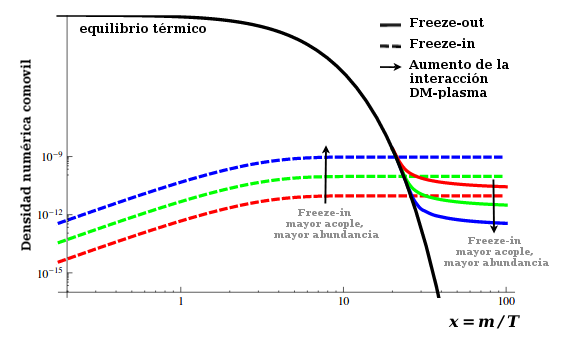}%
\caption{Evolución de la densidad reliquia térmica comóvil para materia oscura mediante el mecanismo de \textit{freeze-out} (curvas en color sólidas) y para el mecanismo de \textit{freeze-in} (curvas en color cortadas). La curva sólida negra indica la abundancia asumiendo que el equilibrio térmico se mantiene. Las flechas representan el efecto que tiene el aumentar la interacción entre la materia oscura y el resto de las partículas del baño térmico, indicando la dirección en que se incrementa el acoplamiento. Figura tomada de Ref.~\cite{Hall:2010}.}
\label{freeze}%
\end{figure}

Si definimos la densidad numérica comovil, $Y\equiv\frac{n_{\chi}}{s}$, y utilizamos la conservación de la densidad de entropía $s$, es decir $s \, a^{3}=\text{cte}$, tenemos que $\frac{d Y}{d t}=\frac{d n_{\chi}}{dt} + 3 H(t) n_{\chi}$, pues $H=\dot{a}/a$. Podemos reescribir Ec.~(\ref{DMboltzmann1}) como
\begin{equation}
\frac{d Y}{d x}=-\frac{x \, s \, \langle \sigma v \rangle}{H(T=m_{\chi})}(Y^2-Y_{eq}^2),
\end{equation}
donde el tiempo $t$ se ha reemplazado por el cociente $x\equiv \frac{m_{\chi}}{T}$. Si el \textit{freeze-out} ocurre durante la época dominada por la radiación entonces $H\sim x^{-2}$. En la Figura~\ref{freeze} se muestra la densidad obtenida para distintos valores de sección eficaz. Se puede observar que la densidad numérica decae exponencialmente hasta que a cierta temperatura ocurre el \textit{freeze-out} y la densidad comóvil permanece constante. Como es esperable, a mayor sección eficaz, la aniquilación de WIMPs es más eficiente por más tiempo, causando que la densidad reliquia resultante sea menor. La temperatura de \textit{freeze-out} típicamente es entre 20 y 25 veces inferior a la masa de materia oscura, es decir que la densidad reliquia está dominada por procesos que ocurren a relativa baja temperatura. Por lo tanto es independiente de la historia temprana del universo y de posibles nuevas interacciones a escalas más altas.

La densidad reliquia puede ser aproximada por 
\begin{equation}
\Omega_{\chi}h^2 \simeq \left\langle \frac{\sigma v}{10^{-26} \, \text{cm}^3 \, s^{-1}} \right\rangle ^{-1},
\end{equation}
por lo tanto, secciones eficaces del orden de la fuerza electrodébil dan como resultado $\Omega_{\chi}h^2 \sim O(1)$. Esta aproximación de primer orden es muy útil para motivar una gran cantidad de candidatos. Si se conserva la paridad-R en los modelos supersimétricos, el neutralino más ligero usualmente estaría incluido en esta categoría.

\subsection{Densidad reliquia por el decaimiento de partículas pesadas}

Existen además otros mecanismos por el cual se puede obtener la densidad reliquia adecuada. Uno de ellos se denomina `\textit{freeze-in}'~\cite{Hall:2010}. Suponemos que a una temperatura $T$ existe un baño térmico de partículas en equilibrio, y otra partícula $X$ la cual posee un tiempo de vida muy largo e interacciones muy débiles con las partículas del baño térmico de manera que $X$ se encuentra térmicamente desacoplada del resto del plasma. Se asume además que en épocas tempranas del universo la abundancia de $X$ es despreciable, por ejemplo por la dilución causada por inflación. Aunque las interacciones con el baño térmico sean débiles, no son nulas, lo que genera la producción de algunas partículas $X$. Si $X$ es más pesada que las partículas del baño térmico con las cual interactúa, el proceso de producción de $X$ es más eficiente cuando $T$ comienza a caer debajo de la masa de $X$. Por esto último, la cantidad de $X$, que nunca alcanza el equilibrio térmico por su baja interacción, queda congelada con el mecanismo de \textit{freeze-in}, con una abundancia que se incrementa para interacciones más fuertes entre $X$ y el baño térmico.

En el proceso de \textit{freeze-out}, al caer la temperatura por debajo de la masa de la partícula candidata,  la materia oscura se está alejando del equilibrio térmico, mientras que en el proceso de \textit{freeze-in} la materia oscura se acerca al equilibrio térmico cuando la temperatura disminuye.

En \textit{freeze-out} se comienza con partículas de materia oscura en equilibrio térmico con el plasma, por lo que reducir su interacción ayuda a mantener su abundancia al ser más difícil aniquilarse. En cambio en \textit{freeze-in} se comienza con una densidad de materia oscura despreciable que incrementa si aumentamos la interacción con el plasma, pues se hace más fácil la producción de materia oscura.

En la Figura~\ref{freeze} se muestra la evolución de la densidad obtenida por el mecanismo de \textit{freeze-in} y el efecto que tiene incrementar el acoplamiento DM-plasma térmico. Como la temperatura de \textit{freeze-in} típicamente es entre 2 y 5 veces inferior a la masa de materia oscura, de manera similar al caso de \textit{freeze-out}, la abundancia de materia oscura al final del proceso está determinada por procesos que ocurren a relativa baja temperatura y por lo tanto es independiente de la historia temprana del universo.

Otro mecanismo importante a tener en cuenta en el cálculo de la abundancia de materia oscura, es la producción a través del decaimiento de partículas más pesadas que, a diferencia de \textit{freeze-in}, no se encuentran en equilibrio térmico. En este caso las densidades reliquias de la materia oscura y de la partícula de la cual decae, denominada partícula madre, se encuentran relacionadas. Si consideramos que solo una partícula de materia oscura se produce en el decaimiento de cada partícula madre, tenemos que sus densidades numéricas son iguales y por lo tanto
\begin{equation}
    \Omega_{\text{DM}} h^2 \, = \, \frac{m_{\text{DM}}}{m_{\text{madre}}} \Omega_{\text{madre}} h^2,
\end{equation}
donde $m_{\text{madre}}$ representa la masa de la partícula madre y, $\Omega_{\text{madre}} h^2$ la densidad reliquia que tendría si no decayese a materia oscura.

Este mecanismo puede ser importante para partículas con interacciones muy débiles, más que para el caso de \textit{freeze-in}, como por ejemplo para gravitinos~\cite{Feng:2003,Feng:2003b} y axinos~\cite{Covi:1999ty}. A este tipo de candidatos se los denomina `\textit{superweakly-interacting massive particles}' o superWIMPs por sus siglas en inglés. Debido a la débil interacción de dichos candidatos, su densidad reliquia resulta independiente del mecanismo por el cual la partícula madre obtiene su abundancia. Por ejemplo, la partícula madre puede ser un WIMP generado por \textit{freeze-out}, mecanismo que a su vez no se ve afectado por la presencia del superWIMP.

\subsection{Detección de materia oscura}
\label{detecDM}

Como trabajaremos en el marco de teorías supersimétricas nos concentraremos en WIMPs, gravitinos y axinos.
Existe actualmente un enorme esfuerzo de la comunidad internacional con el fin de identificar de manera directa o indirecta la materia oscura y sus propiedades. La gran cantidad de experimentos presentes o futuros incluyen tanto experimentos dedicados, como muchos otros conteniendo materia oscura en sus programas de objetivos principales. La base de todos los experimentos es la hipótesis de que la materia oscura tiene algún tipo de interacción (además de la gravitatoria) con la materia bariónica.

\subsubsection{Producción Directa}

Si la materia oscura posee acoples con la materia ordinaria, puede ser producida en colisionadores, como por ejemplo en el LHC~\cite{LHCdarkmatter:2015,LHCdarkmatter:2018}, el cual posee búsquedas dedicadas. El método más común se basa en que luego de la producción de materia oscura, la partícula escapa del detector sin dejar señal ni traza al ser estable y eléctricamente neutra. Por ello se buscan procesos con energía perdida o `\textit{Missing Transverse Energy}' (MET) no atribuible a procesos con neutrinos, los cuales tampoco depositan energía en el aparato.

Uno de los problemas con dicho método es que la producción de partículas en colisionadores es modelo dependiente, al igual que los límites que pueden establecerse. Por otro lado, la identificación de procesos con MET no puede confirmar por si mismos que se tratasen de procesos involucrando partículas de materia oscura, pues podrían resultar ser nuevos estados neutros con tiempos de vida media mayores a las distancias características del detector. Aunque se tratase de un descubrimiento histórico de nueva física, la partícula responsable del MET podría decaer por fuera del detector o incluso ser totalmente estable y aún así no ser responsable de la densidad reliquia de materia oscura presente en el universo. Cualquier descubrimiento en colisionadores debe ser complementado por experimentos astrofísicos de detección directa o indirecta.

\subsubsection{Detección Directa}

Si la materia oscura está formada por partículas, deben formar clusters gravitacionales con las estrellas y demás cuerpos celestes bariónicos en los halos galácticos, incluida la Vía Láctea~\cite{vialactea}, y atravesar la Tierra constantemente. Se espera medir la energía depositada en el detector producto del retroceso de los núcleos o electrones en el scattering elástico de partículas de materia oscura con el material un detector en la Tierra. En particular, WIMPs con masas del orden $O(\text{GeV})$, causarían retrocesos nucleares con energías del orden $O(\text{keV})$, accesibles para los experimentos.

Según los datos experimentales, la densidad de partículas elementales atrapadas gravitacionalmente en nuestra galaxia es $\rho_{0} \approx 5 \times 10^{-25}$ $\text{gr}$ $\text{cm}^{-3}$ $\approx 0.3$ $\text{GeV}$ $\text{cm}^{3}$~\cite{Read:2014}. Para WIMPs con masas del orden de 100 GeV, la densidad sería $n_{0} \approx 3 \times 10^{-3}$ $\text{cm}^{-3}$. Además, su velocidad media sería aproximadamente la velocidad del Sol, ya que se encuentran en el mismo potencial gravitatorio, $v_{0} \approx 220$ km $\text{s}^{-1}$, lo que implica un flujo de partículas de materia oscura incidente sobre la Tierra de $J_{0} = n_{0} v_{0} \approx 10^{5}$ $\text{cm}^{-2}$ $\text{s}^{-1}$. Aunque en principio su detección mediante scattering elástico es posible~\cite{detecta1,detecta2}, existe la enorme dificultad experimental al considerar que la interacción con materia ordinaria es débil.

Experimentos sobre detección de materia oscura comenzaron a fines de la década de 1980~\cite{Ahlen:1987}, estableciendo los primeros límites de la interacción DM-materia mediante detección directa, utilizando germanio como material detector. Algunos experimentos que utilizan distintos tipos de cristales son: Dark Matter (DAMA) en Gran Sasso, Italia~\cite{dama1,dama2}, Cryogenic Dark Matter Search (CDMS) en Estados Unidos~\cite{crio}, EDELWEISS en Laboratorios Subterraneos Modane (Alpes Franco-Italianos)~\cite{edel} y, la nueva generación de detectores SENSEI en Fermi National Accelerator Laboratory en Estados Unidos~\cite{SENSEI:2018}. A partir del 2000, una nueva tecnología permitió el desarrollo de detectores empleando átomos nobles en estado líquido, como xenón o argón. Algunos experimentos son: ZEPLIN en la mina de sal Boudly (Yorkshire, UK)~\cite{zeplin}, PandaX en China Jinping Underground Laboratory~\cite{PandaX:2017}, PICO en SNOLAB underground laboratory, Canadá~\cite{PICO:2019}, XENON1T en Gran Sasso, Italia~\cite{XENON1T:2018}. La próxima generación de detectores incluye a LZ que tomará datos a partir de 2020 en Sanford Underground Research Facility, Estados Unidos~\cite{LZ:2018}, o el experimento planeado DARWIN~\cite{DARWIN:2016}.

Para lograr discriminar señal de ruido se realizan experimentos a grandes profundidades, para blindarse del bombardeo de rayos cósmicos. También deben protegerse de radiactividad natural, tanto de las rocas a su alrededor como de los materiales mismos de los detectores. Por otra parte, pueden considerarse otras maneras de discriminar entre una potencial señal de DM y ruido de fondo. Por ejemplo modulaciones anuales de la señal por el hecho que la Tierra orbita alrededor del Sol~\cite{modulacionsol}, lo que causa una pequeña variación del flujo de DM y por lo tanto también de los eventos en el detector. El experimento DAMA fue la primera colaboración en reportar la detección de una modulación anual, con la forma esperada	considerando materia oscura, y actualmente continúa confirmando sus observaciones luego de veinte ciclos en igual cantidad de años~\cite{dama3:2018}. Sin embargo, el resultado es controversial, pues muchos otros experimentos más sensibles han excluido el espacio de parámetros favorecido~\cite{XENON1T:2018}, aunque con detectores de distinto material.

A pesar de los resultados hasta el momento nulos, gran esfuerzo sigue invirtiéndose en mejorar la sensibilidad de los detectores en particular la región \textit{sub-GeV}, pues en general los experimentos convencionales exploran WIMPs con masas $O(\text{GeV})$. Además de nuevas generaciones de detectores, nuevas señales e interacciones son tenidas en cuenta para los detectores convencionales~\cite{Ibe:2018,XENONmigdal:2019}.

\subsubsection{Detección Indirecta}

Los métodos de detección indirecta se basan en la medición de partículas del modelo estándar originadas a partir del decaimiento o aniquilación de partículas de materia oscura. Se utilizan telescopios de rayos gamma, de rayos-x, de neutrinos y observaciones de rayos cósmicos. En general se estudian las regiones donde hay más densidad de materia oscura, y por lo tanto el flujo de partículas del SM sería mayor, como por ejemplo el centro de la Vía Láctea o galaxias enanas donde la relación materia oscura-materia bariónica es alta.

Algunos excesos han sido medidos pero su interpretación aún se encuentra en discusión, pues existen grandes incertezas de origen astrofísico. Uno de los principales desafíos de los métodos de detección indirecta consiste en poder realizar una substracción del flujo de fondo producto de fuentes astrofísicas tanto identificadas como desconocidas, como por ejemplo pulsares. Además, para la medición de partículas cargadas, como es el caso de algunos rayos cósmicos, la deflexión en su trayectoria producto de campos magnéticos galácticos, o en scatterings con el medio, hacen que la determinación del origen de los mismos sea realmente dificultosa.

En cuanto a la detección de rayos cósmicos podemos mencionar a AMS-02~\cite{AMS:2013,AMS:2016}, el cual ha medido una discrepancia en la proporción de positrones-electrones. Entre los telescopios de neutrinos se encuentran Super-Kamiokande~\cite{superkamio1,superkamio2} e IceCube~\cite{IceCube:2014,IceCube:2018}.

Considerando la detección de gravitinos o axinos, los métodos de detección son puramente indirectos por la naturaleza de la interacción gravitatoria de las partículas. Como desarrollaremos en los siguientes capítulos, una de las interacciones que poseen es responsable de su decaimiento hacia un fotón y un fotino (compañero supersimétrico del fotón), pudiendo el fotón ser detectado en telescopios de rayos gamma~\cite{gamma,fermi,fermihalo}. La señal más clara, al igual que para una gran cantidad de modelos, sería una señal monocromática sobre el espectro esperado.

Existen varios experimentos que han puesto cotas sobre el decaimiento de materia oscura utilizando lineas de emisión de rayos gamma, como por ejemplo el satélite Fermi Gamma-ray Space Telescope (\textit{Fermi}) con uno de sus principales instrumentos, el Large Area Telescope~\cite{fermihalocentro,FERMIdwarfs:2015,Ackermann:2015lka} que se encuentra actualmente en funcionamiento, y misiones previas como los satélites INTEGRAL, SPI, COMPTEL y EGRET~\cite{egret}. Dada la relevancia para el trabajo realizado en esta tesis, se detalla a continuación algunas características de \textit{Fermi} y del instrumento propuesto de rayos gamma de próxima generación e-ASTROGAM.

\subsubsection{Telescopio de rayos gamma Fermi}

El \textit{Fermi Large Area  Telescope} (\Fermi-LAT) es el instrumento principal del telescopio espacial de rayos gamma \textit{Fermi} que comenzó operaciones científicas en 2008. El instrumento \Fermi-LAT posee un campo de visión amplio ($\sim 2.5$ sr) de rayos gamma de alta energía en el siguiente rango energético: desde 20 MeV hasta más de 300 GeV, y puede cubrir la totalidad del cielo una vez cada tres horas. Se encuentra en órbita a $\sim 565$ km de altura con un periodo de $\sim 96$ minutos y una inclinación de $25.6^\circ$. Dicha inclinación implica que el satélite esté 15\% de su tiempo dentro de la anomalía del Atlántico Sur, tiempo durante el cual la toma de datos es suspendida debido al gran flujo de partículas cargadas atrapadas en la región. De todas maneras, el flujo de partículas cargadas que atraviesa \Fermi-LAT es generalmente miles de veces más grande que el flujo de rayos gamma, por lo que diversas etapas de selección son necesarias para el tratamiento de los datos.

\Fermi-LAT consiste en un conversor de pares en frente de un calorímetro, y el conjunto se encuentra rodeado por centelladores plásticos para identificar y rechazar rayos cósmicos. Los rayos gamma individuales son convertidos a pares $e^+ e^-$, los cuales son detectados por el instrumento. Mediante la reconstrucción del par $e^+ e^-$ se puede deducir la energía y dirección del rayo gamma incidente. El análisis de los datos del instrumento se realiza por evento, es decir, que cada partícula incidente se almacena y analiza por separado.

El detector posee tres subsistemas: el rastreador-convertidor (\textit{`tracker/converter'}), el cual consiste en 18 capas de detectores de tiras de silicio y láminas de tungsteno intercaladas que fomentan la conversión de rayos gamma a pares $e^+ e^-$, que luego son detectados mediante la ionización de los detectores de silicio obteniendo la dirección de las partículas incidentes. Luego se encuentra un calorímetro (\textit{`3D imaging calorimeter'}), compuesto por 8 capas de cristales centelladores, donde el par $e^+ e^-$ pierde la mayoría de su energía remanente y se utiliza para la medición de la energía del rayo incidente. Tanto el calorímetro como el rastreador-convertidor se encuentran dispuestos en 16 módulos en una grilla de $4\times 4$. Finalmente, un detector de anticoincidencia (\textit{`anticoincidence detector'}), formado por un conjunto de centelladores plásticos y fibras de cambio de onda, rodea el calorímetro y el rastreador-convertidor para rechazar el fondo producido por rayos cósmicos. Además, un sistema de disparo (\textit{triggering}) y de adquisición de datos selecciona y almacena los candidatos a rayos gamma más probables para su posterior transmisión.


El cielo en rayos gamma de altas energías está dominado por emisión difusa, más del 70\% de los fotones detectados por \Fermi-LAT son producidos en el medio interestelar de nuestra galaxia por la interacción de rayos cósmicos de alta energía con la materia y la radiación de baja energía. Una componente significativa adicional difusa con distribución casi isotrópica proviene de rayos gamma de origen extragaláctico. El resto consiste en varios tipos de fuentes puntuales o extendidas como Nucleos Activos de Galaxias, galaxias normales, pulsares, clusters globulares, sistemas binarios, remanentes de supernova, sistemas solares cercanos, etc.

\subsubsection{Telescopio de rayos gamma e-ASTROGAM}

La misión e-ASTROGAM es un observatorio de rayos gamma de manera análoga al satélite \textit{Fermi}, compuesto por un detector de silicio, un calorímetro y un sistema de anticoincidencia. Se planea que el instrumento detecte fotones con energía desde 0.3 MeV hasta 3 GeV, pudiendo disminuir el límite de energía hasta 150 keV para el \textit{`tracker'}, y hasta 30 keV para el calorímetro~\cite{eAstrogamAngelis:2017}. Además de disminuir el rango energético con respecto a \textit{Fermi}, la nueva generación de instrumentos presentará una mejora en la resolución angular que le permitirá discriminar una mayor cantidad de fuentes puntuales. Al mismo tiempo, presentará una mejora en la sensibilidad de detección de líneas espectrales en el rango energético del MeV de uno o dos órdenes de magnitud con respecto a sus predecesores, convirtiéndolo en un instrumento ideal para la búsqueda de materia oscura por métodos indirectos.

e-ASTROGAM a su vez ayudará a determinar otros procesos astrofísicos, como el origen de ciertos isótopos, la evolución química de la galaxia, los procesos de explosión de supernovas, el movimiento de los flujos astrofísicos en la galaxia, la distribución y abundancia de pulsares, entre otros, que constituyen factores claves en la determinación del flujo de fondo de rayos gamma, disminuyendo las incertezas astrofísicas en la búsqueda de señales de materia oscura.

\begin{figure}
\centering
\includegraphics[scale=0.27]{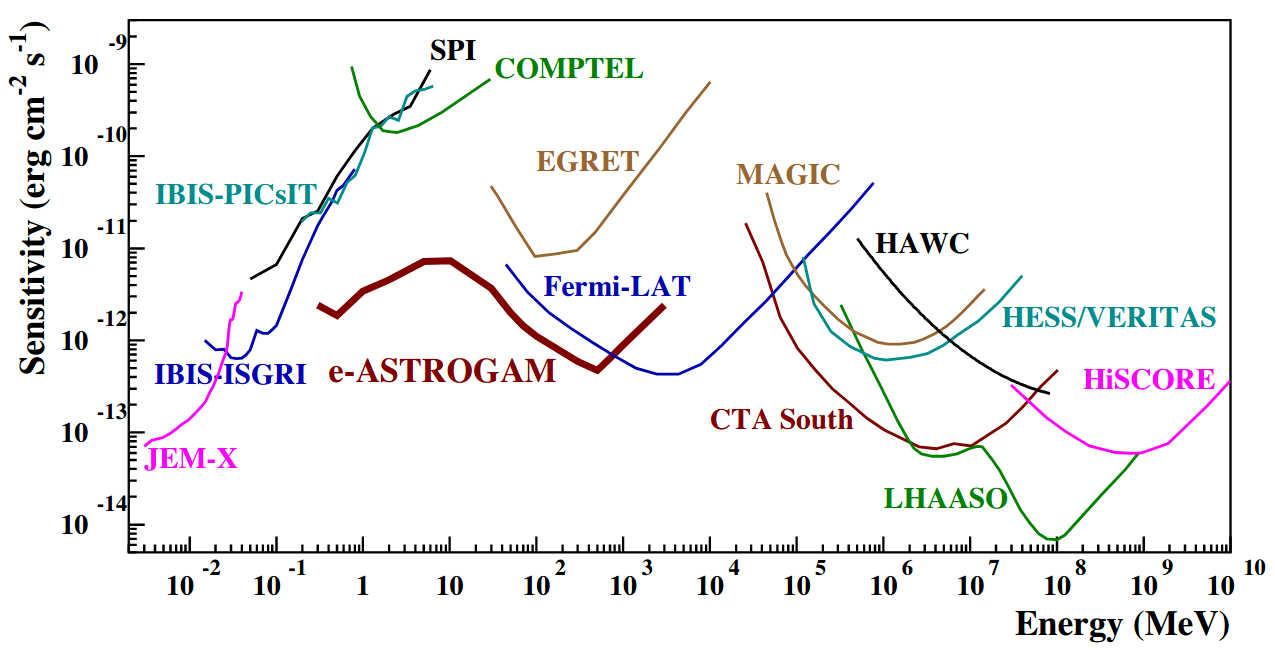}%
\caption{Sensibilidad de instrumentos en rayos-x y rayos gamma en función de la energía. La sensibilidad de \Fermi-LAT corresponde a 10 años de observación del halo galáctico, mientras que la de e-ASTROGAM a una exposición de 1 año de la misma región del cielo. Fig. de Ref.~\cite{eAstrogamAngelis:2017}.}
\label{sensitivity}%
\end{figure}

En la Figura~\ref{sensitivity} mostramos la sensibilidad para distintos instrumentos de rayos-x y rayos gamma. En particular notamos las sensitibidades de \Fermi-LAT y la proyección de e-ASTROGAM.

\newpage

\section{Axiones y el problema CP de QCD}
\label{axiones}

Aunque la motivación teórica para considerar axiones se relaciona con el problema de violación de CP en el sector de QCD, los axiones también pueden ser buenos candidatos a materia oscura. En los últimos años, la búsqueda experimental de axiones ha sido un campo muy activo, pues pueden resolver varios problemas y están bien motivados tanto en las extensiones más simples del SM, como en modelos que pretenden ser más fundamentales a altas energías, por ejemplo en modelos de teoría de cuerdas donde se predice una gran familia de partículas tipo axiones o `\textit{axion-like particles}' (ALPs). En esta tesis nos enfocaremos en los axiones de QCD propuestos como solución al problema CP de la fuerza fuerte en el SM.

\subsection{Ruptura espontánea de simetría} \label{rupturasimetria}

\begin{figure}[t]
 \begin{center}
       \epsfig{file=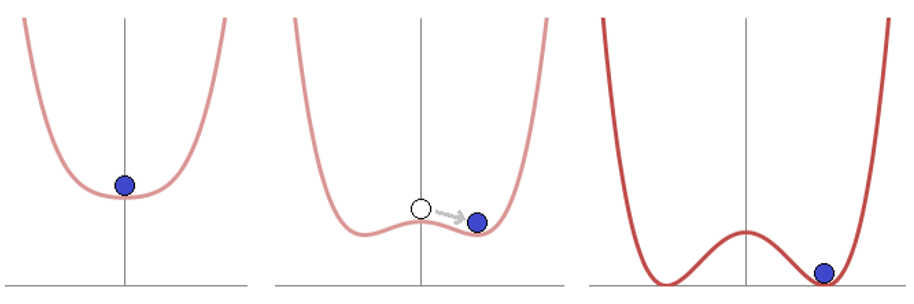,height=4.1cm}
\caption{Potencial $V(|\phi|)$ en función de la coordenada radial para $\mu^2>0$ (izquierda), $\mu^2<0$ (medio), $\mu^2\ll 0$ (derecha). Si $\mu^2=\mu^2(T)$ decrece a medida que la temperatura $T$ decrece, a temperaturas suficientemente chicas, la teoría se minimiza para un valor de $|\phi|=v$.}
    \label{rupturasimetriafig}
 \end{center}
\end{figure}

Antes de continuar ilustraremos brevemente el mecanismo de ruptura espontánea de simetría introduciendo un modelo simple, que además de ilustrar el mecanismo de Higgs, es útil para describir algunas características de los axiones. Consideraremos el siguiente Lagrangiano
\begin{equation}
    L \; = \, \frac{1}{2} (\partial_{\nu} \phi^*) (\partial^{\nu} \phi) - V(|\phi|),
    \label{lagescalarcomplex}
\end{equation}
donde $\phi(x)$ es un campo escalar complejo, cuyo potencial está dado por
\begin{equation}
    V(|\phi|) \; = \; \frac{1}{2} \mu^2 |\phi|^2 + \frac{1}{4} \lambda |\phi|^4.
\end{equation}
El Lagrangiano es independiente de la fase del campo $\phi$, es decir, que la teoría tiene simetría $U(1)$. Si $\mu^2>0$, tenemos la teoría usual $\phi^4$ para un campo escalar con masa $\mu$. Si $\mu^2=0$, obtenemos la teoría de un campo escalar no masivo. El caso interesante se presenta cuando $\mu^2<0$, entonces el potencial $V(|\phi|)$ no se minimiza para $|\phi|=0$, sino para $|\phi|=v=\mu/\sqrt{\lambda}$. De todos modos, es importante remarcar que la teoría respeta la simetría $U(1)$ para cualquier valor de $\mu^2$, por lo tanto el potencial posee simetría azimutal en el espacio de dos dimensiones (las dos coordenadas del campo complejo). En Figura~\ref{rupturasimetriafig} se muestra el potencial en función de la coordenada radial, para varios valores de $\mu^2$.

Si suponemos que $\mu^2=\mu^2(T)$, con $\mu^2(0)<0$ y $\mu^2(T)>0$ para temperaturas suficientemente altas, debajo de la temperatura crítica $T_c$ para la cual $\mu^2(T_c)=0$, $|\phi|=0$ se vuelve un máximo local y el campo evoluciona hacia un nuevo mínimo con $|\phi|=v$. Todos los nuevos mínimos caracterizados por diferentes valores de fase $U(1)$ son energéticamente equivalentes. Si expandimos el campo $\phi(x)$ alrededor del mínimo
\begin{equation}
    \phi(x) \; = \; [v+\sigma(x)] e^{i\pi(x)/f_{\pi}},
    \label{phiminimo}
\end{equation}
donde $f_{\pi}$ es un parámetro con dimensión de masa, y $\sigma(x)$ y $\pi(x)$ son dos campos escalares reales que corresponden a los dos grados de libertad del campo escalar complejo $\phi(x)$. Reemplazando Ec.~(\ref{phiminimo}) en (\ref{lagescalarcomplex}), tenemos
\begin{equation}
    L \; = \; \frac{1}{2} (\partial_{\nu} \sigma)^2 \; + \; \frac{v^2}{2 f_{\pi}^2} (\partial_v \pi)^2 \; + \; \left( \frac{1}{2} \sigma^2 + v \sigma \right) \frac{1}{f_{\pi}^2} (\partial_v \pi)^2 \; + \; V(v+\sigma(x)).
    \label{Lrupturasim}
\end{equation}
Los dos primeros términos representan los términos cinéticos de los campos $\sigma(x)$ y $\pi(x)$; la normalización del segundo término implica que $f_{\pi}=v$. Los términos siguientes corresponden a las interacciones entre $\sigma$ y $\partial_v \pi$. Expandiendo el potencial se puede ver que tenemos términos de masa, $m_{\sigma}^2=-2\mu^2$, y de autointeracción para $\sigma$.

Notamos que no obtenemos ningún término de masa para el campo $\pi$, y que además solo la derivada de $\pi$ aparece en el Lagrangiano. Esto último se debe a que dichos campos se encuentran en la exponencial en Ec.~(\ref{phiminimo}) y solo podemos extraerlos mediante una derivación; por la misma razón siempre están acompañados por un factor $1/f_{\pi}$ (incluso si incluimos en el Lagrangiano interacciones entre $\phi$ y otros campos que respeten la simetría $U(1)$ original). Dichos campos no masivos son característicos en teorías con ruptura espontánea de una simetría continua y son llamados bosones de Goldstone o de Nambu-Goldstone.

En la teoría de campos, la masa de un campo es la frecuencia de las pequeñas oscilaciones alrededor del mínimo de potencial, por lo tanto, el grado de libertad radial $\sigma$ es masivo, mientras que el azimutal (Goldstone) $\pi$ es no masivo. Bajo una transformación de $U(1)$ con parámetro $\alpha$, el campo $\pi$ se transforma de la siguiente manera
\begin{equation}
    \pi(x) \rightarrow \pi(x) + \alpha f_{\pi}.
\end{equation}
Por lo tanto, el valor del campo de Goldstone no es invariante ante transformaciones de $U(1)$, es decir, la simetría está rota. Obtenemos un campo no masivo de Goldstone por cada número de generadores del grupo de simetría que es roto espontáneamente.

Si incluimos en el Lagrangiano original términos que dependan de la fase del campo, es decir que rompan explícitamente la simetría $U(1)$, de manera tal que el potencial solo se modifique levemente, podemos generar un término de masa para el campo $\pi$. En este caso el potencial tendrá una forma similar al mostrado en Figura~\ref{rupturasimetriafig} pero con una leve inclinación. Esto produce que el mínimo de potencial no tenga simetría de revolución, y por lo tanto, la frecuencia de las pequeñas oscilaciones en la dirección azimutal indican que $\pi$ es masivo.

El ejemplo que hemos presentado ha sido tratado de manera clásica. En una teoría cuántica, en vez de un mínimo de un potencial clásico, esperamos un valor de expectación de vacío (VEV) $\langle \phi \rangle =v$ (valor de expectación en el estado de mínima energía sin excitaciones), sin embargo el mecanismo es cualitativamente similar.

Como fue discutido en la Sección~\ref{introduccion}, el modelo estándar de partículas elementales contiene un campo escalar complejo, el bosón de Higgs que es un doblete de $SU(2)_L$, el cual también posee carga de $U(1)_Y$. El potencial para estos escalares es una generalización con más dimensiones del caso anterior con $v=246$ GeV y el grupo de simetría de gauge $SU(2)_L \times U(1)_Y$ se rompe espontáneamente a $U(1)_{em}$. De esta manera, esperamos tres bosones de Goldstone, correspondientes al número de generadores de las simetrías de gauge rotas. Los bosones de Goldstone son absorbidos por los bosones de gauge, los cuales adquieren masa en el denominado mecanismo de Higgs. Finalmente, términos del tipo Yukawa producen términos de masa dependientes de $v$ para los fermiones.

\subsection{El \textit{strong CP problem}} \label{strongCPyaxiones}

El Lagrangiano de QCD del modelo estándar, $L_{QCD-SM}$, permite incluir el siguiente término de violación de CP~\footnote{Términos similares podrían haber sido incluido en el Lagrangiano original con parámetros $\theta_{em}$, y $\theta_W$ en lugar de $\bar{\theta}$, y las constantes de acople para cada simetría de gauge correspondiente. La razón para no incluirlos es que pueden escribirse como una derivada total. Como los observables dependen de la acción $S=\int d^4x \; L$, una derivada total en el Lagrangiano corresponde a términos de superficie. Para campos que se comportan bien en el infinito, dicho término se anula. Sin embargo, los términos de superficie pueden causar efectos observables para teorías de gauge no-Abelianas~\cite{Srednicki:2007qs}. En particular ocurre para QCD.}
\begin{equation}
    L_{\bar{\theta}} \; = \; \frac{g_S^2}{32 \pi^2} \bar{\theta} F_a^{\mu\nu}\widetilde{F}_{\mu\nu a},
    \label{LthetaQCD2}
\end{equation}
donde $-\pi \leq \bar{\theta} < \pi$, $g_S$ es el acople de gauge de $SU(3)_C$, $a$ un índice de color, $F_a^{\mu\nu}$ es el tensor de fuerza del campo de gauge definido como
\begin{equation}
    F_a^{\mu\nu} \; = \; \partial^{\mu} A_a^{\nu} - \partial^{\nu} A_a^{\mu} + g f_{abc} A_b^{\mu} A_c^{\nu},
\end{equation}
y su dual
\begin{equation}
    \widetilde{F}_{\mu\nu a} \; = \; \frac{1}{2} \epsilon_{\mu\nu\alpha\beta}F^{\alpha\beta}_a,
\end{equation}
con $A_a^{\mu}$ el campo de los gluones, $\epsilon_{\mu\nu\alpha\beta}$ el tensor de Levi-Civita antisimétrico en todos los índices, y $f_{abc}$ las constantes de estructura del grupo de gauge~\footnote{El parámetro $\theta$ no puede ser reabsorbido mediante una rotación de la teoría debido a que posee dos contribuciones de origen distinto, una involucrando la topología de QCD y otra debido a las interacciones de Yukawa de los quarks~\cite{Schwartz:2013pla}.}.

La existencia de violación de $CP$ implica que algunos observables dependerán del valor de $\bar{\theta}$, como por ejemplo, el momento dipolar eléctrico del neutrón (el cual puede ser medido experimentalmente con mayor facilidad ya que, comparado con otros bariones, posee un tiempo de vida largo y es eléctricamente neutro). La no observación de dicha cantidad implica que $\bar{\theta}$ debe ser extremadamente pequeño, lo cual representa un problema, llamado \textit{strong CP problem}.

El momento dipolar eléctrico del neutrón se puede aproximar por~\cite{Baluni:1978rf,Crewther:1979pi}
\begin{equation}
    d_N \; \sim \; \bar{\theta}\frac{e}{m_N^2} \frac{m_um_d}{m_u+m_d} \; \sim \; 10^{-16} \; \bar{\theta} \; e \; \text{cm},
    \label{neutronEDM}
\end{equation}
donde $m_N$ es la masa del neutrón y se ha despreciado la contribución de la masa del quark $s$. El límite actual es $d_N < 3 \times 10^{-26} \, e$ cm~\cite{Afach:2015sja}, es decir que $\theta_{QCD} \lesssim 10^{-10}$.


\subsection{El mecanismo de Peccei-Quinn y el axión}\label{mecanismodePQ}

Como fue mostrado por Peccei y Quinn~\cite{Peccei:1977hh,Peccei:1977ur}, el problema se puede resolver de manera natural introduciendo un campo pseudo-escalar denominado axión, $a(x)$~\cite{Weinberg:1977ma,Wilczek:1977pj}, y requiriendo que el Lagrangiano de QCD sea invariante ante una simetría global $U(1)_A$~\footnote{El subíndice $A$ se debe a que $U(1)_A$ representa una simetría \textit{axial o quiral}. En una transformación $U(1)_A$ fermiones derechos e izquierdos con los mismos números cuánticos rotan con una fase de igual magnitud pero distinto signo.}, llamada $U(1)_{PQ}$, la cual se rompe tanto espontánea como explícitamente.

Supongamos que tenemos un par de quarks $q_L$ y $q_R$ (aquí nos referimos a fermiones cargados bajo $SU(3)_C$) los cuales transforman bajo una simetría global $U(1)_A$ como
\begin{equation}
    q_L \rightarrow e^{i\alpha} q_L \hspace{1cm} q_R \rightarrow e^{-i\alpha} q_R.
\end{equation}
Un término de masa $m_q(\bar{q}_Lq_R + h.c)$ en el Lagrangiano es incompatible pues no es invariante (notar la conjugación del primer quark). Sin embargo, si asumimos que poseen un término de interacción del tipo Yukawa con un escalar complejo $\sigma$ que se transforma como $\sigma \rightarrow e^{2i\alpha} \sigma$, podemos generar un término invariante de masa para los quarks
\begin{equation}
    L_{\sigma \bar{q}q} \; = \; \sigma \bar{q}_Lq_R + \sigma^*\bar{q}_R q_L.
\end{equation}
Como vimos en la Sección~\ref{rupturasimetria}, si el potencial tiene un mínimo en $|\sigma|=f_a$, la simetría global puede romperse espontáneamente.  Para temperaturas menores a $f_a$ los quarks adquieren un término de masa $m_q \sim f_a$, mientras la teoría general aún respeta la simetría $U(1)_{PQ}$. Debido al mecanismo de ruptura se genera un pseudo-escalar de Goldstone, el axión.  

Si la simetría $U(1)_{PQ}$ solo estuviera rota espontáneamente, el axión sería el equivalente al bosón de Goldstone $\pi(x)$ con solo interacciones suprimidas por factores de $f_a^{-1}$ (ver Ec.~(\ref{Lrupturasim})). Por lo tanto el axión sería no masivo. Sin embargo, como la simetría global $U(1)_A$ es anómala~\footnote{Una simetría anómala~\cite{Peskin:1995ev,Schwartz:2013pla} es una simetría del Lagrangiano clásico que es violada en la correspondiente teoría cuántica~\cite{Adler:1969gk,Bell:1969ts}. Toda simetría global $U(1)_A$ es anómala.} y los fermiones cargados además tienen carga $SU(3)_C$, se genera un término explícito de ruptura
\begin{equation}
    L_{agg} \; = \; \frac{g_S^2}{32 \pi^2} \frac{a}{f_a} F_a^{\mu\nu}\widetilde{F}_{\mu\nu a},
\end{equation}
y el bosón $a(x)$ adquiere masa. Notar las similitudes con Ec.~(\ref{LthetaQCD2}).

Luego, el Lagrangiano total de QCD resulta
\begin{equation}
L_{QCD} \; = \; L_{QCD-SM} + L_{\bar{\theta}} +  L(\sigma, \bar{q},q) - \frac{1}{2} \partial_{\mu} a \partial^{\mu} a + L_{int}(\partial^{\mu} a/f_a ) + L_{agg},
\end{equation}
donde el último término del lado derecho viola CP y corresponde a la interacción entre el axión y los gluones. Como el potencial del axión se minimiza en $\langle a \rangle = -f_a \bar{\theta}$, donde $\langle a \rangle$ es el valor de expectación del campo, el término $L_{\bar{\theta}}$ se cancela. Entonces, el Lagrangiano total $L_{QCD}$ conserva CP resolviendo dinámicamente el \textit{strong CP problem}. Las interacciones del axión se encuentran suprimidas por la escala de Peccei-Quinn $f_a$ como se remarca en la dependencia de $L_{int}$. Cuando $f_a$ es suficientemente grande, el axión se vuelve `invisible', es decir, con interacciones extremadamente débiles.

Podemos estimar el valor de la masa del axión, $m_a^2$, notando que el Lagrangiano depende solo de la razón $a(x)/f_a$ y que la forma general del término de masa es $\frac{1}{2} m_a^2 a^2$, por lo tanto $m_a^2 \propto f_a^{-2}$. Necesitamos otra cantidad con dimensiones de energía, podemos suponer por análisis dimensional
\begin{equation}
    m_a^2 \sim \frac{\Lambda^4_{QCD}}{f_a^2},
\end{equation}
donde $\Lambda_{QCD}\sim 200$ MeV es la escala de confinamiento de $SU(3)_C$ (donde $g_S(E)$ tiende a infinito).
A bajas energías, el acople axión-gluón-gluón implica interacciones axión-nucleón y axión-mesón. Es de particular importancia la interacción con los piones (pues comparten los mismos números cuánticos). Por lo tanto, términos de mezcla entre axiones y piones existen en el Lagrangiano de bajas energías~\cite{Peccei:2006as} que contribuyen a la masa del axión
\begin{equation}
    m_a^2 \; = \; \frac{m_{\pi}^2 f_{\pi}^2}{f_a^2} \frac{m_um_d}{(m_u+m_d)^2} \; \simeq \; \frac{(77.6 \; \text{MeV})^4}{f_a^2},
    \label{axionmass}
\end{equation}
con $m_u$ y $m_d$ la masa de los quarks up y down respectivamente, $m_{\pi}$ la masa del pión y $f_{\pi}$ su constante de decaimiento.
Notamos que si una de las masas de los quarks es cero la masa del axión se anula; el mecanismo de PQ no es necesario en este caso pues no hay \textit{strong CP problem} (ver Refs.~\cite{Peskin:1995ev,Schwartz:2013pla}).

La mezcla axión-pión implica que el axión comparte la interacción que tenía el pión con dos fotones de la forma
\begin{equation}
    L_{a \gamma\gamma} \; = \; g_{\gamma}\frac{e^2}{16 \pi^2 f_a} a F^{\mu\nu}\widetilde{F}_{\mu\nu},
    \label{axiongammagamma}
\end{equation}
en este caso $F^{\mu\nu}$ y $\widetilde{F}_{\mu\nu}$ corresponden al tensor de fuerza del electromagnetismo, y el coeficiente sin dimensión $g_{\gamma}$ depende del modelo de axión, pues además de la contribución de la mezcla con los piones, pueden existir otras contribuciones de otros campos con carga $U(1)_{PQ}$ que aparecen en $L_{int}$.

\subsection{Modelos y búsquedas de axiones} \label{modelosaxiones}

Originalmente, se trató de vincular la escala de Peccei-Quinn con la escala de ruptura electrodébil. En este caso, el campo escalar que desarrollará luego el potencial para romper espontáneamente la simetría $U(1)_{PQ}$ es el bosón de Higgs del SM (de hecho necesitamos dos dobletes de Higgs). En este caso, los Higgs adquieren VEVs $v_1$ y $v_2$, de modo que $f_a^2=v_1^2 + v_2^2 = v^2 = (174 \text{ GeV})^2$, la escala electrodébil. Entonces la masa del axión, Ec.~(\ref{axionmass}), resulta $m_a \gtrsim 150$ KeV, dependiendo solo de la proporción entre $v_1$ y $v_2$.

Dicho axión puede producir importantes señales experimentales en los decaimientos de mesones pesados, accesibles para los aceleradores de principios de la década de 1980. Los resultados mostraron que dicho modelo, denominado PQWW (Peccei-Quinn-Weinberg-Wilczek) era inconsistente con los datos experimentales~\cite{Edwards:1982zn,Sivertz:1982rb}.

Posteriormente, surgieron modelos donde el mecanismo de PQ era implementado con $f_a \gg v$. Como la masa del axión y todas sus interacciones son inversamente proporcionales a $f_a$, el axión resultante es muy ligero y con acoplamientos muy débiles.

En el modelo KSVZ (Kim-Shifman-Vainshtein-Zakharov)~\cite{Kim:1979if,Shifman:1979if}, se agregan al SM uno o más quarks pesados con masas $\sim f_a$ y un escalar adicional con VEV $\sim f_a \gg v$. Los nuevos campos son singletes de $SU(2)_L$ y poseen carga $U(1)_{PQ}$. Diferentes versiones existen dependiendo de la carga eléctrica de los quarks pesados $q$.

Otro modelo, llamado DFSZ (Dine-Fischler-Srednicki-Zhitnitsky)~\cite{Dine:1981rt,Zhitnitsky:1980tq} es similar al modelo PQWW donde los fermiones del SM poseen carga $U(1)_{PQ}$, pero se agrega un campo escalar adicional con carga $U(1)_{PQ}$. Al igual que en el modelo KSVZ, el escalar debe tener un VEV $\sim f_a \gg v$, y ser singlete de $SU(2)_L$ para no tener acoples directos con los quarks del SM, pero interacciones con los bosones de Higgs son permitidas de acuerdo con la asignación de cargas de $U(1)_{PQ}$. Diferentes versiones de este modelo se caracterizan por distintos acoples del escalar con los fermiones del SM.

Ambos modelos además predicen la existencia de campos muy pesados, con masas $\sim f_a$, energías órdenes de magnitud superiores a lo accesible experimentalmente. A pesar de ello, existen importantes exclusiones experimentales para el sector de los axiones, que en general se presentan en función de la escala de PQ $f_a$, o en el acople entre el axión y partículas del SM. Un resumen de los límites actuales puede ser encontrado en Ref.~\cite{Tanabashi:2018oca}. Algunos de los límites provienen de experimentos en aceleradores buscando determinar generalmente el acople de axiones con fotones, en general para escalas $f_a$ pequeñas. En un rango intermedio de $f_a$ podemos encontrar experimentos diseñados para detectar la conversión de axiones a fotones haciendo uso del acople de la Ec.~(\ref{axiongammagamma}) a través de emisiones del tipo efecto Primakoff empleando un potente campo magnético, y límites provenientes de cotas astrofísicas como el enfriamiento de objetos estelares por la presencia de grados de libertad ligeros adicionales al SM, por ejemplo la duración de las explosiones de supernovas y el ciclo de vida de las enanas blancas brindan datos sobre el acople de axiones con electrones. Para escalas de PQ $f_a$ altas los límites sobre la producción de axiones como materia oscura para no generar sobreabundancia son importantes. De todos modos, un gran rango de $f_a$ es permitido pon las observaciones que general se considera $10^9 \lesssim f_a \lesssim 10^{14}$ GeV, compatible con los modelos de axiones muy livianos y con acoples muy débiles.

\spacing{1}

\chapter{Teorías supersimétricas}\label{SUSY}

\spacing{1.5}

En este capítulo tratamos brevemente el problema de las jerarquías y cómo el marco de teorías supersimétricas brinda una posible solución. Luego describimos cómo construir modelos SUSY en cuatro dimensiones. En la literatura se pueden encontrar muchos reviews sobre SUSY, por ejemplo~\cite{superMartin}, y sus referencias. Introducimos los diferentes supermultipletes y los Lagrangianos SUSY. En particular, discutimos los modelos que mencionamos en la introducción y su fenomenología: el MSSM~\cite{superMartin}, y el $\mu \nu$SSM~\cite{propuvSSM}. El primero es la mínima extensión SUSY fenomenológicamente viable del SM. El segundo es la mínima extensión si se incluyen neutrinos $\textit{\text{right-handed}}$ en el espectro y se soluciona simultáneamente el problema-$\mu$ del MSSM. La fenomenología de cada modelo es muy distinta ya que el primero conserva la paridad-R y el segundo no.

\section{Problema de las jerarquías}
\label{hierarchySUSY}

El problema de las jerarquías consiste en la gran sensibilidad que el espectro de masas del modelo estándar, en particular el bosón de Higgs, presenta frente a extensiones y física nueva a altas energías. De hecho, comprende uno de los pocos ejemplo donde fenómenos con escala característica alta son muy relevantes en fenómenos de escalas muchos órdenes de magnitud inferiores, en contraposición por ejemplo con las interacciones gravitatorias, cuyos efectos son despreciables y no se tienen en cuenta cuando se trabaja en física de colisionadores. Aunque no constituye técnicamente un problema en el SM, sabemos que este modelo no es valido al menos en la escala de Planck, por lo cual el problema de las jerarquías merece tratamiento y puede darnos indicios sobre la física nueva.

El problema reside en la dependencia cuadrática con la escala y acoplamientos de nueva física del cuadrado de la masa del bosón de Higgs, haciendo que dicha escala sea la natural para su masa. Para una teoría conteniendo gravedad, debido a correcciones cuadráticas, sería de esperar que la masa del Higgs fuese del orden de la masa de Planck, anulando de esta manera la ruptura electrodébil tan exitosa del SM.

\begin{figure}
\centering
\includegraphics[width=8cm]{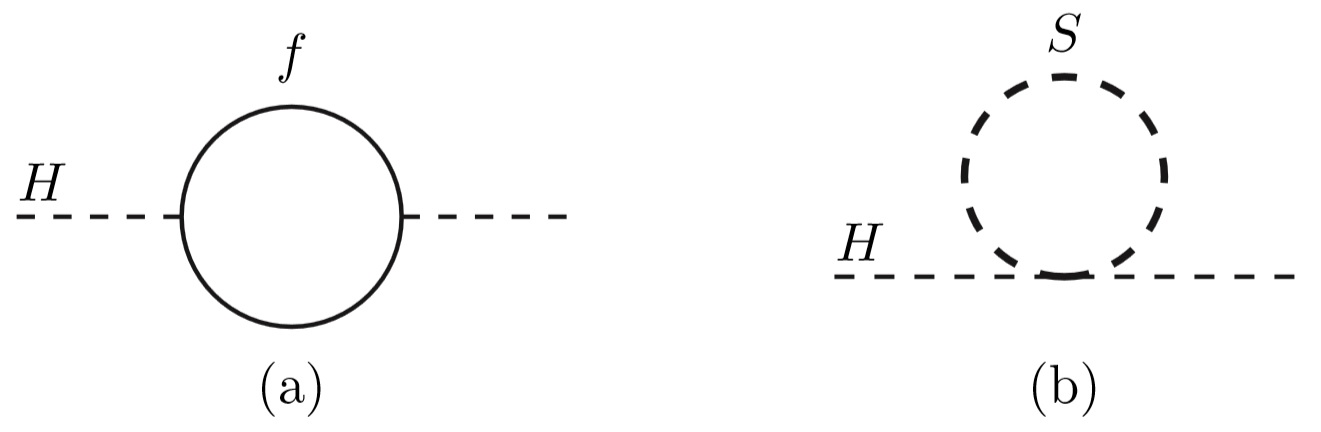}
\caption{Correcciones a un loop para la masa del Higgs debido a (a) un fermión de Dirac $f$, y (b) un escalar $S$.}
\label{higgshierarchy}
\end{figure} 

Por ejemplo, en la Figura~\ref{higgshierarchy} (a) se presenta el diagrama de Feynman de la corrección a un loop de un fermión de Dirac $f$ a la masa del bosón de Higgs $H$, cuyo acople en el Lagrangiano está dado por $-\lambda_f \, H \, \overline{f} \, f$. Por lo tanto, la corrección es
\begin{equation}
	\Delta m_H^2 = -\frac{|\lambda_f|^2}{8\pi^2} \, \left[ \, \Lambda_{UV}^2 \,  + \, 3 m_f^2 \, \text{ln}\left( \frac{\Lambda_{UV}}{m_f} \right) + ... \right]  \, ,
	\label{higgscorrection1}
\end{equation}
donde $\Lambda_{UV}$ es el momento denominado `\textit{cutoff}' usado para regular la integral en el loop, el cual puede ser interpretado como la escala de energía a partir de la cual la nueva física es relevante. Los términos omitidos son proporcionales a $m_f$, el parámetro de masa de los fermiones, y al logaritmo de $\Lambda_{UV}$.

En el modelo estándar, cada leptón y quark (estos últimos multiplicados por 3 para tener en cuenta el grado de libertad de color) pueden tomar el rol de $f$, y la corrección más relevante está dada por el quark más pesado, el top, con $\lambda_f \simeq 0.94$ el acople de Yukawa del top. El problema, como mencionamos, surge cuando consideramos que $\Lambda_{UV}$ puede ser del orden $M_P$ pues las correcciones son órdenes de magnitud superior a la masa del bosón de Higgs experimentalmente medida. Además, el resto de las partículas del SM obtienen su masa a partir de acoples con el Higgs, por lo cual todo el espectro de masas del SM es sensible a dichas correcciones.

Más aún, contribuciones similares a~\ref{higgscorrection1} pueden existir para toda partícula más pesada aún no descubierta involucrando sus masas y no solo la escala de \textit{cutoff}. Por ejemplo, si existen escalares $S$ más pesados que se acoplan al Higgs mediante un término en el Lagrangino dado por $-\lambda_S \, |H|^2 \, |S|^2$, el diagrama de Feynman presentado en la Figura~\ref{higgshierarchy} (b) tiene como correcciones
\begin{equation}
	\Delta m_H^2 = \frac{\lambda_S}{16\pi^2} \, \left[ \, \Lambda_{UV}^2 - 2 m_S^2 \, \text{ln}\left( \frac{\Lambda_{UV}}{m_S} \right) + ... \, \right] \, ,
	\label{higgscorrection2}
\end{equation}
donde $m_S$ es el parámetro de masa del escalar $S$. Por lo tanto, aún si se soluciona el inconveniente del \textit{cutoff}, grandes correcciones estarían dadas por la escala de masa de las nuevas partículas. Asimismo, el problema aún está presente si las nuevas partículas pesadas se acoplan de manera indirecta con el bosón de Higgs, y para cualquier fenómeno físico de alta escala energética como por ejemplo dimensiones extra compactificadas.

La cancelación sistemática de contribuciones a $\Delta m_H^2$ puede ser lograda mediante una simetría. Debido al signo relativo entre las contribuciones de los loops fermiónicos y bosónicos, Ec.~(\ref{higgscorrection1}) y (\ref{higgscorrection2}), se puede pensar en una simetría que relacione fermiones con bosones. Si cada leptón y quark del SM está acompañado por dos escalares con $\lambda_S=|\lambda_f|^2$, las contribuciones con $\Lambda_{UV}$ se cancelan. De hecho, la cancelación también se produce a ordenes superiores de loops. Dicha simetría que relaciona fermiones con bosones se denomina \textit{supersimetría}.

\section{Introducción a modelos supersimétricos}
\label{general SUSY}
Una transformación supersimétrica cambia un estado fermiónico en un estado bosónico y viceversa. El operador que genera tal trasformación debe ser un operador fermiónico de spin 1/2, un spinor que anticonmuta $Q$,
\begin{equation}
    Q|\text{bosón}\rangle \simeq |\text{fermión} \rangle, \hspace{1 cm} Q|\text{fermión}\rangle \simeq |\text{bosón} \rangle.
\end{equation}
Junto con el generador de traslaciones espacio-temporales, el operador de cuadrimomento $P^{\mu}$, conforman el álgebra supersimétrica. Notamos que $Q$ y $P^{\mu}$ conmutan.

En este contexto, los estados de partículas están representados por supermultipletes, los cuales contienen tanto estados fermiónicos como bosónicos, comúnmente llamados supercompañeros, con iguales grados de libertad. Los estados del mismo supermutiplete poseen la misma masa, pues el operador de masas al cuadrado $P^2$ conmuta con $Q$. Del mismo modo, los generadores de las transformaciones supersimétricas también conmutan con los generadores de las transformaciones de gauge, por lo tanto las partículas del mismo supermultiplete poseen la misma representación de grupo de gauge, es decir, misma carga eléctrica, isospín débil y grados de libertad de color.

Existen dos tipos de supermultipletes: supermultiplete quiral y supermultiplete vectorial. El supermultiplete quiral está compuesto por un fermión $\psi$, conformado por un spinor de Weyl, y dos escalares reales que forman un campo escalar complejo $\phi$.

El supermultiplete vectorial contiene un bosón de gauge de spin 1 $A_{\mu}^{a}$ sin masa (cuando la simetría de gauge no se encuentra rota espontáneamente), y un fermión gaugino de Weyl de dos componentes $\lambda^{a}$. El índice $\textit{a}$ recorre la representación adjunta del grupo de gauge ($\textit{a}$ = 1 ... 8
para $SU(3)_{C}$ gluones y gluinos de color; $\textit{a}$ = 1, 2, 3 para $SU(2)_{L}$ isospín débil; $\textit{a}$ = 1 para $U(1)_{Y}$ hipercarga).

Necesitamos que el álgebra supersimétrica sea cerrada. Esto se cumple `\textit{on-shell}', es decir, cuando las ecuaciones de movimiento clásicas se satisfacen. Para que sea cerrada `\textit{off-shell}' necesitamos incluir para cada tipo de supermultiplete un campo escalar complejo que no se propague, es decir, que no tenga término cinético. Para el supermultiplete quiral a dicho campo auxiliar lo denotamos $F_{i}$, y para el supermultiplete vectorial lo llamamos $D^{a}$.

\subsection{Supermultiplete quiral}
Comenzaremos considerando la acción más simple, utilizando un fermión de Weyl y un campo escalar complejo. Entonces, solo tendremos términos cinéticos para cada campo,
\begin{equation}
	S=\int(\textsl{L}_{\text{escalar}}+\textsl{L}_{\text{fermiónico}})d^4x,
	\label{WZ}
\end{equation}

\begin{equation}
	\textsl{L}_{\text{escalar}}=-\partial^{\mu}\phi^*\partial_{\mu}\phi,     \hspace{1cm}      \textsl{L}_{\text{fermiónico}}=-i\psi^{\dag}\overline{\sigma}^{\mu}\partial_{\mu}\psi.
\end{equation}

Una transformación supersimétrica debe convertir un campo bosónico escalar $\phi$ en una expresión que contenga el campo fermiónico $\psi$. La forma más simple de transformación del campo escalar es
\begin{equation}
	\delta\phi=\epsilon\psi    \hspace{1cm}     \delta\phi^*=\epsilon^{\dag}\psi^{\dag}
	\label{tescalar}
\end{equation}
donde $\epsilon^\alpha$, el objeto que parametriza la transformación, es un Weyl de dos componentes infinitesimal que anticonmuta. Como trabajamos en el marco de supersimetría global, $\epsilon^\alpha$ es una constante.

Para que la acción que obtuvimos sea invariante ante transformaciones de supersimetría $\delta\textsl{L}_{\text{escalar}}+\delta\textsl{L}_{\text{fermiónico}}$ debe cancelarse, a menos de una derivada total. Entonces encontramos que la transformación del campo fermiónico debe ser
\begin{equation}
	\delta\psi_{\alpha}=i(\sigma^{\mu}\epsilon^{\dag})_{\alpha}\partial_{\mu}\phi,    \hspace{1cm}     \delta\psi^{\dag}_{\alpha}=-i(\epsilon\sigma^{\mu})_{\alpha}\partial_{\mu}\phi^*.
	\label{tfermion}
\end{equation}

Por lo que arribamos a:
\begin{equation}
	\delta S=\int(\delta\textsl{L}_{\text{escalar}}+\delta\textsl{L}_{\text{fermiónico}})d^4x=0.
\end{equation}

Como anteriormente mencionamos, el álgebra supersimétrica debe ser cerrada, en otras palabras, el conmutador de dos transformaciones supersimétricas distintas debe ser otra simetría de la teoría. Usando Ec.~(\ref{tescalar}) y (\ref{tfermion}) encontramos
\begin{align}
	(\delta_{\epsilon_{2}}\delta_{\epsilon_{1}}-\delta_{\epsilon_{1}}\delta_{\epsilon_{2}})\phi\equiv\delta_{\epsilon_{2}}(\delta_{\epsilon_{1}}\phi)-\delta_{\epsilon_{1}}(\delta_{\epsilon_{2}}\phi)=i(\epsilon_{1}\sigma^{\mu}\epsilon_{2}^{\dag}-\epsilon_{2}\sigma^{\mu}\epsilon_{1}^{\dag})\partial_{\mu}\phi.
\end{align}
Obtenemos que el conmutador de dos transformaciones supersimétricas nos da la derivada del campo original. $\partial_{\mu}$ corresponde al generador de traslaciones espacio temporales $P_{\mu}$.

Debemos ahora encontrar el mismo resultado para el campo fermiónico. Utilizando Ec.~(\ref{tescalar}) y (\ref{tfermion}) junto a la identidad de Fierz encontramos:
\begin{align}
	(\delta_{\epsilon_{2}}\delta_{\epsilon_{1}}-\delta_{\epsilon_{1}}\delta_{\epsilon_{2}})\psi_{\alpha}=i(\epsilon_{1}\sigma^{\mu}\epsilon_{2}^{\dag}-\epsilon_{2}\sigma^{\mu}\epsilon_{1}^{\dag})\partial_{\mu}\psi_{\alpha}-i\epsilon_{1\alpha}\epsilon^{\dag}_{2}\overline{\sigma}^{\mu}\partial_{\mu}\psi+i\epsilon_{2\alpha}\epsilon^{\dag}_{1}\overline{\sigma}^{\mu}\partial_{\mu}\psi.
\label{fermdoble}
\end{align}
Los últimos dos términos en (\ref{fermdoble}) se anulan \textit{on-shell}, es decir, si la ecuación de movimiento se cumple. Para que el álgebra sea cerrada \textit{off-shell}, incluimos un campo escalar complejo F sin término cinético. Dicho campo es denominado \textit{auxiliar}, su densidad Lagrangiana es
\begin{equation}
	\textsl{L}_{\text{auxiliar}}=F^*F.
\end{equation}
Observando los términos de (\ref{fermdoble}) pedimos que la transformación de supersimetría del campo auxiliar sea
\begin{equation}
	\delta F=i\epsilon^{\dag}\overline{\sigma}^{\mu}\partial_{\mu}\psi,   \hspace{1cm}    \delta F^*=-i\partial_{\mu}\psi^{\dag}\overline{\sigma}^{\mu}\epsilon.
\label{taux}
\end{equation}
Sin embargo hemos agregado un término auxiliar a la densidad Lagrangiana total que se transforma como
\begin{equation}
	\delta\textsl{L}_{\text{auxiliar}}=i\epsilon^{\dag}\overline{\sigma}^{\mu}\partial_{\mu}\psi F^*-i\partial_{\mu}\psi^{\dag}\overline{\sigma}^{\mu}\epsilon F,
\end{equation}
que se anula si se cumplen las ecuaciones de movimiento. Agregando un término extra a las leyes de transformación para el campo fermiónico:
\begin{equation}
	\delta\psi_{\alpha}=i(\sigma^{\mu}\epsilon^{\dag})_{\alpha}\partial_{\mu}\phi+\epsilon_{\alpha}F,    \hspace{1cm}     \delta\psi^{\dag}_{\alpha}=-i(\epsilon\sigma^{\mu})_{\alpha}\partial_{\mu}\phi^*+\epsilon_{\alpha}^{\dag}F^*,
\label{ttfermion}
\end{equation}
obtenemos una contribución adicional a $\delta\textsl{L}_{\text{fermiónico}}$ que se cancela con $\delta\textsl{L}_{\text{auxiliar}}$, salvo un término con derivada total. Entonces 
\begin{equation}
	\textsl{L}_{\text{quiral}}=\textsl{L}_{\text{escalar}}+\textsl{L}_{\text{fermiónico}}+\textsl{L}_{\text{auxiliar}}=-\partial^{\mu}\phi^*\partial_{\mu}\phi-i\psi^{\dag}\overline{\sigma}^{\mu}\partial_{\mu}\psi+F^{*}F
\label{Lchiral}
\end{equation}
es invariante ante transformaciones de supersimetría. Procediendo como antes, encontramos que para cada uno de los campos $X=\phi,\phi^*,\psi,\psi^{\dag},F,F^*$,
\begin{equation}
	(\delta_{\epsilon_{2}}\delta_{\epsilon_{1}}-\delta_{\epsilon_{1}}\delta_{\epsilon_{2}})X=i(\epsilon_{1}\sigma^{\mu}\epsilon_{2}^{\dag}-\epsilon_{2}\sigma^{\mu}\epsilon_{1}^{\dag})\partial_{\mu}X.
\end{equation}
usando Ec.~(\ref{tescalar}), (\ref{taux}) y (\ref{ttfermion}) sin recurrir a ninguna ecuación de movimiento. Por lo tanto el álgebra supersimétrica es cerrada.

Hemos descrito el modelo de \textit{Wess-Zumino} no interactuante y sin masa.

\subsubsection{Interacciones entre supermultipletes quirales}\label{intchiral}
En esta sección trabajaremos con varios supermultipletes quirales, cuya componente bosónica notaremos como $\phi_{i}$, y la fermiónica como $\psi_{i}$. El índice $\textit{i}$ corre sobre todos los grados de libertad de familia y sabores.

Los términos más generales de interacciones renormalizables pueden ser escritos en forma simple
\begin{equation}
	\textsl{L}_{\text{int}}=(-\frac{1}{2}W^{ij}\psi_{i}\psi_{j}+W^{i}F_{i})+c.c.,
\end{equation}
donde $W^{ij}$ y $W^{i}$ son polinomios en los campos escalares bosónicos $\phi_{i}$ y $\phi^{i}$ de grados 1 y 2 respectivamente, y "`c.c."' indica complejo conjugado.
Requerimos que $\textsl{L}_{\text{int}}$ sea invariante ante transformaciones de supersimetría, pues $\textsl{L}_{\text{quiral}}$ sin interacciones, lo es por sí mismo. Realizando la transformación de los términos que contienen spinores y utilizando la identidad de Fierz, $\delta\textsl{L}_{\text{int}}$ se anulan si y solo si $\frac{\delta W^{ij}}{\delta \phi_{k}}$ es totalmente simétrico ante el intercambio de i, j, k. También requerimos que $\frac{\delta W^{ij}}{\delta \phi^{*k}}$=0. En otras palabras, $W^{ij}$ es analítica en el campo complejo $\phi_{k}$.
Podemos escribir:
\begin{equation}
	W^{ij}=M^{ij}+y^{ijk}\phi_{k}
\end{equation}
donde $M^{ij}$ es una matriz simétrica de masas de fermiones, y $y^{ijk}$ es el acople de Yukawa de un campo escalar $\phi_{k}$ y dos fermiones $\psi_{i}\psi_{j}$ totalmente simétrico ante el intercambio de i, j, k. Por lo tanto, es conveniente escribir:
\begin{equation}
	W^{ij}=\frac{\delta^2}{\delta\phi_{i}\delta\phi_{j}}W
\end{equation}
donde hemos introducido un objeto muy importante,
\begin{equation}
	W=L^{i}\phi_{i}+\frac{1}{2}M^{ij}\phi_{i}\phi_{j}+\frac{1}{6}y^{ijk}\phi_{i}\phi_{j}\phi_{k},
\label{superpot}
\end{equation}
el cual es llamado \textit{superpotencial}, una función holomorfa del campo escalar $\phi_{i}$. El término con acople $L^{i}$ solo es permitido si $\phi_{i}$ es un singlete de gauge.
Siguiendo con la transformación de supersimetría, los términos que contienen derivadas espacio temporales se anulan, a menos de una derivada total, si:
\begin{equation}
	W^{i}=\frac{\delta W}{\delta \phi_{i}}=M^{ij}\phi_{j}+\frac{1}{2}y^{ijk}\phi_{j}\phi_{k}.
\end{equation}
Los términos restantes de la transformación de $\textsl{L}_{\text{int}}$ son lineales en $F_{i}$ o $F^{*i}$, y se cancelan fácilmente dados los resultados de $W^{i}$ y $W^{ij}$ que encontramos.
Resumiendo, las interacciones más generales entre supermultipletes quirales están determinadas por el superpotencial W, una función analítica de campos escalares complejos.

Es importante remarcar que existe una notación, el lenguaje de supercampos, en la cual los términos del superpotencial están escritos en función de un nuevo ente que relaciona la parte bosónica y la parte fermiónica de un supermultiplete en un solo objeto, el supercampo. La ventaja del lenguaje de supercampos es mostrar de manera manifiesta la invariancia ante supersimetría. El superpotencial se escribe en ese lenguaje en términos de supercampos. Pero tal simetría no se observa a bajas escalas. Por lo tanto, no utilizaremos el lenguaje de supercampos, ya que en esta tesis trataremos por separado la parte fermiónica y la parte bosónica de los supermultipletes.

En $\textsl{L}_{\text{quiral}}+\textsl{L}_{\text{int}}$ los campos auxiliares $F_{i}$ y $F^{*i}$ aparecen en los términos $F_{i}F^{*i}+W^{i}F_{i}+W_{i}^{*}F^{*i}$. Por lo que podemos eliminarlos usando sus ecuaciones clásicas de movimiento:
\begin{equation}
	F_{i}=-W_{i}^{*};   \hspace{1cm}     F^{*i}=-W^{i}
\label{f}
\end{equation}
Entonces los campos auxiliares pueden ser expresados (sin derivadas) en términos de los campos escalares. Luego de reemplazar (\ref{f}) en $\textsl{L}_{\text{quiral}}+\textsl{L}_{\text{int}}$ obtenemos la densidad Lagrangiana
\begin{equation}
	L=-\partial^{\mu}\phi^{*i}\partial_{\mu}\phi_{i}-i\psi^{\dag i}\overline{\sigma}^{\mu}\partial_{\mu}\psi_{i}-\frac{1}{2}(W^{ij}\psi_{i}\psi_{j}+W_{ij}^{*}\psi^{\dag i}\psi^{\dag j})-W^{i}W_{i}^{*}.
\label{lagrangiano1}
\end{equation}
Sin los campos no propagantes $F_{i}$ y $F^{*i}$, de Ec.~(\ref{lagrangiano1}) notamos que el potencial escalar de la teoría está dado en términos del superpotencial por:
\begin{align}
	 &V(\phi,\phi^{*})=W^{k}W_{k}^{*}=F^{*k}F_{k}= \cr
	 &M_{ik}^{*}M^{kj}\phi^{*i}\phi_{j}+\frac{1}{2}M^{in}y_{jkn}^{*}\phi_{i}\phi^{*j}\phi^{*k}+\frac{1}{2}M_{in}^{*}y^{ikn}\phi^{*i}\phi_{j}\phi_{k}+\frac{1}{4}y^{ijn}y_{kln}^{*}\phi_{i}\phi_{j}\phi^{*k}\phi^{*l}.
\end{align}

Este potencial escalar está automáticamente acotado inferiormente, de hecho, al ser una suma de cuadrados, es siempre no negativo.

Finalmente, reemplazando la forma general del superpotencial dada por Ec.~(\ref{superpot}) en Ec.~(\ref{lagrangiano1}), tenemos la densidad Lagrangiana total
\begin{align}
	&L=-\partial^{\mu}\phi^{*i}\partial_{\mu}\phi_{i} - M_{ik}^{*}M^{kj}\phi^{*i}\phi_{j} \cr
	&-\frac{1}{2}M^{in}y_{jkn}^{*}\phi_{i}\phi^{*j}\phi^{*k}-\frac{1}{2}M_{in}^{*}y^{ikn}\phi^{*i}\phi_{j}\phi_{k}-\frac{1}{4}y^{ijn}y_{kln}^{*}\phi_{i}\phi_{j}\phi^{*k}\phi^{*l} \cr
	&-i\psi^{\dag i}\overline{\sigma}^{\mu}\partial_{\mu}\psi_{i}-\frac{1}{2}M^{ij}\psi_{i}\psi_{j}-\frac{1}{2}M_{ij}^{*}\psi^{\dag i}\psi^{\dag j} \cr	 
	&-\frac{1}{2}y^{ijk}\phi_{i}\psi_{j}\psi_{k}-\frac{1}{2}y_{ijk}^{*}\phi^{*i}\psi^{\dag j}\psi^{\dag k}.
\label{langchiral}
\end{align}
Comparando las masas de los fermiones y los escalares en las ecuaciones de movimiento linealizadas, luego de diagonalizar la matriz de masa y redefiniendo los campos, obtenemos una colección de supermultipletes quirales cada uno con un escalar complejo y un fermión de Weyl degenerados en masa.

\subsection{Supermultiplete vectorial}
Los grados de libertad propagantes en un supermultiplete vectorial son un campo bosónico de gauge sin masa $A_{\mu}^{a}$ y un fermión de Weyl gaugino $\lambda^{a}$, donde el indice $\textit{a}$ corre sobre la representación adjunta del grupo de gauge. Al igual que en el caso de los supermultipletes quirales, necesitaremos un campo auxiliar para que el álgebra supersimétrica sea cerrada. Denotamos dicho campo como $D^{a}$, el cual satisface $(D^{a})^{*}=D^{a}$ y no tiene término cinético.
Entonces, la densidad Lagrangiana para un supermultiplete vectorial es:
\begin{equation}
	\textsl{L}_{\text{vectorial}}=-\frac{1}{4}F_{\mu\nu}^{a}F^{\mu\nu a}-i\lambda^{\dag a}\overline{\sigma}^{\mu}D_{\mu}\lambda^{a}+\frac{1}{2}D^{a}D^{a},
\end{equation}
donde
\begin{equation}
	F_{\mu\nu}^{a}=\partial_{\mu}A_{\nu}^{a}-\partial_{\nu}A_{\mu}^{a}+gf^{abc}A_{\mu}^{b}A_{\nu}^{c}
\end{equation}
es la expresión usual de Yang-Mills, y
\begin{equation}
	D_{\mu}\lambda^{a}=\partial_{\mu}\lambda^{a}+gf^{abc}A_{\mu}^{b}\lambda^{c}
\end{equation}
es la derivada covariante del campo gaugino.
De manera similar al caso del supermultiplete quiral, se definen las transformaciones de supersimetría de los campos pidiendo que sean lineales en los parámetros infinitesimales $\epsilon$, $\epsilon^{\dag}$, y que $\delta D^{a}$ sea proporcional a las ecuaciones de movimiento de los gauginos, en analogía con el rol del campo auxiliar F. Luego se puede probar que la transformación es cerrada, llegando a:
\begin{equation}
	(\delta_{\epsilon_{2}}\delta_{\epsilon_{1}}-\delta_{\epsilon_{1}}\delta_{\epsilon_{2}})X=i(\epsilon_{1}\sigma^{\mu}\epsilon_{2}^{\dag}-\epsilon_{2}\sigma^{\mu}\epsilon_{1}^{\dag})D{\mu}X
\end{equation}
para X=$F_{\mu\nu}^{a}$,$\lambda^{a}$,$\lambda^{\dag a}$,$D^{a}$.

\subsubsection{Lagrangianos invariantes de gauge}
Como las transformaciones de gauge y de supersimetría conmutan, el escalar, el fermión y el campo auxiliar de los supermultipletes quirales deben estar en la misma representación del grupo de gauge. Entonces para tener un Lagrangiano invariante de gauge, debemos reemplazar las derivadas ordinarias en Ec.~(\ref{Lchiral}) por derivadas covariantes:
\begin{equation}
	\partial_{\mu}\phi_{i} \rightarrow D_{\mu}\phi_{i}=\partial_{\mu}\phi_{i}-igA_{\mu}^{a}(T^{a}\phi)_{i}
\end{equation}
\begin{equation}
	\partial_{\mu}\phi^{*i} \rightarrow D_{\mu}\phi^{*i}=\partial_{\mu}\phi^{*i}+igA_{\mu}^{a}(\phi^{*}T^{a})^{i}
\end{equation}
\begin{equation}
	\partial_{\mu}\psi_{i} \rightarrow D_{\mu}\psi_{i}=\partial_{\mu}\psi_{i}-igA_{\mu}^{a}(T^{a}\psi)_{i}
\end{equation}
donde $T^{a}$ son matrices hermíticas que cumplen $[T^{a},T^{b}]=if^{abc}T^{c}$ (por ejemplo, si el grupo de gauge es SU(2), $f^{abc}$=$\epsilon^{abc}$ y $T^{a}$ es 1/2 las matrices de Pauli para un supermultiplete quiral que se transforma bajo dicha representación.)
Este procedimiento acopla vectores bosónicos del supermultiplete vectorial con los escalares y fermiones de los supermultipletes quirales. Sin embargo, existen otros tres términos renormalizables permitidos por la invariancia de gauge que involucran gauginos y el campo $D^{a}$, acoplándolos a $\phi$ y $\psi$, que deben ser incluidos en el Lagrangiano
\begin{equation}
	(\phi^{*}T^{a}\psi)\lambda^{a},    \hspace{1cm}      \lambda^{\dag a}(\psi^{\dag}T^{a}\phi),     \hspace{1cm} y \hspace{1cm}    (\phi^{*}T^{a}\phi)D^{a}
\end{equation}
con coeficientes de acople adimensionales. Si pedimos invariancia ante transformaciones de supersimetría, las leyes de transformación para los campos quirales deben ser modificadas incluyendo la derivada covariante en lugar de la derivada ordinaria en Ec.~(\ref{tescalar}), (\ref{taux}) y (\ref{ttfermion}). Además es necesario incluir el término extra $\sqrt{2}g(T^{a}\phi)_{i}\epsilon^{\dag}\lambda^{\dag a}$ en el termino $\delta F_{i}$.
Finalmente:
\begin{equation}
	\textsl{L}=\textsl{L}_{\text{quiral}}+\textsl{L}_{\text{vectorial}}-\sqrt{2}g(\phi^{*}T^{a}\psi)\lambda^{a}-\sqrt{2}g\lambda^{\dag a}(\psi^{\dag}T^{a}\phi)+g(\phi^{*}T^{a}\phi)D^{a}.
	\label{L}
\end{equation}
Aquí $\textsl{L}_{\text{quiral}}$ representa el Lagrangiano del supermultiplete quiral encontrado en la Sección~\ref{intchiral} [es decir Ec.~(\ref{langchiral})] pero con las derivadas ordinarias reemplazadas por derivadas covariantes.

El último término de Ec.~(\ref{L}) junto con el término $1/2 D^{a}D^{a}$ en $\textsl{L}_{\text{vectorial}}$ resultan en la siguiente ecuación de movimiento:
\begin{equation}
	D^{a}=-g(\phi^{*}T^{a}\phi).
\end{equation}
Por lo tanto, el campo auxiliar $D^{a}$ se expresa en términos de los campos escalares, como en el caso de los campos auxiliares $F_{i}$ y $F^{*i}$. Entonces el potencial escalar es (recordando que $\textsl{L}\supset -V$):
\begin{equation}
	V(\phi,\phi^{*})=F^{*i}F_{i}+\frac{1}{2}\sum_{a}D^{a}D^{a}=W_{i}^{*}W^{i}+\frac{1}{2}\sum_{a}g_{a}^{2}(\phi^{*}T^{a}\phi)^{2}.
\end{equation}
Los dos tipos de términos en esta expresión son llamados contribuciones del "`término F"' y "`término D"' respectivamente. $\sum_{a}$ cubre los casos donde los grupos de gauge tengan diferentes acoples de gauge $g_{a}$. (Por ejemplo, en MSSM, los factores $SU(3)_{C}$, $SU(2)_{L}$ y $U(1)_{Y}$ tienen acoples de gauge igual a $g_{3}$, $g$ y $g'$ respectivamente.) Como $V(\phi,\phi^{*})$ es una suma de cuadrados, siempre es mayor o igual a cero para toda configuración de los campos. Notamos que el potencial escalar esta completamente determinado por otras interacciones de la teoría; el término F por los acoples de Yukawa y los términos de masas de los fermiones, y el término D por las interacciones de gauge.

\subsection{Términos de ruptura de supersimetría \textit{soft}}
Esperamos que la supersimetría sea una simetría exacta, la cual es espontáneamente rota, para brindar un modelo fenomenológico realístico. En otras palabras, el modelo debe tener una densidad Lagrangiana invariante ante transformaciones de supersimetría, pero un estado de vacío que no lo es; de manera análoga a la simetría electrodébil. Sin embargo, esperamos que esto suceda en la versión local que incluye gravedad, conocida como SUGRA (ver Ref.~\cite{superMartin} y sus referencias).

La ruptura de supersimetría ocurre en un sector oculto y es mediada al sector visible. Desde un punto de vista práctico es útil introducir términos extra en el Lagrangiano que rompan explícitamente la simetría. Dichos términos deben ser \textit{soft} (con dimensión de masa positiva) para seguir otorgando solución al problema de las jerarquías entre la escala electrodébil y la escala de Planck; en particular, acoples sin dimensión estarán ausentes.
Los posibles términos de ruptura \textit{soft} de supersimetría son:
\begin{equation}
	\textsl{L}_{\text{soft}}=-(\frac{1}{2}M_{a}\lambda^{a}\lambda^{a}+\frac{1}{6}a^{ijk}\phi_{i}\phi_{j}\phi_{k}+\frac{1}{2}b^{ij}\phi_{i}\phi_{j}+t^{i}\phi_{i}) + c.c. - (m^{2})_{j}^{i}\phi^{j*}\phi_{i}.
	\label{Lsoft}
\end{equation}
Contienen la masa de los gauginos $M_{a}$ para cada grupo de gauge, términos de masa al cuadrado para los escalares $(m^{2})_{j}^{i}$ y $b^{ij}$, acoples de $(\textit{\text{escalares}})^{3}$ $a^{ijk}$ y acoples $t^{i}$ solo posible si $\phi_{i}$ es un singlete de gauge.
El término con $M_{a}$ siempre es permitido por las simetrías de gauge; los términos con $(m^{2})_{j}^{i}$ se permiten si $\phi_{i}$, $\phi^{j*}$ se transforman en el complejo conjugado de cada uno, en particular esto es válido para $\textit{i = j}$. Los términos $a^{ijk}$, $b^{ij}$ y $t^{i}$ tienen la misma forma que los términos $y^{ijk}$, $M^{ij}$ y $L^{i}$ del superpotencial, por lo tanto estarán permitidos por invariancia de gauge si cada uno de los correspondientes términos del superpotencial están permitidos.

Por último notamos que la transformación de supersimetría está efectivamente rota por $\textsl{L}_{\text{soft}}$, ya que involucra campos escalares y gauginos pero no sus respectivos supercompañeros. Entonces, los términos en $\textsl{L}_{\text{soft}}$ son capaces de otorgarle masa a todos los escalares y gauginos de la teoría, aún si los bosones de gauge y los fermiones de los supermultipletes quirales no tienen masa (o son relativamente livianos).
Con lo repasado hasta aquí podemos construir cualquier modelo supersimétrico en cuatro dimensiones.

\section{Minimal Supersymmetric Standard Model (MSSM)}
\label{MSSM}
Describiremos brevemente la extensión mínima del SM dentro del marco de las teorías SUSY.

\subsection{Lagrangiano y sus simetrías}
El superpotencial para el MSSM es:
\begin{equation}
	W_{\text{MSSM}}=\epsilon_{ab}(Y_{ij}^u \, \hat H_{u}^{b} \, \hat Q^{a}_i \, \hat u_j^c + Y_{ij}^d \, \hat H_{d}^a \, \hat Q_{i}^b \, \hat d_j^c + Y_{ij}^e \, \hat H_{d}^{a} \, \hat L_i^{b} \, \hat e_j^c) - \epsilon_{ab} \, \mu \, \hat H_{d}^{a} \, \hat H_{u}^{b},
	\label{WMSSM}
\end{equation}
donde $a, b$ son índices $SU(2)_{L}$, $\epsilon_{12}$=1, $i, j$ representan los índices de familia, y se suprimieron los índices de color $SU(3)_{C}$. 
Los objetos $\hat H_{u}$, $\hat H_{d}$, $\hat Q$, $\hat L$, $\hat u$, $\hat d$, $\hat e$ son los campos correspondientes a los supermultipletes quirales que se muestran en la Tabla~\ref{tablachiral}. También presentamos en la Tabla~\ref{tablagauge} los supermultipletes vectoriales. Siguiendo la notación descrita en la sección anterior, al superpotencial mostrado en Ec.~(\ref{WMSSM}) podemos pensarlo como una función de campos escalares. Como hemos visto, a partir del superpotencial podemos hallar el Lagrangiano del modelo que es una función tanto de los campos bosónicos como de los campos fermiónicos.

Como mencionamos en la Sección~\ref{intchiral} podemos escribir el superpotencial y el Lagrangiano en el lenguaje de supercampos exponiendo la invariancia ante supersimetría. Pero al introducir a continuación los términos de ruptura \textit{soft} en el Lagrangiano, trabajaremos en el régimen donde la supersimetría no es una simetría exacta. En consecuencia, no emplearemos el lenguaje de supercampos mostrando explícitamente la distinción entre campos bosónicos y fermiónicos, útil por ejemplo al tratar el sector neutralino-neutrino del $\mu \nu$SSM, más adelante.

\begin{table}
\begin{center}
\begin{tabular}{| c | c | c | c | c | }
     \hline
     Nombre             &       & spin 0                                & spin 1/2                                         & $SU(3)_{C}, SU(2)_{L}, U(1)_{Y}$ \\ \hline
     squarks, quarks    & $\hat Q$     & ($\widetilde{u}_{L}$, $\widetilde{d}_{L}$) & ($u_{L}$, $d_{L}$)                                    & (3, 2, $\frac{1}{6}$)              \\ 
      (x3 familias)      & $\hat u$     & $\widetilde{u}_{R}^{*}$                & $u_{R}^{\dag}$                                     & ($\overline{3}$, 1, $-\frac{2}{3}$)  \\ 
		                   & $\hat d$     & $\widetilde{d}_{R}^{*}$                 & $d_{R}^{\dag}$                                     & ($\overline{3}$, 1, $\frac{1}{3}$)   \\ \hline
		sleptones, leptones & $\hat L$     & ($\widetilde{\nu}$, $\widetilde{e}_{L}$)   & ($\nu, e_{L}$)                                      & (1, 2, $-\frac{1}{2}$)             \\ 
		 (x3 familias)       & $\hat e$     & $\widetilde{e}_{R}^{*}$                 & $e_{R}^{\dag}$                                     & (1, 1, 1)                         \\ \hline
		 Higgs, Higgsinos   & $\hat H_{u}$ & ($H_{u}^{+}$, $H_{u}^{0}$)                 & ($\widetilde{H}_{u}^{\dag}$, $\widetilde{H}_{u}^{0}$) & (1, 2, $+\frac{1}{2}$)              \\
                        & $\hat H_{d}$ & ($H_{d}^{0}$, $H_{d}^{-}$)                 & ($\widetilde{H}_{d}^{0}$, $\widetilde{H}_{d}^{-}$)    & (1, 2, $-\frac{1}{2}$)               \\
     \hline
\end{tabular}
\caption{Supermultipletes quirales del MSSM. Tomamos $Q_{\text{EM}}$=$T_{3}$ + Y}
\label{tablachiral}
\end{center}
\end{table}

\begin{table}
\begin{center}
   \begin{tabular}{| c | c | c | c |}
     \hline
     Nombre             & spin 1/2           & spin 1      & $SU(3)_{C}, SU(2)_{L}, U(1)_{Y}$ \\ \hline
     gluinos, gluones   & $\widetilde{g}$    &  g          & (8, 1, 0)             \\ 
     Winos, bosones W   & $\widetilde{W}^{\pm}$ $\widetilde{W}^{0}$     & $W^{\pm}$ $W^{0}$        &  (1, 3, 0)  \\ 
		 Bino, bosón B      & $\widetilde{B}^{0}$ & $B^{0}$    & (1, 1, 0)   \\
     \hline
   \end{tabular}
	\caption{Supermultipletes vectoriales del MSSM.}
\label{tablagauge} 
\end{center}
\end{table}

Los acoples de Yukawa sin dimensión, $Y^{u}$, $Y^{d}$, $Y^{e}$ son matrices 3x3 en el espacio de las familias. Como en el SM los quarks top y bottom, y el leptón tau son los fermiones más pesados, se considera usualmente la aproximación que solo las componentes de familia (3,3) de cada acople de Yukawa son importantes. En este límite solo las terceras familias, y los campos de Higgs, contribuyen al superpotencial:
\begin{equation}
 W_{\text{MSSM}}\approx y_{t}(\overline{t}tH_{u}^{0}-\overline{t}bH_{u}^{+})-y_{b}(\overline{b}tH_{d}^{-}-\overline{b}bH_{d}^{0})-y_{\tau}(\overline{\tau}\nu_{\tau}H_{d}^{-}-\overline{\tau}\tau H_{d}^{0})+\mu(H_{u}^{+}H_{d}^{-}-H_{u}^{0}H_{d}^{0}).
\label{wmssmaprox}
\end{equation}

Por otra parte, los términos de ruptura \textit{soft} de supersimetría son:
\begin{align}
	&\textsl{L}_{\text{soft}}^{\text{MSSM}}=-\frac{1}{2}(M_{3} \, \widetilde{g} \, \widetilde{g}+M_{2} \, \widetilde{W} \, \widetilde{W}+M_{1} \, \widetilde{B} \, \widetilde{B}+c.c) \cr
	& - \epsilon_{ab}(T^{u}_{ij} \, H_{u}^{b} \, \widetilde{Q}^{a}_i \, \widetilde{u}_j^* + T^{d}_{ij} \, H_{d}^{a} \, \widetilde{Q}^{b}_i \, \widetilde{d}_j^* + T^{e}_{ij} \, H_{d}^{a} \, \widetilde{L}^{b}_i \, \widetilde{e}_j^* + c.c) \cr
	& - (m_{\widetilde{Q}}^{2})_{ij} \, \widetilde{Q}_i^{a*} \, \widetilde{Q}_j^a - (m_{\widetilde{u}}^{2})_{ij} \, \widetilde{u}_i^* \, \widetilde{u}_j - (m_{\widetilde{d}}^{2})_{ij} \, \widetilde{d}_i^* \, \widetilde{d}_j \cr
	& - (m_{\widetilde{L}}^{2})_{ij} \, \widetilde{L}_i^{a*} \, \widetilde{L}_j^a - (m_{\widetilde{e}}^{2})_{ij} \, \widetilde{e}_i^* \, \widetilde{e}_j \cr
	& - m_{H_{u}}^{2} \, H_{u}^{a*} \, H_{u}^a - m_{H_{d}}^{2} \, H_{d}^{a*} \, H_{d}^a - \epsilon_{ab}(b \, H_{d}^{a} \, H_{u}^{b}+c.c.),
\label{MSSMsoft}
\end{align}
donde $M_{3}$, $M_{2}$ y $M_{1}$ son los términos de masa del gluino, del Wino y del Bino respectivamente, y $T_{ij}$ son los parámetros escalares trilineales correspondientes a los acoples trilineales del superpotencial. Notamos que $T_{ij}=A_{ij}Y_{ij}$, sin tener en cuenta la convención de suma de índices repetidos. Dichos términos $(\textit{\text{escalar}})^{3}$ contienen los acoples del tipo $a^{ijk}$ de Ec.~(\ref{Lsoft}), para los cuales vamos a tener en cuenta la aproximación que solo son importantes para las terceras familias, al hacerlos proporcionales a los acoples de Yukawa del superpotencial, por ejemplo $T_{ij}^u=A_{ij}^uY_{ij}^u=A^tY^t$. Los términos de masa al cuadrado para squarks, sleptones y Higgs corresponden a términos del tipo $(m^{2})_{i}^{j}$ de Ec.~(\ref{Lsoft}); donde vamos a considerar que las matrices de masas al cuadrado son invisibles de sabor, es decir, que son proporcionales a la matriz identidad. Por último el término con $b$ corresponde al término del tipo $b^{ij}$ de Ec.~(\ref{Lsoft}).

Todas las aproximaciones mencionadas junto con la asunción de que los parámetros de ruptura \textit{soft} no introducen nuevas fases complejas, evitan cambios de sabor y procesos que violen CP restringidos por los experimentos. Estas tres condiciones son una versión débil de la hipótesis llamada "`universalidad de la ruptura de supersimetría"'.

\subsubsection{Paridad-R}

El superpotencial MSSM de Ec.~(\ref{WMSSM}) es mínimo en el sentido que es suficiente para generar un modelo fenomenológicamente viable. No obstante existen otros términos invariantes de gauge y analíticos que no fueron incluidos porque violan número bariónico (B) o número leptónico (L). La posible presencia de términos de este estilo son preocupantes ya que procesos que violen B o L no han sido observados experimentalmente. Una fuerte cota experimental proviene del no decaimiento del protón. La presencia simultánea de términos que violen B y L podrían dar como resultado que la vida media del protón fuese extremadamente corta.

El Lagrangiano resultante del superpotencial mostrado en Ec.~(\ref{WMSSM}) es simétrico ante multiplicar cada partícula por:
\begin{equation}
	P_{R}=(-1)^{3(B-L)+2s},
\end{equation}
donde $s$ es el espín.
Todas las partículas del SM poseen paridad-R par ($P_{R}$=+1) mientras que todas sus supercompañeras (squarks, sleptones, gauginos y higgsinos) tienen paridad-R impar ($P_{R}$=-1).

El principio de simetría indica que cada término candidato en el Lagrangiano (o superpotencial) solo es permitido si el producto $P_{R}$ de todos sus campos es +1.

Esto implica que cada vértice de la teoría contiene un número par de partículas supersimétricas ($P_{R}$=-1). Entonces:
\begin{itemize}
	\item La partícula supersimétrica más liviana ($P_{R}$=-1), o LSP por sus siglas en inglés, es estable.
	\item Toda partícula supersimétrica debe eventualmente decaer a la LSP.
	\item En experimentos de colisionadores, las partículas supersimétricas solo pueden ser producidas en números pares.
\end{itemize}

\subsubsection{Ruptura espontánea de simetría electrodébil}

En adición a los términos de $\textsl{L}_{\text{soft}}^{\text{MSSM}}$, la parte del potencial escalar que involucra a los campos de Higgs recibe las contribuciones de los términos D y F. Obtenemos:
\begin{align}
	&V=(|\mu|^{2}+m_{H_{u}}^{2})(|H_{u}^{0}|^{2}+|H_{u}^{+}|^{2}) + (|\mu|^{2}+m^{2}_{H_{d}})(|H_{d}^{0}|^{2}+|H_{d}^{-}|^{2}) \cr
	& + [b(H_{u}^{+}H_{d}^{-}-H_{u}^{0}H_{d}^{0}) + c.c.] \cr
	& + \frac{1}{8}(g^{2}+g'^{2})(|H_{u}^{0}|^{2}+|H_{u}^{+}|^{2}-|H_{d}^{0}|^{2}-|H_{d}^{-}|^{2})^{2} \cr
	& + \frac{1}{2}g^{2}|H_{u}^{+}H_{d}^{0*}+H_{u}^{0}H_{d}^{-*}|^{2}.
\label{Vmssm}
\end{align}

Pedimos que el mínimo de potencial rompa la simetría electrodébil, en concordancia con los experimentos. La parte relevante para la ruptura electrodébil del potencial escalar del campo de Higgs neutral es:
\begin{equation}
	V=(|\mu|^{2}+m_{H_{u}}^{2})|H_{u}^{0}|^{2} + (|\mu|^{2}+m^{2}_{H_{d}})|H_{d}^{0}|^{2} - (bH_{u}^{0}H_{d}^{0} + c.c.) + \frac{1}{8}(g^{2}+g'^{2})(|H_{u}^{0}|^{2}-|H_{d}^{0}|^{2})^{2}.
\end{equation}

Entonces los campos $H_{u}^{0}$ y $H_{d}^{0}$ tienen un valor de expectación de vacío diferente de cero; llamamos:
\begin{equation}
	v_{u}=\left\langle H_{u}^{0}\right\rangle , \hspace{1cm} v_{d}=\left\langle H_{d}^{0}\right\rangle,
\end{equation}
y definimos la razón
\begin{equation}
	\tan\beta\equiv\frac{v_{u}}{v_{d}}.
\end{equation}
Los valores de expectación están relacionados con el valor de expectación del Higgs del modelo estándar, $v$, y a su vez con la masa del bosón $Z^{0}$ y los acoples electrodébiles por:
\begin{equation}
	v_{u}^{2}+v_{u}^{2}=v^{2}=2\frac{m_{Z}^{2}}{(g^{2} + g'^{2})}\approx (\text{174 GeV})^{2},
\end{equation}
\begin{equation}
	v_u = v \, \sin \beta \hspace{1cm} v_d= v \, \cos \beta.
\end{equation}

Luego de la ruptura de simetría electrodébil, los autoestados de masa eléctricamente neutros, el fotón $A_{\mu}$, y el bosón $Z$, se definen como
\begin{equation}
	\left( {\begin{array}{c}
   A_{\mu}  \\
   Z_{\mu}  \\
  \end{array} } \right) =
  \left( {\begin{array}{cc}
   \cos \theta_W & \sin \theta_W  \\
   -\sin \theta_W & \cos \theta_W  \\
  \end{array} } \right) \left( {\begin{array}{c}
   B_{\mu}  \\
   W_{\mu}^3  \\
  \end{array} } \right),
\end{equation}
y los bosones eléctricamente cargados $W^{\pm}$ como
\begin{equation}
    W_{\mu}^{\pm} \, = \, \frac{1}{\sqrt{2}} \left( W_{\mu}^1 \mp i W_{\mu}^2 \right).
\end{equation}
 El ángulo de mezcla débil o de Weinberg, $\theta_W$, está definido como
\begin{equation}
    \sin \theta_W \, = \, \frac{g'}{\sqrt{g^2 + g'^2}}, \hspace{0.5cm} \text{y} \hspace{0.5cm} \cos \theta_W \, = \, \frac{g}{\sqrt{g^2 + g'^2}},
\end{equation}
donde $g$ y $g'$ son los acoples de gauge de $SU(2)_L$ y $U(1)_Y$, respectivamente. El acople de gauge de $U(1)_{em}$ entonces es
\begin{equation}
    e\, = \, \sqrt{4 \pi \alpha} \, = \, g \sin \theta_W \, = \, g' \cos \theta_W.
\end{equation}
Mientras que el fotón resulta no masivo, el resto de los bosones de gauge electrodébiles adquieren masa a través del mecanismo de Higgs absorbiendo tres grados de libertad de los dos dobletes de Higgs: los bosones de Goldstone $G$ y $G^+$. Obtenemos de esta manera
\begin{equation}
    m_W \, = \, \frac{g \, v}{\sqrt{2}}, \hspace{0.5cm} \text{y} \hspace{0.5cm} m_Z \, = \, \frac{g \, v}{\sqrt{2} \cos \theta_W}.
\end{equation}

Los autoestados de gauge de los bosones de Higgs pueden ser expresados en función de sus autoestados de masa. Los autoestados neutros como
\begin{equation}
	\left( {\begin{array}{c}
   H_u^0  \\
   H_d^0  \\
  \end{array} } \right) =
  \left( {\begin{array}{c}
   v_u  \\
   v_d  \\
  \end{array} } \right) + \frac{1}{\sqrt{2}}
  \left( {\begin{array}{cc}
   \cos \alpha & \sin \alpha  \\
   -\sin \alpha & \cos \alpha  \\
  \end{array} } \right) 
  \left( {\begin{array}{c}
   h  \\
   H  \\
  \end{array} } \right) + \frac{i}{\sqrt{2}}
  \left( {\begin{array}{cc}
   \cos \beta_0 & \sin \beta_0  \\
   -\sin \beta_0 & \cos \beta_0  \\
  \end{array} } \right) 
  \left( {\begin{array}{c}
   G  \\
   A  \\
  \end{array} } \right),
\end{equation}
mientras los autoestados cargados
\begin{equation}
	\left( {\begin{array}{c}
   H_u^+  \\
   H_d^{-*}  \\
  \end{array} } \right) =
  \left( {\begin{array}{cc}
   \cos \beta_{\pm} & \sin \beta_{\pm}  \\
   -\sin \beta_{\pm} & \cos \beta_{\pm}  \\
  \end{array} } \right) 
  \left( {\begin{array}{c}
   G^+  \\
   H^+  \\
  \end{array} } \right).
\end{equation}
Teniendo en cuenta solamente efectos a orden árbol $\beta_0= \beta_{\pm} = \beta$, y
\begin{align}
    m^2_A \, &= \, 2 |\mu|^2 + m_{H_u}^2 + m_{H_d}^2, \cr
    m^2_{h,H} \, &= \, \frac{1}{2} \left( m_A^2 + m_Z^2 \mp \sqrt{(m_A^2-m_Z^2)^2 + 4 m_Z^2 m_A^2 \sin^2 2 \beta} \right), \cr
    m^2_{H^{\pm}} \, &= \, m_A^2 + m_W^2.
\label{mhiggssector}
\end{align}
En este caso, el ángulo de mezcla $\alpha$ está determinado por
\begin{equation}
    \frac{\sin 2 \alpha}{\sin 2 \beta} \, = \, - \left( \frac{m_H^2 + m_h^2}{m_H^2 - m_h^2} \right) \hspace{0.5cm} \text{y} \hspace{0.5cm}  \frac{\tan 2 \alpha}{\tan 2 \beta} \, = \, - \left( \frac{m_A^2 + m_Z^2}{m_A^2 - m_Z^2} \right).
\end{equation}
En el límite $|\mu|$ grande, o de desacople, tenemos que $m_A \gg m_Z$. Los estados $H$, $A$, y $H^{\pm}$ resultan muy pesados, mientras que $h$ es del orden de la escala electrodébil. Bajo esta condición el ángulo de mezcla es $\alpha \simeq \beta - \pi/2$, y el bosón más ligero $h$ tiene los acoples del bosón de Higgs del SM.

\subsection{Espectro de masas} \label{MSSMespectro}
Los Higgsinos y los gauginos electrodébiles se mezclan por efectos de la ruptura de simetría electrodébil [ver Ec.~(\ref{L}) cuando los bosones de Higgs adquieren VEVs]. Los Higgsinos neutros ($\widetilde{H}_{u}^{0}$ y $\widetilde{H}_{d}^{0}$) y los gauginos neutros ($\widetilde{B}$ y $\widetilde{W}^{0}$) se combinan para formar cuatro autoestados de masa llamados $\textit{neutralinos}$, denotados por $\widetilde{N}_{i}$ (i=1,2,3,4). Los Higgsinos cargados ($\widetilde{H}_{u}^{+}$ y $\widetilde{H}_{d}^{-}$) y los Winos ($\widetilde{W}^{+}$ y $\widetilde{W}^{-}$) se mezclan para formar dos autoestados de masa, con carga $\pm$1, llamados $\textit{charginos}$, denotados por $\widetilde{C}^{\pm}_{i}$ (i=1,2). Por convención, sus índices se encuentran en orden ascendente, siguiendo a la masa de las partículas. Usualmente se asume al neutralino más liviano, $\widetilde{N}_{1}$, como la partícula supersimétrica más liviana, ya que es la única partícula del MSSM que no está seriamente restringida experimentalmente en el modelo para ser candidato a materia oscura fría.

En la base de autoestados de gauge $\psi^{0}=(\widetilde{B},\widetilde{W}^{0},\widetilde{H}_{d}^{0},\widetilde{H}_{u}^{0})$, la parte del Lagrangiano para la masa de los neutralinos es:
\begin{align}
	\textsl{L}_{\text{masa-neutralinos}}=-\frac{1}{2}(\psi^{0})^{T}M_{\widetilde{N}}\psi^{0} + c.c.,
\end{align}
donde
\begin{equation}
	M_{\widetilde{N}}=
  \left( {\begin{array}{cccc}
   M_{1} & 0 & -\cos\beta \sin\theta_{w} m_{Z} & \sin\beta \sin\theta_{w} m_{Z} \\
   0 & M_{2} & \cos\beta \cos\theta_{w} m_{Z} & -\sin\beta \cos\theta_{w} m_{Z} \\
	 -\cos\beta \sin\theta_{w} m_{Z} & \cos\beta \cos\theta_{w} m_{Z} & 0 & -\mu \\
	 \sin\beta \sin\theta_{w} m_{Z} & -\sin\beta \cos\theta_{w} m_{Z} & -\mu & 0 \\
  \end{array} } \right).
\end{equation}
$M_{\widetilde{N}}$ puede ser diagonalizada por una matriz unitaria para obtener los autoestados de masa.

El espectro de los charginos puede ser analizado de una manera similar. En base de autoestados de gauge $\psi^{\pm}=(\widetilde{W^{+}},\widetilde{H}_{u}^{+},\widetilde{W}^{-},\widetilde{H}_{d}^{-})$, la parte del Lagrangiano para la masa de los charginos es:
\begin{align}
	\textsl{L}_{\text{masa-charginos}}=-\frac{1}{2}(\psi^{\pm})^{T}M_{\widetilde{C}}\psi^{\pm} + c.c.,
\end{align}
donde, en bloques de 2 x 2,
\begin{equation}
	M_{\widetilde{C}}=
  \left( {\begin{array}{cc}
   0 & X^{T}  \\
   X & 0  \\
  \end{array} } \right),
\end{equation}
con
\begin{equation}
	X=
  \left( {\begin{array}{cc}
   M_{2} & \sqrt{2} \sin\beta m_{W} \\
   \sqrt{2} \cos\beta m_{W} & \mu  \\
  \end{array} } \right).
\end{equation}

En cuanto a los campos escalares, todos los que posean la misma carga eléctrica, paridad-R y número de color, se pueden mezclar entre ellos. Por lo tanto, los autoestados de masa se obtienen al diagonalizar tres matrices de masas 6 x 6, para squarks tipo up, squarks tipo down y leptones cargados, y una matriz 3 x 3 para sneutrinos.
Sin embargo, si consideramos las hipótesis de universalidad, la mayoría de los ángulos de mezcla serán muy chicos. Solo las terceras familias tienen masas muy diferentes comparadas con las dos primeras familias debido a los acoples de Yukawa ($y_{t}$, $y_{b}$, $y_{\tau}$) y los acoples \textit{soft} ($a_{t}$, $a_{b}$, $a_{\tau}$), y sus ángulos de mezcla pueden ser significativos.
Entonces consideraremos como buena aproximación matrices de masas de 2 x 2 involucrando las partes derechas e izquierdas de cada una de las terceras familias.
En la base de autoestados de gauge $(\widetilde{t}_{L},\widetilde{t}_{R})$, la parte del Lagrangiano para la masa de los stop es:
\begin{align}
	\textsl{L}_{\text{masa-stop}}=-(\widetilde{t}_{L}^{*},\widetilde{t}_{R}^{*}) m_{\widetilde{t}}^{2} (\widetilde{t}_{L},\widetilde{t}_{R})^{T},
\end{align}
donde
\begin{equation}
	m_{\widetilde{t}}^{2}=
  \left( {\begin{array}{cc}
   m^{2}_{\widetilde{Q}} + m^{2}_{t} + \frac{1}{6}(4m^{2}_{W} - m^{2}_{Z}) \cos2\beta & m_{t}(A^{t} - \mu\cot\beta)  \\
   m_{t}(A^{t} - \mu\cot\beta) & m^{2}_{\widetilde{u}} + m^{2}_{t} - \frac{2}{3}(m^{2}_{W} - m^{2}_{Z}) \cos2\beta  \\
  \end{array} } \right).
\end{equation}

De manera similar, para los squarks bottom y los sleptones stau tenemos:
\begin{equation}
	m_{\widetilde{b}}^{2}=
  \left( {\begin{array}{cc}
   m^{2}_{\widetilde{Q}} + m^{2}_{b} + \frac{1}{6}(4m^{2}_{W} - m^{2}_{Z}) \cos2\beta & m_{b}(A^{b} - \mu\cot\beta)  \\
   m_{b}(A^{b} - \mu\cot\beta) & m^{2}_{\widetilde{d}} + m^{2}_{b} - \frac{2}{3}(m^{2}_{W} - m^{2}_{Z}) \cos2\beta  \\
  \end{array} } \right),
\end{equation}
\begin{equation}
	m_{\widetilde{\tau}}^{2}=
  \left( {\begin{array}{cc}
   m^{2}_{\widetilde{L}} + m^{2}_{\tau} - \frac{1}{2}(2m^{2}_{W} - m^{2}_{Z}) \cos2\beta & m_{\tau}(A^{\tau} - \mu\tan\beta)  \\
   m_{\tau}(A^{\tau} - \mu\tan\beta) & m^{2}_{\widetilde{e}} + m^{2}_{\tau} + (m^{2}_{W} - m^{2}_{Z}) \cos2\beta  \\
  \end{array} } \right).
\end{equation}

\subsection{Problemas del MSSM}

El modelo MSSM posee dos problemas principales que discutiremos aquí. El primero consiste en un problema de naturalidad, el denominado "`problema-$\mu$"', proveniente del término de masa para el campo de Higgs en el superpotencial, $\mu H_{u} H_{d}$ en Ec.~(\ref{WMSSM}). 

Dicho término es la versión supersimétrica de la masa del bosón de Higgs en el SM y es único ya que términos del tipo $H_{u}^{*}H_{u}$ o $H_{d}^{*}H_{d}$ se encuentran prohibidos en el superpotencial, el cual debe ser analítico en los supermultipletes quirales (o de forma equivalente, en los campos escalares) como se mostró en la Sección~\ref{intchiral}. De manera análoga podemos ver que necesitamos tanto a $H_{u}$ como a $H_{d}$, pues el término $\overline{u}QH_{u}$ con acople de Yukawa en Ec.~(\ref{WMSSM}) no puede ser reemplazado por algo que involucre $\overline{u}QH_{d}^{*}$ sin violar la condición que el superpotencial sea analítico. De manera similar no es posible reemplazar $\overline{d}QH_{d}$ y $\overline{e}LH_{d}$ por $\overline{d}QH_{u}^{*}$ y $\overline{e}LH_{u}^{*}$. Por lo tanto es preciso contar con dos (en el modelo mínimo) supermultipletes de Higgs para obtener una fenomenología viable.

Retomando el primero de los problemas del MSSM, como hemos visto el parámetro $\mu$ está involucrado en los términos de masa de los higgsinos [último término de Ec.~(\ref{wmssmaprox})] y en la masa de los Higgs [dos primeros términos de Ec.~(\ref{Vmssm})].

El término $\mu$ está presente en el potencial escalar neutral que debemos minimizar para romper la simetría electrodébil, sin embargo requerimos que $\mu$ sea del orden de la ruptura electrodébil, para permitir que el Higgs tenga VEV del orden $10^{2}$ GeV sin una cancelación fina entre $\left|\mu\right|^{2}$ y las masas $m_{H_{u}}^{2}$ y $m_{H_{d}}^{2}$. Entonces surge la pregunta de naturalidad: por qué $\left|\mu\right|^{2}$ debe ser distinto de cero o demasiado pequeño comparado con, por ejemplo, la masa de Planck al cuadrado. Tal vez aún más intrigante resulte que $\left|\mu\right|^{2}$ deba ser del mismo orden de $m_{soft}^{2}$, es decir, de los parámetros con dimensión involucrados en los términos de ruptura \textit{soft} de supersimetría, Ec.~(\ref{MSSMsoft}). El potencial escalar del MSSM, Ec.~(\ref{Vmssm}), depende de dos tipos de parámetros con dimensión que son conceptualmente distintos: la masa $\mu$ proveniente de un origen supersimétrico, y los términos de masa \textit{soft} provenientes del mecanismo de ruptura de supersimetría. Aún así, los experimentos sugieren una coincidencia entre escalas de masa que en principio no tienen relación.

El segundo problema del MSSM, compartido con el SM, es que los neutrinos tienen masa cero, sin embargo se ha demostrado experimentalmente lo contrario. Esto se manifiesta en el superpotencial, Ec.~(\ref{WMSSM}), al observar que no existe un término del tipo $Y_{\nu}H_{u}L\nu^{c}$.

\section{\texorpdfstring{$\mu$-from-$\nu$}{mu-from-nu} Supersymmetric Standard Model (\texorpdfstring{$\mu\nu$}{munu}SSM)}
\label{uvMSSM}

Al verificarse experimentalmente en los últimos años que los neutrinos poseen masa, debemos incluir dicho resultado en nuestro modelo si pretendemos reproducir todos los fenómenos físicos. En este sentido, el modelo denominado $\mu$-from-$\nu$ Supersimétrico ($\mu \nu$SSM) cumple este requisito, al incluir el campo de neutrinos derechos, y por lo tanto, un término en el superpotencial de la forma $Y_{\nu}H_{u}L\nu^{c}$.

Si además permitimos que los compañeros supersimétricos de los neutrinos derechos adquieran valores de expectación de vacío, VEVs, un término del tipo  $\lambda \nu^{c} H_{d}H_{u}$ genera un término con $\mu_{\text{eff}}$, resolviendo de forma dinámica el problema-$\mu$ de naturalidad.

Aunque no es la única propuesta viable para resolver los problemas del MSSM, la ventaja de $\mu \nu$SSM reside en brindar solución simultáneamente tanto a la física de neutrinos como al problema-$\mu$, utilizando solo supermultipletes derechos de neutrinos. Relacionando de esta forma la ruptura de paridad-R con la física de neutrinos, y brindando una rica fenomenología accesible en un futuro próximo.

\subsection{Lagrangiano y sus simetrías}
El superpotencial para el $\mu \nu$SSM es:
\begin{equation}
\begin{split}
	W_{\mu \nu\text{SSM}} =& \epsilon_{ab}(Y_{ij}^u \, \hat H_{u}^{b} \, \hat Q^{a}_i \, \hat u_j^c + Y_{ij}^d \, \hat H_{d}^a \, \hat Q_{i}^b \, \hat d_j^c + Y_{ij}^e \, \hat H_{d}^{a} \, \hat L_i^{b} \, \hat e_j^c + Y_{ij}^{\nu} \, \hat H_{u}^{b} \, \hat L^{a}_i \, \hat \nu^{c}_j)\\
	& - \epsilon_{ab}\lambda_i \, \hat \nu^{c}_i \, \hat H_{d}^{a} \, \hat H_{u}^{b} + \frac{1}{3}\kappa_{ijk} \, \hat \nu^{c}_i \, \hat \nu^{c}_j \, \hat \nu^{c}_k,
\end{split}
	\label{WNMSSM}
\end{equation}
donde hemos explicitado en $\nu^{c}_i$ que corresponde a los neutrinos derechos, $i,j,k$ van a de 1 a 3 corresponden a los índices de familias, $a,b=1,2$ son los índices de la representación $SU(2)$ y $\epsilon_{ab}$ es el tensor antisimétrico tal que $\epsilon_{12}=1$. Los índices de color no han sido escritos.

Como hemos mencionado solo acoples trilineales sin dimensión están presentes, $Y$ es la matriz de Yukawa, $\lambda$ un vector y $\kappa$ un tensor totalmente antisimétrico.
Remarcamos las características importantes del modelo:
\begin{itemize}
	\item Se encuentran presentes los acoples de Yukawa usuales para los quarks y leptones cargados, y también un término de Yukawa para los neutrinos $Y_{\nu}H_{u}L\nu^{c}$.
	\item Además del término de masa de Dirac antes mencionado,  $Y_{\nu}H_{u}L\nu^{c}$, existe un término de masa de Majorana $\frac{1}{3}\kappa_{ijk} \, \hat \nu^{c}_i \, \hat \nu^{c}_j \, \hat \nu^{c}_k$ para los neutrinos.
	\item Los términos del tipo $\lambda \nu^{c} H_{d}H_{u}$ producen un término con $\mu$ efectivo por los valores de expectación de vacío de los sneutrinos derechos.
	\item Hay ruptura explicita de paridad-R por los últimos dos términos de Ec.~(\ref{WNMSSM}), los cuales violan número leptónico.
	\item La ruptura de paridad-R produce mezcla de neutralinos con neutrinos derechos, dando a orden árbol, tres valores de masa ligeros de los neutrinos.
	\item En el límite donde los Yukawas para los neutrinos se anulan, los $\nu^{c}$ son singletes sin conexión con los neutrinos, y la paridad-R se conserva.
\end{itemize}

Notamos que si los acoples de Yukawa de los neutrinos son distintos de cero, como hemos dicho, tenemos ruptura de paridad-R, la cual será chica si los acoples de Yukawa de neutrinos son del orden $\approx$ $10^{-6}$, como el acople de Yukawa del electrón. En el caso de ruptura de paridad-R, podríamos enfrentarnos al problema del decaimiento rápido del protón, sin embargo la elección de paridad-R es $\textit{ad hoc}$. Existen simetrías que solo prohíben operadores que violan número bariónico, evitando que el protón decaiga. El superpotencial mostrado en Ec.~(\ref{WNMSSM}) es el mínimo posible fenomenológicamente viable, solucionando el problema-$\mu$ e incluyendo neutrinos derechos.

Los términos de ruptura \textit{soft} de supersimetría compatibles con (\ref{WNMSSM}) son:
\begin{align}
	&\textsl{L}_{\text{soft}}^{\text{$\mu \nu$SSM}}=-\frac{1}{2}(M_{3} \, \widetilde{g} \, \widetilde{g}+M_{2} \, \widetilde{W} \, \widetilde{W}+M_{1} \, \widetilde{B} \, \widetilde{B}+c.c) \cr
	& - \epsilon_{ab}(T^{u}_{ij} \, H_{u}^{b} \, \widetilde{Q}^{a}_i \, \widetilde{u}_j^* + T^{d}_{ij} \, H_{d}^{a} \, \widetilde{Q}^{b}_i \, \widetilde{d}_j^* + T^{e}_{ij} \, H_{d}^{a} \, \widetilde{L}^{b}_i \, \widetilde{e}_j^* + T^{\nu}_{ij} \, H_{u}^{b} \, \widetilde{L}^{a}_i \, \widetilde{\nu}^{c*}_j + c.c) \cr
	& -\epsilon_{ab}(-T^{\lambda}_i \, \widetilde{\nu}^{c*}_i \, H_{d}^{a} \, H_{u}^{b} + \frac{1}{3} T^{\kappa}_{ijk} \, \widetilde{\nu}^{v*}_i \, \widetilde{\nu}^{v*}_j \, \widetilde{\nu}^{v*}_k + c.c. ) \cr
	& - (m_{\widetilde{Q}}^{2})_{ij} \, \widetilde{Q}_i^{a*} \, \widetilde{Q}_j^a - (m_{\widetilde{u}}^{2})_{ij} \, \widetilde{u}_i^* \, \widetilde{u}_j - (m_{\widetilde{d}}^{2})_{ij} \, \widetilde{d}_i^* \, \widetilde{d}_j \cr
	& - (m_{\widetilde{L}}^{2})_{ij} \, \widetilde{L}_i^{a*} \, \widetilde{L}_j^a - (m_{\widetilde{e}}^{2})_{ij} \, \widetilde{e}_i^* \, \widetilde{e}_j - (m_{\widetilde{\nu}^{c}}^{2})_{ij} \, \widetilde{\nu}^{c*}_i \, \widetilde{\nu}^{c}_j - m_{H_{u}}^{2} \, H_{u}^{a*} \, H_{u}^a - m_{H_{d}}^{2} \, H_{d}^{a*} \, H_{d}^a,
	\label{MUNUMSSMsoft}
\end{align}
donde $m^2_{\widetilde{x}}$ son las matrices de masas \textit{soft} para los escalares y $T^x$ son los parámetros escalares trilineales correspondientes a los acoples trilineales del superpotencial. Notamos que $T_{ij}^x=A_{ij}^xY_{ij}^x$, con $n=u,d,e,\nu$, $T_i^{\lambda}=A_{i}^{\lambda}\lambda_i$, $T_i^{\kappa}=A_{ijk}^{\kappa}\kappa_{ijk}$, sin tener en cuenta la convención de suma de índices repetidos
En la primera línea de Ec.~(\ref{MUNUMSSMsoft}) los campos representan los compañeros fermionicos de los bosones de gauge. Las últimas cuatro líneas denotan las componentes escalares de los supercampos. 

\subsubsection{Ruptura espontánea de simetría electrodébil}

En adición a los términos de $\textsl{L}_{\text{soft}}^{\text{$\mu \nu$SSM}}$, la parte del potencial escalar que involucra a los campos de Higgs recibe las contribuciones de los términos D y F. Pedimos que el mínimo de potencial rompa la simetría electrodébil. La parte relevante a minimizar es la que involucra al campo de Higgs neutral, cuando los escalares adquieren valor de expectación de vacío:
\begin{align}
	&V= \frac{1}{8}(g^{2}+g'^{2}) \, (|v_{\nu}|^{2}+|v_{d}|^{2}-|v_{u}|^{2})^{2} + |\lambda|^{2} \, (|v_{\nu^c}|^{2} \, |v_{d}|^{2} + |v_{\nu^c}|^{2} \, |v_{u}|^{2} + |v_{u}|^{2} \, |v_{d}|^{2}) \cr
	& + |Y^{\nu}|^{2} \, (|v_{\nu^c}|^{2} \, |v_{d}|^{2} + |v_{\nu^c}|^{2} \, |v_{u}|^{2} + |v_{u}|^{2} \, |v_{d}|^{2}) \cr
	& + |\kappa|^{2} \, |v_{\nu^c}|^{4} +  m_{H_{d}}^{2} \, |v_{d}|^{2} + m_{H_{u}}^{2} \,|v_{u}|^{2} + m_{\nu^{c}}^{2} \, |v_{\nu^c}|^{2} + m_{\nu}^{2} \, |v_{\nu}|^{2} \cr
	& + (-\lambda \, \kappa^{*} \, v_{u} \, v_{d} \, v_{\nu^c}^{*2} - \lambda \, Y^{\nu *} \, |v_{\nu^c}|^{2} \, v_{d} \, v_{\nu}^{*} - \lambda \, Y^{\nu *} \, |v_{u}|^{2} \, v_{d} \, v_{\nu}^{*} \cr
	& + \kappa \, Y^{\nu *} \, v_{u}^{*} \, v_{\nu}^{*} \, v_{\nu^2}^{2} - \lambda \, A^{\lambda} \, v_{\nu^c} \, v_{u} \, v_{d} + Y^{\nu} \, A^{\nu} \, v_{\nu^c} \, v_{\nu} \, v_{u} + \frac{1}{3} \, \kappa \, A^{\kappa} \, v_{\nu^c}^{3} + c.c.),
	\label{Velectro}
\end{align}

donde
\begin{equation}
\langle H_d^0 \rangle = v_d, \, \quad \langle H_u^0 \rangle = v_u, \,
\quad \langle \widetilde \nu_i \rangle = v_{\nu_i}, \,  \quad
\langle \widetilde \nu_i^c \rangle = v_{\nu^c_i}\ ,
\label{vevs}
\end{equation}
\\
En el apéndice~\ref{minimizacionapendice} se presentan las ocho condiciones de minimización del potencial escalar neutro, Ec.~(\ref{Velectro}), con respecto al módulo de $v_{d}$, $v_{u}$, $v_{\nu^c_i}$ y $v_{\nu_i}$. Allí podemos ver que tanto $Y^{\nu}_{ij}$ como  $v_{\nu_i}$ deben ser pequeños.

\subsection{Espectro de masas}
Como consecuencia de la ruptura de paridad-R, los neutrinos (izquierdos y derechos) se mezclan entre sí y con los fermiones neutros restantes, los neutralinos. Esto da como resultado un total de diez neutralinos-neutrinos (reservamos el nombre neutralino para la resultante de la mezcla Bino, Wino, Higgsino y neutrinos derechos).

En la base de autoestados de gauge $\psi^{0}=(\widetilde{B},\widetilde{W}^{0},\widetilde{H}_{d}^{0},\widetilde{H}_{u}^{0},\nu^{c},\nu)$, la parte del Lagrangiano para la masa de los neutralinos-neutrinos es:
\begin{align}
	\textsl{L}_{\text{masa-neutralinos-neutrinos}}=-\frac{1}{2}(\psi^{0})^{T}M_{\widetilde{N}}\psi^{0} + c.c.,
\end{align}
para valores de expectación reales tenemos (para el caso complejo ver Ref.~\cite{neutrinocp}):
\begin{equation}
	M_{\widetilde{N}}=
  \left( {\begin{array}{cc}
	M & m \\
	m^{T} & O_{3x3}
	\end{array} } \right),
\label{MuvMSSM}
\end{equation}
con
\begin{equation}
\hspace{-1.5cm}  M=
   \left( {\begin{array}{ccccccc}
   M_{1} & 0 & -A\cos\beta  & A\sin\beta  & 0 & 0 & 0 \\
   0 & M_{2} & B\cos\beta  & -B\sin\beta  & 0 & 0 & 0 \\
	 -A\cos\beta  & B\cos\beta  & 0 & -\lambda_{i}v_{\nu^c_i} & -\lambda_{1} v_{u} & -\lambda_{2} v_{u} & -\lambda_{3} v_{u} \\
	 A\sin\beta  & -B\sin\beta  & -\lambda_{i}v_{\nu^c_i} & 0 & -\lambda_{1} v_{d} + Y^{\nu}_{i1}v_{\nu_i} & -\lambda_{2} v_{d} + Y^{\nu}_{i2}v_{\nu_i} & -\lambda_{3} v_{d} + Y^{\nu}_{i3}v_{\nu_i} \\
	0 & 0 & -\lambda_{1} v_{u} & -\lambda_{1} v_{d} + Y^{\nu}_{i1}v_{\nu_i} & 2\kappa_{11j} v_{\nu^c_j} & 2\kappa_{12j} v_{\nu^c_j} & 2\kappa_{13j} v_{\nu^c_j} \\
	0 & 0 & -\lambda_{2} v_{u} & -\lambda_{2} v_{d} + Y^{\nu}_{i2}v_{\nu_i} & 2\kappa_{21j} v_{\nu^c_j} & 2\kappa_{22j} v_{\nu^c_j} & 2\kappa_{23j} v_{\nu^c_j} \\
	0 & 0 & -\lambda_{3} v_{u} & -\lambda_{3} v_{d} + Y^{\nu}_{i3}v_{\nu_i} & 2\kappa_{31j} v_{\nu^c_j} & 2\kappa_{32j} v_{\nu^c_j} & 2\kappa_{33j} v_{\nu^c_j}
  	\end{array} } \right),
\label{MM}
\end{equation}
A=$\sin\theta_{w} m_{Z}$, B= $\cos\theta_{w} m_{Z}$ y
\begin{equation}
	m^{T}=
  \left( {\begin{array}{ccccccc}
	 -\frac{g}{\sqrt{2}}v_{\nu_1} & \frac{g'}{\sqrt{2}}v_{\nu_1} & 0 & Y^{\nu}_{1i} v_{\nu^c_i} & Y^{\nu}_{11} v_{u} & Y^{\nu}_{12} v_{u} & Y^{\nu}_{13} v_{u} \\
	-\frac{g}{\sqrt{2}}v_{\nu_2} & \frac{g'}{\sqrt{2}}v_{\nu_2} & 0 & Y^{\nu}_{2i} v_{\nu^c_i} & Y^{\nu}_{21} v_{u} & Y^{\nu}_{22} v_{u} & Y^{\nu}_{23} v_{u} \\
	 -\frac{g}{\sqrt{2}}v_{\nu_3} & \frac{g'}{\sqrt{2}}v_{\nu_3} & 0 & Y^{\nu}_{3i} v_{\nu^c_i} & Y^{\nu}_{31} v_{u} & Y^{\nu}_{32} v_{u} & Y^{\nu}_{33} v_{u}
	\end{array} } \right).
	\label{mm}
\end{equation}

Por simplicidad se utilizó la convención de suma sobre los índices repetidos. Notamos que el bloque 4 x 4 superior izquierdo de la matriz $\textit{M}$, Ec.~(\ref{MM}), es la matriz de masas del sector neutralino en el modelo MSSM.

La matriz (\ref{MuvMSSM}) es del tipo \textit{seesaw} y da origen a masas muy chicas para los neutrinos. Este es el caso, pues las entradas de la matriz $\textit{M}$ son del orden de la escala electrodébil, mientras que las entradas de la matriz $\textit{m}$ son del orden de la masa de Dirac ($10^{-4}$ GeV). Entonces, para obtener la masa física de los estados neutralinos-neutrinos, es necesario diagonalizar la matriz $M_{\widetilde{N}}$ de 10 x 10, Ec.~(\ref{MuvMSSM}). Siete autoestados serán pesados, es decir, del orden de la escala electrodébil. Los tres restantes autoestados serán livianos, del orden de la masa de los neutrinos. Cuatro de los estados más pesados serán similares a los neutralinos del modelo MSSM.

Si estamos interesados en el sector de los neutrinos, ya que la matriz $M_{\widetilde{N}}$ es del tipo \textit{seesaw}, en primera aproximación la matriz de masa efectiva de mezcla de neutrinos esta dada por:
\begin{equation}
	m_{eff}=-m^{T}M^{-1}m.
	\label{mmeff}
\end{equation}

Esta es una ecuación muy útil que permite conocer masas y mezclas del sector en buena aproximación. Debemos notar que la matriz (\ref{MuvMSSM}) es una matriz mal condicionada, su diagonalización debe realizarse con cuidado si requerimos obtener una buena aproximación de los autovalores y autovectores ligeros.

\subsection{Supergravedad}

Si supersimetría es una simetría local, es decir si el parámetro en las transformaciones supersimétricas depende de las coordenadas, la teoría debe incluir gravedad. Esto se debe a que para lograr invariancia local ante transformaciones supersimétricas es necesario incluir un nuevo supermultiplete a la teoría, el cual incluye el gravitón de spin 2 y el gravitino de spin $3/2$, como se muestra en la Tabla~\ref{tablagrav}. El resultado es una teoría llamada supergravedad (SUGRA)~\cite{Deser:1976}. Una introducción a supergravedad puede encontrarse en Ref.~\cite{munozSUGRA} y sus referencias.

\begin{table}
\begin{center}
\begin{tabular}{| c | c | c | c | }
     \hline
     Nombre           & spin 2                                & spin 3/2                                         & $SU(3)_{C}, SU(2)_{L}, U(1)_{Y}$ \\ \hline
     Gravitón, gravitino     & $g_{\mu\nu}$ &   $\psi_{\mu}$              & (1, 1, 0)              \\ 
     \hline
\end{tabular}
\caption{Supermultiplete de gravedad en teorías con supersimetría local.}
\label{tablagrav}
\end{center}
\end{table}

Las teorías de supergravedad están descritas en general por tres funciones de los campos escalares de la teoría: el superpotencial $W(\psi)$, el potencial de K$\ddot{\text{a}}$hler $K(\psi,\psi^*)$, y la función cinética de gauge $f_{ab}(\psi)$. SUGRA resulta en una teoría no renormalizable, y por lo tanto el Lagrangiano completo contiene términos de interacción con acoples de dimensión de masa negativa. Se asume entonces que supergravedad es la aproximación de bajas energías de una teoría más general como por ejemplo teoría de cuerdas, del mismo modo que la teoría de interacciones de Fermi es una aproximación a las interacciones débiles a bajas energías. El Lagrangiano más general y completo puede ser visto en el libro de Wess y Bagger~\cite{Wess:1992}, o para el caso más conveniente del punto de vista fenomenológico en cuatro dimensiones y supergravedad $N=1$ en Ref.~\cite{Moroi:1995}.

\subsubsection{Ruptura de supersimetría y el super mecanismo de Higgs}\label{rupturaSUSY}

Como hemos mencionado, supersimetría debe ser una simetría rota para poder explicar la diferencia de masas entre las partículas de un mismo supermultiplete. Generalmente, se asume la existencia de un sector oculto donde la supersimetría se rompe, y luego este efecto es mediado al sector visible a través de interacciones que se encuentran suprimidas por alguna escala.

De manera análoga al mecanismo de Higgs en la ruptura de simetría electrodébil, en supergravedad existe un super mecanismo de Higgs de ruptura de supersimetría~\cite{Deser:1977}. Como los generadores de supersimetría son fermiónicos, la ruptura de supersimetría genera un bosón de goldstone fermiónico, el goldstino. Sin embargo, debido a términos de mezcla de masa entre el goldstino y el gravitino, el primero es absorbido por el gravitino, el cual se vuelve masivo. Entonces, la masa del gravitino está dada por
\begin{equation}
    m_{3/2} \, \sim \, \frac{\langle F \rangle}{M_{Pl}},
\end{equation}
donde $\langle F \rangle$ es el valor de expectación de vacío de un campo auxiliar del sector oculto responsable por la ruptura espontánea de supersimetría. Por lo tanto, el valor de su masa depende del modelo particular que se considere para el mediador y la ruptura de supersimetría, pudiendo ubicarse la masa del gravitino desde el eV hasta más allá del TeV.

Uno de los mecanismos mejores estudiados para mediar la ruptura de supersimetría es el mediado por gravedad. Se asume que la mediación entre el sector oculto y el visible se da a través de términos de interacción no renormalizables. En el caso de mediación por gravedad, por razones dimensionales se espera que los términos \textit{soft} sean del orden
\begin{equation}
    m_{\text{soft}} \, \sim \, \frac{\langle F \rangle}{M_{Pl}} \, \sim \, m_{3/2},
    \label{gravrange1}
\end{equation}
ya que los términos \textit{soft} se anulan en los límites $\langle F \rangle \rightarrow 0$, y $M_{Pl} \rightarrow \infty $. Si requerimos que el gravitino esté `cerca' de la escala electrodébil para resolver el problema de las jerarquías, la escala de ruptura de supersimetría debería ser $\sqrt{\langle F \rangle} \sim 10^{10} - 10^{11}$ GeV.

Otro mecanismo importante es el mediado por interacciones de gauge, donde la interacción entre el sector oculto y el visible esta dado por partículas mensajeras con acoples del tipo gauge. En estos modelos los parámetros de masa \textit{soft} son generados por loops. Por análisis dimensional se espera
\begin{equation}
    m_{\text{soft}} \, \sim \, \frac{\alpha}{4 \pi} \frac{\langle F \rangle}{M_{\text{mens}}},
    \label{gravrange2}
\end{equation}
donde el primer factor de la derecha se debe a un loop y $M_{\text{mens}}$ es la masa del mensajero, cuya escala se considera típicamente debajo de la escala de Planck para que la contribución de los términos mediados por gravedad sean despreciables. Por lo tanto, la ruptura de supersimetría debería ocurrir a una escala inferior que en el caso mediado por gravedad, lo que a su vez genera que la masa del gravitino también resulte menor a la estimada con el caso anterior. Entonces, en este caso, el gravitino siempre es más ligero que los compañeros supersimétricos de las partículas del modelo estándar.

\subsection{Axiones en supersimetría}\label{axionesenSUSY}

En los modelos supersimétricos el axión $a$, con paridad-R par, es parte de un supermultiplete. Los campos que lo completan son un fermión de Majorana de spin $1/2$ con paridad-R impar, el axino $\widetilde{a}$, y un escalar de paridad-R par llamado saxión $s$~\cite{Nilles:1981py,Kim:1983dt}.

Al igual que para los axiones, los acoples de las nuevas partículas están suprimidos por la escala de PQ, $f_a$. Pero mientras que la masa del axión está dada por mecanismo de Peccei-Quinn (ver Sección~\ref{mecanismodePQ}), la masa de sus compañeros está relacionada con la escala de ruptura espontánea de supersimetría.

Para energías superiores a dicha escala, todos los miembros del supermultiplete del axión son degenerados en masa, dada por el mismo mecanismo que le otorga masa al axión. Luego de la ruptura de SUSY, el saxión desarrolla un término de masa de ruptura \textit{soft} $m_s^2 \; s^2$ pues es un escalar, donde esperamos que $m_s \sim m_{\text{soft}}$~\cite{Tamvakis:1982mw}. En cambio, el axino al ser un fermión no desarrolla un término de masa \textit{soft}, y por lo tanto no está limitado por dicha escala. El rango que puede tener la masa del axino varía dependiendo los detalles de cada modelo de axiones y los campos del superpotencial junto con la asignación de cargas de $U(1)_{PQ}$. Por ejemplo en Ref.~\cite{Rajagopal:1990yx,Chun:1992zk} se puede ver que en forma general
\begin{equation}
    m_{\widetilde{a}}\sim \frac{m_{\text{soft}}^2}{f_a}\sim O(\text{keV-MeV}),
\end{equation}
es decir que el axino puede ser la partícula supersimétrica más ligera, con otras contribuciones extra dependientes del modelo tales que
\begin{equation}
    m_{\widetilde{a}}\sim  O(\text{keV-GeV}).
    \label{axinomassrange}
\end{equation}

Por lo tanto, incluir a los axiones en modelos supersimétricos resulta muy atractivo, pues además de resolver el \textit{strong CP problem} tratado en la Sección~\ref{axiones}, tenemos dos candidatos a materia oscura: el axión y el axino~\cite{Covi:1999ty,Covi:2001nw} (el saxión generalmente más pesado que el resto de las partículas supersimétricas resulta en un tiempo de vida muy corto y por lo tanto no constituye un buen candidato a materia oscura).

Luego de la era de inflación, los axinos pueden ser producidos por diversos mecanismos. Un proceso eficiente es mediante producción térmica a través de colisiones y decaimientos de partículas en el plasma primordial. También existen otros mecanismos no térmicos como por ejemplo el decaimiento de partículas supersimétricas pesadas luego de que estas se hayan desacoplado del baño térmico.

Existen varios paralelismos entre el axino y el gravitino, como por ejemplo que ambas partículas son fermiones neutros con masas que dependen fuertemente del modelo particular que se considere, pudiendo ser tan livianas como $m\sim O(\text{keV})$. Asimismo, la escala característica de interacción típicamente es órdenes de magnitud superior tanto a la escala electrodébil como las escalas actualmente accesibles por los colisionadores.

Sin embargo, los acoples del axino con la materia ordinaria son mayores a los del gravitino por un factor $M_{Pl}/f_a \sim 10^7$. Esto significa que la producción termal de axinos es importante para temperaturas de \textit{reheating} más bajas que las del gravitino. Otra consecuencia importante con respecto al acople más fuerte, es que otras partículas supersimétricas más pesadas decaen al axino con tiempos característicos más chicos, generalmente decaen mucho antes del comienzo de nucleosíntesis. Por lo tanto, no existe inyección de energía por decaimientos que pueda destruir la producción de elementos ligeros y las predicciones exitosas de BBN (para el caso del gravitino esto puede suceder y se denomina `\textit{gravitino problem}').

Finalmente, considerando Ec.~(\ref{gravrange1}) y (\ref{gravrange2}) con (\ref{axinomassrange}), vemos que el axino y el gravitino pueden ser de forma natural las partículas más ligeras del espectro supersimétrico.

\spacing{1}



\chapter{Gravitino como materia oscura en el \texorpdfstring{$\mu\nu$}{munu}SSM y su detección}
\label{gravitinoresults1}

\spacing{1.5}

En este capítulo comenzamos nuestro análisis del espacio de parámetros del $\mu\nu$SSM buscando reproducir la física de neutrinos, y estudiamos las condiciones para que el gravitino constituya la materia oscura. Remarcamos que una de las consecuencias de la ruptura de paridad-R en el $\mu\nu$SSM es la existencia de una relación directa entre el sector de neutrinos y el decaimiento del gravitino. A partir de la predicción del flujo de rayos gamma imponemos límites sobre la masa y el tiempo de vida del gravitino. Para ello se tienen en cuenta los distintos regímenes y características de los decaimientos a dos y tres cuerpos. Este capítulo está basado en Ref.~\cite{Gomez-Vargas:2016ocf}.

\section{Matriz de masa efectiva de neutrinos} \label{seccionmeff}

Como consecuencia de la violación de la paridad-R en el $\mu \nu$SSM, los neutralinos del MSSM se mezclan con los neutrinos izquierdos y derechos, resultando en la matriz de masas mostrada en Ec.~(\ref{MuvMSSM}). Dicha matriz es del tipo \textit{seesaw} ya que los elementos de la matriz $M$ son del orden de la escala electrodébil, mientras que las entradas de la matriz $m$ son del orden de la masa de Dirac. En primera aproximación, podemos escribir la matriz de masa efectiva de mezcla de neutrinos como $m_{eff}=-m^{T}M^{-1}m$, Ec.~(\ref{mmeff}).

Para obtener los autoestados de masa de los neutrinos diagonalizamos la matriz de masas $m_{eff}$ con la matriz unitaria $U$ de Pontecorvo–Maki–Nakagawa–Sakata (PMNS) cuya parametrización se muestra en Ec.~(\ref{PMNSmatrix}) en función de los ángulos de mezcla $\theta_{ij}$, con $i,j$ corriendo desde 1 a 3, y de la fase de Dirac $\delta$, la cual viola CP y asumiremos nula en el resto del trabajo.
Entonces
\begin{equation}
	U^{-1}m_{eff}U=\text{diag}(m_{1},m_{2},m_{3}),
\end{equation}
donde $m_{i}$ son las masas de los neutrinos.

Los parámetros independientes en el sector de neutrinos del $\mu \nu$SSM son el siguiente conjunto de variables:
\begin{equation}
	\lambda_{i},\hspace{0.1cm} \kappa_{ijk},\hspace{0.1cm} \tan\beta=\frac{v_{u}}{v_{d}},\hspace{0.1cm} v_{\nu_{i}},\hspace{0.1cm} v_{\nu^{c}_{i}},\hspace{0.1cm} Y^{\nu}_{ij}\hspace{0.1cm} y \hspace{0.1cm} M.
\end{equation}

Por simplicidad trabajaremos bajo las siguientes consideraciones:
\begin{itemize}
	\item Utilizaremos parámetros reales.
	\item Despreciaremos los términos que contengan $(Y_{\nu}v_{\nu})^{2}$, $(Y^{\nu})^{3}v_{\nu}$ y $Y^{\nu}(v_{\nu})^{3}$, debido a que $Y^{\nu}$ y $v_{\nu}$ son chicos (del orden de $10^{-6}$ y $10^{-4}$ GeV respectivamente).
	\item Asumiremos que no existe mezcla intergeneracional entre los parámetros del modelo. Entonces los acoples están dados por $\lambda_{i}\equiv\lambda$, $\kappa_{iii}\equiv\kappa_{i}\equiv\kappa$ y 0 en cualquier otro caso, tomaremos los acoples de Yukawa diagonales $Y^{\nu}_{ii}\equiv Y^{\nu}_{i}$ y VEVs $v_{\nu_{i}^{c}}\equiv v_{\nu^{c}}$; donde $\textit{i}$ corresponde al índice de familia.
\end{itemize}
  
Por lo tanto, los parámetros independientes en nuestro análisis del sector de neutrinos resultan: 
\begin{equation}
	\lambda,\hspace{0.1cm} \kappa,\hspace{0.1cm} \tan\beta=\frac{v_{u}}{v_{d}},\hspace{0.1cm} v_{\nu_{i}},\hspace{0.1cm} v_{\nu^{c}},\hspace{0.1cm} Y^{\nu}_{i}\hspace{0.1cm} y \hspace{0.1cm} M.
\label{param}
\end{equation}
	
Bajo dichas condiciones podemos obtener una expresión analítica para la matriz de masa efectiva de neutrinos~\cite{neutrinoindio,neutrinocp,LopezFogliani:2010bf}:
\begin{equation}
	(m_{eff})_{ij}\approx \frac{v_{u}^{2}}{6\kappa v_{\nu^c}}Y^{\nu}_{i}Y^{\nu}_{j}(1-3\delta_{ij}) - \frac{1}{2M_{eff}}\left[v_{\nu_i}v_{\nu_j}+\frac{v_{d}(Y^{\nu}_{i}v_{\nu_j} + Y^{\nu}_{j}v_{\nu_i})}{3\lambda} + \frac{Y^{\nu}_{i}Y^{\nu}_{j}v_{d}^{2}}{9\lambda^{2}}\right],
\label{mefff}
\end{equation}
con
\begin{equation}
	M_{eff}\equiv M\left[1 - \frac{v^{2}}{2M(\kappa v_{\nu^c}^2 + \lambda v_{u} v_{d}) 3\lambda v_{\nu^c}}\left(2\kappa v_{\nu^c}^2\frac{v_{u} v_{d}}{v^{2}} + \frac{\lambda v^{2}}{2}\right)\right],
\end{equation}
donde $v^{2}=v_{u}^{2}+v_{d}^{2}+\sum v_{\nu_i}^{2} \approx v_{u}^{2}+v_{d}^{2} \approx (174 \text{GeV})^{2}$ ya que $v_{\nu_i}<<v_{u},v_{d}$, y definimos $\frac{1}{M}=\frac{g_{1}^{2}}{M_{1}}+\frac{g_{2}^{2}}{M_{2}}$.

Para ilustrar cómo funciona el mecanismo $seesaw$, analizaremos cómo se comporta la matriz de masas efectiva en distintos límites.

Si consideramos el límite donde los gauginos se desacoplan~\cite{neutrinocp,neutrinoindio}, es decir $M\rightarrow\infty$, entonces solo el primer término de Ec.~(\ref{mefff}) es relevante:
\begin{equation}
	(m_{eff})_{ij}\approx \frac{v_{u}^{2}}{6\kappa v_{\nu^c}}Y^{\nu}_{i}Y^{\nu}_{j}(1-3\delta_{ij}).
	\label{limite1}
\end{equation}
Notamos que aún obtenemos entradas fuera de la diagonal en la matriz de masas de los neutrinos, con matrices de Yukawa diagonales. Esto se debe a la mezcla de neutrinos derechos con Higgsinos. Si además consideramos el límite donde los neutrinos derechos también se desacoplan, es decir $v_{\nu^c}\rightarrow\infty$, la masa de los neutrinos es cero.

Otro límite interesante es el desacople de los neutrinos derechos~\cite{neutrinocp,neutrinoindio}, $v_{\nu^c}\rightarrow\infty$. En este caso Ec.~(\ref{mefff}) se reduce a:
\begin{equation}
	(m_{eff})_{ij}\approx - \frac{1}{2M}\left[v_{\nu_i}v_{\nu_j}+\frac{v_{d}(Y^{\nu}_{i}v_{\nu_j} + Y^{\nu}_{j}v_{\nu_i})}{3\lambda} + \frac{Y^{\nu}_{i}Y^{\nu}_{j}v_{d}^{2}}{9\lambda^{2}}\right].
\end{equation}
Tomando también el limite $v_{d}\rightarrow 0$, es decir $\tan\beta\rightarrow\infty$ obtenemos:
\begin{equation}
	(m_{eff})_{ij}\approx - \frac{v_{\nu_i}v_{\nu_j}}{2M},
\end{equation}
donde nuevamente tenemos entradas fuera de la diagonal. Este caso nos indica que al considerar desacoplados los neutrinos derechos y los Higgsinos, el mecanismo $seesaw$ se genera a partir de la mezcla de neutrinos izquierdos con los gauginos.
Si $v_{\nu_i} \gg \frac{Y^{\nu}_{i}v_{d}}{3\lambda}$, el mismo resultado puede ser obtenido como se muestra en Ref.~\cite{neutrinocp}.

El límite que mejor ilustra el mecanismo \textit{seesaw} en el $\mu\nu$SSM corresponde a una mezcla de los dos casos límites considerados anteriormente. Consiste en suprimir parte de la mezcla de Higgsinos y gauginos~\cite{neutrinocp}.
Tomaremos $v_{d}\rightarrow 0$ (ó $v_{\nu_i} \gg \frac{Y^{\nu}_{i}v_{d}}{3\lambda}$), entonces Ec.~(\ref{mefff}) resulta:
\begin{equation}
	(m_{eff})_{ij}\approx \frac{v_{u}^{2}}{6\kappa v_{\nu^c}}Y^{\nu}_{i}Y^{\nu}_{j}(1-3\delta_{ij}) - \frac{v_{\nu_i}v_{\nu_j}}{2M_{eff}},
	\label{mefflimite}
\end{equation}
con
\begin{equation}
	M_{eff}\equiv M\left(1 - \frac{v^{4}}{12M\kappa v_{\nu^c}^3}\right).
\end{equation}
La masa efectiva $M_{eff}$ representa que la mezcla entre gauginos con Higgsinos y neutrinos derechos no esta completamente suprimida, por lo que el resultado es más general que en los casos anteriores. Sin embargo, para los valores típicos de los parámetros involucrados en el modelo, tenemos que $M_{eff}\approx M$ y Ec.~(\ref{mefflimite}) se reduce a:
\begin{equation}
	(m_{eff})_{ij}\approx \frac{v_{u}^{2}}{6\kappa v_{\nu^c}}Y^{\nu}_{i}Y^{\nu}_{j}(1-3\delta_{ij}) - \frac{v_{\nu_i}v_{\nu_j}}{2M},
	\label{meffsimple}
\end{equation}
donde la simplificación reside en haber despreciado por completo la mezcla entre gauginos con Higgsinos y neutrinos derechos. A modo ilustrativo en el apéndice~\ref{maximal} se muestra trabajando en este límite, cómo de manera simple puede encontrarse la jerarquía normal e invertida del sector de neutrinos, en el régimen trimaximal (donde se considera $\sin\theta_{13}=0$).

En resumen, notamos que en nuestro trabajo emplearemos para el estudio de la física de neutrinos los parámetros libres mostrados en Ec.~(\ref{param}) y la matriz de masas efectiva sin considerar las simplificaciones de los tres límites arriba mencionados, es decir Ec.~(\ref{mefff}).

En este escenario veremos mediante un análisis numérico que podemos encontrar a orden árbol el correcto patrón de mezclas y masas del sector, para tres familias de neutrinos y matrices de Yukawa diagonales, lo cual torna tanto al modelo, su estudio y comprensión más simple al reducir el número de parámetros necesarios para reproducir los datos experimentales.

\section{Gravitino y su relación con la física de neutrinos}
\label{gravitinooscuro}

En primer lugar vamos a mostrar que el gravitino cumple las condiciones que un candidato a materia oscura debe cumplir, nombradas en la Sección~\ref{candiDM}, en modelos de paridad-$R$ rota.

\begin{figure}[t]
 \begin{center}
       \epsfig{file=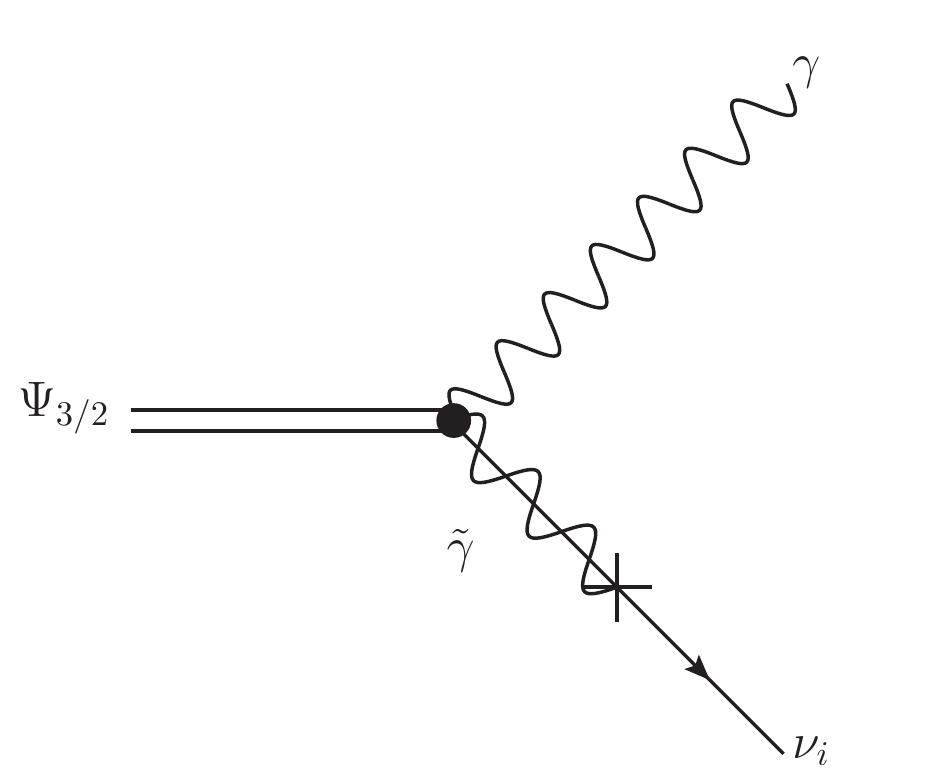,height=4.1cm}
\caption{Diagrama a orden árbol para el decaimiento de dos cuerpos del gravitino a un fotón y neutrino, vía el término de mezcla fotino-neutrino.}
    \label{fig_gfn}
 \end{center}
\end{figure}

Veamos que su vida media es típicamente varios órdenes de magnitud superior a la edad del universo. En el Lagrangiano de supergravedad existe un término de interacción entre el gravitino, el campo del fotón y el fotino (el compañero supersimétrico del fotón, una mezcla del Bino y Wino). Como en $\mu\nu$SSM la paridad-R está rota, los neutralinos se mezclan con los neutrinos. En particular los neutrinos izquierdos se mezclan con los Bino y Wino, por lo tanto el gravitino puede decaer a un fotón y un neutrino como se muestra en la Figura~\ref{fig_gfn}, debido al término de interacción fotón-fotino~\cite{gravitinonoR}\footnote{A continuación veremos que las búsquedas de rayos gamma permiten $m_{3/2}<m_W$ donde $m_W$ es la masa del bosón W. Por lo tanto, el gravitino no puede decaer a estados finales con bosones Z, W, o de Higgs que satisfagan las ecuaciones clásicas de movimiento, estados denominados \textit{on-shell}.}. El punto grande en el vértice gravitino-fotón-fotino representa el término de orden 5 suprimido por la masa de Planck, y la cruz la mezcla entre el fotino y el neutrino.
Como el decaimiento es a dos cuerpos, el fotón producido es monocromático con energía igual a la mitad de la masa del gravitino, despreciando la masa del neutrino pues $m_{\nu} \ll m_{3/2}$. Obtenemos:
\bea
\Gamma\left(\Psi_{3/2}\rightarrow\sum_i\gamma\nu_i\right)\simeq\frac{m_{3/2}^3}{64\pi M_{Pl}^2}|U_{\tilde{\gamma} \nu}|^2\ ,
\label{decay2body}
\eea
donde $\Psi_{3/2}$ denota al gravitino, $m_{3/2}$ la masa del gravitino, $M_{Pl}$ la masa de Planck, $\gamma$ al fotón, $\widetilde{\gamma}$ al fotino, y $|U_{\widetilde{\gamma}\nu}|^{2}$ el contenido de fotino en el neutrino
\begin{equation}
	|U_{\widetilde{\gamma}\nu}|^{2}=\sum_{i=1}^{3}|N_{i1}\cos\theta_{w} + N_{i2}\sin\theta_{w}|^{2}.
\label{photino}
\end{equation}
$N_{i1}$ ($N_{i2}$) corresponde a la componente Bino (Wino) del neutrino $\textit{i}$-ésimo. El mismo resultado se obtiene para el proceso conjugado $\Psi_{3/2}\rightarrow\gamma\bar{\nu}_i$.

Asumiendo que Ec.~(\ref{decay2body}) es el único canal de decaimiento para el gravitino, podemos escribir su vida media como:
\begin{equation}
{\tau}_{3/2}
(\Psi_{3/2}\rightarrow\sum_i\gamma\nu_i
)
=\frac{1}{2\Gamma\left(\Psi_{3/2}\rightarrow\sum_i\gamma\nu_i\right)}
\simeq 3.8\times 10^{27}\, {s}
\left(\frac{10^{-16}}{|U_{\widetilde{\gamma}\nu}|^2}\right)
\left(\frac{10\, \mathrm{GeV}}{m_{3/2}}\right)^{3}\ ,
\label{lifetimegamma}
\end{equation}
donde el factor 2 tiene en cuenta los procesos conjugados en el estado final. Recordamos que la edad del universo es del orden de $10^{17}$s.

Sin embargo, como fue establecido en Ref.~\cite{Choi:2010xn}, gravitinos con masas menores a la masa del bosón $W$, como es en nuestro caso, pueden también decaer con un BR significativo a estados finales de tres cuerpos. De este modo se produce un espectro continuo en el rango de los rayos gamma que también puede ser detectado en \Fermi LAT.

\begin{figure}[t]
\begin{center}
\begin{tabular}{lll}
\hspace*{-2mm}\epsfig{file=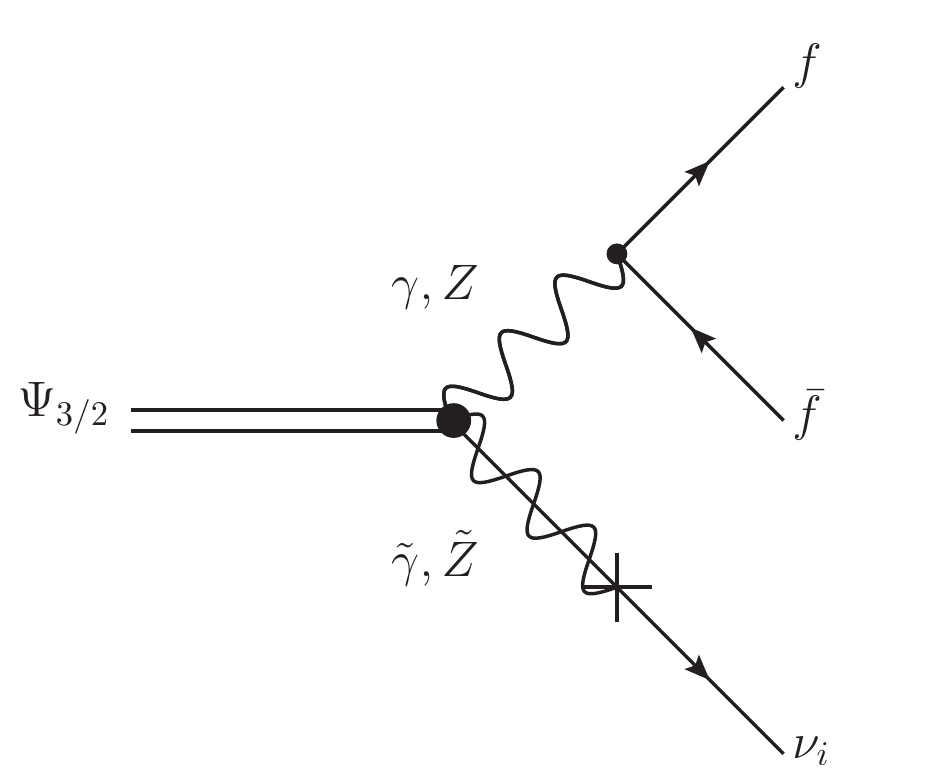,height=4.1cm} & \epsfig{file=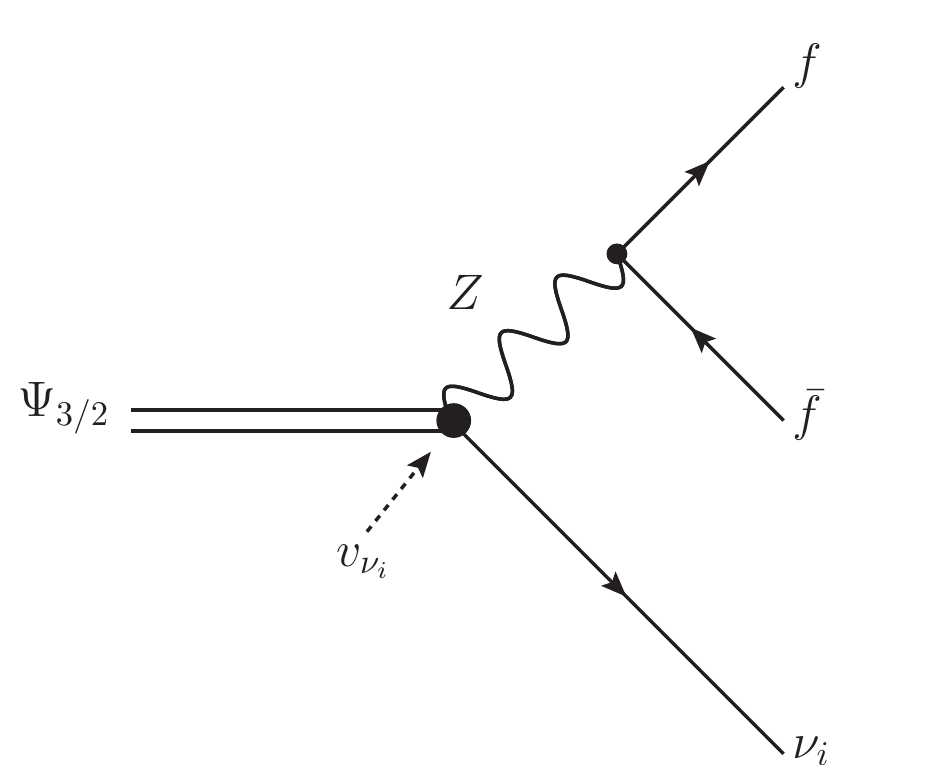,height=4.1cm} & \epsfig{file=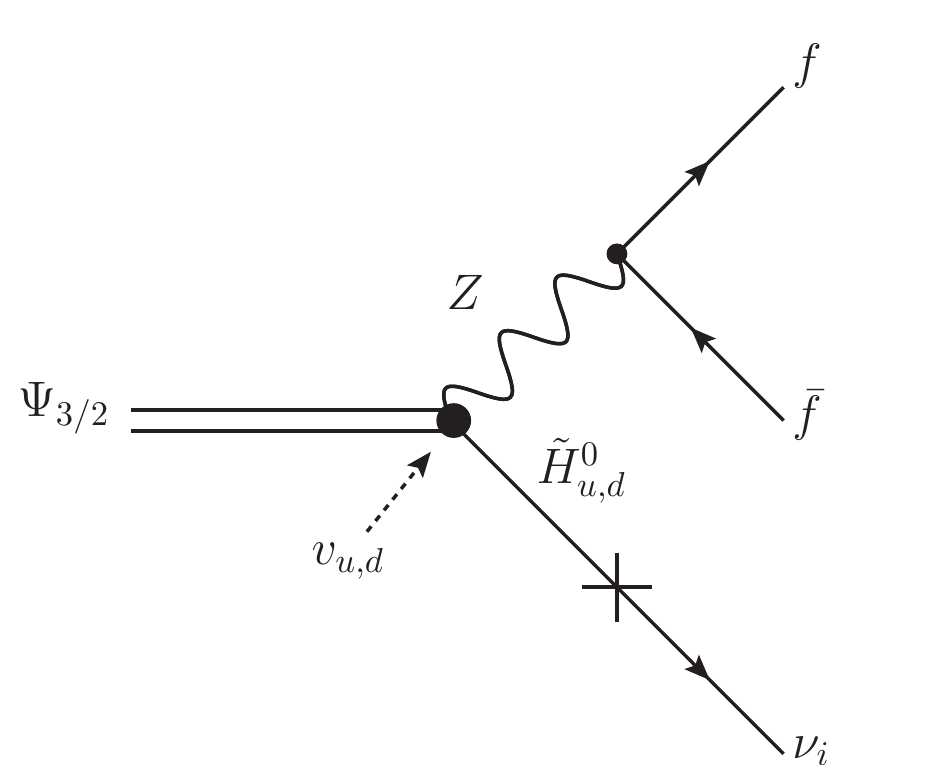,height=4.1cm} \\
\hspace*{2cm}(a) & \hspace*{2cm}(b) & \hspace*{2cm}(c)
\end{tabular}
\end{center}
\caption{Diagrama a orden árbol para el decaimiento del gravitino a un par fermión-antifermión y un neutrino, vía un fotón o un bosón $Z$ intermedio.}
    \label{fig_gz}
\end{figure}

\begin{figure}[t]
\begin{center}
\begin{tabular}{lll}
\hspace*{-2mm}\epsfig{file=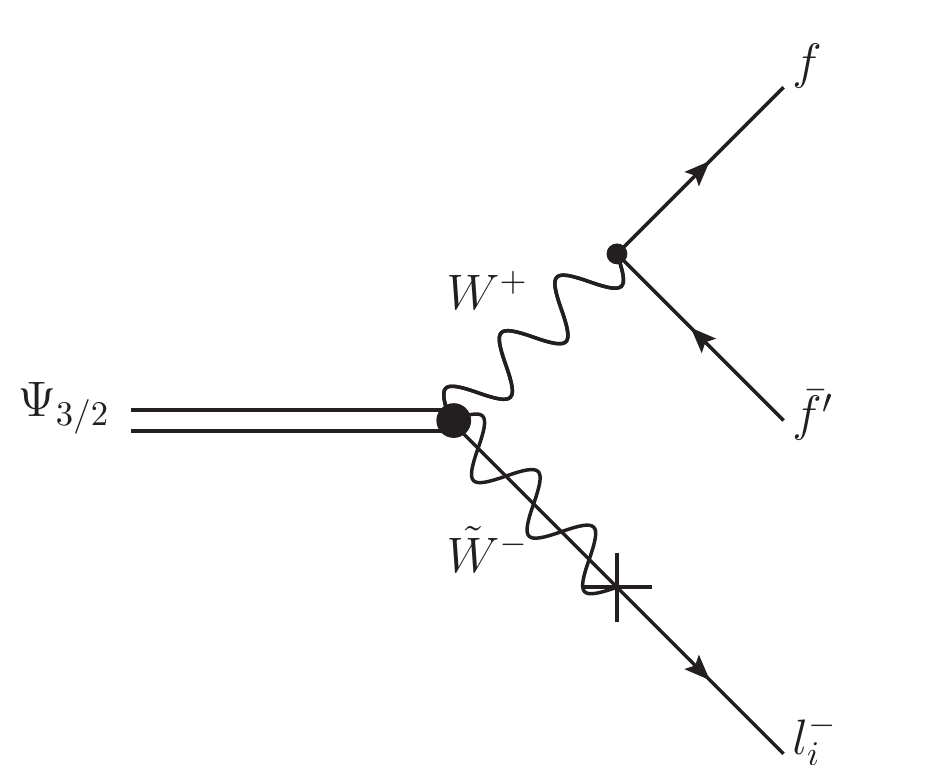,height=4.1cm} & \epsfig{file=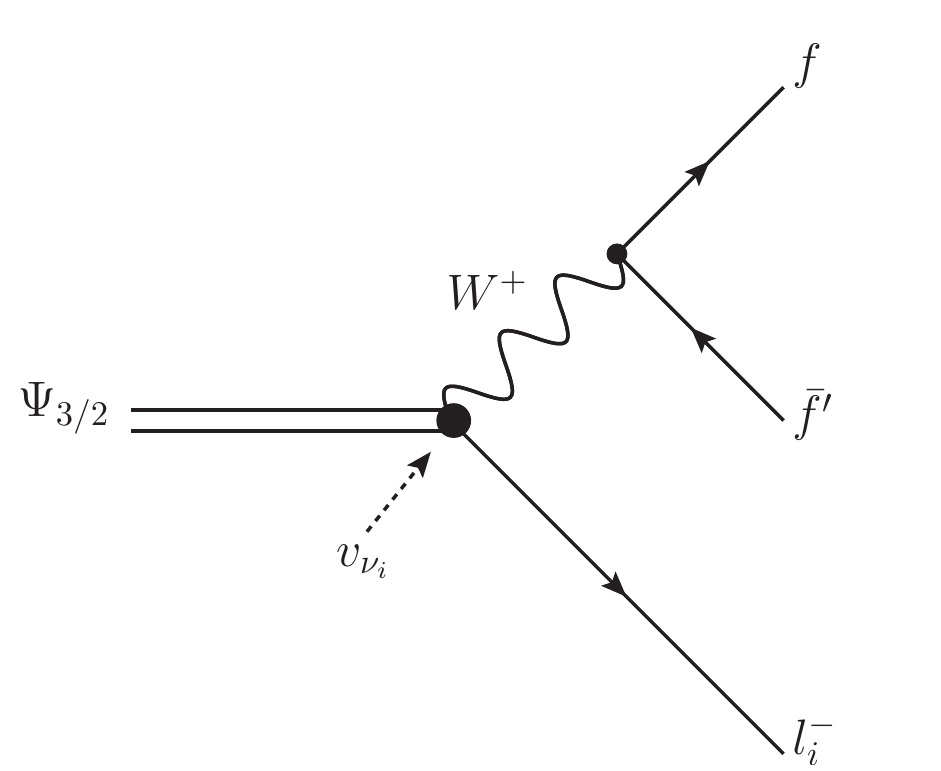,height=4.1cm} & \epsfig{file=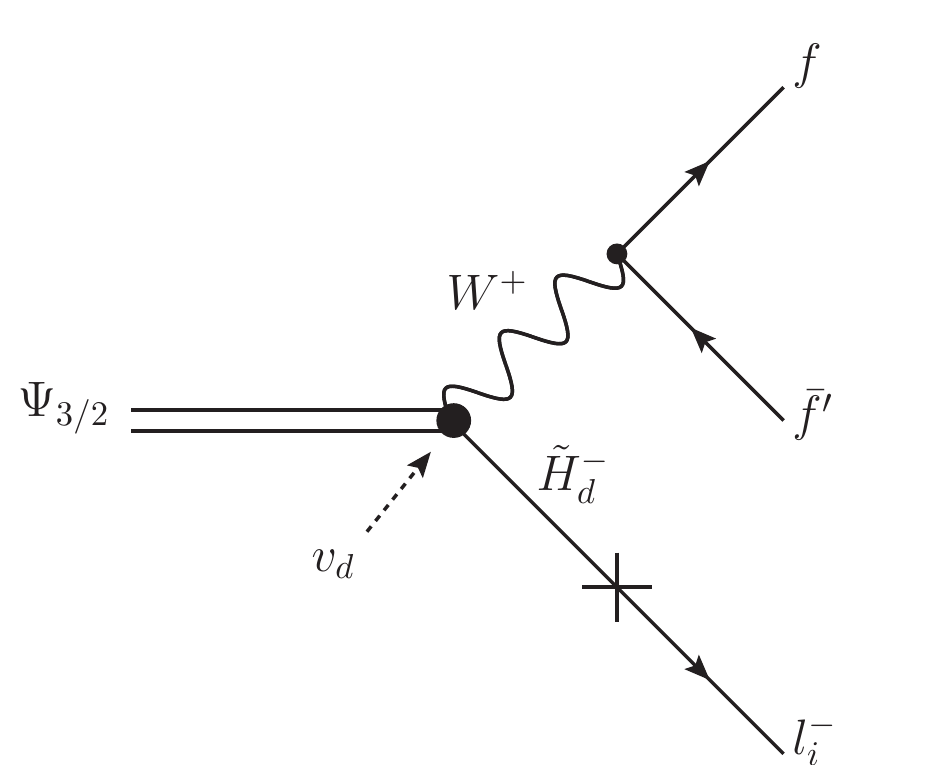,height=4.1cm}\\
\hspace*{2cm}(a) & \hspace*{2cm}(b) & \hspace*{2cm}(c)
\end{tabular}
\end{center}
\caption{Diagrama a orden árbol para el decaimiento del gravitino a dos fermiones y un leptón cargado, vía un bosón $W$ intermedio.}
    \label{fig_gw}
\end{figure}

Estos canales son $\Psi_{3/2}\rightarrow\gamma^*/Z^* \, \nu_i\rightarrow f \,  \bar{f} \, \nu_i$ vía un fotón o un bosón $Z$ intermedio, y $\Psi_{3/2}\rightarrow W^* \, l\rightarrow f  \, \bar{f}' \, l$ vía un bosón $W$ intermedio, donde $f$ denota a fermiones y $l$ leptones. Ambos canales son mostrados en las Figuras~\ref{fig_gz} y~\ref{fig_gw} respectivamente, donde las flechas punteadas indican términos de interacción donde un bosón adquiere el VEV especificado. Los decaimientos fueron calculados en Refs.~\cite{Choi:2010xn,Choi:2010jt,Grefe:2011dp,Diaz:2011pc}, y por completitud mostramos los resultados de las anchuras de decaimiento en el apéndice~\ref{3body}.

Entonces, la anchura de decaimiento total del gravitino está dada por:
\bea
\Gamma_{\text{total}}(\Psi_{3/2})=2\left(\Gamma(\Psi_{3/2} \rightarrow \gamma\nu) +\Gamma(\Psi_{3/2} \rightarrow f\bar{f}\nu)+\Gamma(\Psi_{3/2} \rightarrow f\bar{f'}l) \right)
\ .
\label{total_width}
\eea
Si el canal $\gamma\nu$ no es siempre el canal de decaimiento del gravitino dominante, el tiempo de vida media estimado en Ec.~(\ref{lifetimegamma}) se ve modificado. Naturalmente, la formula completa es
\begin{equation}
\tau_{3/2}(\Psi_{3/2})
=\frac{1}{\Gamma_{\text{total}}(\Psi_{3/2})}\ .
\label{lifetimetotal}
\end{equation}

A partir de la discusión anterior, notamos que el parámetro de mezcla entre el fotino y el neutrino $|U_{\tilde{\gamma} \nu_i}|$ cumple un papel crucial en el análisis del gravitino como materia oscura y su detección mediante rayos gamma. Considerando Ec.~(\ref{decay2body}), es fácil ver que para valores cada vez más grandes (chicos) de $|U_{\tilde{\gamma} \nu_i}|$, la anchura de decaimiento es más grande (chica), y como consecuencia los límites sobre el espacio de parámetros del modelo son más (menos) restrictivos por la no observación de líneas espectrales en los datos de \Fermi LAT.

Por otra parte, en regiones del espacio de parámetros donde $|U_{\tilde{\gamma} \nu_i}|$ se encuentra suprimido, el BR a $\gamma\nu$ también está suprimido, y los BRs a tres cuerpos se vuelven más importantes.
A partir de las Figuras~\ref{fig_gz} y~\ref{fig_gw} (ver también apéndice~\ref{3body}) podemos ver que solo un canal a tres cuerpos depende de $|U_{\tilde{\gamma} \nu_i}|$. Los diagramas en los paneles (a) y (c) de la Figuras~\ref{fig_gz} involucran $|U_{\widetilde{\chi}\nu}|$, donde $\widetilde{\chi}$ representa un neutralino, y se obtienen a partir de la matriz de fermiones neutros (neutralinos-neutrinos). Los diagramas en los paneles (a) y (c) de la Figura~\ref{fig_gw} involucran $|U_{\widetilde{\chi}^-l^-}|$, donde $\widetilde{\chi}^-$ representa un chargino, y se calcula a partir de la matriz de fermiones cargados (charginos-leptones). Por otro lado, los diagramas en los paneles (b) son independientes de los parámetros de mezcla, proporcionales a  $v_{\nu_{i}}$, y pueden dominar en algunos límites. Esto puede tener dos efectos. Primero, puede afectar significativamente el resultado del tiempo de vida media del gravitino, como ya hemos mencionado. Segundo, los límites del espacio de parámetros provenientes de líneas espectrales son menos restrictivos, pero nuevos límites pueden aparecer a partir del análisis del espectro continuo generado por los decaimientos con tres cuerpos en el estado final, utilizando datos de \Fermi LAT.
En la próxima sección, estudiaremos estas cuestiones cruciales en el análisis de gravitino como materia oscura y su detección.

Por último es necesario ver que la densidad de reliquia necesaria puede ser reproducida por el gravitino. Por el proceso inflacionario, los gravitinos primordiales son diluidos como consecuencia de la expansión exponencial del universo. Sin embargo, después del proceso de inflación, los gravitinos son producidos durante el periodo llamado $reheating$, a través de colisiones y decaimientos dominados por procesos de QCD entre las partículas en equilibrio térmico con el plasma primordial. Entonces la densidad reliquia será proporcional a la temperatura de \textit{reheating} $T_{R}$, estimada en~\cite{producciongravitino}:
\begin{equation}
	\Omega_{3/2}h^{2}\cong0.27\left(\frac{T_{R}}{10^{10}\text{GeV}}\right)\left(\frac{100\text{GeV}}{m_{3/2}}\right)\left(\frac{m_{\widetilde{g}}}{1\text{TeV}}\right)^{2},
	\label{relicgravitinos1}
\end{equation}
donde $m_{\overline{g}}$ es la masa del gluino. Ajustando $T_{R}$ podemos reproducir la densidad reliquia correcta para todos el rango de posibles masas de gravitinos. Por ejemplo, para $m_{3/2}$ entre 1-1000 GeV tendremos $\Omega_{3/2}h^{2} \cong$0.1 para $T_{R} \approx 10^{8}-10^{11}$GeV y $m_{\widetilde{g}}\approx 1$TeV.

\subsection{El parámetro de mezcla fotino-neutrino}

Una vez establecida la conexión entre la física de neutrinos y el decaimiento del gravitino, dado por la presencia de $|U_{\widetilde{\gamma}\nu}|^{2}$ en Ec.~(\ref{decay2body}) consecuencia de la ruptura de paridad-R, analizaremos a continuación el espacio de parámetros del sector de neutrinos con la finalidad de cumplir las cotas experimentales actuales y obtener el rango permitido de $|U_{\widetilde{\gamma}\nu}|^{2}$. Dichos valores a su vez están involucrados en los rangos de validez a la masa y vida media del gravitino dentro del marco del $\mu\nu$SSM.

Como estimación de $|U_{\widetilde{\gamma}\nu}|^{2}$ consideremos una matriz 2x2,
\begin{equation}
	  \left( {\begin{array}{cc}
	a & c \\
	c & b
	\end{array} } \right),
\end{equation}
cuyo ángulo de mezcla está dado por $\tan 2\theta = 2c/(a-b)$. Para el $\mu\nu$SSM dicha matriz representa la matriz de masas del sector neutralinos-neutrinos para un modelo con Bino y una sola familia de neutrinos izquierdos: $a$ $\sim$ 1TeV (masa \textit{soft} del Bino $M_{1}$), $b$=0 (término de masa del neutrino izquierdos) y $c$ $\sim$ $g_{1}v_{\nu}$ (mezcla del Bino con el neutrino izquierdo). Es decir,
\begin{equation}
|U_{\widetilde{\gamma}\nu}|\sim \frac{g_{1}v_{\nu}}{M_{1}}\ .
\label{representa}
\end{equation}
Para valores típicos de la escala electrodébil, $M_1 \sim 10^{3}$ GeV, y para reproducir los valores observados de masas de neutrinos y sus ángulos de mezcla, $v_{\nu}\sim 10^{-4}$ GeV, el parámetro de mezcla fotino-neutrino tiene que estar en el siguiente rango,
\begin{equation}
10^{-15} \lesssim |U_{\widetilde{\gamma}\nu}|^{2} \lesssim 10^{-14}.
\label{representative}
\end{equation}
Esto fue confirmado realizando un scan en el espacio de parámetros de baja energía del modelo en Ref.~\cite{Choi:2009ng}. Si bien la paridad-R está rota y el LSP no es estable, el decaimiento del gravitino se encuentra suprimido tanto por la débil interacción gravitatoria, manifestado en Ec.~(\ref{decay2body}) por la masa de Planck, como por lo chico de la mezcla del fotino con los neutrinos $|U_{\widetilde{\gamma}\nu}|^{2}$.

A partir de la no observación de líneas espectrales en las mediciones de rayos gamma reportadas por la colaboración \Fermi LAT, se obtuvieron límites sobre el espacio de parámetros del modelo en Refs.~\cite{Choi:2009ng,Albert:2014hwa}, utilizando el rango mostrado en Ec.~(\ref{representative}). En un trabajo más reciente~\cite{Albert:2014hwa}, se obtuvieron límites para la masa y el tiempo de vida del gravitino,  $m_{3/2}\lesssim 2.5$ GeV y $\tau_{3/2}\gtrsim 10^{28}$ s, donde el primero (último) surge tomando el límite inferior (superior) en Ec.~(\ref{representative}).

Sin embargo, se puede inferir a partir de Ec.~(\ref{photino}) que valores inferiores del parámetro de mezcla  podrían ser alcanzables mediante la cancelación entre las contribuciones de Bino y Wino. Se ha sugerido en Ref.~\cite{Choi:2009ng} que es posible relajar el límite inferior un orden de magnitud
\begin{equation}
10^{-16} \lesssim |U_{\widetilde{\gamma}\nu}|^{2} \lesssim 10^{-14}\ ,
\label{relaxing}
\end{equation}
extendiendo los valores permitidos de $m_{3/2}$. El análisis en Ref.~\cite{Albert:2014hwa} bajo dichas suposiciones obtuvo como límite $m_{3/2}\lesssim 5$ GeV.

En este trabajo queremos comprobar dicha suposición de manera cuantitativa, dada la importancia que tiene el parámetro de mezcla al restringir al gravitino como materia oscura. Para entender la situación de una manera cualitativa, utilizaremos el resultado obtenido en Refs.~\cite{Ibarra:2007wg,Grefe:2011dp,Diaz:2011pc}, donde una aproximación más precisa del valor del parámetro de mezcla ha sido llevada a cabo. Podemos recuperar el resultado utilizando nuevamente las entradas de la matriz de masas de los fermiones neutros mostrada en Ec.~(\ref{MuvMSSM}), para reemplazar
$N_{i1} \approx \frac{-g_1 v_{\nu_i}}{\sqrt 2 M_1} $ y 
$N_{i2} \approx \frac{ g_2 v_{\nu_i}}{\sqrt 2 M_2}$
en Ec.~(\ref{photino}). Por lo tanto, obtenemos:
\begin{equation}
	U_{\tilde{\gamma} \nu_i}\approx -\frac{g_1}{\sqrt 2} v_{\nu_i} \cos \theta_W \frac{M_2-M_1}{M_1M_2}\ ,
	\label{UAP1}
\end{equation}
es decir que se puede suprimir el decaimiento del gravitino a $\gamma\nu$ cancelando el numerador, simplemente tomando $M_2 \to M_1$.

A partir de Ec.~(\ref{UAP1}) (o Ec.~(\ref{representa})), se puede deducir otra manera de obtener una menor composición de fotino en los neutrinos, incrementando los valores de $|M_1|$ y $|M_2|$. Sin embargo, remarcamos que los parámetros involucrados en esta ecuación, las masas de los gauginos $M_i$ y los VEVs de los neutrinos levógiros o izquierdos $v_{\nu_i}$, también están involucrados en el mecanismo generalizado de $seesaw$ de escala electrodébil que genera las masas de los neutrinos en el $\mu \nu$SSM (mirar por ejemplo Ec.~(\ref{meffsimple})). Como los valores elegidos para los parámetros deben reproducir los datos actuales sobre la masas y ángulos de mezcla del sector de los neutrinos, debemos tener en cuenta las posibles correlaciones entre ellos.

Por ejemplo, considerando el segundo término en Ec.~(\ref{meffsimple}) podemos ver que dado un set de parámetros que reproduce la física de neutrinos, si incrementamos los valores de $|M_{1}|$ y $|M_{2}|$ por dos órdenes de magnitud, tenemos que incrementar también los VEVs de los neutrinos izquierdos $v_{\nu_{i}}$ un orden de magnitud. De lo contrario, la física de neutrinos se modificaría. Entonces el contenido de fotino en los neutrinos, $|U_{\tilde{\gamma} \nu_i}|$,  disminuye solo un orden de magnitud de acuerdo con Ec.~(\ref{UAP1}) (o Ec.~(\ref{representa})). Esto indica que para disminuir el valor de $|U_{\tilde{\gamma} \nu_i}|$, la estrategia de hacer las masas de los gauginos similares es más eficiente que incrementar su valor absoluto.

\begin{figure}[t!]
 \begin{tabular}{cc}
 \hspace*{-4mm}
 \epsfig{file=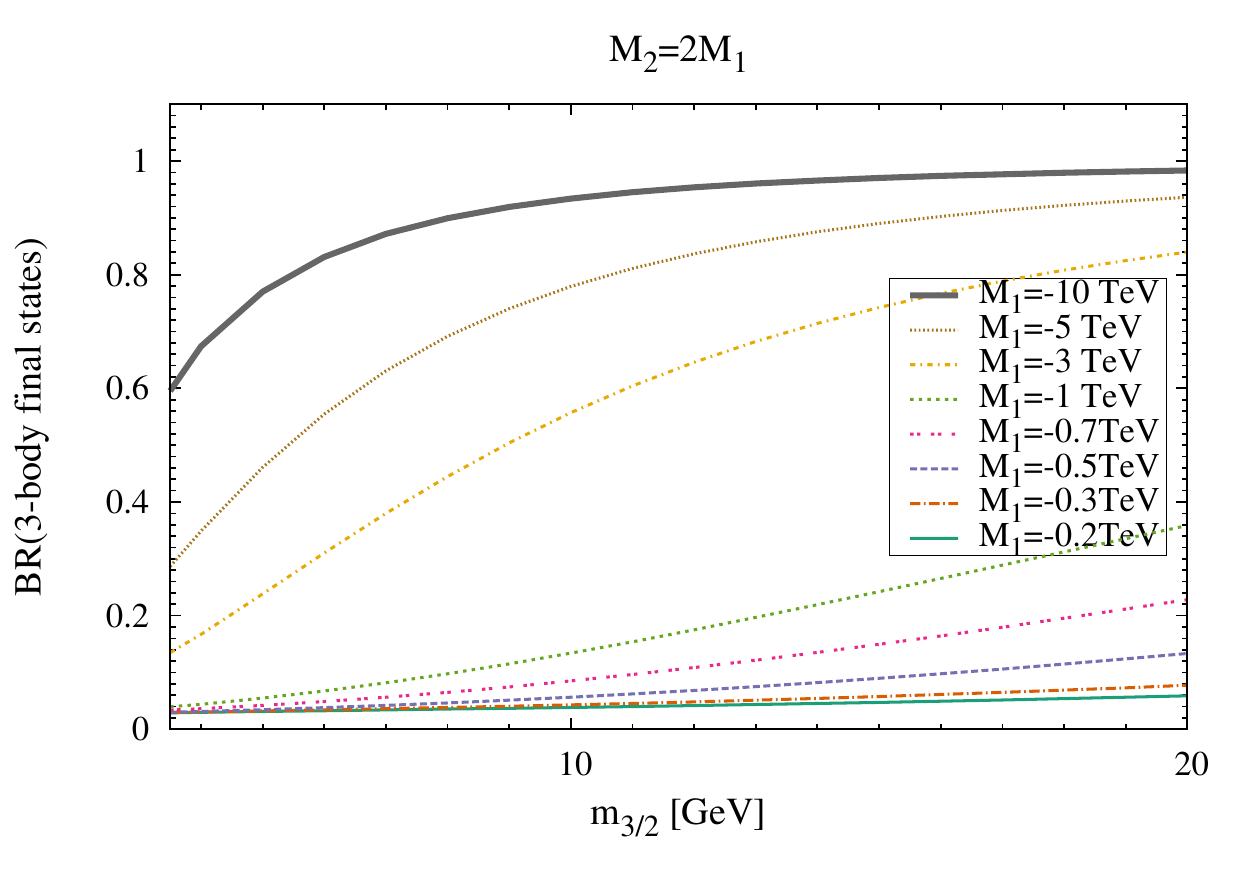,height=5.4cm} 
       \epsfig{file=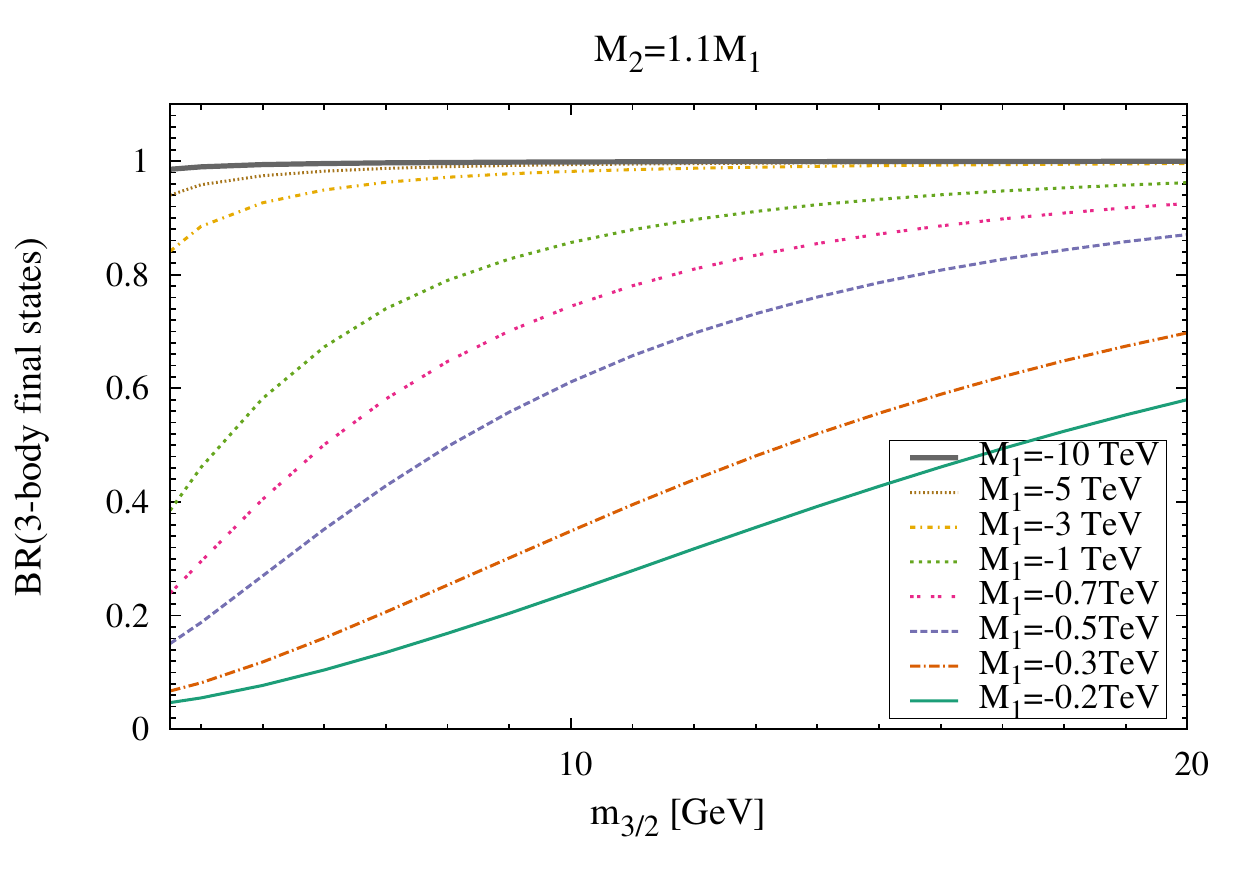,height=5.4cm}   \vspace*{-0.4cm}\\
       \vspace*{-0.1cm}       
     (a)\hspace*{3cm} & \hspace*{-2cm} (b)
    \end{tabular}
    \caption{BR del gravitino a tres cuerpos en función de la masa del gravitino, para varios valores de baja energía de $M_1$ considerando (a) $M_2 = 2 \, M_1$, (b) $M_2 = 1.1 \, M_1$. En ambos casos, los siguientes valores representativos se utilizaron para los parámetros a baja energía: $\lambda = 0.1 $, $\kappa = 0.1$, $\tan\beta = 10$, $v_{\nu^c}$ = 1750 GeV. 
}
    \label{fig2}
\end{figure}

Una vez que dichas estrategias nos permiten suprimir la composición de fotino en los neutrinos, como hemos mencionado la anchura de decaimiento del proceso $\psi \rightarrow\gamma\nu$ también es suprimida y la contribución de procesos a tres cuerpos se vuelve más relevante. Mostramos este comportamiento en la Figura~\ref{fig2} para dos relaciones diferentes entre $M_1$ y $M_2$, donde graficamos la razón entre la anchura de decaimiento de cada canal y la anchura de decaimiento total (BR$_i=\Gamma_i/\Gamma_{tot}$) llamado `\textit{branching ratio}', en función de la masa del gravitino.

El panel (a) presenta el BR para estados finales con tres cuerpos para varios valores de $M_1$ para el espectro de baja energía, suponiendo la aproximación de la relación entre las masas de los gauginos dada por las teorías de gran unificación (GUT), $M_2=2M_1$. Para otros parámetros del modelo utilizamos valores típicos $\lambda = 0.1 $, $\kappa = 0.1$, $\tan\beta = 10$, $v_{\nu^c}$ = 1750 GeV. Variaciones en estos valores no modifican las conclusiones significativamente. Como podemos ver, las contribuciones al decaimiento del gravitino de los canales a tres cuerpos puede ser importante para varios rangos de masas de gravitino y los  gauginos, especialmente para  $M_1\gtrsim 1$ TeV.

En el panel (b) mostramos los mismos casos que en el panel (a) pero utilizando la siguiente relación a bajas energías $M_2=1.1M_1$. Este ejemplo ilustra el caso límite discutido anteriormente, $M_2 \to M_1$, para obtener la cancelación del parámetro de mezcla. Los resultados en la figura confirman nuestra discusión, y podemos ver que los canales a tres cuerpos son más importantes que en el panel (a). En particular, ya para $M_1=-200$ GeV su BR es más grande que 0.5 cuando $m_{3/2}\gtrsim 17$ GeV, y por ejemplo, para $M_1=-1$ TeV este valor se obtiene cuando $m_{3/2}\gtrsim 4$ GeV.

Para calcular numéricamente el rango del parámetro de mezcla fotino-neutrino, y al mismo tiempo reproducir el patrón experimental de los neutrinos~\cite{Tortola:2012te} dentro del rango de 3$\sigma$ C.L., se ha realizado un barrido del espacio de parámetros de baja energía del $\mu \nu$SSM, considerando los siguientes rangos:
\begin{center}
   \begin{tabular}{ r  c  l }
	$0.1 \leq$ & $\lambda$ & $\leq 0.4$, \cr
	$0.1 \leq$ & $\kappa$ & $\leq 0.55$, \cr
	$5 \leq$ & $\tan\beta$ & $\leq 30$, \cr
	$10^{-6}\  \text{GeV} \leq$ & $v_{\nu_{1}}$ & $\leq 10^{-3}\ \text{GeV}$, \cr
	$10^{-6}\ \text{GeV}\leq$ & $v_{\nu_{2,3}}$ & $\leq 10^{-4}\ \text{GeV}$, \cr
	$500\ \text{GeV} \leq$ & $v_{\nu^{c}}$ & $\leq 5\ \text{TeV}$, \cr
	$10^{-8} \leq$ & $Y_{\nu_{1}}$ & $\leq 10^{-6}$, \cr
	$10^{-7} \leq$ & $Y_{\nu_{2,3}}$ & $\leq 10^{-5}$, \cr
	$-75\ \text{TeV} \leq$ & $M_{1}$ & $\leq -200\ \text{GeV}$, \cr
	$0.5 \leq$ & $M_{2}/M_{1}$ & $\leq 2$, \cr
	\end{tabular}
\end{center}
donde tomamos acoples de Yukawa diagonales, y los límites en las masas de gauginos siguiendo la discusión de las secciones anteriores. Para hallar soluciones a la física de neutrinos permitidas por los resultados experimentales  de la manera más fácil posible, se eligieron valores negativos para las masas de los gauginos, como se ha discutido en Ref.~\cite{neutrinocp} y en el apéndice~\ref{maximal} para el caso de jerarquía normal. Entonces obtenemos el siguiente resultado:
\bea
10^{-20} \lesssim|U_{\tilde{\gamma} \nu}|^2\lesssim 10^{-14}\ ,
\label{new_range}
\eea
el cual extiende el límite inferior de las previas estimaciones (\ref{relaxing}). Este límite inferior es alcanzado utilizando los casos límites para algunos parámetros, como por ejemplo $M_1=-75$ TeV, $M_2/M_1=1$. En el estudio del gravitino como materia oscura, no necesitamos incluir en el análisis dichos valores extremos para las masas de los gauginos, debido a que los resultados no cambian de manera significativa si consideramos $|M_1|\leq 10$ TeV como asumiremos a continuación, . 

Por lo tanto, dado el rango de $|U_{\tilde{\gamma} \nu_i}|$, y las contribuciones al decaimiento del gravitino por estados finales con tres cuerpos, se ha extendido el análisis del espacio de parámetros del $\mu \nu$SSM empleando datos de \Fermi LAT. Como veremos estos efectos causan gran impacto en la determinación de las restricciones experimentales.

\subsection{Barrido del espacio de parámetros del \texorpdfstring{$\mu\nu$}{munu}SSM}

Antes de continuar, daremos algunos detalles sobre el barrido del espacio de parámetros del $\mu \nu$SSM. Como primer paso se buscó reproducir las cotas experimentales de la física de neutrinos mostradas en la Tabla~\ref{tablaneutrinos} en el rango de 3$\sigma$. Para ello se realizó mediante métodos numéricos un barrido de los parámetros independientes del modelo. Se utilizó la matriz de masa efectiva de neutrinos, $m_{eff}$, Ec.~(\ref{mefff}), la cual es una matriz simétrica y puede ser diagonalizada por una matriz unitaria para luego calcular las diferencias de masa. Se seleccionaron los conjuntos de parámetros tales que las diferencias de masas cumplan las cotas experimentales. Con ellos se calcularon los ángulos de mezcla del sector, utilizando nuevamente $m_{eff}$, y se los comparó con los datos experimentales. Notamos que es posible hallar de manera eficiente tanto las diferencias de masas como los ángulos de mezcla de este modo gracias a el mecanismo de $seesaw$ y las características del $\mu \nu$SSM permiten trabajar con la forma simplificada $m_{eff}$ en vez de la matriz de masas completa del sector de neutralinos-neutrinos $M_{\widetilde{N}}$, Ec.~(\ref{MuvMSSM}).

Una vez hallados los parámetros que cumplen la física actual de neutrinos, se los utilizó para calcular el parámetro de mezcla $|U_{\tilde{\gamma} \nu}|^2$. Para obtener la componente Bino y Wino (fotino) en los neutrinos, es necesario utilizar la matriz de masas de los neutralinos-neutrinos completa, $M_{\widetilde{N}}$, la cual se diagonaliza con una matriz unitaria. En primer lugar se verificó que las diferencias de masa y los ángulos de los neutrinos calculados mediante la matriz efectiva no varían de manera significativa. Es decir, se verificó que aún cumplen las cotas experimentales. En segundo lugar se extrajo cada componente a partir de los autovectores correspondientes a los neutrinos y se calculó $|U_{\overline{\gamma}\nu}|^{2}$ según Ec.~(\ref{photino}).


\begin{figure}
\makebox[\textwidth][c]{\centering
\begin{tabular}{c c}
               \includegraphics[scale=0.85]{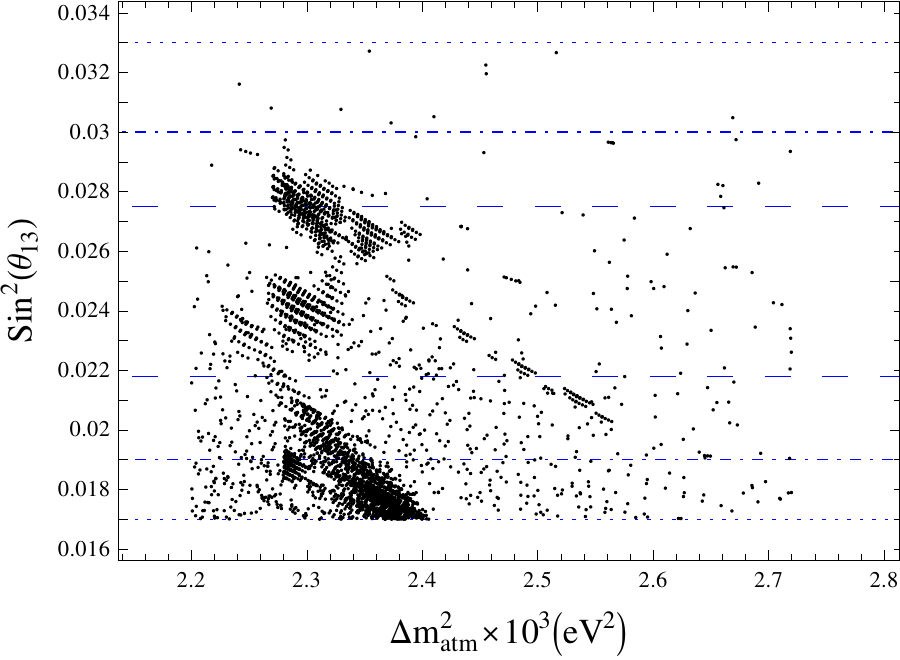}      & \includegraphics[scale=0.85]{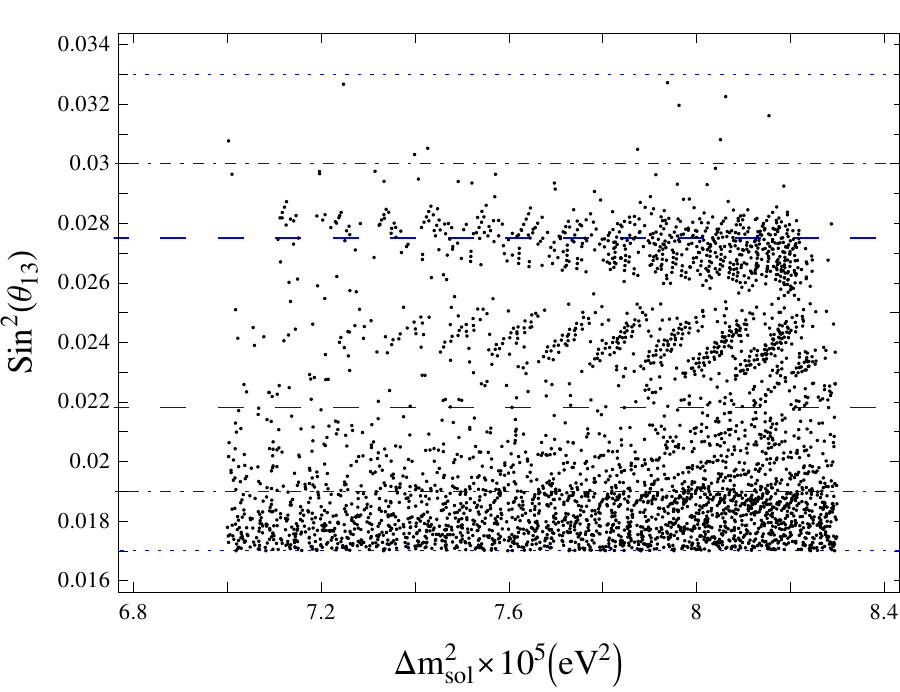}  
		
   \end{tabular}}
\caption{$\sin^{2}\theta_{13}$ en función de las diferencias de masas de los neutrinos de mediciones atmosféricas y solares. Se muestran también las cotas experimentales para el ángulo de mezcla calculado; las rectas de lineas, lineas con puntos y de puntos corresponden al rango 1$\sigma$, 2$\sigma$ y 3$\sigma$ respectivamente.}%
\label{so13}%
\end{figure}

Remarcamos que estudios sobre el espacio de parámetros en el $\mu \nu$SSM han sido llevados a cabo en trabajos anteriores, ver por ejemplo Refs.~\cite{gamma,neutrinocp,neutrinoindio}, sin embargo, las cotas experimentales hasta el momento permitían un ángulo de mezcla $\theta_{13}$ nulo. Entonces, se consideraban regímenes maximales tomando $Y_{\nu_{2}}$=$Y_{\nu_{3}}$ y $\nu_{2}$=$\nu_{3}$, simplificando el análisis al reducir el número de parámetros libres. En este trabajo, se utilizaron los regímenes maximales como condiciones iniciales y se realizaron variaciones alrededor de ellos considerando los datos actuales.

A continuación se presentan ejemplos y resultados del barrido del espacio de parámetros considerando la relación GUT entre las masas de gauginos, $M_2/M_1=2$. En la Figura~\ref{so13} se muestra el ángulo de mezcla $\sin^{2}\theta_{13}$ en función de las diferencias de masa entre los neutrinos junto con líneas discontinuas que indican las cotas experimentales de 1$\sigma$, 2$\sigma$ y 3$\sigma$. En general las soluciones favorecen un ángulo de mezcla $\sin^{2}\theta_{13}$ y una diferencia de masa $\Delta m^{2}_{\text{atm}}$, en el extremo inferior de las cotas experimentalmente permitidas.

\begin{figure}
\makebox[\textwidth][c]{\centering
\begin{tabular}{c c}
               \includegraphics[scale=0.85]{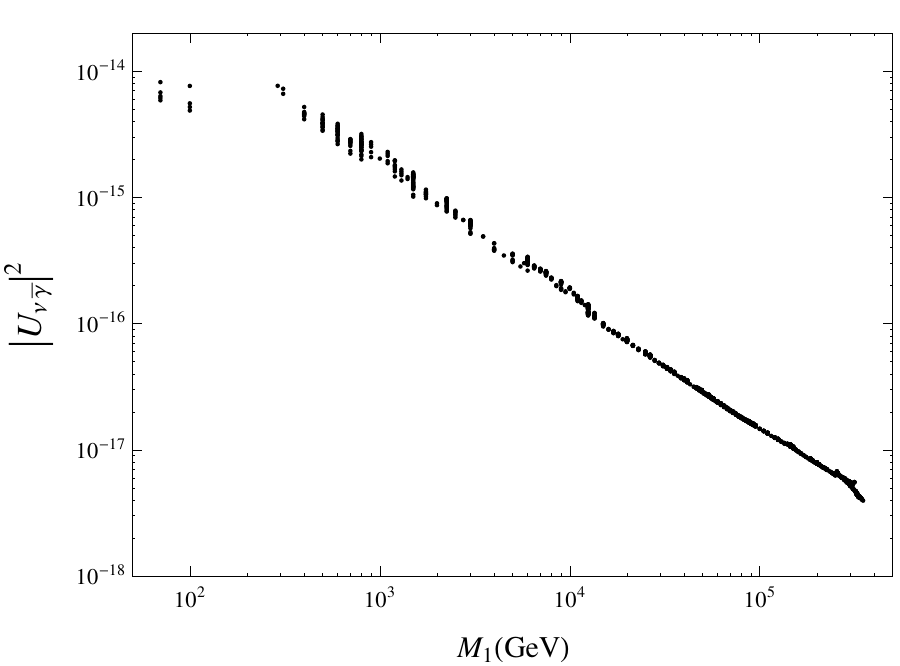}      & \includegraphics[scale=0.85]{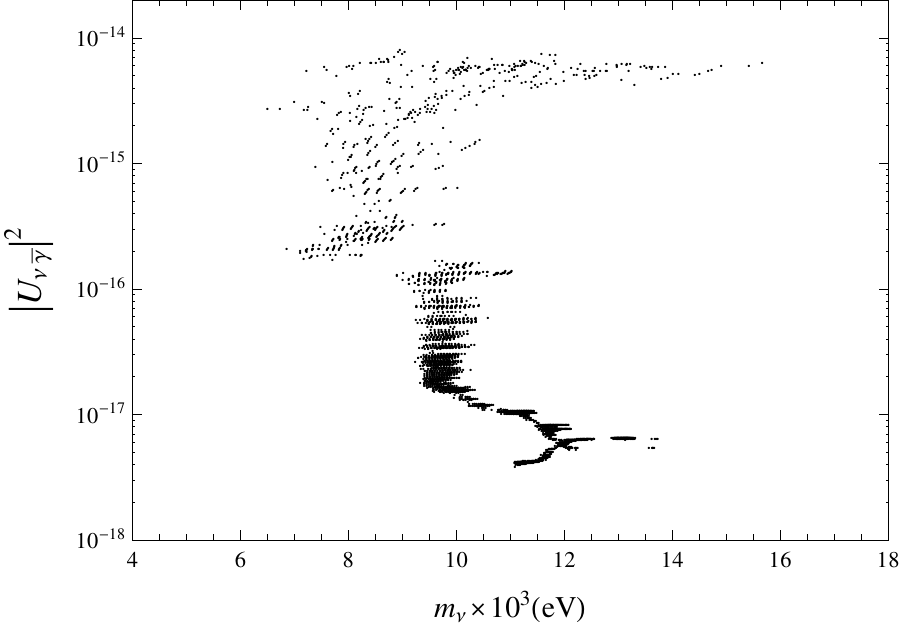}  
		
   \end{tabular}}
\caption{$|U_{\overline{\gamma}\nu}|^{2}$ en función de la masa de los gauginos $|M_{1}|$ (izquierda) y en función de la masa del neutrino mas ligero (derecha), para los parámetros analizados que reproducen la física de neutrinos, con $M_2/M_1=2$.}%
\label{neutripuntos}%
\end{figure}

En la Figura~\ref{neutripuntos} (panel izquierdo) se presenta la contaminación de fotino en función de la masa de los gauginos $|M_1|$. Un amplio rango de valores para $|U_{\overline{\gamma}\nu}|^{2}$ puede ser hallado, entre $10^{-18}-10^{-14}$, a medida que se aumenta el valor de $M_1$. 
Sin embargo, como hemos fijado $M_2/M_1=2$, es necesario tomar valores de $|M_1|$ muy grandes para extender el rango de $|U_{\overline{\gamma}\nu}|^{2}$, por ejemplo necesitamos $|M_1|>10^5$ GeV para $|U_{\overline{\gamma}\nu}|^{2}<10^{-17}$. Como discutimos en la sección anterior, es más eficiente considerar casos con $M_2 \rightarrow M_1$ y valores de $|M_i|$ más cercanos a la escala electrodébil.

En la Figura~\ref{neutripuntos} (panel derecho) se muestra la contaminación de fotino en función de la masa del neutrino más ligero. La disminución de la dispersión de las soluciones con valores decrecientes de $|U_{\overline{\gamma}\nu}|^{2}$ es consecuencia de la estrategia utilizada para explorar el rango de la contaminación fotino-neutrino. Los puntos hallados con $|U_{\overline{\gamma}\nu}|^{2}$ del orden $10^{-15}-10^{-14}$ corresponden a una excursión amplia del espacio de parámetros, obteniendo diferentes valores para la masa del neutrino más ligero. 
A medida que se incrementa el valor de $|M_1|$, se introduce una jerarquía al paulatinamente ir desacoplando los gauginos del resto de los neutralinos. Entonces empezamos a trabajar en el primer caso límite presentado en la Sección~\ref{seccionmeff} al estudiar el comportamiento del mecanismo de $seesaw$ en el $\mu\nu$SSM, donde $m_{eff}$ está dada por Ec.~(\ref{limite1}). Bajo esas condiciones, podemos separar los parámetros libres que consideramos en el análisis en dos grupos:



i. el conjunto de parámetros que no están involucrados en Ec.~(\ref{limite1}), es decir, la masa de los gauginos $M_i$ y los VEVs de los neutrinos izquierdos $v_{\nu}$; 

ii. el conjunto de parámetros involucrados en Ec.~(\ref{limite1}), es decir, los acoples de Yukawa $Y_{\nu_{i}}$, los neutrinos derechos $\nu^{c}$, el parámetro $\kappa$ y $\beta$ o $v_u$.

Partiendo de un punto del espacio de parámetros que reproduzca la física de neutrinos, si los gauginos están desacoplados, al aumentar $M_1$ no es necesario realizar un cambio significativo de los parámetros del segundo conjunto para continuar reproduciendo la física de neutrinos hallada en la iteración previa, por lo tanto la masa de los neutrinos varía levemente.

\section{Flujo de rayos gamma y \Fermi LAT}

A continuación veremos qué tipo de señal se espera que los experimentos detecten si el gravitino conforma la materia oscura. Para simular el espectro de rayos gamma debemos tener en cuenta la anchura de decaimiento de los distintos canales, la región del cielo que se analiza, el perfil de densidad de materia oscura que se emplea en el modelo, y la respuesta del detector.

Las contribuciones a las emisiones de rayos gamma observadas por \Fermi LAT, debido al decaimiento de gravitino como candidato a materia oscura, pueden tener tres orígenes: i) halo galáctico de la Vía Láctea, ii) sub-halos que orbiten al halo galáctico, iii) estructuras extragalácticas. La señal proveniente de origen extragaláctico se espera que sea isotrópica, y en contraste con el caso de aniquilación de materia oscura, independiente de la cantidad de clustering de materia oscura en cada valor de corrimiento al rojo~\cite{2012MNRAS.421L..87S,2002PhR...372....1C,Sefusatti:2014vha}.  Consideraremos las emisiones de i) y iii) en este análisis, lo cual constituye el caso más conservativo y al mismo tiempo, el más independiente de modelo. Entonces,   
\begin{equation}\label{eq:decayFluxTotal}
 \frac{d\Phi_{\gamma}^{\text{total}}}{dEd\Omega}= \frac{d\Phi_{\gamma}^{\text{halo}}}{dEd\Omega} +  \frac{d\Phi_{\gamma}^{\text{extragal}}}{dEd\Omega}\ .
\end{equation}

El flujo de rayos gamma del decaimiento de gravitinos en el halo galáctico se calcula integrando la distribución de materia oscura a través de la línea de visión tierra-fuente.
\begin{equation}\label{eq:decayFlux}
 \frac{d\Phi_{\gamma}^{\text{halo}}}{dEd\Omega}=\frac{1}{4\,\pi\,\tau_{3/2}\,m_{3/2}}\,\frac{1}{\Delta\Omega}\,\frac{dN^{\text{total}}_{\gamma}}{dE} \int_{\Delta\Omega}\!\!\cos
 b\,db\,d\ell\int_0^{\infty}\!\! ds\,\rho_{\text{halo}}(r(s,\,b,\,\ell))\ ,
\end{equation}
donde $b$ y $\ell$ denotan la latitud y longitud galácticas respectivamente, y $s$ representa la distancia a el sistema solar, el parámetro $\Delta \Omega$ es la región de interés, o ROI por sus siglas en inglés, y el radio $r$ en el perfil de densidad de materia oscura del halo de la Vía Láctea, $\rho_{\text{halo}}$, se expresa en función de las coordenadas galácticas como
\begin{equation}
 r(s,\,b,\,\ell)=\sqrt{s^2+R_{\odot}^2-2\,s\,R_{\odot}\cos{b}\cos{\ell}}\ ,
\end{equation}
donde $R_{\odot}\simeq 8.5$ kpc es el radio de la órbita solar alrededor del centro galáctico.
El número total de fotones producidos en el decaimiento del gravitino puede ser expresado como
\begin{equation}
\frac{dN_{\gamma}^{\text{total}}}{dE}=\sum_{i} BR_i\frac{dN_{i}}{dE}\ ,
\label{eq:dndephotona}
\end{equation}
\noindent donde $dN_{i}/dE$ es el espectro de energía de los fotones producidos por los diferentes canales de decaimiento estudiados en las secciones anteriores. Para calcular $dN_{i}/dE$ se empleó Pythia 8.205~\cite{Sjostrand:2014zea}. Se creó una resonancia con energía igual a la masa del gravitino que solo permite que decaiga en un canal particular $i$. Luego Pythia hadroniza las partículas producidas y hace decaer a los hadrones principalmente a leptones que luego generan fotones a través de procesos de QED. Los eventos son almacenados en un histograma, a partir del cual se crea una tabla involucrando todos los posibles canales de decaimiento del gravitino para una masa fija. Para realizar lo descrito se utilizó Monash tune~\cite{Skands:2014pea} para correr Pythia.

Definimos el factor-J, el cual contiene los parámetros astrofísicos involucrados en Ec.~(\ref{eq:decayFlux}) que describe el flujo de rayos gamma proveniente del halo galáctico por decaimiento de materia oscura:
\begin{equation}
	J_{\text{dec}}=\int_{los}\rho_{\text{halo}}(\overline{l}) \ d\overline{l}=\int_{\Delta\Omega}\cos b\  db\  dl \int_{0}^{\infty}\rho_{\text{halo}}(r(s,b,l)) \ ds.
\end{equation}
El ángulo sólido que subtiende la región analizada, definida por la Colaboración Fermi en Ref.~\cite{Ackermann:2015lka} como R180 o halo galáctico, es $\Delta \Omega=$10.4. El factor-J para distintos modelos de perfiles de densidades de materia oscura es

$\hspace{2.1cm} \text{NFW} \hspace{4.15cm} \rightarrow \hspace{0.5cm} 21.9 \times 10^{22} \text{ GeV cm}^{-2}$,

$\hspace{2.1cm} \text{Einasto} \hspace{3.8cm} \rightarrow \hspace{0.5cm} 21.9 \times 10^{22} \text{ GeV cm}^{-2}$,

$\hspace{2.1cm} \text{Isotermal} \hspace{3.5cm} \rightarrow \hspace{0.5cm} 22.7 \times 10^{22} \text{ GeV cm}^{-2}$,

$\hspace{2.1cm} \text{NFW contraido } (\gamma =1.3) \hspace{0.8cm} \rightarrow \hspace{0.5cm} 21.1 \times 10^{22} \text{ GeV cm}^{-2}$.

Como los perfiles difieren en la vecindad del centro galáctico pero son similares en las regiones exteriores, los factores-J que se obtienen para distintos perfiles de densidades solo introducen una incertidumbre del orden del 10\% en el flujo de rayos gamma producto del decaimiento de materia oscura en el halo galáctico. En el resto de esta sección emplearemos el perfil NFW como referencia.

Por otro lado, la contribución al espectro de rayos gamma de estructuras extragalácticas puede ser modelada por
\begin{equation}\label{eq:decayFlux2}
 \frac{d\Phi_{\gamma}^{\text{extragal}}}{dEd\Omega}=\frac{c}{4\pi}\frac{\Omega_{DM} \; \rho_{c}}{m_{3/2} \; \tau_{3/2}}\frac{E_{\gamma}}{H_0}\int^{\infty}_{E_{\gamma}}{dE'_{\gamma}\frac{E_{\gamma}}{E'_{\gamma}}\frac{Q_{\gamma}(E_{\gamma},E'_{\gamma})}{\sqrt{\Omega_{\Lambda}+\Omega_M(E_{\gamma},E'_{\gamma})^{3}}}}\ ,
\end{equation}
donde $c$ es la velocidad de la luz, y utilizamos los valores de los parámetros cosmológicos medidos por la colaboración Planck combinados con WMAP~\cite{Ade:2013zuv}: $H_0 = 67.04$ km s$^{-1}$ Mpc$^{-1}$, $\Omega_M = 0.3183$, $\Omega_{DM} = 0.2678$, $\Omega_{\Lambda} = 0.6817$, y $\rho_c = 1.054\times 10^{-5}$ h$^2$ GeV cm$^{-3}$.
Además,
$E'_{\gamma}=(1+z)E_{\gamma}$ es la energía de los rayos $\gamma$ cuando son producidos en un valor del corrimiento al rojo, o redshift, $z$, y
\begin{equation}\label{eq:Q}
 Q_{\gamma}(E_{\gamma},E'_{\gamma})=e^{-\tau(z,E_{\gamma})}(1+z)\frac{dN_{\gamma}^{\text{total}}}{dE}
 \ .
\end{equation}
\noindent Donde $\tau(z,E_{\gamma})$ es la profundidad óptica, para la cual tomamos el resultado dado en Ref.~\cite{2012MNRAS.422.3189G}. 

Usando estas fórmulas, podemos calcular la línea espectral y el flujo esperado producido en el decaimiento del gravitino como materia oscura. Como nuestros resultados y predicciones deben ser comparados con las observaciones de rayos gamma, en la próxima subsección discutiremos las observaciones de \Fermi LAT relevantes para nuestro trabajo.

\subsection{Observaciones de \Fermi LAT}

Las observaciones del cielo de $\gamma$-ray realizadas por \Fermi LAT poseen un detalle sin precedente y constituyen un avance experimental crucial de los últimos años. La mayoría de los rayos gamma detectados provienen de estructuras puntuales o de fuentes con extensiones muy pequeñas, y se identificó una emisión difusa prominente correlacionada con estructuras galácticas~\cite{diffuse2}. Además, una tenue componente difusa también ha sido detectada, el fondo isotrópico de rayos gamma (IGRB por sus siglas en inglés)~\cite{2010PhRvL.104j1101A}. El origen del IGRB pueden ser fuentes que permanecen por debajo del límite de  detección de \Fermi LAT, por ejemplo, materia oscura que decae o se aniquila a rayos gamma~\cite{Bergstrom:2001jj,Ullio:2002pj} puede conformar una contribución significativa del IGRB\footnote{Las contribuciones más importantes al IGRB provienen de blazares, galaxias con importante actividad en la formación de estrellas, procesos difusos como shocks intergalácticos~\cite{Colafrancesco:1998us,Loeb:2000na,Zandanel:2013wea}, interacciones de rayos cósmicos ultra energéticos con los fotones de fondo (EBL)~\cite{Berezinsky:1975zz}, e interacciones de rayos cósmicos con sistemas estelares~\cite{Moskalenko:2009tv}.}.

El IGRB observado depende del límite de detección de fuentes puntuales del instrumento que se utilice. En cambio, la cantidad física es la totalidad de la luz de fondo o EGB por sus siglas en inglés, definida como la combinación de fuentes identificadas y el IGRB. La colaboración \Fermi LAT ha determinado el EGB usando datos tomados en 50 meses reprocesados con el método Pass 8 event-level analysis, que cubre un rango desde los 100 MeV a 820 GeV~\cite{Ackermann:2014usa}. En la figura 3 de dicho trabajo, se muestra la cantidad integrada de las cuentas del instrumento LAT por encima de 100 MeV que han sido utilizadas  y las regiones alrededor del plano galáctico que han sido enmascaradas en el análisis.

\begin{figure}[t!]
 \begin{tabular}{cc}
 \hspace*{-4mm}
       \epsfig{file=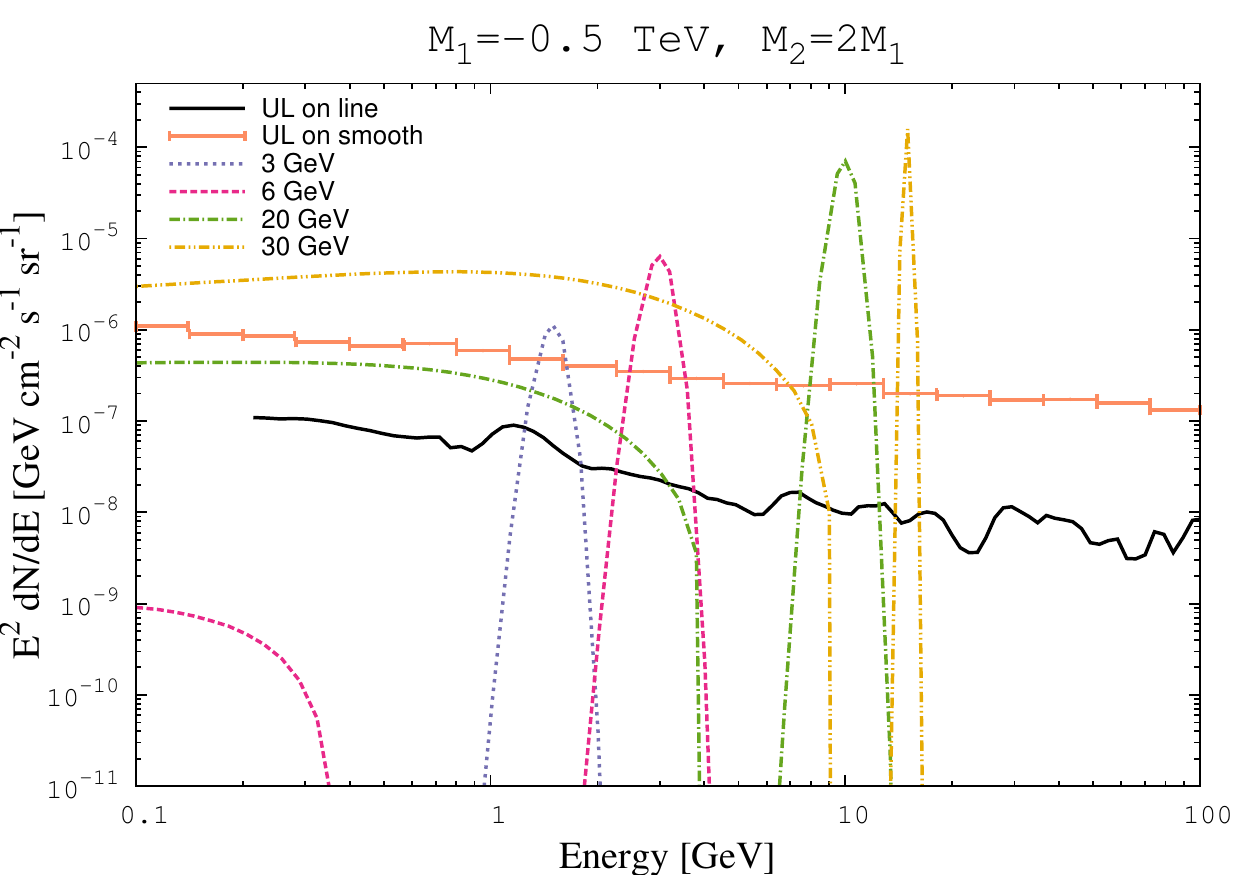,height=5.4cm}
       \epsfig{file=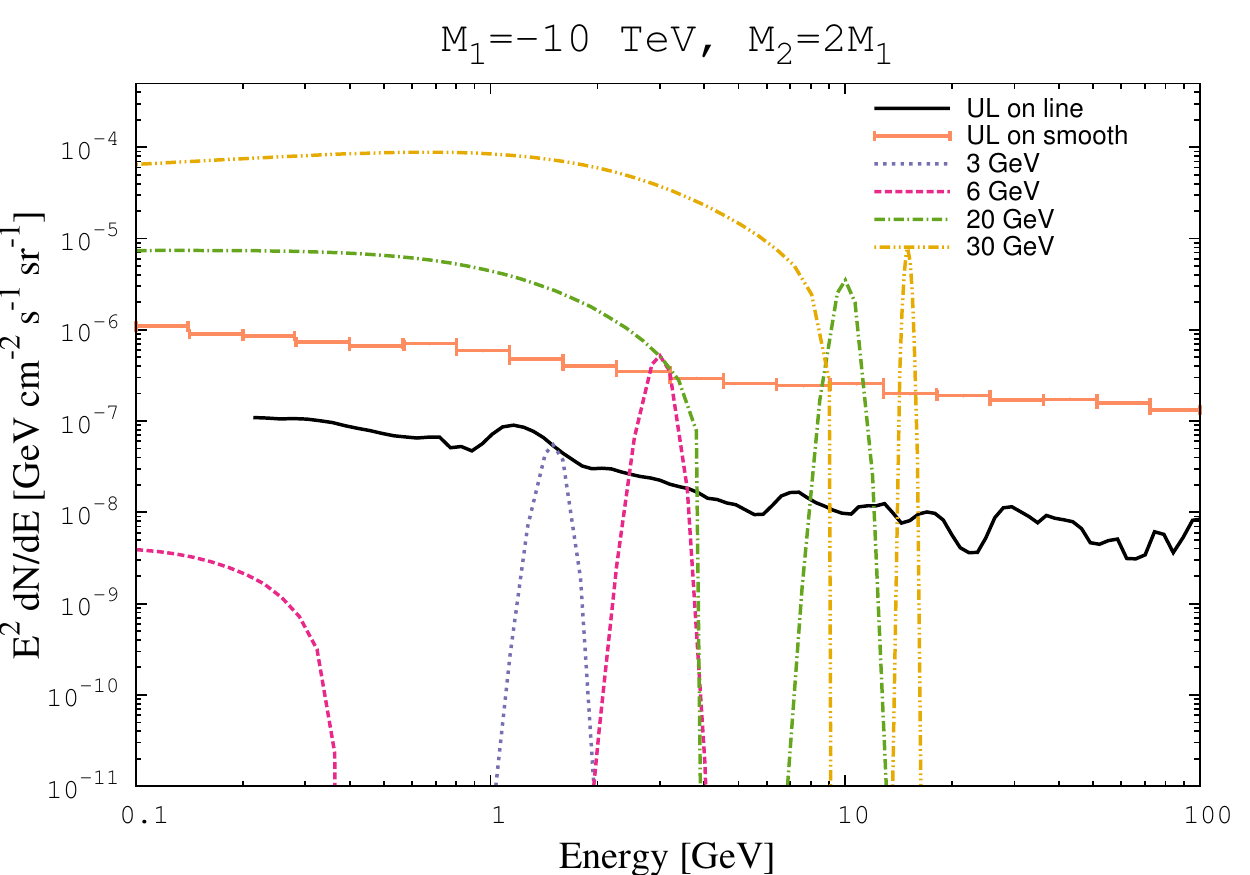,height=5.4cm} \vspace*{0.cm}\\
     (a)\hspace*{3cm} & \hspace*{-2cm} (b)\\
\hspace*{-4mm}
       \epsfig{file=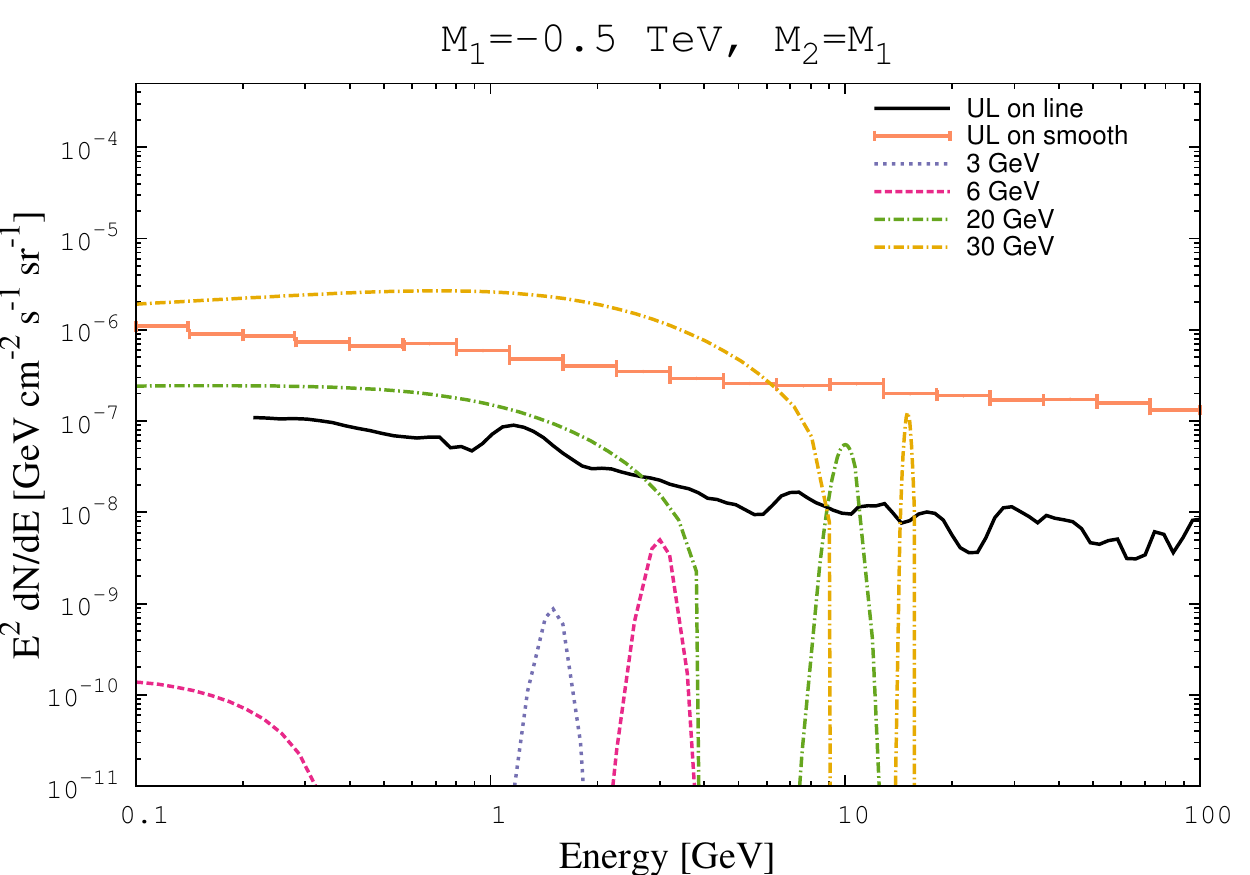,height=5.4cm}
       \epsfig{file=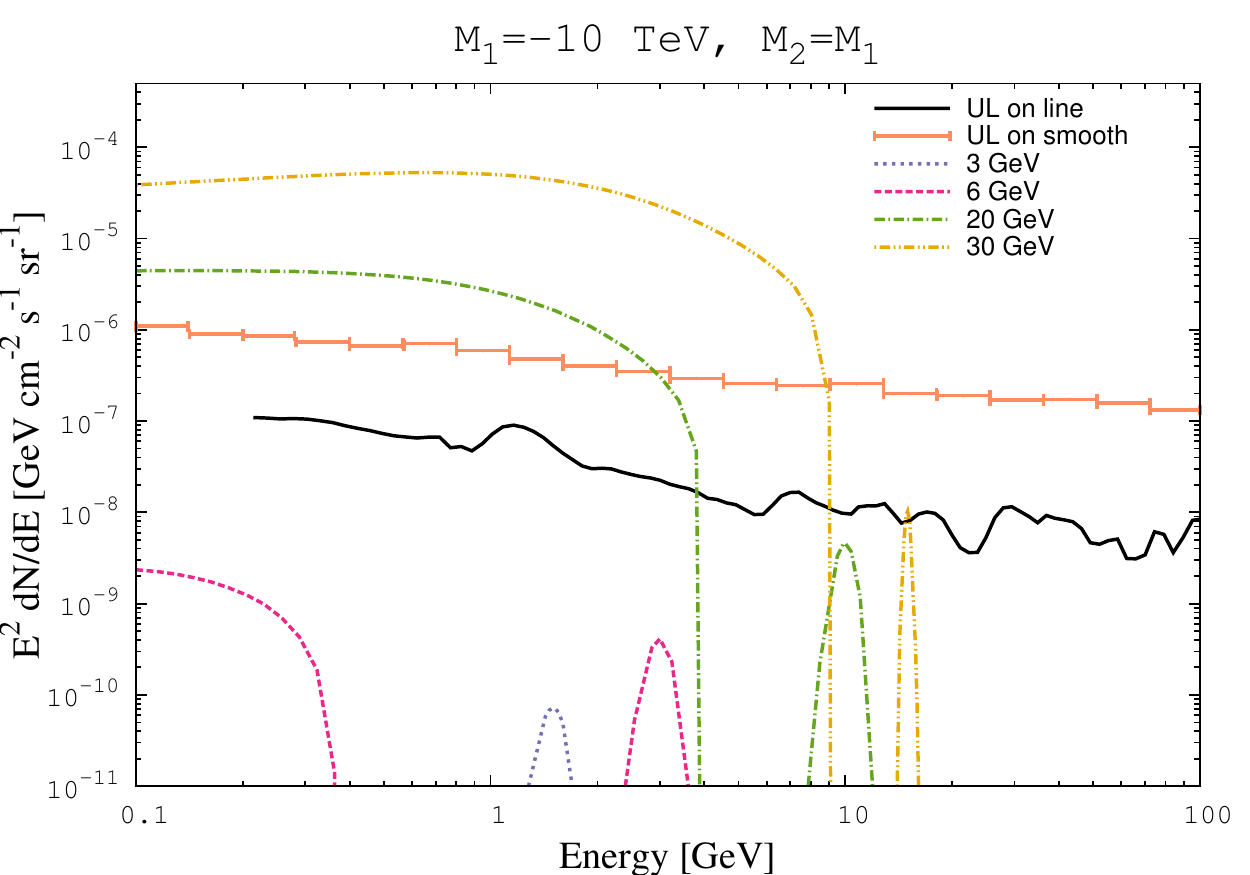,height=5.4cm} \vspace*{0.cm}\\
     (c)\hspace*{3cm} & \hspace*{-2cm} (d) \vspace*{0.1cm} \\
    \end{tabular}
\caption{Espectro de rayos gamma esperado para distintos ejemplos del decaimiento de gravitino como materia oscura, $m_{3/2}=3,6,20,30$ GeV, usando cuatro combinaciones de masas de gauginos, (a) $M_2=2M_1$ con $M_1=-0.5$ TeV, (b) $M_2=2M_1$ con $M_1=-10$ TeV, 
(c) $M_2=M_1$ con $M_1=-0.5$ TeV, (d) $M_2=M_1$ con $M_1=-10$ TeV.
Las curvas negras corresponden a los límites superiores (UL) provenientes de la búsqueda de líneas espectrales, mientras que los puntos naranjas corresponden a los límites superiores de la determinación del EGB, luego de sustraer los modelos de las contribuciones conocidas. Ambos límites emplean datos de \Fermi LAT.
}
    \label{fig_EGBa}
\end{figure}

En la Figura~\ref{fig_EGBa} mostramos los límites superiores con 95\% C.L. en la determinación del EGB (puntos naranjas). Para obtener estos límites el promedio de las emisiones de las contribuciones provenientes de fuentes no exóticas han sido sustraídas\footnote{Asumiendo errores Gaussianos, donde el 95\% del área de una distribución Gaussiana se encuentra dentro de 1,64 desviaciones estándares de la media.}. Las contribuciones no exóticas en el EGB que se han considerado son: galaxias con formación de estrellas~\cite{Ackermann:2012vca}, galaxias que emiten fuertemente en radio~\cite{Inoue:2011bm} y la emisión integrada de blazars con la absorción del EBL modelada en~\cite{AjelloBlazars}. Los límites han sido tomados de~\cite{Carquin:2015uma}. Utilizaremos estos límites para sondear y restringir la señal espectral continua producto del decaimiento a tres cuerpos del gravitino como materia oscura.

Por otro lado, la línea espectral generada por decaimientos a dos cuerpos posee una energía igual a la mitad de la masa del gravitino, y en general, es la señal dominante en el espectro de rayos gamma producidos por el candidato considerado. Utilizamos los límites en la emisión de líneas (curva negra) a partir de un análisis realizado por la colaboración \Fermi LAT~\cite{Ackermann:2015lka}, donde han utilizado la denominada R180 ROI. Dicha ROI está definida como la región circular de radio 180$^{\text{o}}$ centrada en el centro galáctico. Además, la región del plano galáctico con longitud superior a 6$^{\text{o}}$ del centro galáctico y latitud más chica que 5$^{\text{o}}$ ha sido removida (ver Figura~4 de Ref.~\cite{Ackermann:2015lka}).
Los límites son establecidos utilizando el método de análisis desarrollado en Ref.~\cite{Albert:2014hwa}, que tiene en cuenta las incertezas sistemáticas en la parte de bajas energías de la banda de medición de \Fermi LAT. Se han tenido en cuenta 69.9 meses de datos.

Por lo tanto, en la próxima sección se usaran los límites del EGB~\cite{Carquin:2015uma} y las búsquedas de líneas espectrales realizadas por \Fermi LAT~\cite{Ackermann:2015lka} para establecer restricciones en el espacio de parámetros del gravitino como materia oscura en el $\mu\nu$SSM.

\section{Límites de gravitino DM con los datos de \Fermi LAT}

En esta sección aplicaremos las fórmulas anteriores para calcular el flujo de rayos gamma a partir de gravitinos como candidatos a materia oscura en el $\mu \nu$SSM, siguiendo el análisis de Ref.~\cite{Albert:2014hwa} respecto a la aplicación de los límites de exclusión de \Fermi LAT.

En la Figura~\ref{fig_EGBa} presentamos la forma del espectro de rayos gamma para cuatro masas diferentes de gravitino DM, $m_{3/2}=3, 6, 20, 30$ GeV. Para cada caso, la línea espectral correspondiente a la mitad de la masa del gravitino proviene del canal de decaimiento a dos cuerpos, convolucionado con la dispersión debido a la resolución de \Fermi LAT \footnote{http://www.slac.stanford.edu/exp/glast/groups/canda/lat\_Performance.htm}. Como es esperado a partir de Ec.~(\ref{decay2body})
y~(\ref{eq:decayFlux}), cuanto más grande sea $m_{3/2}$ más grande es la anchura de decaimiento y el flujo, produciendo una señal más brillante. Por otra parte, la señal continua y extendida corresponde a los canales de decaimiento a tres cuerpos. Para establecer las restricciones comparamos la señal angosta con los límites superiores provenientes de la búsqueda de líneas (curvas negras) y la señal extendida con los límites superiores del EGB (puntos naranjas), calculando la señal producida por el gravitino de acuerdo a la ROI utilizada para calcular dichos límites.

Los paneles (a) y (b) de la Figura~\ref{fig_EGBa} corresponden a escenarios con relaciones del tipo GUT para los parámetros $M_2 = 2 \, M_1$. Como podemos ver, los flujos correspondientes a las masas de gravitino ejemplificadas se encuentran excluidos por los límites de la búsqueda de líneas espectrales en ambos paneles, salvo el caso con $m_{3/2}=3$ GeV en el panel (b) donde la masa de los gauginos ha sido incrementada más de un orden de magnitud respecto a los valores tomados en el panel (a). Esto es consistente con los resultados en Ref.~\cite{Albert:2014hwa}. Sin embargo, en los paneles (c) y (d), donde se consideran los casos límite para la relación entre las masas de gauginos, $M_2 = M_1$, es evidente la supresión de la intensidad de la línea con respecto a los paneles previos. Como es esperado, para evadir los límites impuestos por \Fermi LAT, es más eficiente hacer los valores de las masas de gauginos similares que incrementar sus valores absolutos. En estos casos, masas de gravitinos de 3 y 6 GeV son permitidas para ambos paneles (c) y (d). Es importante notar que un gravitino de masa igual a 20 GeV es permitido por los límites de líneas para el caso $M_1=-10$ TeV, sin embargo este punto se encuentra excluido debido a los límites de EGB, indicando que los decaimientos a tres cuerpos son relevantes en el análisis.

En conclusión, aumentar la masa de los gauginos $M_{1,2}$ y/o reducir la relación $M_2/M_1$ suprime la intensidad de la línea. Además, los límites de líneas espectrales son lo suficientemente restrictivos como para excluir casi todos los escenarios excepto los extremos, es decir, con masas de gauginos muy altas y cercanas entre sí.

\begin{figure}[ht!]
 \begin{center}
       \epsfig{file=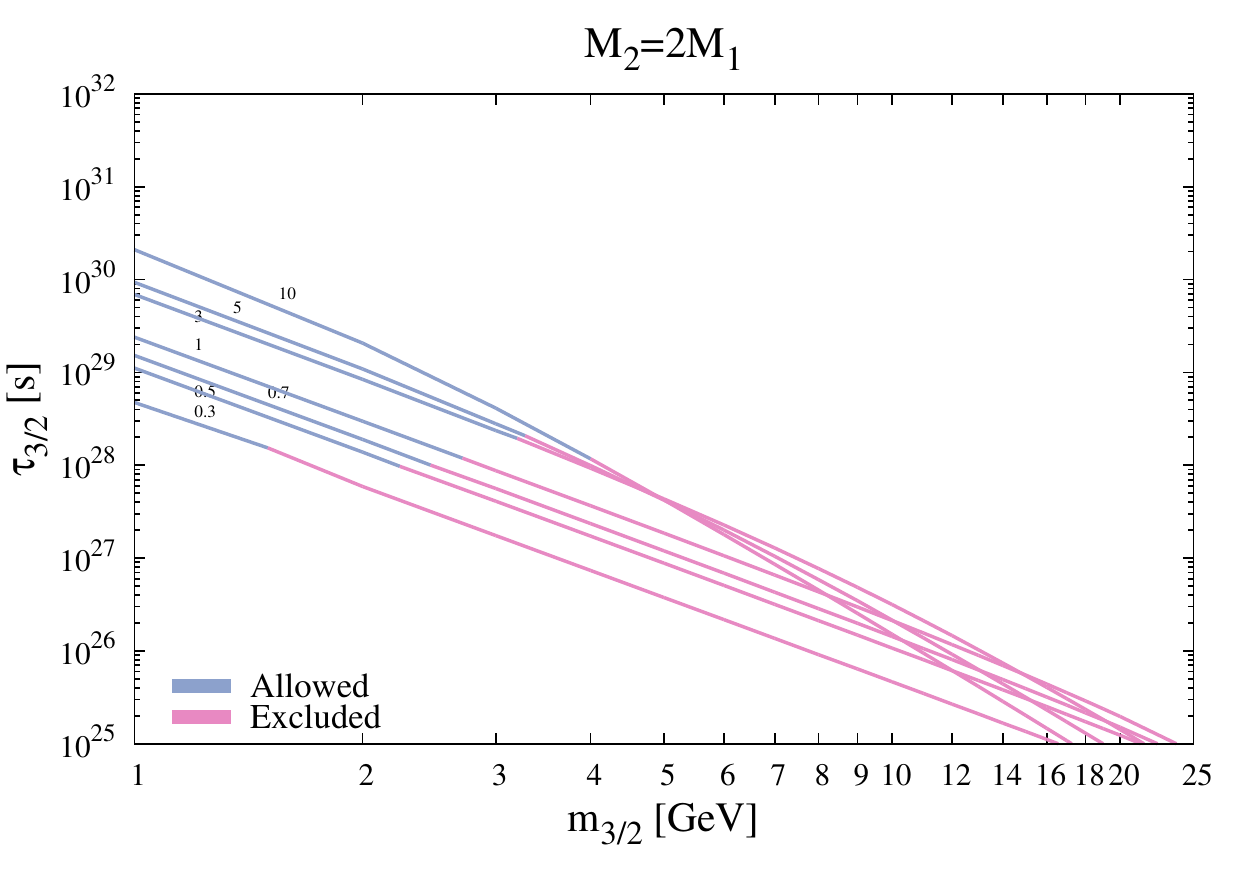,height=6.5cm}
       \epsfig{file=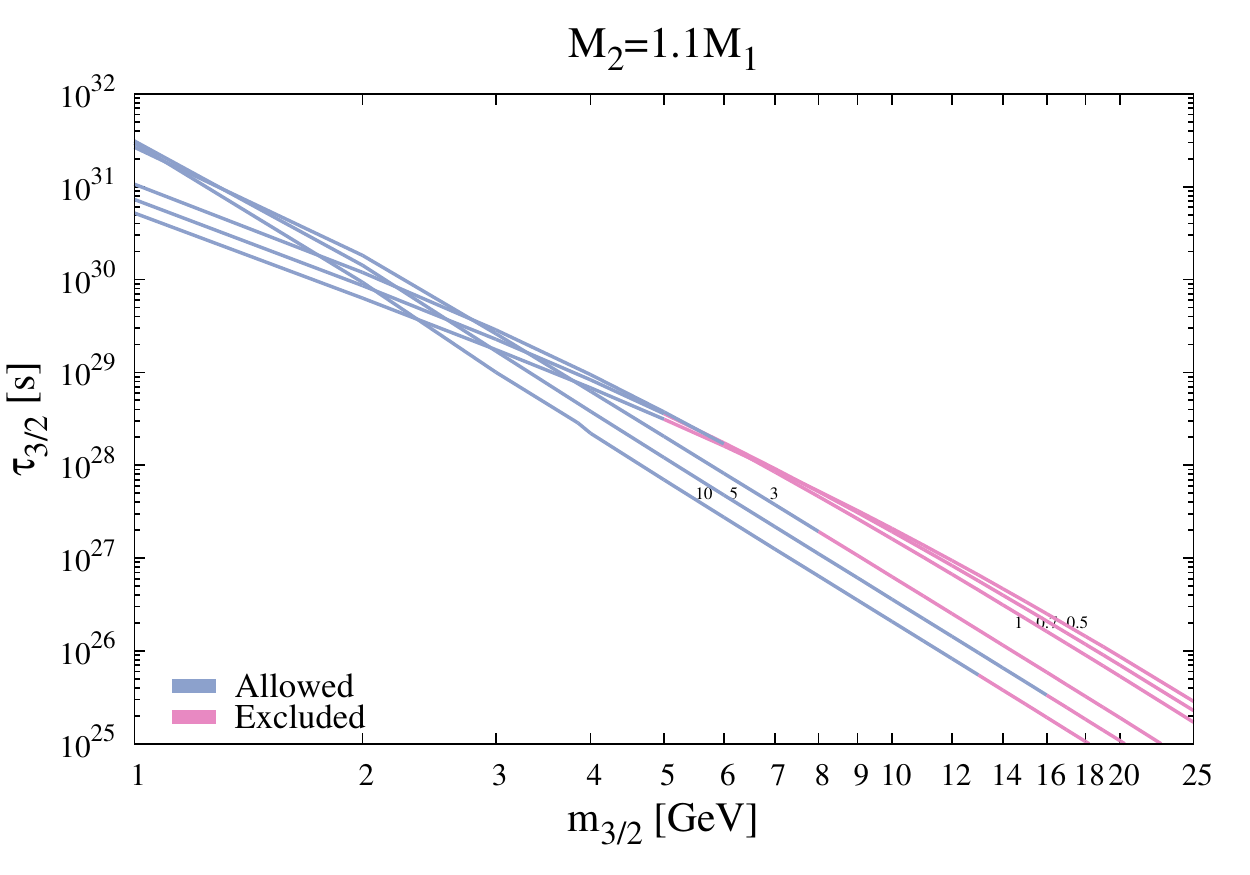,height=6.5cm}
       \epsfig{file=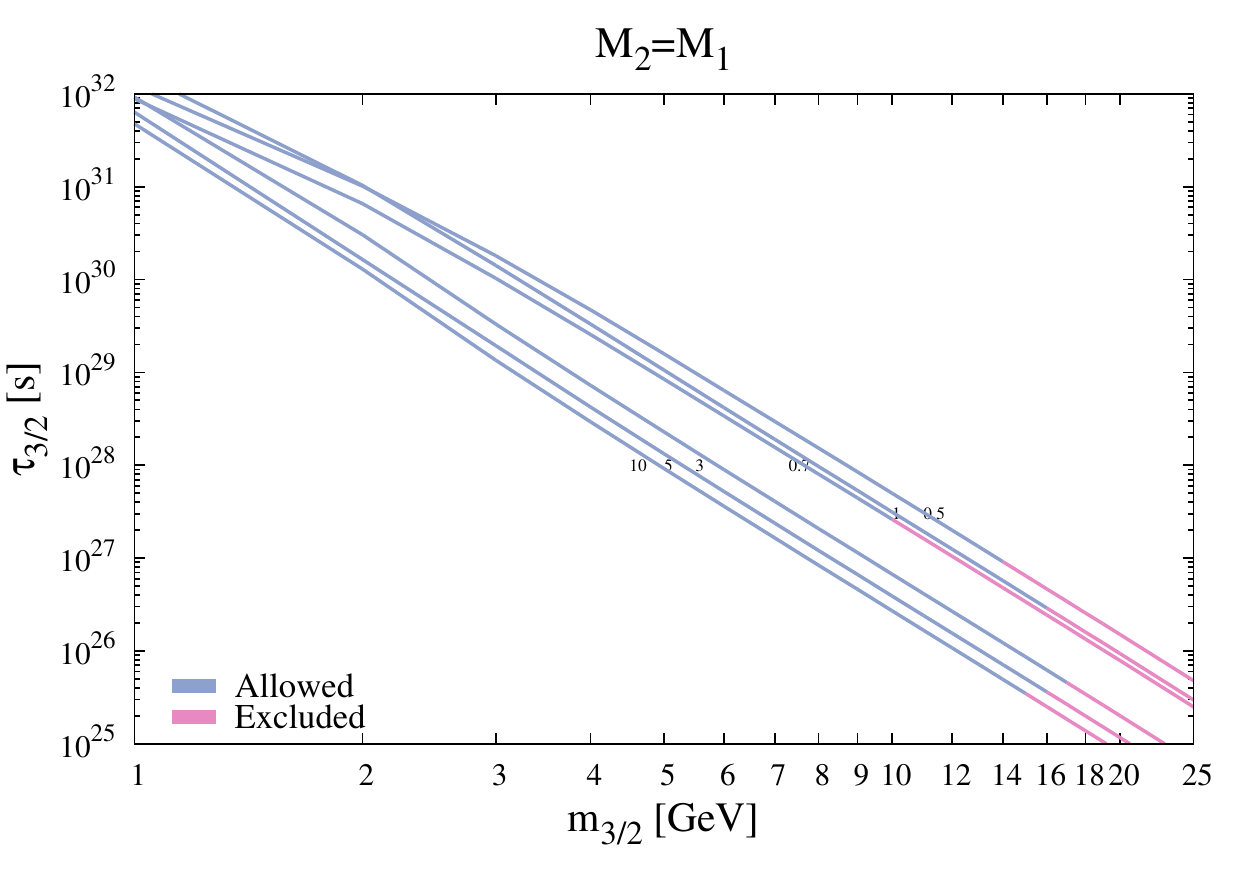,height=6.5cm}
  \caption{Tiempo de vida en función de la masa del gravitino como candidato a materia oscura. Se muestran tres relaciones para la masa de los gauginos $M_2=2M_1$, $M_2=1.1 M_1$ y $M_2=M_1$ en el panel superior, medio e inferior correspondientemente. Para cada relación, las curvas representan distintos valores de $M_1=-0.5, -0.7, -1, -3, -5, -10$ TeV, identificadas con una etiqueta con $|M_1|$. La sección azul de cada curva se encuentra permitida, mientras que la sección magenta se encuentra excluida por los datos de \Fermi LAT. 
}
    \label{figconstrains}
\end{center}
\end{figure}

En la Figura~\ref{figconstrains} mostramos los límites en el plano tiempo de vida media del gravitino versus su masa, para tres relaciones diferentes entre $M_1$ y $M_2$: $M_2=2M_1$, $M_2=1.1 M_1$ y $M_2=M_1$. Para cada uno de los tres casos presentamos curvas con los siguientes valores de $M_1$: $M_1=-0.5, -0.7, -1, -3, -5, -10$ TeV, y los mismos valores representativos para los parámetros de bajas energías que se utilizaron en la Figura~\ref{fig2}:
$\lambda = 0.1 $, $\kappa = 0.1$, $\tan\beta = 10$, $v_{\nu^c} = 1750$~GeV (recordamos que $\mu_{eff}=3\lambda v_{\nu^c}$). En las Tablas~\ref{param2}, \ref{param11} y \ref{param1} mostramos para cada caso, los parámetros relevantes para reproducir el patrón de masas de los neutrinos y sus ángulos de mezcla, junto con los valores correspondientes de $|U_{\tilde{\gamma} \nu}|^2$. Podemos observar que el parámetro de mezcla alcanza su valor inferior límite solo para los casos más extremos con $M_2 \to M_1$.

La sección azul de cada curva en la Figura~\ref{figconstrains} indica los valores permitidos de $m_{3/2}$ y $\tau_{3/2}$, mientras que las secciones magenta indican los valores excluidos por los datos de \Fermi LAT. Como hemos mencionado anteriormente, en general la exclusión dominante proviene de los límites de búsquedas de líneas espectrales, generado por los canales de decaimiento a dos cuerpos al producir la señal más brillante y dominante. Más aún, las restricciones impuestas por \Fermi LAT sobre la búsqueda de líneas son más restrictivas que los límites del EGB, pues son más fáciles de identificar.

\newpage
\thispagestyle{empty}
\clearpage

\begin{table}
\begin{center}
   \begin{tabular}{| c | c | c | c | }
    \hline
       $M_1=-0.3 \text{TeV}$  &  $v_{\nu_1}=1.7\times10^{-4} \text{GeV}$  & $v_{\nu_2}=0.2\times10^{-4} \text{GeV}$   &  $v_{\nu_3}=0.8\times10^{-4} \text{GeV}$     \\ \hline
       $|U_{\tilde{\gamma} \nu}|^2=7.85\times10^{-15}$ &  $Y_{\nu_{1}}=5\times10^{-7}$  &  $Y_{\nu_{2}}=0.8\times10^{-6}$  &   $Y_{\nu_{3}}=0.9\times10^{-6}$ \\ \hline \hline
       $M_1=-0.5 \text{TeV}$  &  $v_{\nu_1}=1.5\times10^{-4} \text{GeV}$  & $v_{\nu_2}=0.2\times10^{-4} \text{GeV}$   &  $v_{\nu_3}=1.5\times10^{-4} \text{GeV}$     \\ \hline
       $|U_{\tilde{\gamma} \nu}|^2=3.35\times10^{-15}$ &  $Y_{\nu_{1}}=3\times10^{-7}$  &  $Y_{\nu_{2}}=0.7\times10^{-6}$  &   $Y_{\nu_{3}}=0.9\times10^{-6}$ \\ \hline \hline
       $M_1=-0.7 \text{TeV}$  &  $v_{\nu_1}=1.7\times10^{-4} \text{GeV}$  & $v_{\nu_2}=2\times10^{-4} \text{GeV}$   &  $v_{\nu_3}=0.3\times10^{-4} \text{GeV}$     \\ \hline
       $|U_{\tilde{\gamma} \nu}|^2=2.44\times10^{-15}$ &  $Y_{\nu_{1}}=3\times10^{-7}$  &  $Y_{\nu_{2}}=0.9\times10^{-6}$  &   $Y_{\nu_{3}}=0.7\times10^{-6}$ \\ \hline \hline
       $M_1=- 1\text{TeV}$  &  $v_{\nu_1}=2.3\times10^{-4} \text{GeV}$  & $v_{\nu_2}=2\times10^{-4} \text{GeV}$   &  $v_{\nu_3}=0.6\times10^{-4} \text{GeV}$     \\ \hline
       $|U_{\tilde{\gamma} \nu}|^2=1.55\times10^{-15}$ &  $Y_{\nu_{1}}=3\times10^{-7}$  &  $Y_{\nu_{2}}=0.9\times10^{-6}$  &   $Y_{\nu_{3}}=0.7\times10^{-6}$ \\ \hline \hline
       $M_1=-3 \text{TeV}$  &  $v_{\nu_1}=4.3\times10^{-4} \text{GeV}$  & $v_{\nu_2}=4\times10^{-4} \text{GeV}$   &  $v_{\nu_3}=0.8\times10^{-4} \text{GeV}$     \\ \hline
       $|U_{\tilde{\gamma} \nu}|^2=5.35\times10^{-16}$ &  $Y_{\nu_{1}}=3\times10^{-7}$  &  $Y_{\nu_{2}}=0.9\times10^{-6}$  &   $Y_{\nu_{3}}=0.7\times10^{-6}$ \\ \hline \hline
       $M_1=-5 \text{TeV}$  &  $v_{\nu_1}=6.1\times10^{-4} \text{GeV}$  & $v_{\nu_2}=6.2\times10^{-4} \text{GeV}$   &  $v_{\nu_3}=0.4\times10^{-4} \text{GeV}$     \\ \hline
       $|U_{\tilde{\gamma} \nu}|^2=3.91\times10^{-16}$ &  $Y_{\nu_{1}}=3\times10^{-7}$  &  $Y_{\nu_{2}}=0.9\times10^{-6}$  &   $Y_{\nu_{3}}=0.8\times10^{-6}$ \\ \hline \hline
       $M_1=-10 \text{TeV}$  &  $v_{\nu_1}=8\times10^{-4} \text{GeV}$  & $v_{\nu_2}=8.1\times10^{-4} \text{GeV}$   &  $v_{\nu_3}=1.4\times10^{-4} \text{GeV}$     \\ \hline
        $|U_{\tilde{\gamma} \nu}|^2=1.65\times10^{-16}$ &  $Y_{\nu_{1}}=3\times10^{-7}$  &  $Y_{\nu_{2}}=0.9\times10^{-6}$  &   $Y_{\nu_{3}}=0.7\times10^{-6}$ \\ \hline
   \end{tabular}
    \caption{Parámetros relevantes para reproducir el patrón de masas observado y los ángulos de mezcla del sector de los neutrinos para los casos con $M_2=2M_1$ de la Figura~\ref{figconstrains}. También se muestran los valores correspondientes al parámetro de mezcla fotino-neutrino $|U_{\tilde{\gamma} \nu}|^2$.
}
\label{param2} 
\end{center}
\end{table}

\begin{table}
\begin{center}
   \begin{tabular}{| c | c | c | c | }
    \hline
       $M_1=-0.3 \text{TeV}$  &  $v_{\nu_1}=1.3\times10^{-4} \text{GeV}$  & $v_{\nu_2}=0.6\times10^{-4} \text{GeV}$   &  $v_{\nu_3}=0.09\times10^{-4} \text{GeV}$     \\ \hline
       $|U_{\tilde{\gamma} \nu}|^2=1.25\times10^{-16}$ &  $Y_{\nu_{1}}=5\times10^{-7}$  &  $Y_{\nu_{2}}=0.9\times10^{-6}$  &   $Y_{\nu_{3}}=0.8\times10^{-6}$ \\ \hline \hline
       $M_1=-0.5 \text{TeV}$  &  $v_{\nu_1}=1.9\times10^{-4} \text{GeV}$  & $v_{\nu_2}=0.8\times10^{-4} \text{GeV}$   &  $v_{\nu_3}=0.03\times10^{-4} \text{GeV}$     \\ \hline
       $|U_{\tilde{\gamma} \nu}|^2=7.10\times10^{-17}$ &  $Y_{\nu_{1}}=5\times10^{-7}$  &  $Y_{\nu_{2}}=0.9\times10^{-6}$  &   $Y_{\nu_{3}}=0.8\times10^{-6}$ \\ \hline \hline
       $M_1=-0.7 \text{TeV}$  &  $v_{\nu_1}=2.2\times10^{-4} \text{GeV}$  & $v_{\nu_2}=1.2\times10^{-4} \text{GeV}$   &  $v_{\nu_3}=0.2\times10^{-4} \text{GeV}$     \\ \hline
       $|U_{\tilde{\gamma} \nu}|^2=5.05\times10^{-17}$ &  $Y_{\nu_{1}}=5\times10^{-7}$  &  $Y_{\nu_{2}}=0.9\times10^{-6}$  &   $Y_{\nu_{3}}=0.8\times10^{-6}$ \\ \hline \hline
       $M_1=-1 \text{TeV}$  &  $v_{\nu_1}=2.8\times10^{-4} \text{GeV}$  & $v_{\nu_2}=1.3\times10^{-4} \text{GeV}$   &  $v_{\nu_3}=0.3\times10^{-4} \text{GeV}$     \\ \hline
       $|U_{\tilde{\gamma} \nu}|^2=3.40\times10^{-17}$ &  $Y_{\nu_{1}}=5\times10^{-7}$  &  $Y_{\nu_{2}}=0.9\times10^{-6}$  &   $Y_{\nu_{3}}=0.8\times10^{-6}$ \\ \hline \hline
       $M_1=-3 \text{TeV}$  &  $v_{\nu_1}=4.7\times10^{-4} \text{GeV}$  & $v_{\nu_2}=3\times10^{-4} \text{GeV}$   &  $v_{\nu_3}=1.3\times10^{-4} \text{GeV}$     \\ \hline
       $|U_{\tilde{\gamma} \nu}|^2=1.09\times10^{-17}$ &  $Y_{\nu_{1}}=5\times10^{-7}$  &  $Y_{\nu_{2}}=0.9\times10^{-6}$  &   $Y_{\nu_{3}}=0.8\times10^{-6}$ \\ \hline \hline
       $M_1=-5 \text{TeV}$  &  $v_{\nu_1}=6.3\times10^{-4} \text{GeV}$  & $v_{\nu_2}=4\times10^{-4} \text{GeV}$   &  $v_{\nu_3}=1.5\times10^{-4} \text{GeV}$     \\ \hline
       $|U_{\tilde{\gamma} \nu}|^2=6.50\times10^{-18}$ &  $Y_{\nu_{1}}=5\times10^{-7}$  &  $Y_{\nu_{2}}=0.9\times10^{-6}$  &   $Y_{\nu_{3}}=0.8\times10^{-6}$ \\ \hline \hline
       $M_1=-10 \text{TeV}$  &  $v_{\nu_1}=7.2\times10^{-4} \text{GeV}$  & $v_{\nu_2}=7.1\times10^{-4} \text{GeV}$   &  $v_{\nu_3}=0.4\times10^{-4} \text{GeV}$     \\ \hline
       $|U_{\tilde{\gamma} \nu}|^2=2.67\times10^{-18}$ &  $Y_{\nu_{1}}=3\times10^{-7}$  &  $Y_{\nu_{2}}=0.9\times10^{-6}$  &   $Y_{\nu_{3}}=0.8\times10^{-6}$ \\ \hline
   \end{tabular}
    \caption{Igual que en la Tabla~\ref{param2} pero para $M_2=1.1M_1$.
}
\label{param11} 
\end{center}
\end{table}

\begin{table}
\begin{center}
   \begin{tabular}{| c | c | c | c | }
    \hline
       $M_1=-0.3 \text{TeV}$  &  $v_{\nu_1}=1.3\times10^{-4} \text{GeV}$  & $v_{\nu_2}=0.5\times10^{-4} \text{GeV}$   &  $v_{\nu_3}=0.01\times10^{-4} \text{GeV}$     \\ \hline
       $|U_{\tilde{\gamma} \nu}|^2=6.05\times10^{-18}$ &  $Y_{\nu_{1}}=5\times10^{-7}$  &  $Y_{\nu_{2}}=0.9\times10^{-6}$  &   $Y_{\nu_{3}}=0.8\times10^{-6}$ \\ \hline \hline
       $M_1=-0.5 \text{TeV}$  &  $v_{\nu_1}=1.3\times10^{-4} \text{GeV}$  & $v_{\nu_2}=0.8\times10^{-4} \text{GeV}$   &  $v_{\nu_3}=0.06\times10^{-4} \text{GeV}$     \\ \hline
       $|U_{\tilde{\gamma} \nu}|^2=2.62\times10^{-18}$ &  $Y_{\nu_{1}}=3\times10^{-7}$  &  $Y_{\nu_{2}}=0.9\times10^{-6}$  &   $Y_{\nu_{3}}=0.7\times10^{-6}$ \\ \hline \hline
       $M_1=-0.7 \text{TeV}$  &  $v_{\nu_1}=2.2\times10^{-4} \text{GeV}$  & $v_{\nu_2}=1\times10^{-4} \text{GeV}$   &  $v_{\nu_3}=0.1\times10^{-4} \text{GeV}$     \\ \hline
       $|U_{\tilde{\gamma} \nu}|^2=3.35\times10^{-18}$ &  $Y_{\nu_{1}}=5\times10^{-7}$  &  $Y_{\nu_{2}}=0.9\times10^{-6}$  &   $Y_{\nu_{3}}=0.8\times10^{-6}$ \\ \hline \hline
       $M_1=-1 \text{TeV}$  &  $v_{\nu_1}=1.8\times10^{-4} \text{GeV}$  & $v_{\nu_2}=1.3\times10^{-4} \text{GeV}$   &  $v_{\nu_3}=0.5\times10^{-4} \text{GeV}$     \\ \hline
       $|U_{\tilde{\gamma} \nu}|^2=1.45\times10^{-18}$ &  $Y_{\nu_{1}}=3\times10^{-7}$  &  $Y_{\nu_{2}}=0.9\times10^{-6}$  &   $Y_{\nu_{3}}=0.7\times10^{-6}$ \\ \hline \hline
       $M_1=-3 \text{TeV}$  &  $v_{\nu_1}=4.7\times10^{-4} \text{GeV}$  & $v_{\nu_2}=2.8\times10^{-4} \text{GeV}$   &  $v_{\nu_3}=0.9\times10^{-4} \text{GeV}$     \\ \hline
       $|U_{\tilde{\gamma} \nu}|^2=9.58\times10^{-19}$ &  $Y_{\nu_{1}}=5\times10^{-7}$  &  $Y_{\nu_{2}}=0.9\times10^{-6}$  &   $Y_{\nu_{3}}=0.8\times10^{-6}$ \\ \hline \hline
       $M_1=-5 \text{TeV}$  &  $v_{\nu_1}=6.1\times10^{-4} \text{GeV}$  & $v_{\nu_2}=4\times10^{-4} \text{GeV}$   &  $v_{\nu_3}=1.1\times10^{-4} \text{GeV}$     \\ \hline
       $|U_{\tilde{\gamma} \nu}|^2=6.11\times10^{-19}$ &  $Y_{\nu_{1}}=5\times10^{-7}$  &  $Y_{\nu_{2}}=0.9\times10^{-6}$  &   $Y_{\nu_{3}}=0.8\times10^{-6}$ \\ \hline \hline
       $M_1=-10 \text{TeV}$  &  $v_{\nu_1}=6.2\times10^{-4} \text{GeV}$  & $v_{\nu_2}=6.1\times10^{-4} \text{GeV}$   &  $v_{\nu_3}=1.2\times10^{-4} \text{GeV}$     \\ \hline 
       $|U_{\tilde{\gamma} \nu}|^2=2.16\times10^{-19}$ &  $Y_{\nu_{1}}=3\times10^{-7}$  &  $Y_{\nu_{2}}=0.9\times10^{-6}$  &   $Y_{\nu_{3}}=0.7\times10^{-6}$ \\ \hline
   \end{tabular}
    \caption{Igual que en la Tabla~\ref{param2} pero para $M_2=M_1$.
}
\label{param1} 
\end{center}
\end{table}

Sin embargo, la señal extendida y continua cumple un papel muy importante aún cuando no constituya el espectro más restrictivo. Esto se debe a la presencia de canales de decaimiento con BR no despreciables que cambian el tiempo de vida media del gravitino, y por lo tanto modifican los límites para su espacio de parámetros. Por ejemplo, se puede observar dicho efecto comparando los distintos paneles de la Figura~\ref{figconstrains}, para $M_2=2 M_1$ se obtuvo $\tau_{3/2} \geq 10^{28}$ s, en cambio para $M_2=1.1 M_1$ el límite inferior del tiempos de vida media del gravitino es órdenes de magnitud más chico $\tau_{3/2} \geq 4\times 10^{25}$ s.

Además, notamos que las curvas se cruzan entre si. Para entender esto, necesitamos tener en cuenta que la pendiente de cada curva es diferente si el decaimiento dominante es a través de canales con estado final de dos o tres cuerpos. La zona en el plano para la cual este cambio de régimen sucede es distinta para cada una de las curvas mostradas en las figuras.

Este cambio de pendiente, que depende de $M_1$ y de su relación con $M_2$, nos conduce a otro punto importante para el análisis. Como hemos mencionado, al considerar masas de gaugino cada vez más grandes, el parámetro de mezcla fotino-neutrino disminuye, lo que a su vez implica menos restricciones por parte de los datos de \Fermi LAT en la búsqueda de líneas. Por lo tanto uno esperaría que al incrementar $M_1$, el límite superior de $m_{3/2}$ aumente. Sin embargo, para $M_1\gtrsim 10$ TeV el espectro difuso y extendido domina completamente la pendiente de las curvas en el rango de $m_{3/2}$ estudiado y la exclusión por decaimientos a tres cuerpos hace que para valores crecientes de $M_1$, el límite superior de $m_{3/2}$ disminuya, como podemos ver en la Figura~\ref{figconstrains} para los casos extremos $M_2=1.1 M_1$ y $M_2=M_1$.

En conclusión, para $M_2=2 M_1$ los límites de línea son los dominantes, y la masa del gravitino tiene que ser inferior a $4$ GeV para evitar las restricciones de \Fermi LAT. Esto se encuentra en acuerdo con los resultados de Ref.~\cite{Albert:2014hwa}.

Para $M_2=1.1M_1$ y $M_1$ debajo de 1 TeV, la exclusión está dada por el decaimiento a dos cuerpos, como en el caso anterior, y los valores permitidos para la masa de los gravitinos son similares ($\lesssim 6$ GeV). Sin embargo, por encima de 1 TeV los límites de línea son evadidos cuando se suprime fuertemente $|U_{\tilde{\gamma} \nu_i}|$ por una combinación del incremento de la masa de los gauginos y la cercanía de $M_1$ y $M_2$. Masas de gravitino de hasta $16$ GeV son permitidas.

Para $M_2=M_1$, los límites de línea todavía son relevantes y muy importantes, pero aún para valores bajos de $M_1$ es posible alcanzar masas de gravitino no excluidas que superen los $10$ GeV.

\begin{figure}[t]
 \begin{center}
       \epsfig{file=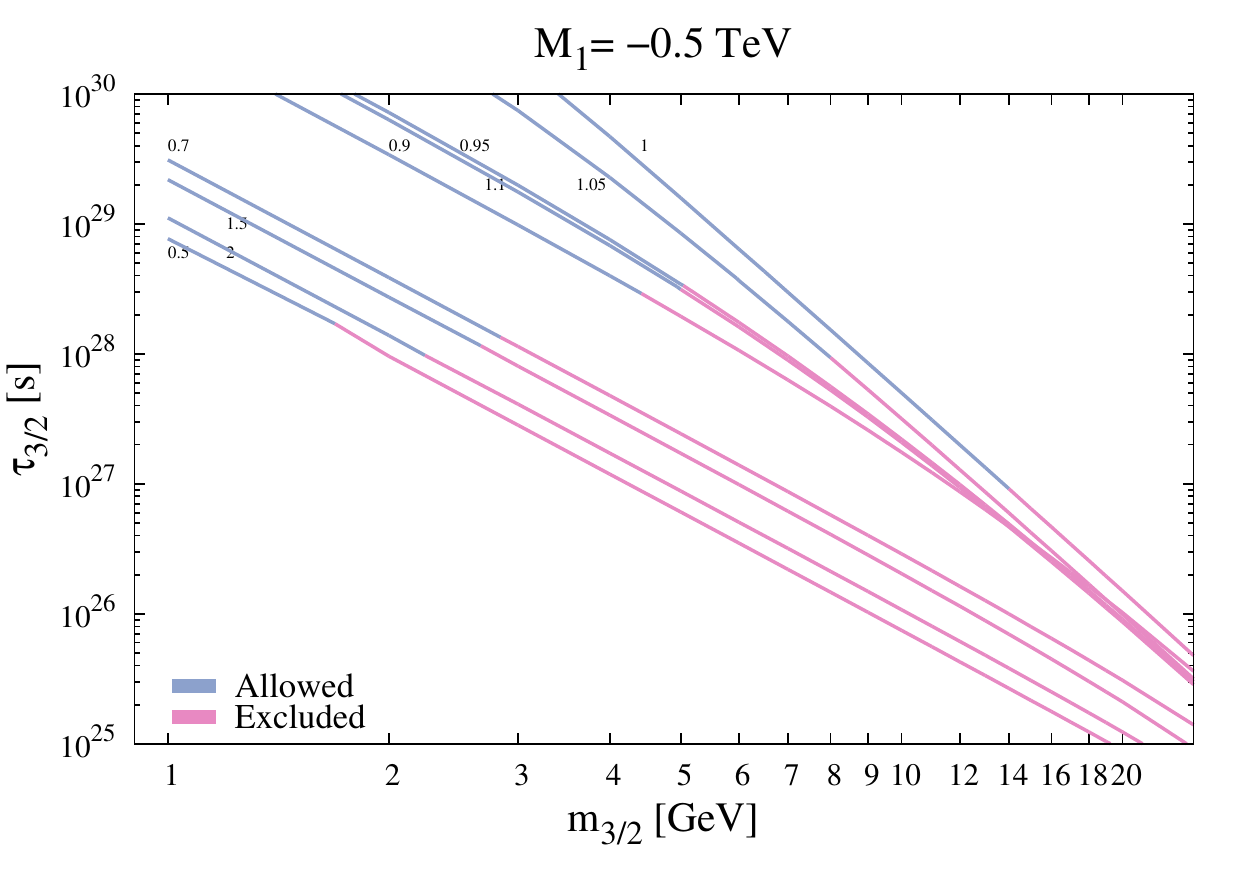,height=6.5cm}
\caption{Mismas referencias que en la Figura~\ref{figconstrains} pero fijando $M_1=-0.5$ TeV y 
$M_2 = x M_1$ con $x=0.5,0.7,0.9,0.95,1,1.05,1.1,1.5,2$. Cada curva se identifica con una etiqueta que corresponde al valor de $x$.
}
    \label{figsymmetric12}
 \end{center}
\end{figure}

Por lo tanto, utilizando los datos de \Fermi LAT, hemos obtenido los siguientes límites para la masa y el tiempo de vida media del gravitino como materia oscura en el $\mu \nu$SSM:
$m_{3/2}\lesssim 17$ GeV y $\tau_{3/2}\gtrsim 4\times 10^{25}$ s.

Finalmente, podemos considerar la posibilidad de modificar dichos límites usando que $M_2 = x M_1$ con $x<1$, además de $x>1$ como hemos hecho hasta ahora. Empero, a partir de la aproximación mostrada en Ec.~(\ref{UAP1}), la mezcla fotino-neutrino y por ende el espectro producido por el decaimiento a estados finales de dos cuerpos es proporcional a $|(1-x)/x|$, lo cual implica que mientras más grande sea esta relación, más fuertes serán las restricciones sobre $m_{3/2}$. Por lo tanto, los resultados para $x<1$ no van a diferir significativamente de aquellos con $x>1$. Este argumento no es del todo exacto debido al espectro generado por el decaimiento a tres cuerpos. Por ejemplo, los canales que involucran un bosón $W$ intermedio dependen de parámetros de mezcla que provienen de la matriz de masas de los charginos-leptones cargados, donde solo $M_2$ aparece. No obstante, un análisis numérico muestra que el argumento puede ser utilizado como primera aproximación. En la Figura~\ref{figsymmetric12} mostramos cómo se modifican los límites de $m_{3/2}$ y $\tau_{3/2}$ impuestos por \Fermi LAT considerando $M_1=-0.5$ TeV y diferentes valores de $x$. Los límites más (menos) restrictivos para la masa de los gravitinos se obtienen para $x=0.5$ ($x=1$).

\section{Conclusiones del capítulo}

El $\mu\nu$SSM es un modelo supersimétrico que resuelve el problema $\mu$ y reproduce la física de los neutrinos, simplemente empleado acoples con tres familias de neutrinos dextrógiros. Como dichos acoples rompen la paridad-$R$, el gravitino es un candidato interesante a materia oscura en el modelo ya que puede producir una señal.

En este capítulo hemos descrito un análisis completo de la detección del gravitino como candidato a materia oscura en el $\mu\nu$SSM. Además de considerar canales de decaimiento con estados finales de dos cuerpos (ver Figura~\ref{fig_gfn}) los cuales producen una línea espectral angosta con una energía igual a la mitad de la masa del gravitino, hemos incluido en el análisis los canales de decaimiento con estados finales de tres cuerpos (ver Figuras~\ref{fig_gz} y~\ref{fig_gw}) que producen una señal extendida, continua y suave. Hemos calculado el flujo de rayos gamma predicho por el modelo, considerando el área de observación (ROI) y los perfiles de densidades utilizados por los experimentos. Finalmente, hemos comparado los flujos totales con observaciones de \Fermi LAT para imponer restricciones al modelo teórico utilizado. En particular, empleamos los límites superiores del fondo difuso de rayos gamma extragaláctico (EGB) con 95$\%$ C.L. a partir de 50 meses de datos para restringir la señal espectral continua producto del decaimiento a tres cuerpos, y los límites superiores en la búsqueda de líneas espectrales a partir de 69.9 meses de datos para la emisión proveniente del decaimientos a dos cuerpos.

Primero hemos realizado una exploración numérica exhaustiva del espacio de parámetros de baja energía del $\mu \nu$SSM teniendo en cuenta que los datos de la física de neutrinos deben ser reproducidos. Esto impone importantes restricciones, principalmente determinando el rango del parámetro de mezcla fotino-neutrino, cantidad crucial en el cálculo de la señal con forma de línea. Para valores decrecientes de $|U_{\tilde{\gamma} \nu_i}|^2$ la anchura de decaimiento es cada vez más chica, y por lo tanto, son cada vez menos restrictivos los límites para la masa del gravitino (ver Ec.~(\ref{decay2body})) a partir de la no observación de líneas espectrales en los datos de \Fermi LAT. Se comprobó que el parámetro de mezcla fotino-neutrino  debe estar en el rango $10^{-20} \lesssim|U_{\tilde{\gamma} \nu}|^2\lesssim 10^{-14}$. El límite inferior ha sido extendido en comparación con estimaciones de trabajos previos, y puede ser obtenido para valores grandes de las masas de los gauginos $M_{1,2}$ y/o en el límite $M_2 \to M_1$ (ver la fórmula aproximada en Ec.~(\ref{UAP1})).

Hemos encontrado que para escenarios estándares, es decir para relaciones entre los parámetros de baja energía $M_2 = 2 \, M_1$ del tipo GUT, los límites provenientes de líneas son cruciales y solo permiten valores de masa de gravitinos $\lesssim 4$ GeV (y tiempos de vida media $\gtrsim 10^{28}$ s), aún para valores de $|M_1|$ tan grandes como 10 TeV. En el caso de $M_2 \to M_1$, aunque el tamaño de la línea se encuentra suprimido restringiendo menos la masa de gravitino ($\lesssim 17$ GeV), aún los límites de línea son importantes (ver Figura~\ref{fig_EGBa}). Los límites de EGB pueden establecer las restricciones solo en los escenarios extremos, es decir, con masas de gauginos grandes y muy cercanas entre si. De todos modos, la señal extendida proveniente de los decaimientos a tres cuerpos cumple un rol muy importante, ya que puede dominar la anchura de decaimiento del gravitino en una región muy amplia del espacio de parámetros (ver Figura~\ref{fig2}), modificando los límites de exclusión para su tiempo de vida media.

Los resultados se resumen en las Figuras~\ref{figconstrains} y~\ref{figsymmetric12}, donde podemos ver que usando los datos de \Fermi LAT los siguientes límites para la masa y tiempo de vida del gravitino han sido obtenidos: $m_{3/2}\lesssim 17$ GeV y $\tau_{3/2}\gtrsim 4\times 10^{25}$ s.

\spacing{1}


\chapter{Axino como materia oscura en el \texorpdfstring{$\mu\nu$}{munu}SSM y su detección}
\label{axinoDMchapter}

\spacing{1.5}

En este capítulo describimos las condiciones para que el axino, compañero supersimétrico del axión, sea un buen candidato a materia oscura en el $\mu\nu$SSM. De manera similar al gravitino, modelos SUSY con ruptura de paridad-R implican que el axino LSP puede decaer a partículas del SM, en particular a un fotón y un neutrino en el $\mu\nu$SSM. Calculamos el flujo de rayos gamma producto del decaimiento del axino, e imponemos límites sobre su masa y tiempo de vida media mediante la utilización de datos actuales sobre las búsquedas de líneas espectrales. Finalmente mostramos las perspectivas de detección para la nueva generación de detectores de rayos gamma al considerar la sensibilidad del proyecto e-ASTROGAM como referencia. Estudiamos los dos modelos más populares de axiones: KSVZ y DFSZ. Este capítulo está basado en Ref.~\cite{Gomez-Vargas:2019vci}.

\section{Axino como materia oscura}
\label{axinocomoDM}

Como hemos visto en la Sección~\ref{axiones} el axión $a$ es un pseudo-escalar que surge a partir la solución al `\textit{strong CP problem}'. El axino $\widetilde{a}$ es el compañero supersimétrico del axión, y como vimos en la Sección~\ref{axionesenSUSY} puede ser el LSP y un candidato a materia oscura.

En el marco de teorías de supergravedad, el axino tiene un término de interacción en el Lagrangiano con el fotón y el fotino. De manera análoga al gravitino tratado en la sección anterior, en presencia términos de ruptura de paridad-R, el fotino y los neutrinos izquierdos se mezclan en la matriz de masa de fermiones neutros (neutralino-neutrino). Por consiguiente, el axino LSP puede decaer mediante esta interacción a un fotón y un neutrino. Esto tiene consecuencias muy importantes, ya que el fotón puede ser detectado como una señal de rayos gamma en forma de línea monocromática de energía ${m_{\tilde{a}}}/{2}$, donde $m_{\tilde{a}}$ es la masa del axino. Análisis similares para el axino en el marco de modelos con violación de paridad-R mediante términos bilineales/trilineales han sido llevados a cabo en Refs.~\cite{Kim:2001sh,Hooper:2004qf,Chun:2006ss,Endo:2013si,Kong:2014gea,Choi:2014tva,Liew:2014gia,Colucci:2015rsa,Bae:2017tqn,Colucci:2018yaq}.

Instrumentos actuales como \Fermi LAT, poseen la sensibilidad y el rango de energías adecuado para contrastar está hipótesis, mientras que futuros experimentos planeados como e-ASTROGAM mejoraran la sensibilidad de forma significativa. De esta manera, en modelos con paridad-R rota, podemos encontrar una señal medible proveniente de una partícula con interacciones extremadamente débiles que la hacen imposible de detectar mediante experimentos de detección directa de materia oscura.

\subsection{Decaimiento de axinos}
\label{axinodecay}

El decaimiento del axino a un fotón y un neutrino utilizando acoples que violan paridad-R está dado por~\cite{Covi:2009pq}:
\bea
\Gamma(\tilde{a}\rightarrow
\gamma
\nu_i)
\simeq\frac{m_{\tilde{a}}^3}{128\pi^3 f_a^2}\alpha_{em}^2C_{a\gamma\gamma}^2|U_{\tilde{\gamma} \nu}|^2,
\label{decay2bodyaxino}
\eea
donde 
$\Gamma(\tilde{a}\rightarrow \gamma \nu_i)$ representa la suma de las anchuras de decaimiento parciales a  $\nu_i$ y $\bar\nu_i$,
$C_{a\gamma\gamma}$ es una constante de orden uno que depende del modelo, $\alpha_{em}=e^2/4\pi$, y $f_a$ es la escala de PQ. Notamos que $|U_{\tilde{\gamma} \nu}|$ es el mismo parámetro de mezcla involucrado en el decaimiento a dos cuerpos del gravitino definido en la Ec.~(\ref{photino}) que determina la composición de fotino en los neutrinos. Como fue discutido anteriormente, relajando algunas suposiciones como la relación GUT entre las masas de los gauginos y/o escalas $O(\text{TeV})$ podemos tomar como rango
\begin{equation}
10^{-10} \lesssim |U_{\widetilde{\gamma}\nu}| \lesssim 10^{-6}.
\label{relaxingmore}
\end{equation}

Como se puede ver en Ec.~(\ref{decay2bodyaxino}), el decaimiento del axino está suprimido por dos factores: por el valor pequeño del parámetro de mezcla $|U_{\tilde{\gamma} \nu}|$ relacionado con la ruptura de paridad-R, y por el valor grande de la escala de PQ $f_a \gtrsim 10^{9}$ GeV obtenido a partir de observaciones de la supernova SN1987A~\cite{Kawasaki:2013ae}. Esto hace que el tiempo de vida media del axino sea mayor a la edad del universo ${\tau}_{\tilde{a}}\gg t_{hoy}\sim 10^{17}$ s, con 
\begin{equation}
{\tau}_{\tilde{a}} = \Gamma^{-1}(\tilde{a}\rightarrow
\gamma
\nu_i) 
\simeq
3.8\times 10^{28}\, {s}
\left(\frac{f_a}{10^{13}\, \mathrm{GeV}}\right)^2
\left(\frac{10^{-8}}{|U_{\widetilde{\gamma}\nu}|}\right)^2
\left(\frac{0.1\, \mathrm{GeV}}{m_{\tilde{a}}}\right)^{3},
\label{axinolifetime}
\end{equation}
donde en la última igualdad hemos asumido $C_{a\gamma\gamma}=1$.

\subsection{Densidad reliquia de los axinos}
\label{axinorelic}

A pesar de que los axinos decaen, en la sección anterior hemos visto que su tiempo de vida es mayor a la edad del universo  $\tau_{\tilde a}\gg t_{hoy}$, y por lo tanto como una buena aproximación podemos considerar que su densidad reliquia coincide con la densidad reliquia que los axinos hubiesen tenido si fuesen estables y no decayesen. Para los axinos, su abundancia depende del modelo de axiones considerado, pudiendo diferir varios órdenes de magnitud.
En el marco del modelo de axiones KSVZ~\cite{Kim:1979if,Shifman:1979if}, la producción de axinos está dominada por interacciones entre gluones y gluinos, y su densidad reliquia a partir de producción termal resulta ser~\cite{Brandenburg:2004,Strumia:2010}
\begin{equation}
\Omega^{\text{TP}}_{\tilde{a}}h^2\simeq 0.3 \ (g_3 (T_R))^4 \ \left(\frac{F(g_3(T_R))}{23}\right) \left( \frac{m_{\tilde{a}}}{1 \text{ GeV}} \right) \left(\frac{T_R}{10^4 \text{ GeV}}\right) \left(\frac{10^{12}\text{ GeV}}{f_a}\right)^2,
\label{relicaxinos}
\end{equation}
donde $T_R$ es la temperatura de \textit{reheating} luego de inflación, $g_3$ es el acople de $SU(3)$ que depende de la energía, y la función $F(g_3(T_R))$ describe la tasa de producción de axinos con $F\simeq 24-21.5$ para $T_R\simeq 10^4-10^6$~GeV~\cite{Strumia:2010}. Para los cálculos numéricos que realizaremos a continuación usaremos $F\simeq 23$. Si consideramos otros valores los resultados finales no cambian de forma significativa.

Antes de continuar debemos mencionar la producción de axiones. Proviene del mecanismo de desalineación o `\textit{misalignment mechanism}', y por lo tanto podemos escribir la abundancia reliquia de axiones como materia oscura fría de la siguiente manera
\begin{equation}
\Omega_{a}h^2\simeq 0.18\ \theta^2_i \left(\frac{f_a}{10^{12}\text{~GeV}}\right)^{1.19},
\label{relicaxions}
\end{equation}
donde $\theta_i$ es el ángulo inicial de $misalignment$. Como estamos interesados en estudiar el escenario con axino como el único componente de la materia oscura, podemos hacer despreciable la densidad reliquia de axiones eligiendo de ser necesario un valor apropiado para el ángulo $\theta_i$, es decir, cuando $f_a\gtrsim 10^{12}$~GeV. Sin embargo, en nuestro trabajo es conveniente considerar el límite superior $f_a\leq 10^{13}$ GeV para evitar demasiado ajuste fino en los parámetros.

Por lo tanto, si asumimos que el axino es el único componente de la materia oscura, la densidad reliquia dada por Ec.~(\ref{relicaxinos}) es proporcional a la temperatura de \textit{reheating}. Claramente, si fijamos la masa del axino y la escala de PQ, ajustando $T_R$ se puede obtener el valor de densidad reliquia medido por la colaboración Planck~\cite{Aghanim:2018eyx},
$\Omega_{cdm}^{\text{Planck}}h^2\simeq 0.12$.
En particular, se obtiene
\begin{equation}
T_R
\simeq \frac{0.4}{(g_3 (T_R))^4}\times {10^4\ \text{GeV}}\ \left(\frac{1\ \text{GeV}}{m_{\tilde{a}}}\right) 
\left(\frac{f_a}{10^{12}\ \text{GeV}}\right)^2.
\label{reheating}
\end{equation}
Por ejemplo, para $m_{\tilde{a}}=0.1$ GeV y $f_a=10^{12}$ GeV se necesita $T_R\simeq 7.2 \times 10^4$ GeV. Además, si asumimos el límite conservativo $T_R \gtrsim 10^4$ GeV, un límite superior para $m_{\tilde{a}}$ se obtiene a partir de Ec.~(\ref{reheating}) para cada valor de $f_a$:
\begin{equation}
m_{\tilde{a}}\lesssim 
0.5\ \text{GeV}\ \left(\frac{f_a}{10^{12}\ \text{GeV}}\right)^2.
\label{reheating2}
\end{equation}
Por ejemplo, para $f_a=10^{13}, 10^{12}, 10^{11}$ GeV se obtienen los siguientes límites superiores $m_{\tilde{a}}\lesssim 50, 0.5, 0.005$ GeV, respectivamente.
Notamos que es conveniente utilizar el límite inferior $f_a\geq 10^{11}$ GeV para evitar masas de axino demasiado pequeñas, por debajo del alcance de los detectores propuestos (ver Figura~\ref{figaxinoalone1} abajo). Entonces, para estudiar el axino como materia oscura consideraremos el siguiente rango para la escala de PQ:
\begin{equation}
10^{11} \leq f_a \leq 10^{13}\ \text{GeV}.
\label{pqscale}
\end{equation}

Por otro lado, en el caso del modelo de axiones DFSZ~\cite{Dine:1981rt,Zhitnitsky:1980tq} la producción de axinos está dominada por interacciones entre axino-Higgs-Higgsino y/o axino-quark-squark. Para temperaturas de \textit{reheating} superiores a $10^4$~GeV como las utilizadas en esta tesis, es una buena aproximación considerar que la densidad reliquia de axinos es independiente de la temperatura de \textit{reheating}, y resulta ser~\cite{Bae:2011}
\begin{equation}
\Omega^{\text{TP}}_{\tilde{a}}h^2\simeq 20.39 \left( \frac{m_{\tilde{a}}}{1 \text{ GeV}} \right) \left(\frac{10^{12}\text{ GeV}}{f_a}\right)^2.
\label{relicaxinosDFSZ}
\end{equation}
Por lo tanto, si el axino es el único componente de la materia oscura obtenemos un valor de masa fijo para un dado valor de $f_a$:
\begin{equation}
m_{\tilde{a}}\simeq 6\ \text{MeV}\ \left(\frac{f_a}{10^{12}\ \text{GeV}}\right)^2.
\label{maDFSZ}
\end{equation}
Por ejemplo, para $f_a=10^{13}, 10^{12}, 10^{11}$ GeV tenemos $m_{\tilde{a}}\simeq 600, 6, 0.06$ MeV, respectivamente.

A continuación discutiremos las perspectivas de detección para líneas espectrales provenientes del decaimiento de axinos. Veremos que el modelo DFSZ abarca un subconjunto del espacio de parámetros predicho por el modelo KSVZ.

\section{Resultados}
\label{resultsaxinochapter}

Como en modelos con violación de paridad-R el axino decae a un fotón y un neutrino, se produce un espectro monocromático de rayos gamma con energía ${m_{\tilde{a}}}/{2}$. El flujo producido por las partículas de nuestra galáctico está dado por la Ec.~(\ref{eq:decayFlux}). Para este caso, $\tau_{DM}$ y $m_{DM}$ corresponden al tiempo de vida media y masa del axino, respectivamente, mientras que el número total de fotones producidos en el decaimiento del axino puede ser expresado como
\begin{equation}
\frac{dN_{\gamma}^{\text{total}}}{dE} \; = \; \sum_{i} BR_i\frac{dN_{i}}{dE}\; = \; \frac{2}{m_{DM}} \delta\left( 1 - \frac{2E}{m_{DM}} \right) ,
\label{eq:dndephotonaaxino}
\end{equation}
ya que consideramos que el decaimiento del axino a fotón-neutrino es el único canal relevante, y además despreciamos la masa de los neutrinos.

\begin{figure}[t!]
\begin{center}
 \begin{tabular}{cc}
 \hspace*{-4mm}
 \epsfig{file=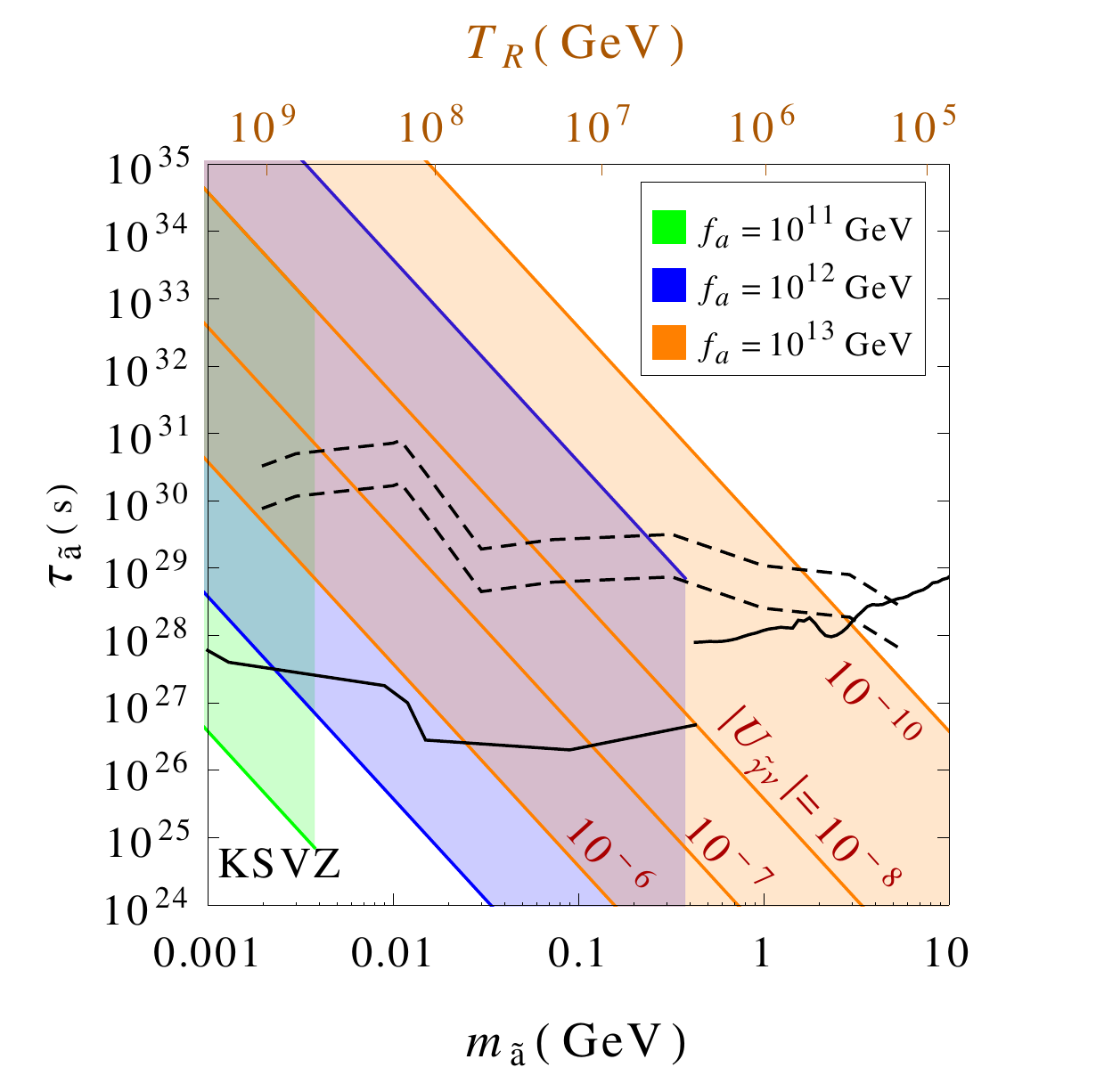,height=8cm} 
       \epsfig{file=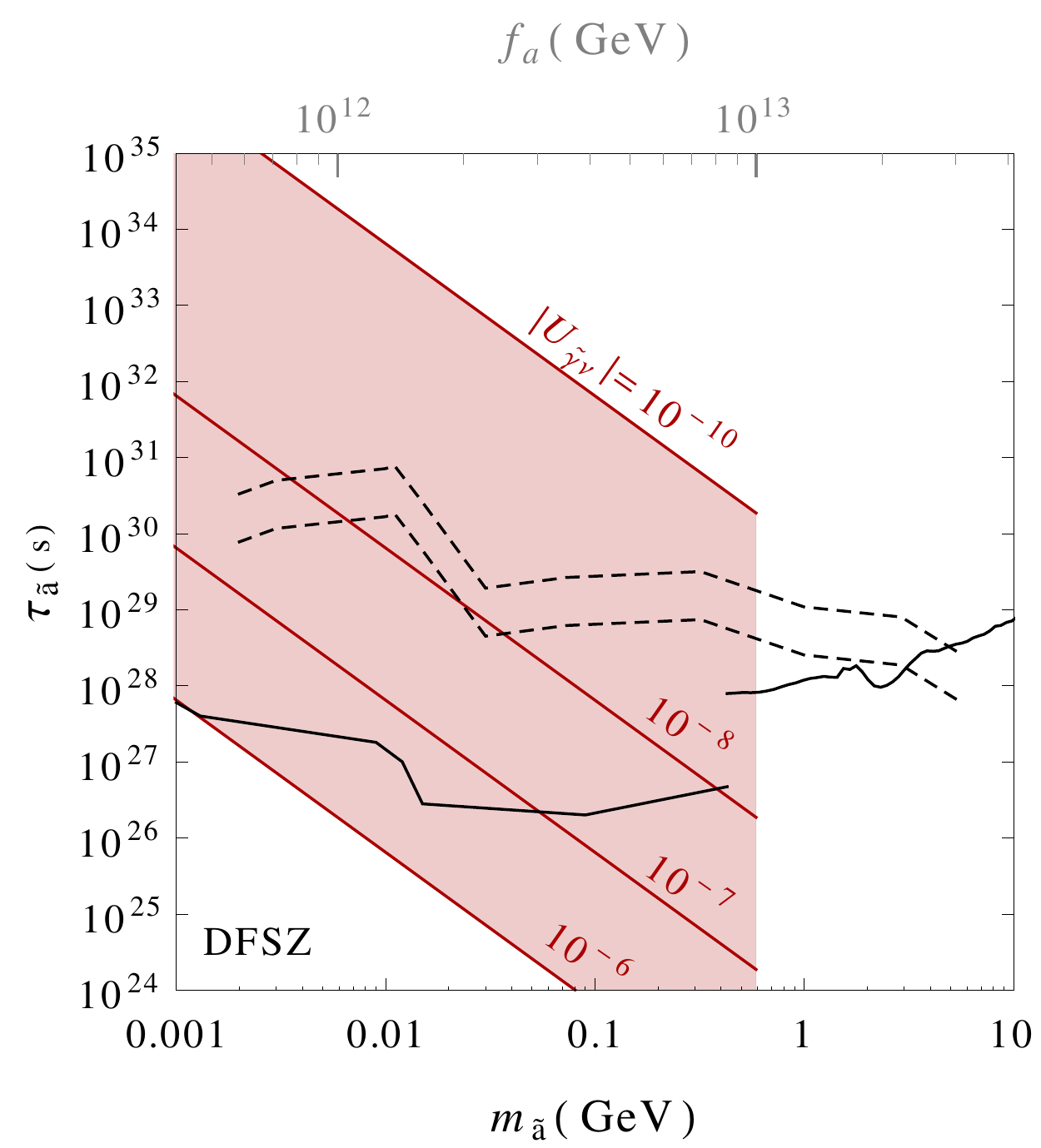,height=7.8cm}   
    \end{tabular}
    \caption{Límites actuales y perspectivas de detección para el tiempo de vida media y la masa del axino como único candidato a materia oscura. La región por debajo de la línea sólida negra para $m_{\tilde{a}} < (>) 0.4$ GeV se encuentra excluida por la búsqueda de líneas espectrales en el halo galáctico por COMPTEL (\Fermi LAT~\cite{Ackermann:2015lka}). La región debajo de las líneas punteadas negras superiores (inferiores) podrán ser exploradas por e-ASTROGAM~\cite{eAstrogamAngelis:2017}, observando el centro galáctico y asumiendo el perfil de densidad de materia oscura dado por Einasto B (Burkert). Panel izquierdo (modelo de axiones KSVZ): La zona naranja corresponde a las predicciones del $\mu\nu$SSM para varios valores representativos de $|U_{\tilde{\gamma} \nu}|$, para el caso con $f_a=10^{13}$ GeV.
    Para los demás casos $f_a=10^{12}$ y $10^{11}$ GeV, regiones azules y verdes, se muestra $10^{-10} \leq|U_{\tilde{\gamma} \nu}|\leq 10^{-6}$. 
    La temperatura de $reheating$ en función de la masa del axino para el caso con $f_a=10^{13}$ GeV se muestra en el eje superior naranja. Para otros valores de $f_a$, $T_R$ puede ser hallada con Ec.~(\ref{reheating}). El límite superior en $m_{\tilde{a}}$ para cada región se puede obtener a partir de Ec.~(\ref{reheating2}). Panel de la derecha (modelo de axiones DFSZ): La región roja corresponde a los valores predichos por el $\mu\nu$SSM para el mismo conjunto de valores representativos de $|U_{\tilde{\gamma} \nu}|$ empleados en el panel izquierdo. En el eje superior de la figura se muestra la escala de PQ en función de la masa del axino usando Ec.~(\ref{maDFSZ}).
}
    \label{figaxinoalone1}
\end{center}
\end{figure}

Para restringir el espacio de parámetros del axino con observaciones de rayos gamma, consideramos los límites determinados por la búsqueda de líneas espectrales realizadas por COMPTEL y \Fermi LAT~\cite{Ackermann:2015lka}, los cuales se presentan independientes del modelo. En la Figura~\ref{figaxinoalone1}, las regiones por debajo de las curvas negras se encuentran excluidas por las colaboraciones antes mencionadas, tomando como ROI el halo galáctico R180 en~\cite{Ackermann:2015lka} y empleando el perfil de materia oscura NFW. También se presenta el poder de exclusión que tendrá e-ASTROGAM proyectando la sensibilidad a la búsqueda de líneas espectrales propuesta en Ref.~\cite{eAstrogamAngelis:2017}. Para el caso de e-ASTROGAM, se considera como región de interés una sección del cielo de 10$^\text{o} \times$10$^\text{o}$ alrededor del centro galáctico, de acuerdo con los datos publicados por la colaboración. Como vimos en la Sección~\ref{perfilesDM} los perfiles de densidad de DM pueden diferir significativamente en la zona del centro galáctico, por lo tanto distintos perfiles han sido utilizados para el caso de e-ASTROGAM: NFW, Moore, Einasto, Einasto B y Burkert. En particular, Einasto B (Burkert) es el perfil que predice el límite más (menos) restrictivo y corresponde en la figura a la línea punteada superior (inferior). Si empleamos el perfil NFW, generalmente considerado el perfil de referencia, la sensibilidad de e-ASTROGAM se ubicaría equidistante entre las líneas punteadas.

Utilizando los resultados de las secciones anteriores, también mostramos en la Figura~\ref{figaxinoalone1} los valores predichos del tiempo de vida media y masa del axino para el $\mu\nu$SSM. El panel izquierdo corresponde al modelo de axiones KSVZ y las curvas naranjas representan el espacio de parámetros usando Ec.~(\ref{axinolifetime}) con $f_a=10^{13}$ GeV para varios valores representativos de $|U_{\tilde{\gamma} \nu}|$. Para los casos con $f_a~=~10^{11}$ y $10^{12}$ GeV mostramos solamente las líneas correspondientes a los límites superiores e inferiores dados por la física de neutrinos en Ec.~(\ref{relaxingmore}). La temperatura de \textit{reheating} en función de la masa del axino se muestra en el eje superior de la figura en naranja ya que corresponde al caso con $f_a=10^{13}$ GeV. Para otros valores de  $f_a$, $T_R$ puede ser obtenida a partir de Ec.~(\ref{reheating}).

Como podemos ver, masas de axino mayores a 3 GeV se encuentran desfavorecidas por los datos actuales de \Fermi LAT. Además, una región significativa del espacio de parámetros del axino como candidato único a materia oscura podrá ser explorada por futuras misiones de rayos gamma, como e-ASTROGAM, permitiendo estudiar masas y tiempos de vida media en los rangos $2$ MeV$-3$ GeV y $2 \times 10^{26}-8 \times 10^{30}$ s, respectivamente.

Remarcamos nuevamente que el límite superior graficado de la masa de los axinos para cada valor de $f_{a}$, es decir $m_{\tilde{a}}\lesssim 0.005, 0.5$ GeV para $f_a=10^{11}, 10^{12}$ GeV, respectivamente, se obtiene a partir de la Ec.~(\ref{reheating2}) considerando el límite conservativo para la temperatura de \textit{reheating} $T_R \gtrsim 10^4$ GeV.

Por otro lado, la región roja del panel de la derecha de la Figura~\ref{figaxinoalone1} corresponde a las predicciones realizadas con el modelo de axiones DFSZ. A diferencia del modelo KSVZ, la densidad reliquia resulta independiente de $T_R$, como hemos discutido en Ec.~(\ref{relicaxinosDFSZ}). Esto permite que la figura para el modelo DSFZ sea más simple. En el eje superior gris se muestra la escala de PQ en función de la masa del axino utilizando Ec.~(\ref{maDFSZ}).

En este caso, masas superiores a 0.6 GeV son desfavorecidas debido al límite conservativo considerado en la escala de PQ $f_a \leq 10^{13}$ GeV. Sin embargo, una importante región del espacio de parámetros podrá ser explorado por e-ASTROGAM: $2$ MeV$-0.6$ GeV y $2 \times 10^{26}-8 \times 10^{30}$ s.

Es importante notar que la región roja obtenida con el modelo de axiones DFSZ está incluida en la región predicha por el modelo KSVZ. Para cada valor de $m_{\tilde{a}}$ dado debajo de Ec.~(\ref{maDFSZ}), podemos identificar el rango predicho de $\tau_{\tilde{a}}$ en el $\mu\nu$SSM a partir de la correspondiente región coloreada para un $f_a$ en el panel de la izquierda. También se pueden extrapolar los resultado para valores intermedios de $f_a$. Por lo tanto, superponiendo ambos paneles podemos ver que la región permitida del modelo DFSZ representa un subconjunto de la región del modelo KSVZ, de modo que este último es más general. Esto constituye un resultado esperado, ya que el primer modelo tiene un grado de libertad menos, $T_R$, para obtener la densidad reliquia correcta.

\section{Conclusiones del capítulo}

En este capítulo estudiamos al axino, compañero supersimétrico del axión, como único candidato a materia oscura en el contexto de modelos con paridad-R rota, en particular hemos analizado el espacio de parámetros del $\mu\nu$SSM.

Primero discutimos las condiciones para considerar al axino como un buen candidato a materia oscura. A pesar que el axino LSP puede decaer a partículas del SM, su anchura de decaimiento a fotón-neutrino se encuentra suprimida tanto por lo grande de la escala de PQ $10^{11} \leq f_a \leq 10^{13}$ GeV, como por lo pequeño de la mezcla fotino-neutrino $10^{-10} \lesssim |U_{\widetilde{\gamma}\nu}| \lesssim 10^{-6}$ parámetro que involucra violación de paridad-R. Estos dos factores generan que el tiempo de vida media del axino sea varios órdenes de magnitud superior al tiempo de vida del universo. Notamos también la similitud del decaimiento del axino a fotón-neutrino con el correspondiente decaimiento del gravitino. Ambos candidatos están suprimidos por escalas superiores a la escala electrodébil y por $|U_{\widetilde{\gamma}\nu}|$ relacionado con la física de neutrinos vía el mecanismo de \textit{seesaw} generalizado a escala electrodébil que le da masas a los neutrinos a orden árbol en el $\mu\nu$SSM.

La abundancia reliquia de axinos ha sido discutida, considerando los dos modelos más populares de axiones: KSVZ y DFSZ. Una de las diferencias importantes entre ambos, es que la densidad reliquia de axinos depende de la temperatura de $\textit{reheating}$ solo en el primer modelo. Si imponemos el limite inferior conservativo de $T_R \gtrsim 10^4$ GeV, en el modelo KSVZ hallamos un límite superior para la masa del axino de $m_{\tilde a} \lesssim 50$ GeV. Considerando en cambio el modelo DFSZ el límite resulta $m_{\tilde a} \lesssim 0.6$ GeV, independiente de $T_R$.

Por último, estudiamos el flujo de rayos gamma y mostramos los límites de exclusión mediante búsquedas de líneas espectrales con datos de \Fermi LAT en el halo galáctico. También empleamos la proyección de la sensibilidad del instrumento propuesto e-ASTROGAM observando el centro galáctico con distintos perfiles de densidades de materia oscura.

Mostramos que la región predicha del modelo DFSZ al tener un grado de libertad menos, $T_R$, representa un subconjunto de la región del modelo KSVZ que resulta más general.

Finalmente, encontramos que masas $m_{\tilde a} \gtrsim 3$ GeV se encuentran excluidas por \Fermi LAT, y que misiones propuestas especializadas en la exploración del rango energético MeV-GeV como e-ASTROGAM, serán capaces de explorar importantes sectores del espacio de parámetros del modelo:
$2$ MeV $\lesssim m_{\tilde a} \lesssim 3$ GeV, 
$2\times 10^{26}\lesssim \tau_{\tilde a} \lesssim 8\times 10^{30}$ s.

\spacing{1}


\chapter{Axino y gravitino como candidatos simultáneos a materia oscura en el \texorpdfstring{$\mu\nu$}{munu}SSM}
\label{axinogravitinoDMchapter}

\spacing{1.5}

En este capítulo estudiamos escenarios con múltiples candidatos a materia oscura, el axino y el gravitino como LSP y Next-to-LSP (NLSP). Como hemos visto en capítulos anteriores, de forma individual son buenos candidatos aún cuando pueden decaer a fotón-neutrino y producir una señal. Esto se debe a que su tiempo de vida tiende a ser mayor que la edad del universo, al estar sus decaimientos suprimidos por el parámetro de mezcla fotino-neutrino que viola paridad-R y la escala característica de cada partícula: $M_{Pl}$ y $f_a$. Sin embargo, existe un término de interacción entre el axino y gravitino que conserva paridad-R y produce otro tipo de decaimiento de crucial importancia para el análisis: el NLSP al LSP más un axión.

En Ref.~\cite{Hamaguchi:2017}, los autores estudiaron la cosmología de un modelo con paridad-R conservada y materia oscura formada por gravitinos y axinos, que puede relajar las potenciales tensiones del modelo cosmológico estándar $\Lambda$CDM entre observaciones del universo temprano y el universo reciente~\cite{Berezhiani:2015,Chudaykin:2016,Poulin:2016,Chudaykin:2017,Bringmann:2018}. Las masas del gravitino $m_{3/2}$ y axino $m_{\tilde a}$ dependen del modelo pudiendo ser del mismo orden o algunos órdenes de magnitud diferentes en escenarios realistas~\cite{Kawasaki:2013ae,Goto:1991gq,Chun:1992zk,Chun:1995hc,Kim:2012bb} como en supergravedad. Por lo tanto, si el axino (gravitino) es el LSP el gravitino (axino) puede tomar el rol del NLSP.

Primero analizamos el espacio de parámetros y establecemos las condiciones necesarias para la coexistencia de los dos candidatos, mediante la relación entre la densidad reliquia y la temperatura de \textit{reheating} favorecida. Como el NLSP decae al LSP más un axión ultrarelativista, empleamos límites sobre la cantidad de especies ultrarelativistas que puede inyectarse al universo en distintas etapas de su evolución. Finalmente estudiamos las señales en rayos gamma imponiendo los límites actuales y las perspectivas de detección para la próxima generación de telescopios de rayos gamma. En la región donde ambos candidatos pueden coexistir en el presente, concluimos
que es posible obtener una señal característica de este escenario muy difícil de emular, dos líneas espectrales simultáneas provenientes de cada candidato. Este capítulo está basado en Refs.~\cite{Gomez-Vargas:2019vci,Gomez-Vargas:2019mqk}.

\section{Gravitino y axino como candidatos múltiples
}
\label{sec:DM multiplecomp}

Como hemos discutido, tanto el axino como el gravitino tienen términos de interacción en el Lagrangiano con fotones y fotinos. Por lo tanto, en presencia de violación de paridad-R, ambos pueden decaer a un fotón y un neutrino produciendo una línea espectral monocromática con energía ${m_{3/2}}/{2}$ o ${m_{\tilde{a}}}/{2}$. Las anchuras de decaimiento de cada uno de los candidatos se puede ver en Ec.~(\ref{decay2body}) para el gravitino y en Ec.~(\ref{decay2bodyaxino}) para el axino, y notamos que ambos dependen del parámetro $|U_{\tilde{\gamma} \nu}|$. Podemos compara ambos decaimientos como:
\begin{equation}
\frac{\Gamma(\tilde{a}\rightarrow\gamma\nu_i)}
{{\Gamma}(\psi_{3/2}\rightarrow\gamma\nu_i)}
\simeq 10^{5}\,
\left(\frac{10^{13}\, \mathrm{GeV}}{f_a}\right)^2
r_{\tilde a}^3\ ,
\label{relation}
\end{equation}
donde
\bea 
r_{\tilde{a}}\equiv \frac{m_{\tilde{a}}}{{m_{3/2}}}.
\label{ra}
\eea
Entonces, 
$\Gamma(\tilde{a}\rightarrow\gamma\nu_i)$ es típicamente más grande que 
$\Gamma(\psi_{3/2}\rightarrow\gamma\nu_i)$ a menos qué $r_{\tilde a}$ sea muy chico.
En particular, para $f_a=10^{13}, 10^{12}, 10^{11}$ GeV tiene que ser más chico que aproximadamente 0.02, 0.004, 0.001, respectivamente. Este resultado será útil para nuestra discusión en la Sección~\ref{subsec: Prospects}.

Como veremos en las próximas secciones, la región donde el gravitino y el axino pueden coexistir y dar señal múltiple corresponde a $m_{3/2} \lesssim 1$ GeV. En dicha región consideramos que el decaimiento a dos cuerpos del gravitino es el dominante. Para masas superiores, veremos que el término de interacción entre gravitino-axino-axión causa que generalmente el tiempo de vida del NLSP sea inferior a la edad del universo, es decir, el NLSP no produce una señal en el espectro de rayos gamma pues no se encuentra en el presente. Por lo tanto, para estudiar las posibilidades de detección de líneas espectrales, es suficiente considerar solo decaimientos a fotón-neutrino.

Con respecto al rango de masas, en modelos de supergravedad se relacionan con el mecanismo de ruptura espontánea de supersimetría. En particular, en modelos de ruptura mediados por gravedad, donde la escala de masas \textit{soft} están típicamente determinadas por la masa del gravitino (ver Sección~\ref{rupturaSUSY}). Por lo tanto, esperamos que el rango de masas del gravitino sea $O(\text{GeV-TeV})$~\cite{Brignole:1997dp}, es decir alrededor de la escala electrodébil. Sin embargo, potenciales de Kahler y/o superpotenciales específicos en supergravedad pueden generar diferentes situaciones, produciendo gravitinos con masas varios órdenes de magnitud más chicos que la escala electrodébil. Por ejemplo, esto sucede en el caso de modelos de supergravedad `\textit{no-scale}', donde la masa del gravitino se desacopla del resto del espectro de partículas supersimétricas~\cite{Ellis:1984kd}. Por otra parte, masas de gravitino mucho más chicas que la escala electrodébil son obtenidas en modelos con ruptura de supersimetría mediada por acoples gauge~\cite{Giudice:1998bp}. También, en teorías-F con unificación GUT y el último modelo de ruptura de supersimetría mencionado pueden obtenerse masas de gravitino de $10-100$ MeV~\cite{Heckman:2009mn}. Dada la dependencia de la masa del gravitino y axino con el modelo, consideramos que no elegir un modelo especifico de supergravedad es apropiado para nuestro enfoque fenomenológico, y por lo tanto, tratamos a las masas de nuestros candidatos a materia oscura como parámetros libres.

\section{Caso 1: axino LSP - gravitino NLSP
}
\label{sec:axinoLSPgravitinoNLSP}

En el marco de supergravedad existe un término de interacción que no viola paridad-R entre el gravitino, el axino y el axión. En este caso donde el gravitino es el NLSP y el axino el LSP, tenemos que~\cite{Hamaguchi:2017}
\begin{equation}
\Gamma(\psi_{3/2} \rightarrow \tilde{a} \, a) \simeq \frac{m_{3/2}^3}{192\pi M_P^2}(1-r_{\tilde{a}})^2
(1-r_{\tilde{a}}^2)^3, 
\label{decaytoaa}
\end{equation}
donde despreciamos la masa del axión. Este canal de decaimiento domina sobre el decaimiento a dos cuerpos mostrado en Ec.~(\ref{decay2body}) y que involucra parámetros que viola paridad-R, por lo tanto el tiempo de vida del gravitino que fue estimado en Ec.~(\ref{lifetimegamma}), en este escenario resulta
\begin{equation}
{\tau}_{3/2}\simeq {\Gamma}^{-1}
(\psi_{3/2} \rightarrow a \, \tilde{a})\simeq  2.3 \times 10^{15} s \: \left(\frac{1\, \text{GeV}}{m_{3/2}} \right)^3,
\label{gravitinolifetime1}
\end{equation}
donde en la última fórmula hemos despreciado la contribución de $r_{\tilde{a}}$ en Ec.~(\ref{decaytoaa}), aproximación válida si $m_{\tilde a}\ll m_{3/2}$.

En este punto es importante notar que si bien ${\Gamma^{-1}}(\psi_{3/2}\rightarrow\gamma\nu_i) \gg t_{hoy}$, esto no sucede para ${\tau}_{3/2}$, indicando que este resultado va a afectar las ecuaciones de la densidad reliquia que a continuación estudiaremos.

\subsection{Abundancia de axino LSP y gravitino NLSP
}
\label{subsec: relic}

Para calcular la densidad reliquia del axino en el caso con múltiples candidatos a materia oscura, tenemos que considerar el mecanismo de producción térmica, dado en Ec.~(\ref{relicaxinos}), y producción no térmica. Este último se relaciona con el decaimiento del gravitino NLSP involucrando su abundancia y tiempo de vida media. Los axinos producidos de esta manera no volverán a reaniquilarse ya que las interacciones entre gravitino y/o axinos se encuentran suprimidas por la masa de Planck.

Contrario al caso de los axinos LSP, cuyo tiempo de vida media es mayor a la edad del Universo, los gravitinos NLSP tienen un tiempo de vida menor y por lo tanto su densidad reliquia depende de dicho tiempo, resultando
\bea
\Omega_{3/2}h^2 &=& \Omega_{3/2}^{\text{TP}}h^2 e^{-(t_{\text{hoy}}-t_0)/ \tau_{3/2}},
\label{NLSPrelicgeneral0}
\eea
donde $t_0$ es la edad del universo correspondiente al momento de producción térmica de gravitinos, y
$\Omega_{3/2}^{\text{TP}}h^2$ corresponde a la densidad reliquia que los gravitinos NLSP tendrían si fuesen completamente estables y no decayesen, dado por~\cite{Pradler:2006qh,Rychkov:2007uq}
\begin{equation}
\Omega^{\text{TP}}_{3/2}h^2\simeq 0.02\left(\frac{T_R}{10^5 \text{ GeV}}\right)\left(\frac{1 \text{ GeV}}{m_{3/2}}\right)\left(\frac{M_3(T_R)}{3\text{ TeV}}\right)^2\left(\frac{ \gamma(T_R) / (T_R^6/M_P^2)}{0.4}\right),
\label{relicgravitinos}
\end{equation}
donde, $M_3(T_R)$ es la masa del gluino dependiente de la escala de energía, y el último factor parametriza la producción efectiva cuya rango es $\gamma(T_R) / (T_R^6/M_P^2)\simeq 0.4-0.35$ para $T_R\simeq 10^4-10^6$~GeV~\cite{Rychkov:2007uq}. Para nuestros cálculos numéricos usaremos $M_3(T_R) \simeq 3$ TeV y $\gamma(T_R)/(T_R^6/M_P^2) \simeq 0.4$. Otros valores no modifican nuestros resultados significativamente. Utilizamos esta expresión para la densidad reliquia de gravitinos en vez de Ec.~(\ref{relicgravitinos1}) ya que es más precisa para el rango de temperaturas de \textit{reheating} que consideramos a continuación.

Usando el límite conservativo $T_R \gtrsim 10^4$ GeV y el valor medido de la densidad reliquia total de materia oscura, se obtiene el siguiente límite inferior para la masa del gravitino $m_{3/2} \gtrsim 0.017 \text{ GeV}$. Recordamos que en la Sección~\ref{axinorelic} del capítulo anterior encontramos límites superiores para la masa del axino, dependientes del modelo y de la escala de PQ, que también aplicaremos en esta sección.

Teniendo en cuenta lo discutido arriba, la densidad reliquia para el axino LSP resulta
\bea
\Omega_{\tilde{a}}h^2 &=& \Omega_{\tilde{a}}^{\text{TP}}h^2  + \Omega_{\tilde{a}}^{\text{NTP}}h^2,
\label{LSPrelicgeneral}
\eea
donde $\Omega_{\tilde{a}}^{\text{TP}}h^2$ está dada por Ec.~(\ref{relicaxinos}), y el término $\Omega_{\tilde{a}}^{\text{NTP}}h^2$ tiene en cuenta la producción no termal a través del decaimiento del gravitino:
\bea
\Omega_{\tilde{a}}^{\text{NTP}}h^2  
= 
r_{\tilde a}\
\Omega_{3/2}^{\text{TP}}h^2\ \left(1-e^{-(t_{\text{hoy}}-t_0)/ \tau_{3/2}}\right).
\label{LSPrelicgeneral2}
\eea
Notamos que el factor $r_{\tilde a}$ tiene en cuenta si el LSP producido mediante el mecanismo no termal es relativista o no relativista, ya que solo estamos interesados en materia oscura fría. Como es esperado, si $\tau_{3/2}\ll t_{hoy}$, obtenemos las relaciones usuales~\cite{Covi:1999ty,Choi:2011yf,Roszkowski:2014}:
\bea
\Omega_{3/2}h^2 &\approx& 0,\\
\Omega_{\tilde{a}} h^2 \; &\approx& \Omega_{\tilde{a}}^{\text{TP}}h^2 + 
r_{\tilde a}\
\Omega_{3/2}^{\text{TP}}h^2.
\label{usual}
\eea

Con respecto a la producción de axiones, en este escenario con candidatos múltiples, además del mecanismo de \textit{misalignment} discutido en la Sección~\ref{axinorelic}, tenemos la producción de axiones provenientes del decaimiento del gravitino. De todas maneras, estos axiones constituyen radiación oscura o \textit{dark radiation}, es decir especies invisibles y ultrarelativistas. La cantidad de radiación oscura se encuentra bajo límites experimentales muy estrictos~\cite{Poulin:2016,Berezhiani:2015,Chudaykin:2016,Chudaykin:2017,Bringmann:2018}, y en consecuencia solo representa una pequeña contribución a la cantidad de materia oscura total.

Una cantidad que será útil a lo largo de este capítulo es la fracción de gravitinos NLSP que puede decaer a radiación oscura. Para ello definimos
\bea
f_{ddm}^{\text{DR}} = 
f_{3/2} \left( 1 - 
r_{\tilde a}
\right),
\label{ddmfraction}
\eea
con
\bea
f_{3/2}=\frac{\Omega_{3/2}^{\text{TP}}}{\Omega_{cdm}^{\text{Planck}}}
\label{fracgra}
\eea
la fracción de gravitino NLSP con respecto a la densidad de materia oscura total.
El subíndice $ddm$ y DR se refieren a las siglas en inglés de materia oscura que puede decaer y radiación oscura, respectivamente. Es importante aclarar lo siguiente:
\begin{itemize}
\item La colaboración Planck obtiene $\Omega_{cdm}^{\text{Planck}}h^2\simeq 0.12$ hoy a partir de datos provenientes de fotones generados en la era de recombinación usando la evolución del universo descrita por $\Lambda$CDM. Estamos trabajando con materia oscura que puede decaer, por lo tanto la cantidad de materia oscura fría tiene una dependencia temporal ya que parte de la densidad energética del gravitino NLSP puede `perderse' como radiación oscura. De todas formas, la última cantidad tiene que ser pequeña, como fue discutido arriba.
\item Materia oscura que decae o DDM, y la fracción de radiación oscura $f_{ddm}^{\text{DR}}$, se refieren a la contribución proveniente del decaimiento de gravitino NLSP a axino LSP y axiones. Es importante remarcar que no se refiere a los decaimientos del axino LSP y gravitino NLSP a fotones y neutrinos, cuyos tiempos de vida característicos son órdenes de magnitud mayores a la edad actual del universo.
\end{itemize}

\noindent
Finalmente debemos mencionar que debido a la mezcla entre axiones y fotones, los axiones emitidos por el decaimiento del gravitino pueden ser convertidos en fotones en presencia de campos magnéticos, produciendo potencialmente una señal. Sin embargo, el trabajo realizado en Ref.~\cite{Bae:2019} muestra que si se consideran axiones de QCD, como en nuestro caso, la probabilidad de conversión es muy chica como para ser observada.

\subsection{Flujo de rayos gamma}
\label{subsec: gamma-ray flux}

Los límites a materia oscura por emisiones de rayos gamma usualmente consideran que está compuesta solo por una especie de partícula. Para los casos de decaimientos de materia oscura, las restricciones establecen límites inferiores al tiempo de vida de la partícula. Si el axino y el gravitino pueden coexistir, uno como el LSP y otro como el NLSP, ambos pueden ser fuentes de rayos gamma. Sin embargo, es fácil normalizar la señal considerando que una fuente específica es una fracción de $\Omega_{cdm}^{\text{Planck}}h^2$.

El flujo diferencial de rayos gamma producidos por el decaimiento de materia oscura del halo galáctico está dado por Ec.~(\ref{eq:decayFlux}), donde $\rho_{\text{halo}}$ es una cantidad crucial. Asumiendo que la materia oscura es multicomponente, y que la distribución de cada una de las especies es homogénea a lo largo de la distribución de materia oscura, simplemente tenemos
\begin{equation}
\rho_{\text{halo}}=\sum_i \rho_{\text{DM}_i},
 \label{rhohalo}
\end{equation}
donde la componente $i$-ésima de la densidad de materia oscura $\rho_{\text{DM}_i}$ puede expresarse como
\begin{equation}
\rho_{\text{DM}_i}=
f_{\text{DM}_i}\ \rho_{\text{halo}},
 \label{rhoDMj}
\end{equation}
con 
\begin{equation}
f_{\text{DM}_i}\equiv  
\frac{\Omega_{\text{DM}_i}}{\Omega_{cdm}^{\text{Planck}}}.
 \label{rhoDMj2}
\end{equation}
Para calcular el flujo de rayos gamma de la componente $i$-ésima de materia oscura que decae a fotones, solo tenemos que reemplazar $\rho_{\text{halo}}\rightarrow \rho_{\text{DM}_i}$ en Ec.~(\ref{eq:decayFlux}) obteniendo el siguiente flujo de rayos gamma:
\begin{equation}
\frac{d\Phi_{\gamma}^{\text{DM}_i}}{dEd\Omega}=f_{\text{DM}_i} \frac{d\Phi_{\gamma}^{\text{100\% DM}_i}}{dEd\Omega},
\label{fluxcase1}
\end{equation}
donde $\frac{d\Phi_{\gamma}^{\text{100\% DM}_i}}{dEd\Omega}$ es el flujo que tendríamos si consideramos que $\rho_{\text{DM}_i}=\rho_{\text{halo}}$. Finalmente, teniendo en cuenta que los límites al flujo de rayos gamma se presentan como límites inferiores al tiempo de vida media de la materia oscura considerando solo un componente, en un escenario multicomponente es útil emplear para cada componente un tiempo de vida efectivo
\begin{equation}
\tau_{\text{DM}_i \text{-eff}}
= f^{-1}_{\text{DM}_i}\
\tau_{\text{DM}_i},
 \label{newtime1}
\end{equation}
donde $\tau_{\text{DM}_{i} \text{-eff}}$ puede ser contrastado contra los límites inferiores presentados por las colaboraciones experimentales.

Sin embargo, no podemos aplicar directamente las formulas anteriores para nuestro caso de materia oscura multicomponente compuesto por axino LSP (DM$_1$) y gravitino NLSP (DM$_2$). La razón es que sus fracciones cambian con el tiempo debido al decaimiento del gravitino a axino-axión. Entonces, teniendo en cuenta las Ecs.~(\ref{NLSPrelicgeneral0}) y~(\ref{LSPrelicgeneral}), debemos realizar los siguientes reemplazos en Ec.~(\ref{fluxcase1}) para gravitinos y axinos respectivamente:
\bea
f_{\text{DM}_2}  &\rightarrow & 
f_{3/2}\ 
e^{-(t_{\text{hoy}}-t_0)/\tau_{3/2}},
\\
\label{primera}
f_{\text{DM}_1} & \rightarrow & 
f_{\tilde a} + 
r_{\tilde a}\
f_{3/2}\ \left(1 - e^{-(t_{\text{hoy}}-t_0)/\tau_{3/2}}\right),
\label{fractionstime}
\eea
con $f_{3/2}$ definido en Ec.~(\ref{fracgra}) y
\begin{equation}
f_{\tilde a}=  
\frac{\Omega_{\tilde a}^{\text{TP}}}{\Omega_{cdm}^{\text{Planck}}}.
 \label{rhoDMj222}
\end{equation}
Como es esperado, si el decaimiento del gravitino NLSP a axino LSP y un axión no está permitido, se obtienen los mismos resultados que en Ec.~(\ref{fluxcase1}). 

Finalmente, en la misma forma que establecimos antes, es más fácil para el análisis considerar un tiempo de vida efectivo en nuestro escenario multicomponente de materia oscura que decae. Por lo tanto, Ec.~(\ref{newtime1}) resulta
\bea
\tau_{3/2\text{-eff}} & = & 
\left(
f_{3/2}\ 
e^{-(t_{\text{hoy}}-t_0)/\tau_{3/2}}\right)^{-1} 
\Gamma^{-1}(\psi_{3/2}\rightarrow\gamma\nu_i),
 \label{newtime2}
 \\
 \tau_{\tilde a\text{-eff}} & = & 
\left[
f_{\tilde a} + 
r_{\tilde a}\
f_{3/2}\ \left(1 - e^{-(t_{\text{hoy}}-t_0)/\tau_{3/2}}\right)\right]^{-1} 
\Gamma^{-1}(\tilde a\rightarrow\gamma\nu_i).
 \label{newtime22}
\eea

De esta manera, podemos aplicar el análisis de esta sección para estudiar las restricciones actuales y las perspectivas de detección del espacio de parámetros de nuestro escenario. Tomamos a continuación la siguiente simplificación $t_0=0$ para nuestros cálculos numéricos.


\subsection{Resultados 
}
\label{subsec: Results}

En el Capítulo~\ref{axinocomoDM} notamos que una de las más importantes diferencias entre los dos modelos más populares de axiones, el DFSZ y el KSVZ, reside en la producción térmica de axinos ya que el primero es independiente de la temperatura de \textit{reheating}. En cambio, para ambos modelos la anchura de decaimiento a fotón-neutrino es la misma. Como consecuencia hemos visto que la región predicha por el modelo de axiones DFSZ representa un subconjunto de la región predicha por el modelo KSVZ, al tener un grado de libertad menos. Entonces, en el resto de esta tesis nos enfocamos en el modelo de axiones KSVZ para explorar el escenario gravitino-axino DM desde el punto de vista fenomenológico más amplio posible.

\subsubsection{Límites a partir de observaciones cosmológicas}

\begin{figure}[t!]
\begin{center}
 \begin{tabular}{cc}
 \hspace*{-4mm}
 \epsfig{file=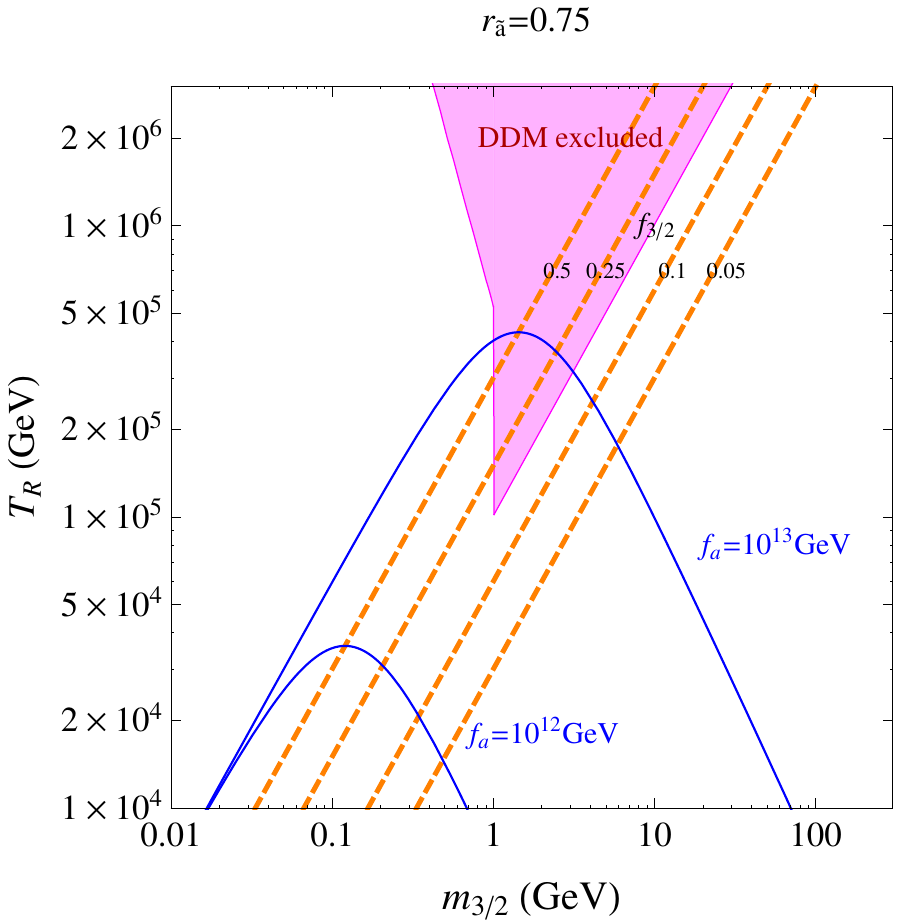,height=7cm} 
       \epsfig{file=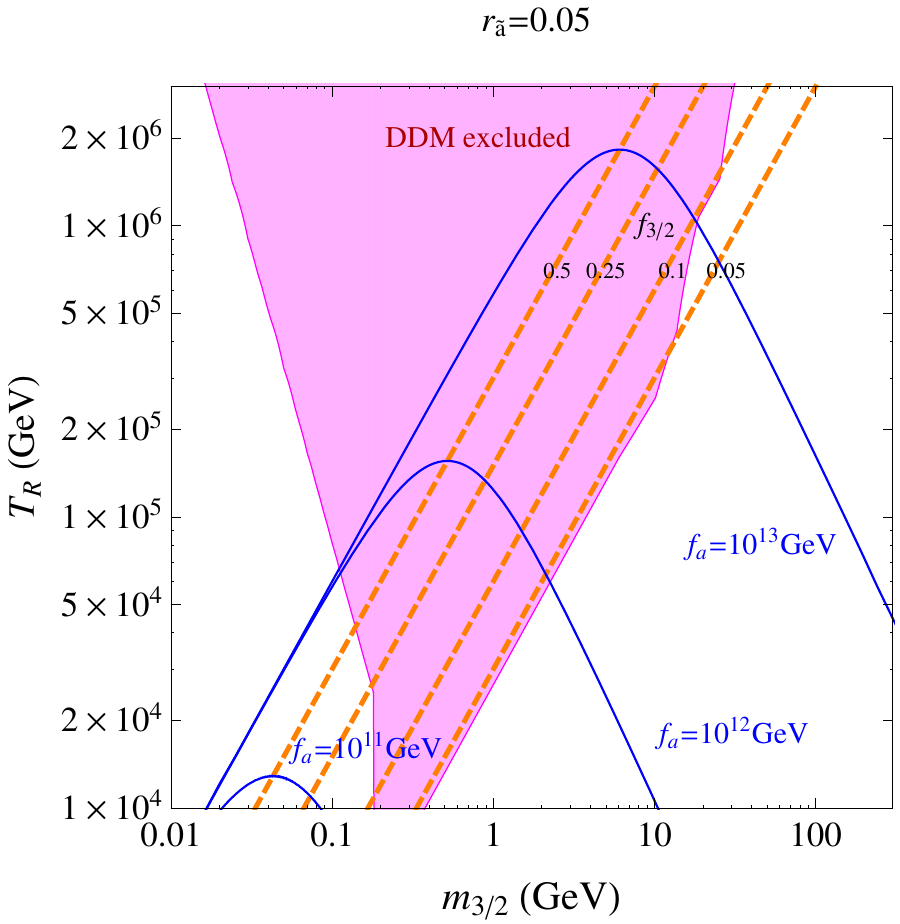,height=7cm}   
    \end{tabular}
    \caption{Límites sobre la temperatura de \textit{reheating} en función de la masa del gravitino NLSP para el caso multicomponente con axino LSP, y relaciones entre las masas $r_{\tilde a} = 0.75, 0.05$ en el panel izquierdo y derecho respectivamente. Las líneas azules corresponden a puntos con $\Omega_{3/2}h^2+\Omega_{\tilde{a}}h^2$ igual a  $\Omega_{cdm}^{\text{Planck}}h^2$ en la era de recombinación de acuerdo con las observaciones de Planck, para varios valores de la escala de Peccei-Quinn $f_a=10^{11}, 10^{12}, 10^{13}$ GeV. Para un dado $f_a$, la región encima de la correspondiente curva azul se encuentra excluida por exceso de producción de materia oscura fría. La región magenta se encuentra excluida por observaciones cosmológicas para modelos con materia oscura que decae~\cite{Poulin:2016,Berezhiani:2015,Chudaykin:2016,Chudaykin:2017,Bringmann:2018}, considerando las restricciones sobre $f_{ddm}^{\text{DR}}$. Las líneas discontinuas naranjas corresponden a la fracción de gravitino NLSP, $f_{3/2}=
0.5, 0.25, 0.1, 0.05$. El límite inferior $m_{3/2}\gtrsim 0.017$ GeV se obtuvo a partir de Ec.~(\ref{relicgravitinos}) asumiendo el límite conservativo $T_R \gtrsim 10^4$ GeV.
}
    \label{figaxinoLSP1}
\end{center}
\end{figure}

Para determinar la región del espacio de parámetros que cumple los límites experimentales sobre los modelos de materia oscura que decae, de forma similar a lo realizado en Ref.~\cite{Hamaguchi:2017} en la Figura~\ref{figaxinoLSP1} mostramos $T_R$ en función de $m_{3/2}$ para nuestro escenario con axino LSP y gravitino NLSP. En el panel izquierdo utilizamos la relación entre las masas $r_{\tilde a}=0.75$, mientras que en el panel derecho $r_{\tilde a}=0.05$.

Las líneas azules representan los puntos del espacio de parámetros con $\Omega_{3/2}h^2+\Omega_{\tilde{a}}h^2$, satisfaciendo las observaciones de la colaboración Planck en la era de recombinación, para diferentes escalas de PQ. Es decir, el valor de densidad reliquia determinado por Planck igual al valor que obtendríamos si una vez alcanzada la era de recombinación, el gravitino NLSP fuese completamente estable sin decaer a axino-axión, y posteriormente se continuase la evolución del universo según $\Lambda$CDM.

Para un dado valor de $f_a$, la región por encima de la línea azul correspondiente se encuentra excluida por exceso de producción de materia oscura fría. La región por debajo se encuentra permitida, pero necesitamos asumir otra contribución a la DM, por ejemplo axiones producidos mediante el mecanismo de \textit{misalignment}. A pesar que este último puede ser un escenario interesante, un tercer candidato a materia oscura fría se encuentra más allá del alcance de este trabajo, por lo tanto nos enfocaremos en los puntos del espacio de parámetros que cumplan con el contorno graficado en azul. Por otra parte, las líneas discontinuas naranjas corresponden a diferentes valores de la fracción de gravitino NLSP $f_{3/2}$.

Notamos que utilizando Ec.~(\ref{decaytoaa}) podemos definir tres regiones distintas en la figura, dependiendo si el decaimiento del gravitino NLSP a axino LSP más axión tiene lugar después de la era actual, entre recombinación y la era actual, o antes de recombinación. Por ejemplo, en el panel de la derecha tenemos la región con masas chicas o tiempos de vida grandes de gravitino NLSP para $m_{3/2}\lesssim 0.2$ GeV, una zona de masas intermedias para $0.2 \lesssim m_{3/2}\lesssim 10$ GeV, y una región de masas grandes o tiempos de vida pequeños de gravitino NLSP para $m_{3/2} \gtrsim 10$ GeV. 

Finalmente, la región magenta en ambos paneles se encuentra excluida por observaciones cosmológicas~\cite{Poulin:2016,Berezhiani:2015,Chudaykin:2016,Chudaykin:2017,Bringmann:2018}, teniendo en cuenta los importantes límites sobre la fracción de gravitino NLSP que puede decaer a radiación oscura, $f_{ddm}^{\text{DR}}$. Generalmente, estas restricciones son presentadas como límites superiores para dicha fracción. Para la región intermedia de masas de gravitino NLSP tenemos~\cite{Poulin:2016,Berezhiani:2015,Chudaykin:2016,Chudaykin:2017,Bringmann:2018} $f_{ddm}^{\text{DR}} \lesssim 0.042$. Los límites correspondientes para las regiones de masas chicas y grandes puede encontrarse en Ref.~\cite{Poulin:2016}. Como puede verse en la Figura~\ref{figaxinoLSP1}, estos límites permites diferentes valores de $f_{3/2}$ dependiendo de la relación entre las masas del axino y el gravitino $r_{\tilde a}$ (ver Ec. (\ref{ddmfraction})). Recordamos que $f_{ddm}^{\text{DR}}$ corresponde a la cantidad de densidad reliquia fría perdida como especies ultrarelativistas. Para el decaimiento del gravitino NLSP a axino LSP más un axión, este último siempre será ultrarelativista, pero el comportamiento del axino producido de forma no termal dependerá de $r_{\tilde a}$.

\begin{figure}[t!]
\begin{center}
 \begin{tabular}{cc}
 \hspace*{-4mm}
 \epsfig{file=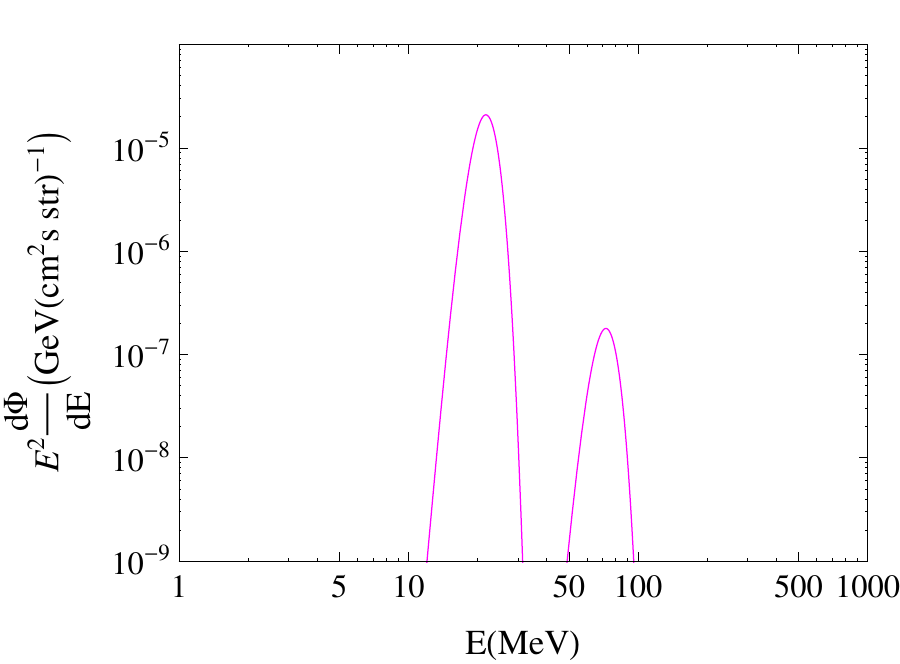,height=7cm} 
    \end{tabular}
    \caption{Señal de doble línea de rayos gamma generada por el decaimiento de axino y gravitino como componentes múltiples de materia oscura del universo. El espectro ha sido generado con un axino LSP de 43.5 MeV y un gravitino NLSP de 145 MeV decayendo a un par fotón-neutrino para $f_a=10^{13}$ GeV y $\left|U_{\tilde{\gamma} \nu}\right|=10^{-6}$. Las líneas espectrales fueron convolucionadas con una Gaussiana asumiendo un instrumento con una resolución energética del 10\%. La doble línea graficada corresponde a la emisión de un cuadrado de $10^\circ \times 10^\circ$ alrededor del centro galáctico. Para determinar los límites al espacio de parámetros utilizamos los resultados estándares de búsquedas de líneas espectrales, aplicándolos a cada una de las líneas predichas por separado.
}
    \label{flux_lines}
\end{center}
\end{figure}

\begin{figure}[t!]
\begin{center}
 \begin{tabular}{cc}
 \hspace*{-4mm}
 \epsfig{file=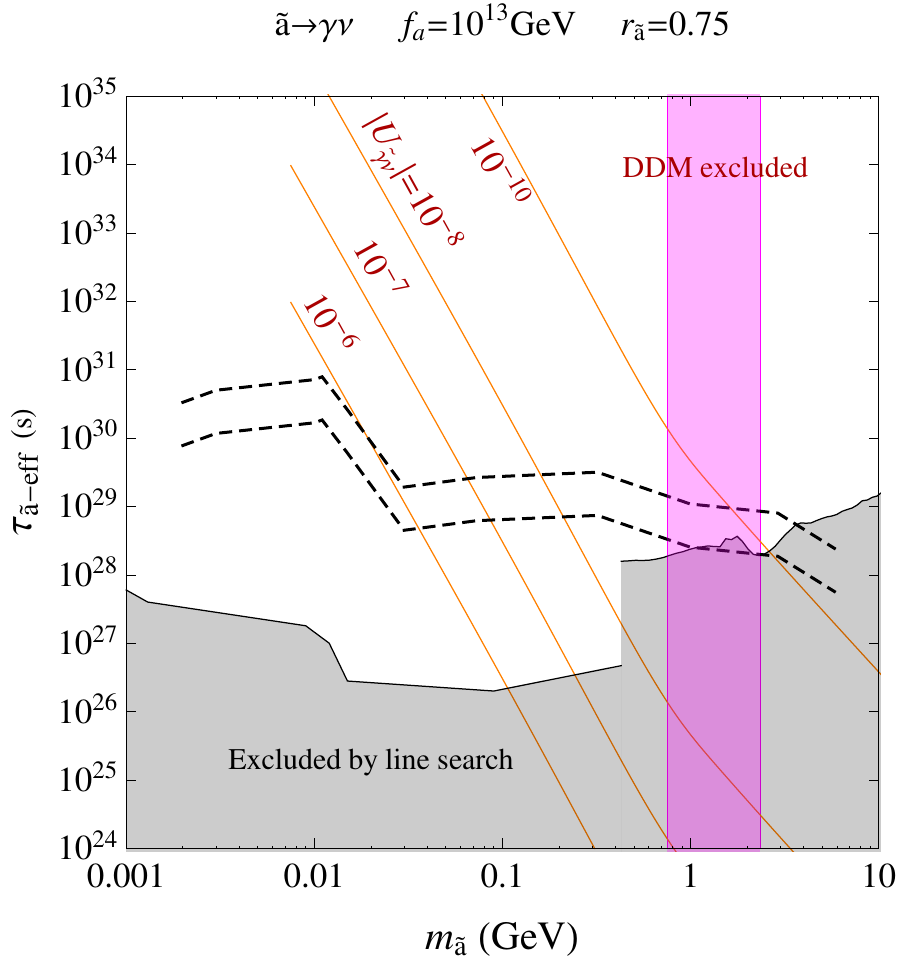,height=7cm} 
       \epsfig{file=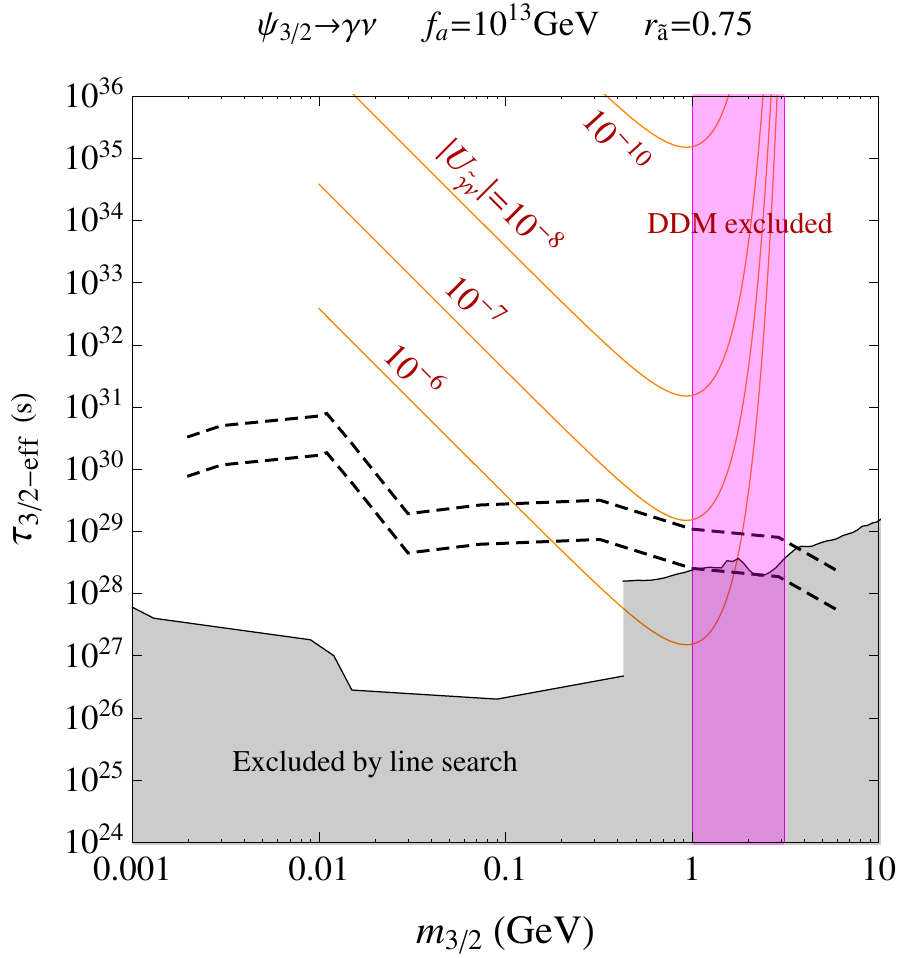,height=7cm}
       \vspace*{-0.8cm}       
   \\ & 
    \end{tabular}
    \caption{
    Limites sobre el tiempo de vida efectivo en función de la masa del axino LSP (panel izquierdo) y de la masa del gravitino NLSP (panel derecho). Las señales de rayos gamma provenientes de axinos y gravitinos son analizadas por separado en los paneles de la izquierda y derecha, respectivamente. La región gris debajo de las curvas negras se encuentran excluidas por observaciones del halo galáctico realizadas por COMPTEL y \Fermi LAT~\cite{Ackermann:2015lka}. La región por debajo de la línea discontinua negra superior (inferior) puede ser explorada por e-ASTROGAM~\cite{eAstrogamAngelis:2017} con observaciones del centro galáctico asumiendo el perfil Einasto B (Burkert) de DM. Las líneas naranjas corresponden a las predicciones del $\mu\nu$SSM para varios valores representativos de $|U_{\tilde{\gamma} \nu}|$, para el caso con $f_a=10^{13}$ GeV y $r_{\tilde a}=0.75$. El límite inferior $m_{3/2}\gtrsim 0.017$ GeV se obtuvo a partir de Ec.~(\ref{relicgravitinos}) asumiendo el límite conservativo $T_R \gtrsim 10^4$ GeV. La región magenta está excluida por observaciones cosmológicas para modelos DDM~\cite{Poulin:2016,Berezhiani:2015,Chudaykin:2016,Chudaykin:2017,Bringmann:2018}, considerando el límite sobre $f_{ddm}^{\text{DR}}$. 
}
    \label{figaxinoLSP2}
\end{center}
\end{figure}

\subsubsection{Límites por observaciones de rayos gamma y perspectivas de detección}
\label{subsec: Prospects}

Para analizar el efecto de considerar un escenario con DDM multicomponente sobre la señal de rayos gamma, en la Figura~\ref{flux_lines} mostramos un ejemplo del espectro generado por gravitinos y axinos coexistiendo. Utilizamos las líneas por separado para determinar los límites sobre el modelo, considerando que el flujo resultante de cada línea no resulta afectado por la otra señal. Como veremos a continuación, en el caso de una doble línea detectable, las dos señales resultan estar localizadas a energías lo suficientemente distintas como para no superponerse, dando lugar a dos líneas que pueden resolverse por separado. Por esta razón, en la Figura~\ref{figaxinoLSP2} presentamos dos paneles, uno para cada candidato a materia oscura. En el panel de la izquierda (derecha) mostramos los límites al espacio de parámetros considerando la producción de una línea espectral proveniente de un axino LSP (gravitino NLSP) decayendo a  $\gamma \nu$. En este ejemplo hemos fijado $f_a=10^{13}$ GeV y $r_{\tilde{a}}=0.75$, es decir, los mismos valores que fueron utilizados para graficar la curva azul superior en el panel de la izquierda de la Figura~\ref{figaxinoLSP1}.  Por lo tanto, es importante remarcar que los paneles izquierdo y derechos de la Figura~\ref{figaxinoLSP2} corresponden al mismo escenario con axino LSP y gravitino NLSP, y que los límites obtenido en ambos paneles tienen que ser tomados en cuenta para cada punto del espacio de parámetros. Por ejemplo, el punto con $m_{3/2}=0.5$ GeV y $|U_{\tilde{\gamma} \nu}|=10^{-7}$ no parece estar excluido por búsqueda de líneas en el panel derecho, sin embargo para la correspondiente masa de axino $m_{\tilde{a}}=r_{\tilde a}\times 0.5=0.375$ GeV y $|U_{\tilde{\gamma} \nu}|=10^{-7}$ se puede ver claramente en el panel izquierdo que dicho punto se encuentra excluido.

Se pueden ver los efectos que tiene el decaimiento de gravitinos NLSP a axinos LSP y axiones en el tiempo efectivo de vida media de los dos candidatos a DM, comparando el panel izquierdo de la Figura~\ref{figaxinoLSP2} con la región naranja de la Figura~\ref{figaxinoalone1}, donde solo se considera al axino LSP como único candidato a materia oscura. Para $m_{\tilde{a}}\leq 0.8$ GeV, el tiempo efectivo es más grande que la vida media sin la presencia de gravitino NLSP, debido principalmente a la baja fracción de axino $f_{\tilde a}$ contribuyendo al primer término de la Ec.~(\ref{newtime22}). Una situación similar ocurre para $0.8\leq m_{\tilde{a}}\leq 1.2$ GeV, donde la contribución del segundo término de la Ec.~(\ref{newtime2}) es significativa debido a la contribución de axinos producto del decaimiento de gravitinos. Por otra parte, para $m_{\tilde{a}} \geq$ $1.2$ GeV, el decaimiento del gravitino ocurre a tiempos lo suficientemente tempranos y/o la fracción de gravitinos $f_{3/2}$ es tan chica que el tiempo efectivo de vida es similar al escenario mostrado en la Figura~\ref{figaxinoalone1} con inicialmente 100\% axino DM.

El panel derecho de la Figura~\ref{figaxinoLSP2} muestra el mismo espacio de parámetros, pero para el tiempo efectivo de vida media del gravitino NLSP. Podemos ver el efecto que produce la reducción de la densidad energética de gravitinos debido a su decaimiento a axino LSP para $1\leq m_{3/2}\leq 3$ GeV, como se puede deducir de Ec.~(\ref{newtime2}). Masas de gravitino más ligeras implican un tiempo de vida más grande, por lo tanto en dicha región podemos tener en la época actual una distribución de materia oscura con ambos candidatos produciendo una doble línea espectral.

\begin{figure}[H]
 \begin{center}
  \begin{tabular}{cc}
 \hspace*{-0mm}
  	\includegraphics[height=6cm]{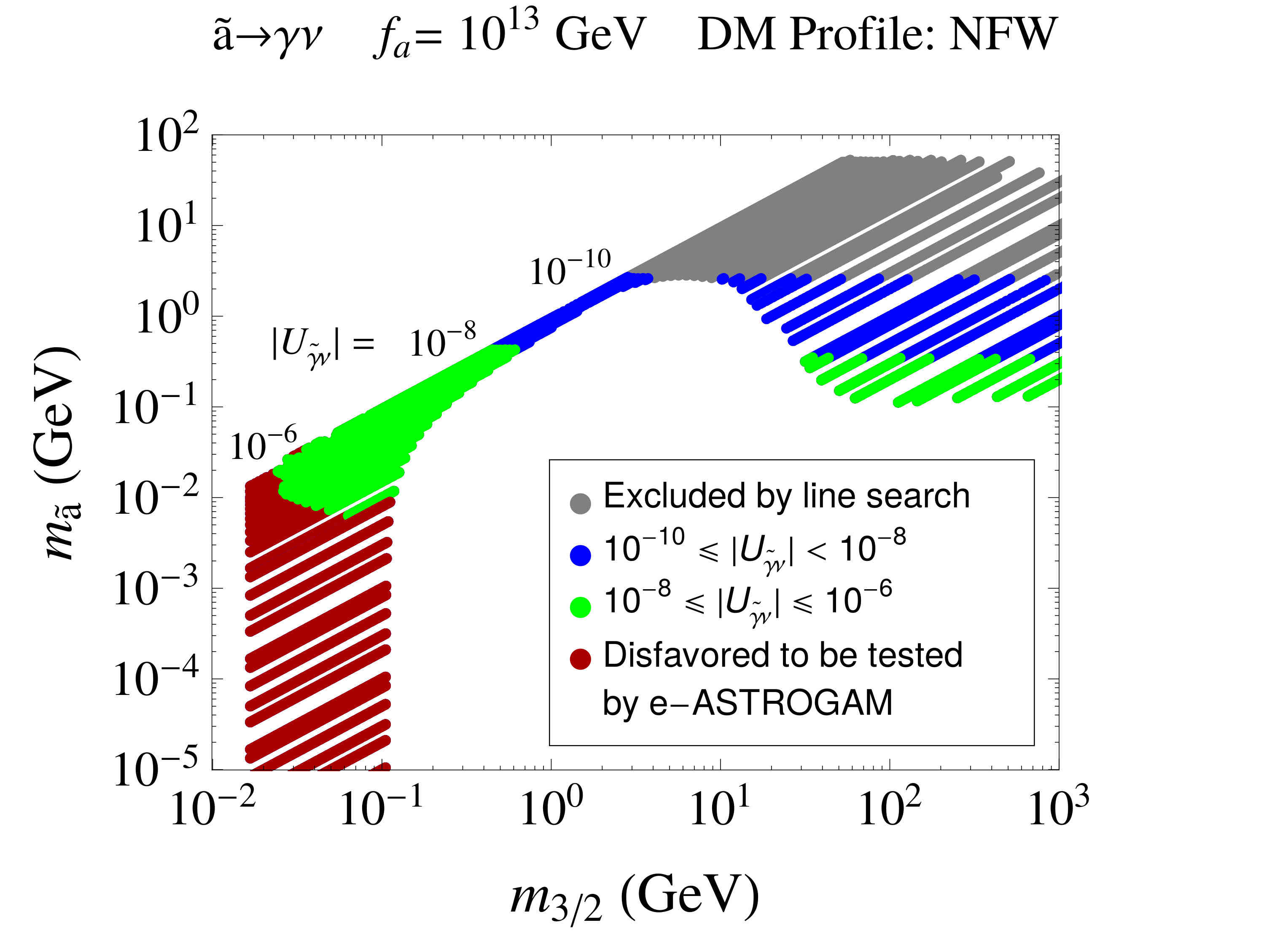} 
\hspace*{-0.5cm} \includegraphics[height=6cm]{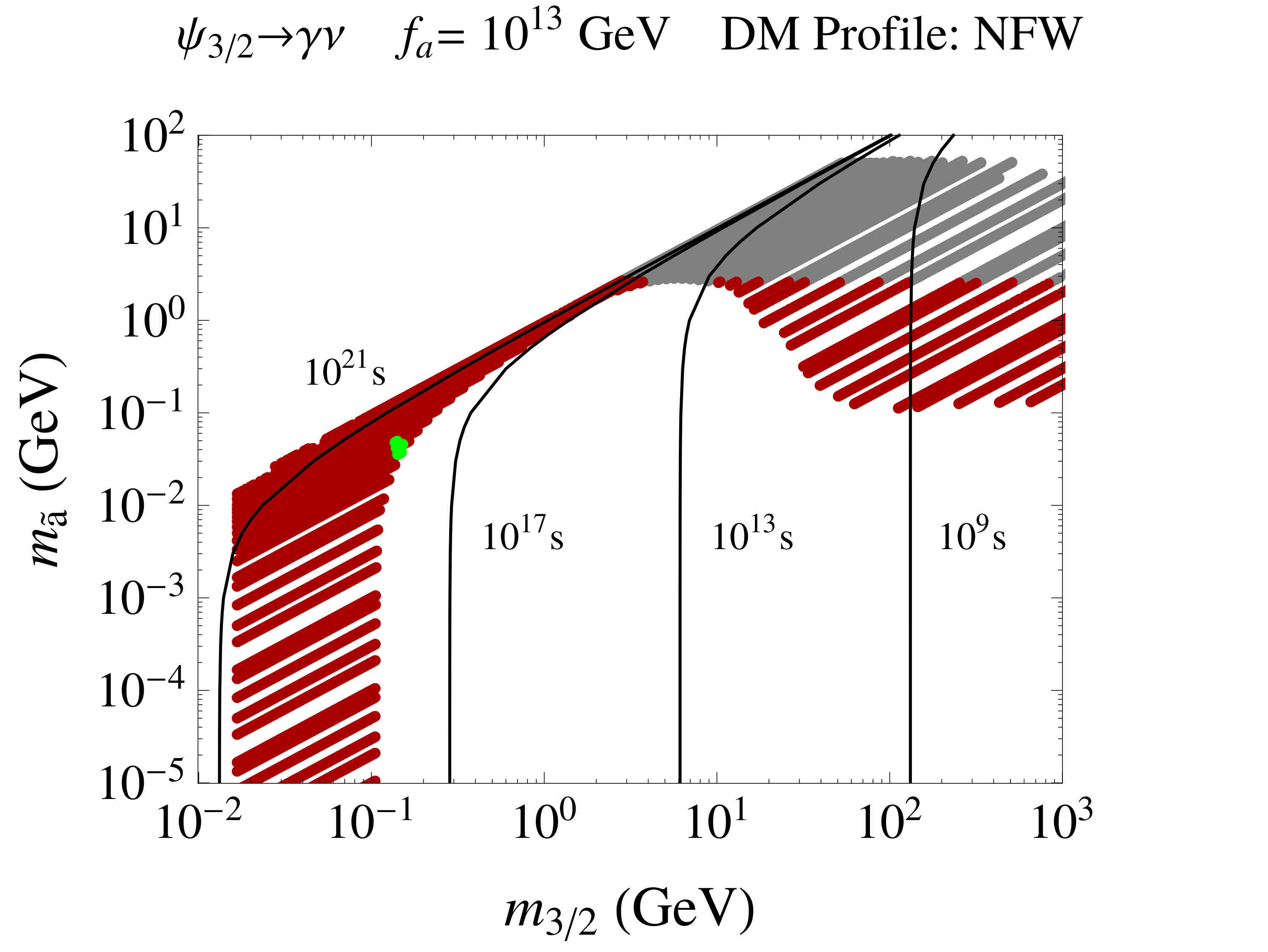}
   \\ \hspace*{-0mm} \includegraphics[height=6cm]{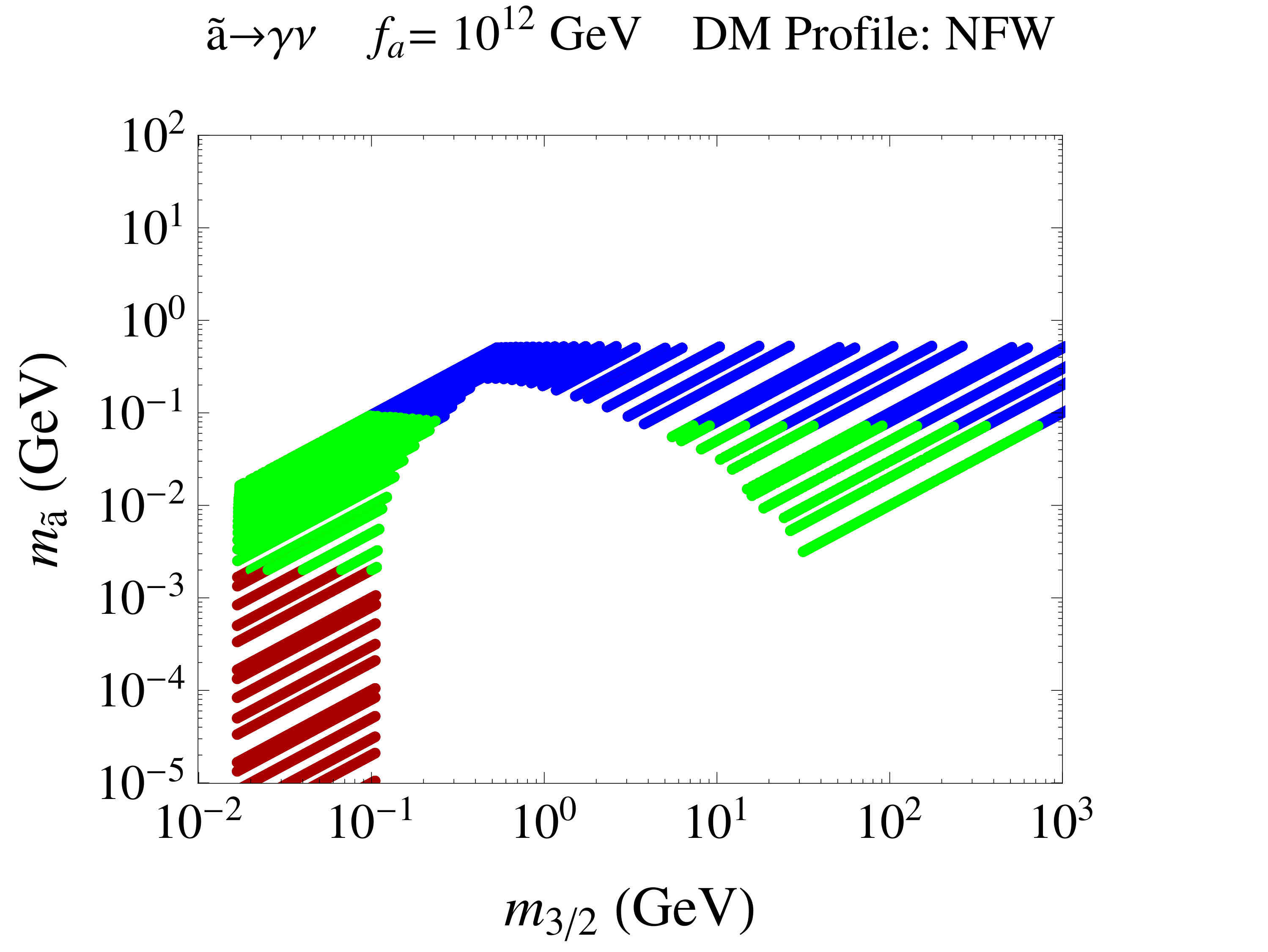} \hspace*{-0.5cm} \includegraphics[height=6cm]{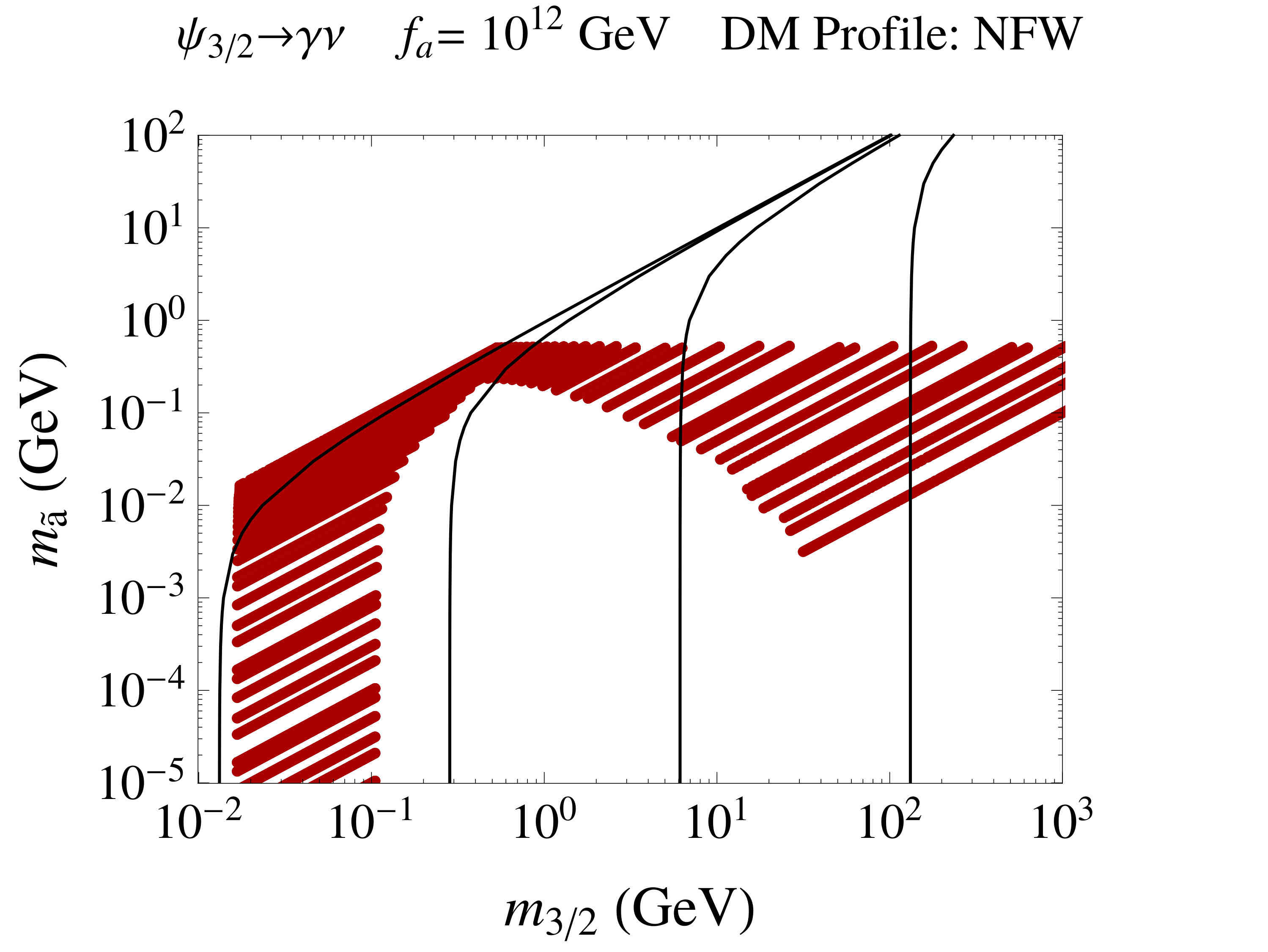}\\     
	\hspace*{-0mm}  \includegraphics[height=6cm]{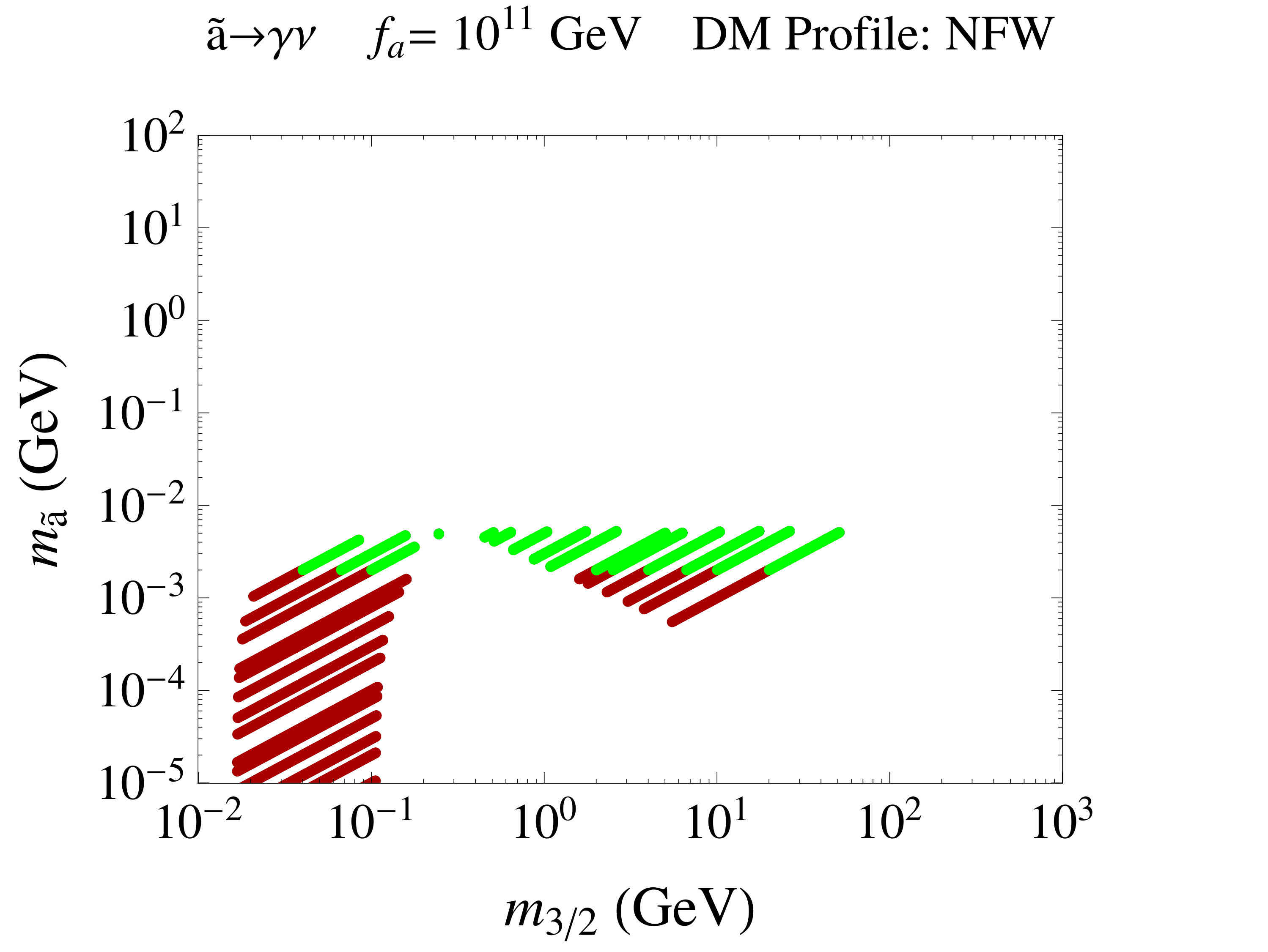} \hspace*{-0.5cm}
       \includegraphics[height=6cm]{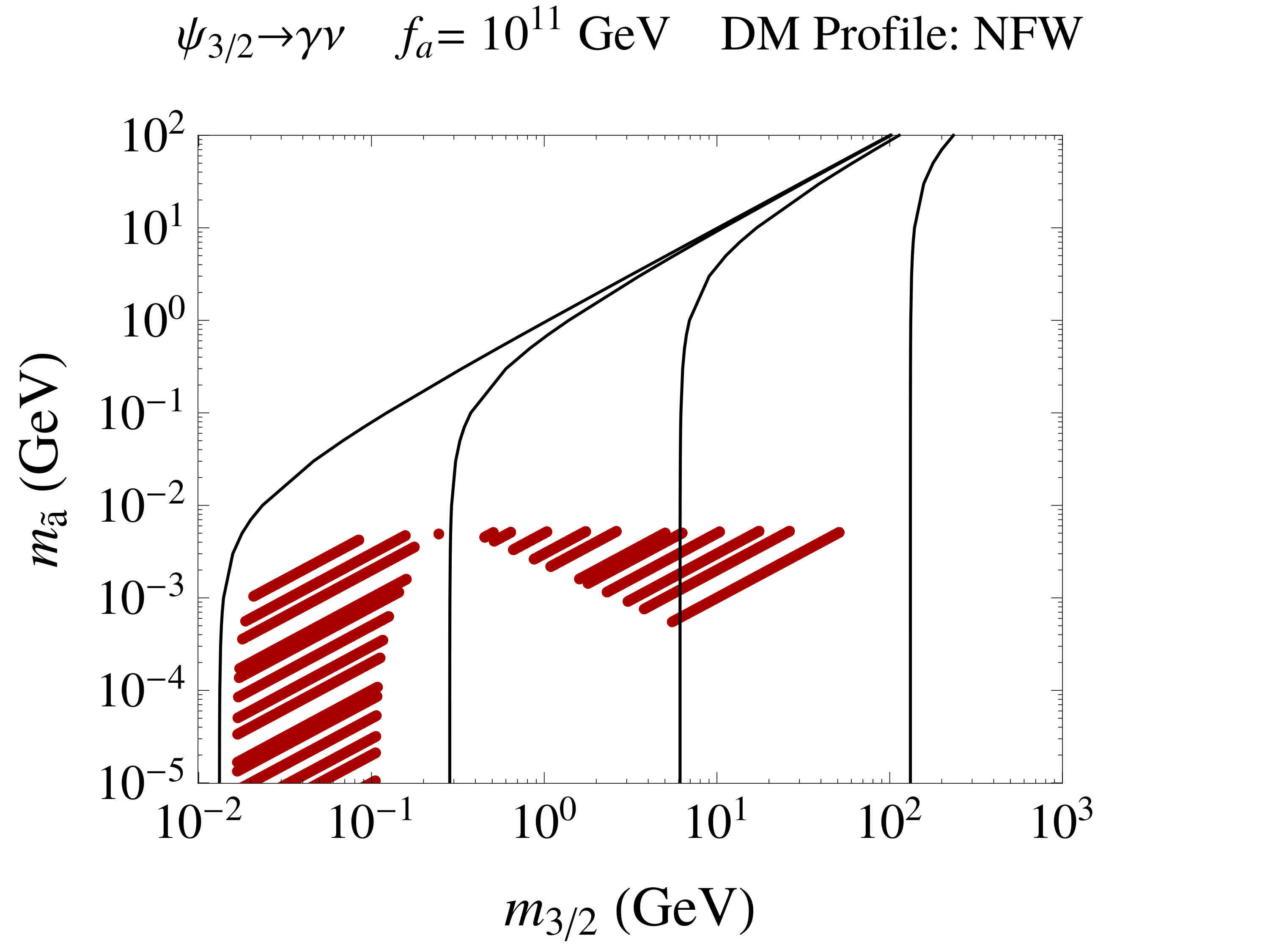} 
  \end{tabular}
  \caption{\small 
  Perspectivas de detección axino LSP con gravitino NLSP. Las señales de rayos gamma se analizaron por separado para axinos (izquierda) y gravitinos (derecha), asumiendo un perfil NFW de DM. La región gris está excluida por búsqueda de líneas en el halo galáctico por COMPTEL y \Fermi LAT~\cite{Ackermann:2015lka}. Las regiones verdes y azules podrían ser exploradas por e-ASTROGAM para dos rangos representativos de $|U_{\tilde{\gamma}\nu}|$ en el $\mu\nu$SSM, en el panel superior izquierdo se muestran algunos valores representativos, el resto de los paneles es similar. Si el mismo punto puede ser explorado en ambas regiones, una señal de doble línea podría ser medida. La región roja no podría ser explorada por e-ASTROGAM. En los paneles derechos, las líneas negras muestran distintos valores de $\tau_{3/2}\simeq {\Gamma}^{-1} (\psi_{3/2} \rightarrow a \, \tilde{a})$. Todos los puntos satisfacen $\Omega_{3/2}h^2+\Omega_{\tilde{a}}h^2$ igual a $\Omega_{cdm}^{\text{Planck}}h^2$ en recombinación, y los límites para modelos con DDM para $f_{ddm}^{\text{DR}}$.
}
    \label{figconstrains1}
\end{center}
\end{figure}

Para realizar un análisis completo del espacio de parámetros permitido, hemos hecho un barrido sobre el siguiente rango
\begin{equation}
10^{-4} \leq r_{\tilde a} \leq 0.95.
\end{equation}
El resultado se presenta en la Figura~\ref{figconstrains1}, donde la señal de rayos gamma del decaimiento del axino y gravitino fueron analizadas por separado en los paneles izquierdos y derechos, respectivamente\footnote{Si paridad-R se conserva, todos los puntos mostrados están permitidos ya que satisfacen las cotas sobre la abundancia de materia oscura y los límites establecidos por radiación oscura.}. Las regiones verdes y azules corresponden a los puntos que podrían ser explorados por un instrumento con las características proyectadas de e-ASTROGAM asumiendo el perfil NFW de DM y una región de interés de 10$^\text{o} \times$10$^\text{o}$ alrededor del centro galáctico, para diferentes valores del parámetro de mezcla fotino-neutrino $|U_{\tilde{\gamma} \nu}|$. En particular, los puntos verdes corresponden al rango más natural de $|U_{\tilde{\gamma} \nu}|$, como fue discutido en Ec.~(\ref{relaxing}). Mencionamos también que dicho rango incluye el espacio de parámetros típico que puede reproducir la física de neutrinos en modelos de violación de paridad-R con términos bilineales, por lo tanto, los límites encontrados en este trabajo también pueden aplicarse a esos modelos.

Como podemos ver en la figura, para valores de $r_{\tilde a}$ cercanos a 1, es decir, la región superior permitida, recuperamos el espacio de parámetros permitido que se obtuvo cuando se consideró que el axino LSP era el único candidato a materia oscura (sin el efecto del gravitino NLSP). Esto se debe a que las restricciones a modelos DDM para $f_{ddm}^{\text{DR}}$ se relajan ya que ${\Gamma} (\psi_{3/2} \rightarrow a \, \tilde{a})  \rightarrow 0$ cuando $r_{\tilde a} \rightarrow 1$ (ver Ec.~(\ref{decaytoaa})). Los efectos restantes sobre los flujos de rayos gamma son solamente determinados por la fracción de densidad reliquia del LSP y del NLSP. Para valores más pequeños de  $r_{\tilde a}$, el espacio de parámetros permitido se modifica empleando los límites a modelos DDM, y generando las dos regiones separadas de masas permitidas mostradas en la Figura~\ref{figconstrains1}.

\newpage
\thispagestyle{empty}

En ese sentido, notamos que la exclusión por DDM en la Figura~\ref{figaxinoLSP1} para $r_{\tilde a}=0.75$ y $f_a=10^{13}$~GeV, dejan dos ramas de soluciones permitidas para la curva azul e ($m_{3/2} \lesssim 1$ GeV y $m_{3/2} \gtrsim 3$ GeV) con la cantidad correcta de densidad reliquia.

A partir de la Figura~\ref{figconstrains1}, podemos concluir que en el escenario DDM considerado, una región significativa del espacio de parámetros, dentro del rango de masas $7 \, \text{MeV} \lesssim m_{\tilde a} \lesssim 3$ GeV y $20 \, \text{MeV} \lesssim m_{3/2} \lesssim 1$ TeV, podrá ser explorado por la próxima generación de telescopios de rayos gamma, lo cual se debe principalmente gracias a la línea espectral proveniente del axino LSP (paneles de la izquierda). Notemos que las anchuras de decaimientos del axino y gravitino a fotón-neutrino, que son cantidades relevantes en el cálculo del flujo de fotones (ver Ecs.~(\ref{eq:decayFlux}),~(\ref{newtime2}) y~(\ref{newtime22})), cumplen $\Gamma(\tilde{a}\rightarrow\gamma\nu_i) > \Gamma(\psi_{3/2}\rightarrow\gamma\nu_i)$ dentro del rango de masas mencionado para los valores de $r_{\tilde a}$ que discutimos en la Sección~\ref{sec:DM multiplecomp}.

Para valores menores de $r_{\tilde a}$, esperaríamos un flujo importante atribuido al decaimiento del gravitino. Sin embargo, debido a la región excluida por exceso de radiación oscura, la masa del gravitino que cumple lo antes dicho tiene que ser muy grande, lo que a su vez implica un tiempo de vida media del gravitino muy chico, es decir que ha decaído a axino-axión antes de $t_{\text{hoy}}$ (en los paneles derechos que corresponden a los puntos a la derecha de la línea negra etiquetada con $10^{17}$ s), o la masa del gravitino tiene que ser muy chica $m_{3/2} \lesssim 0.1$ GeV, lo que implica que el flujo de fotones es muy pequeño (ver Ec.~(\ref{lifetimegamma})). Por lo tanto, esperamos que una línea espectral proveniente de gravitino NLSP sea detectable en una región muy pequeña. Esta se puede ver como la región verde en el panel superior derecho que corresponde a $f_a=10^{13}$~GeV, con masas $m_{3/2}\sim 150$ MeV y $m_{\tilde a}\sim 40$ MeV. Como los mismos puntos pueden ser explorados por e-ASTROGAM en ambos paneles, una señal conformada por dos líneas espectrales puede ser medida como una característica clara de este modelo al ser muy difícil de obtener mediante procesos astrofísicos. Las líneas negras en los paneles de la derecha nos indican que la señal detectable proveniente del decaimiento del gravitino NLSP se encuentra que en la región del espacio de parámetros con $\tau_{3/2}\simeq {\Gamma}^{-1} (\psi_{3/2} \rightarrow a \, \tilde{a})~>~t_{\text{hoy}}$, como es esperado.

Notemos también que en este escenario existe una importante región del espacio de parámetros detectable (señal proveniente del axino LSP) donde se permiten gravitinos pesados, $m_{3/2}> 10$ GeV. En esta región, además de los canales a fotón-neutrino, otros modos de decaimiento se vuelven relevantes, como aquellos que involucran bosones de gauge $Z$, $W$ y al bosón de Higgs en estados finales de dos y tres cuerpos~\cite{Gomez-Vargas:2016ocf}. Esto resulta en un incremento en la anchura de decaimiento total del gravitino a partículas visibles, produciendo una inyección energía mediante especies hadrónicas y electromagnéticas en el universo temprano que pueden alterar el proceso de nucleosíntesis (BBN) o el espectro del fondo cósmico de microondas (CMB). Sin embargo, la región de masas de gravitino para lo cual esto ocurre, el proceso de decaimiento dominante es gravitino NLSP a axino LSP y axión, como se puede estimar comparando los paneles derechos de las Figuras~\ref{figconstrains1} y~\ref{figaxinoLSP2}. Otro factor importante a tener en cuenta es que en la región de gravitinos pesados, su densidad reliquia de origen térmico es muy chica, pues domina la producción de axinos (ver las curvas naranjas discontinuas en la Figura~\ref{figaxinoLSP1}). Por lo tanto, no hay una cantidad significativa de energía depositada en el sector visible durante el universo temprano debido a la reducción de la densidad energética del gravitino, y para nuestro análisis del espectro de rayos gamma es suficiente considerar que el gravitino solo decae mediante la interacción a dos cuerpos a fotón-neutrino.

\begin{figure}[t!]
 \begin{center}
  \begin{tabular}{cc}
 \hspace*{-4mm}
 \includegraphics[height=6cm]{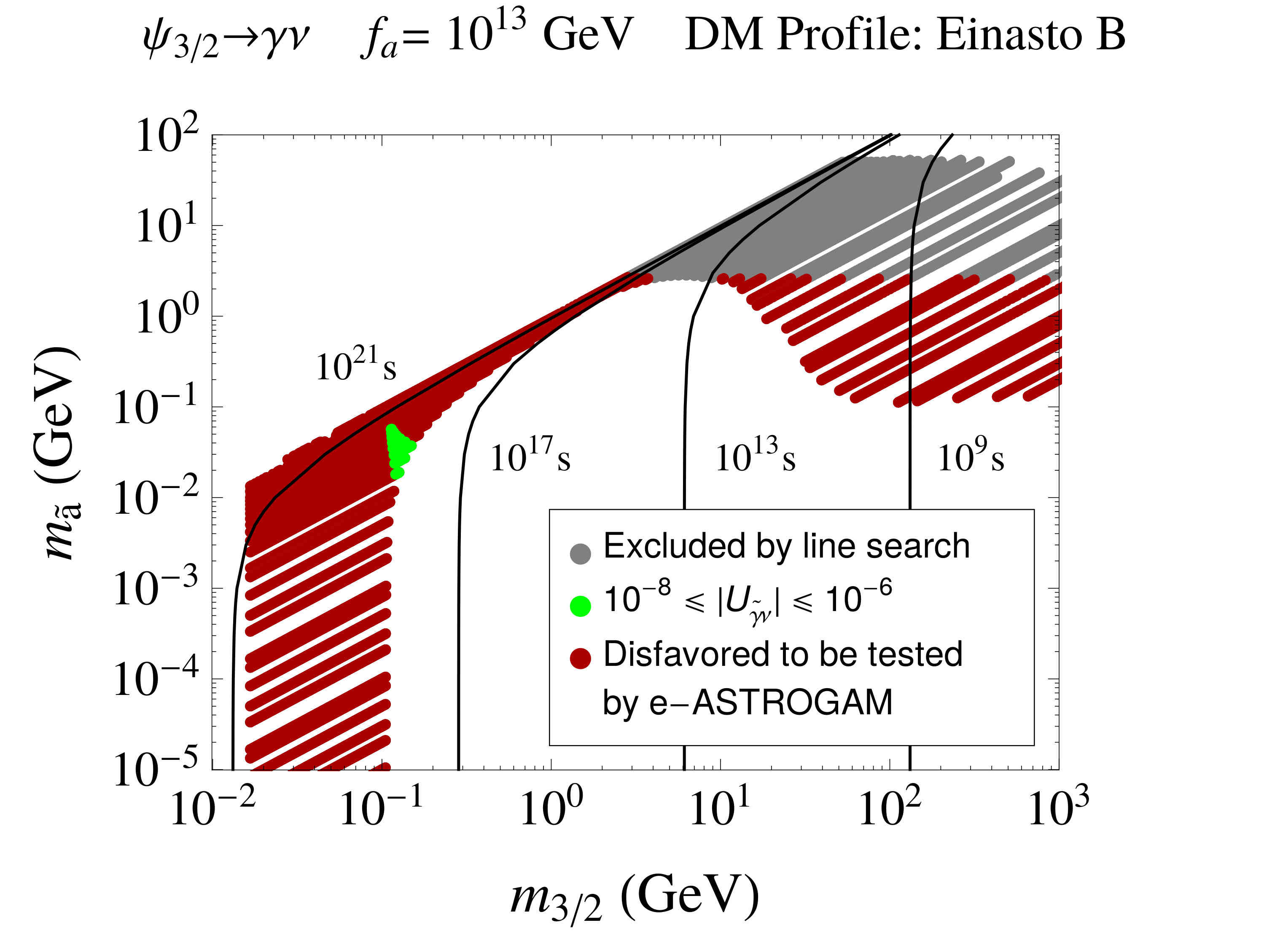} 
  \end{tabular}
  \caption{Mismas referencias que en la Figura~\ref{figconstrains1}, pero solo mostrando el espacio de parámetros para una señal de rayos gamma proveniente del decaimiento de gravitinos a fotón-neutrino asumiendo un perfil Einasto B de materia oscura
}
    \label{figprofiles11}
\end{center}
\end{figure}

Por otro lado, como hemos mencionado, para obtener los resultados mostrados en la Figura~\ref{figconstrains1} utilizamos la sensibilidad propuesta para el instrumento e-ASTROGAM asumiendo un perfil NFW de materia oscura. En la Figura~\ref{figprofiles11}, mostramos los efectos en las perspectivas de detección al considerar un perfil de distribución de materia oscura distinto, Einasto B. En este caso, solamente presentamos el espacio de parámetros considerando la señal proveniente de gravitino NLSP decayendo a fotón-neutrino, y una escala PQ $f_a=10^{13}$~GeV. Comparando esta figura con el panel superior derecho de la Figura~\ref{figconstrains1}, podemos ver que la región verde que corresponde a la zona que puede ser explorada por e-ASTROGAM mediante la detección de una doble línea espectral se extiende levemente al rango $100\lesssim m_{3/2}\lesssim 200$ MeV y $10 \lesssim m_{\tilde a}\lesssim 60$ MeV. Hemos verificado que el resto de los paneles de la Figura~\ref{figconstrains1} no son modificados de manera significativa por esta fuente de incerteza astrofísica.

\begin{figure}[t!]
 \begin{center}
  \begin{tabular}{cc}
  \hspace*{-0mm}
  \includegraphics[height=6cm]{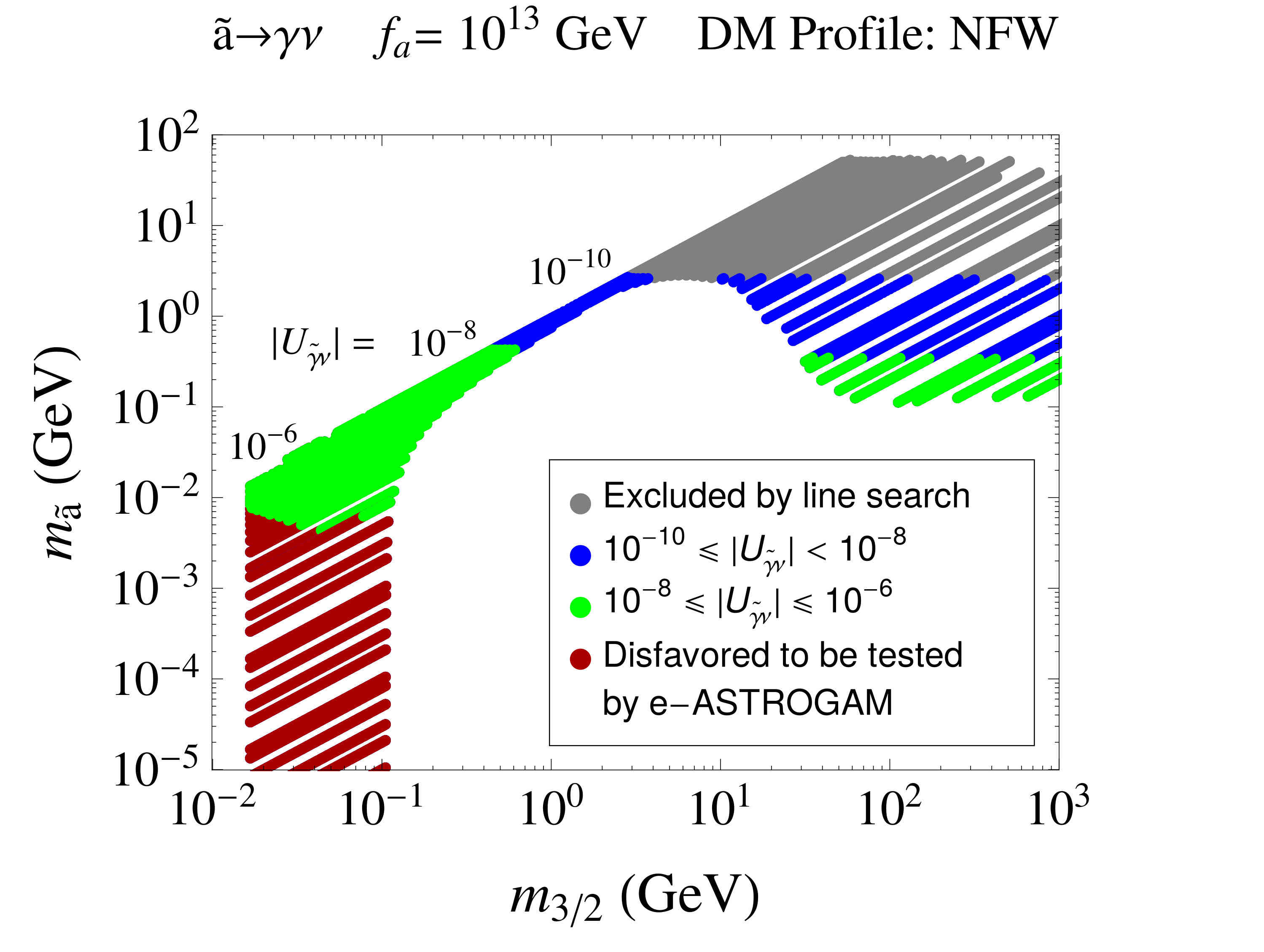} 
\hspace*{-0.5cm} \includegraphics[height=6cm]{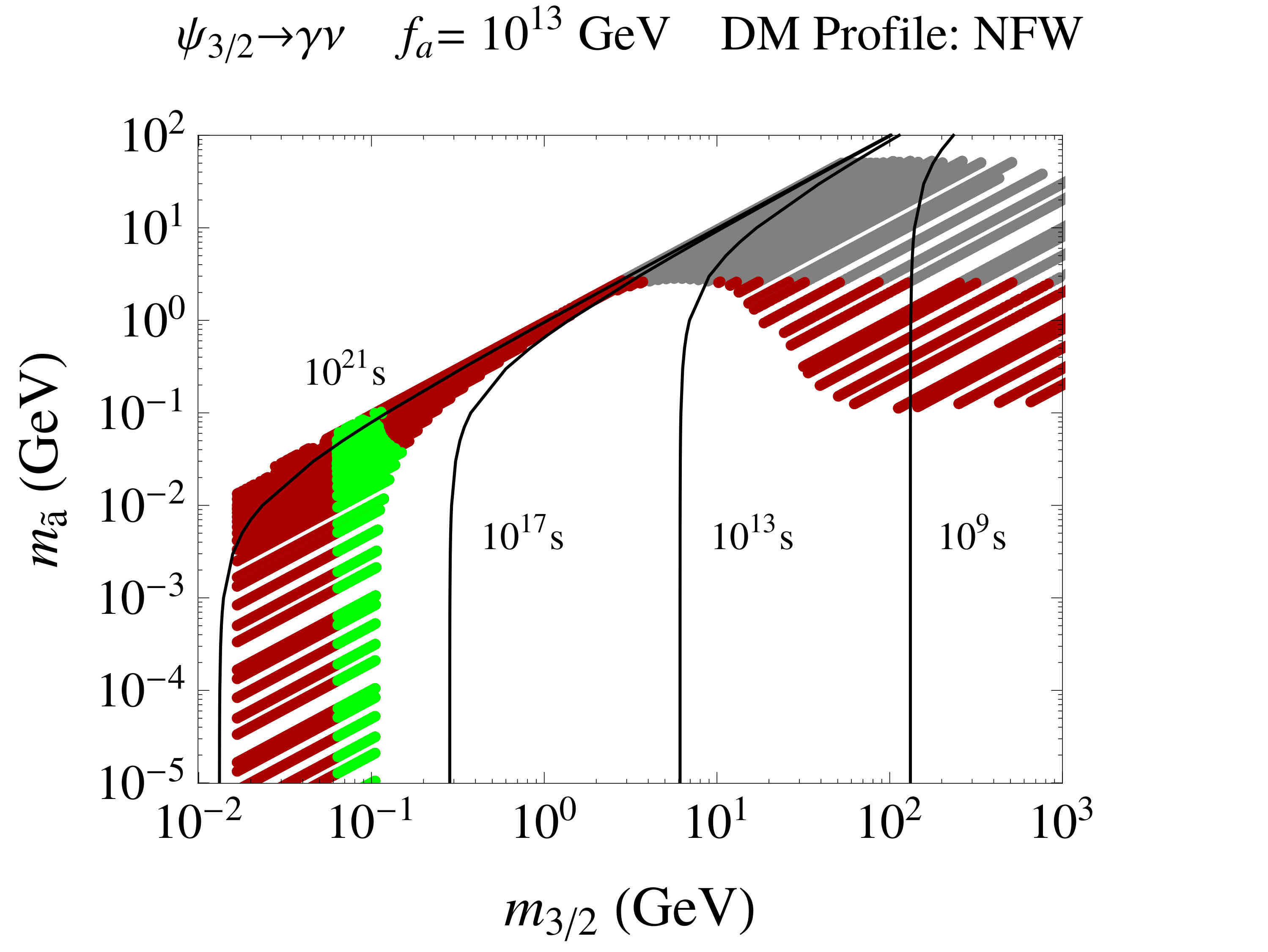}
\\     
\hspace*{-0mm} 
\\ \hspace*{-0mm} \includegraphics[height=6cm]{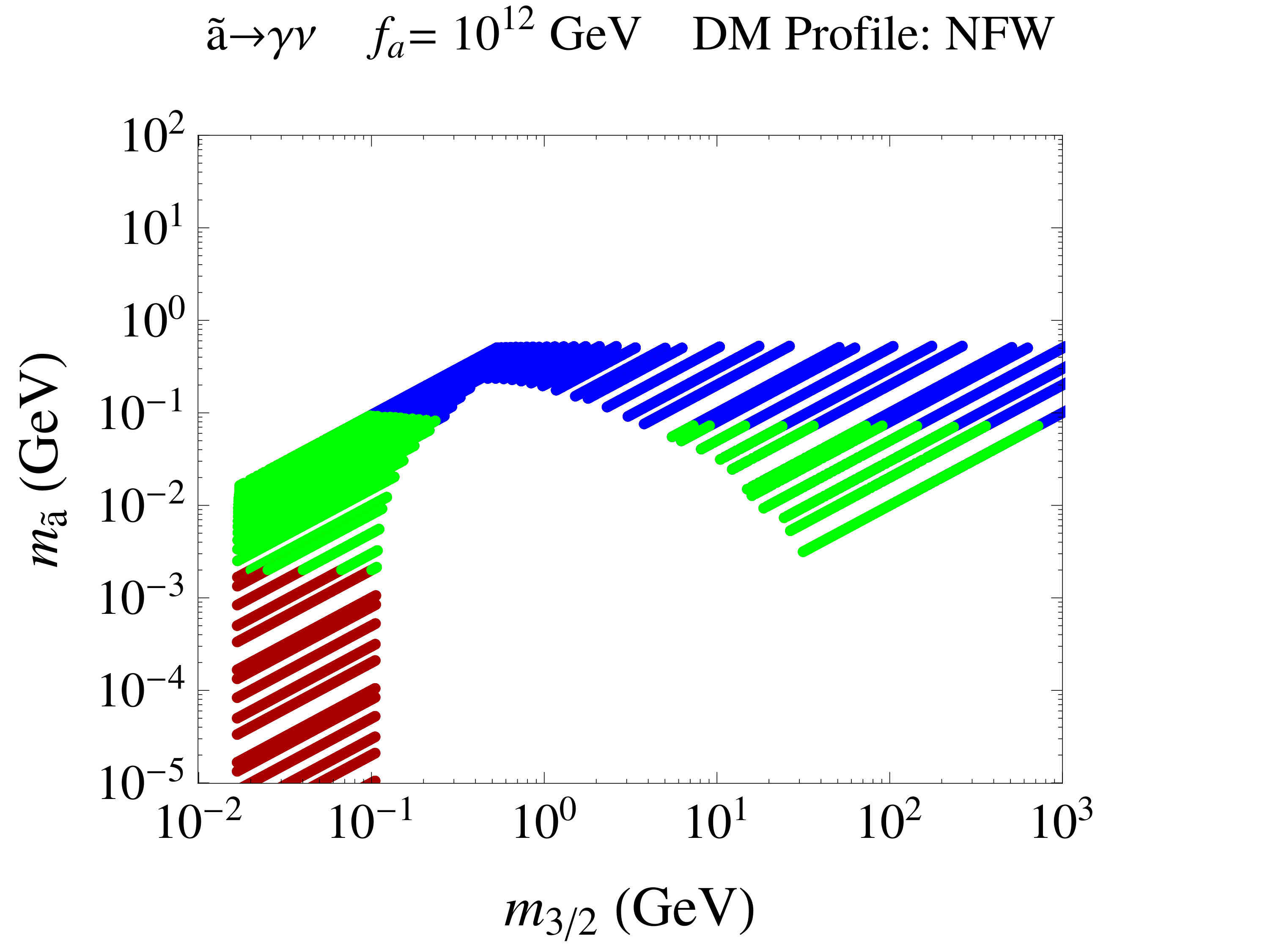} 
\hspace*{-0.5cm}\includegraphics[height=6cm]{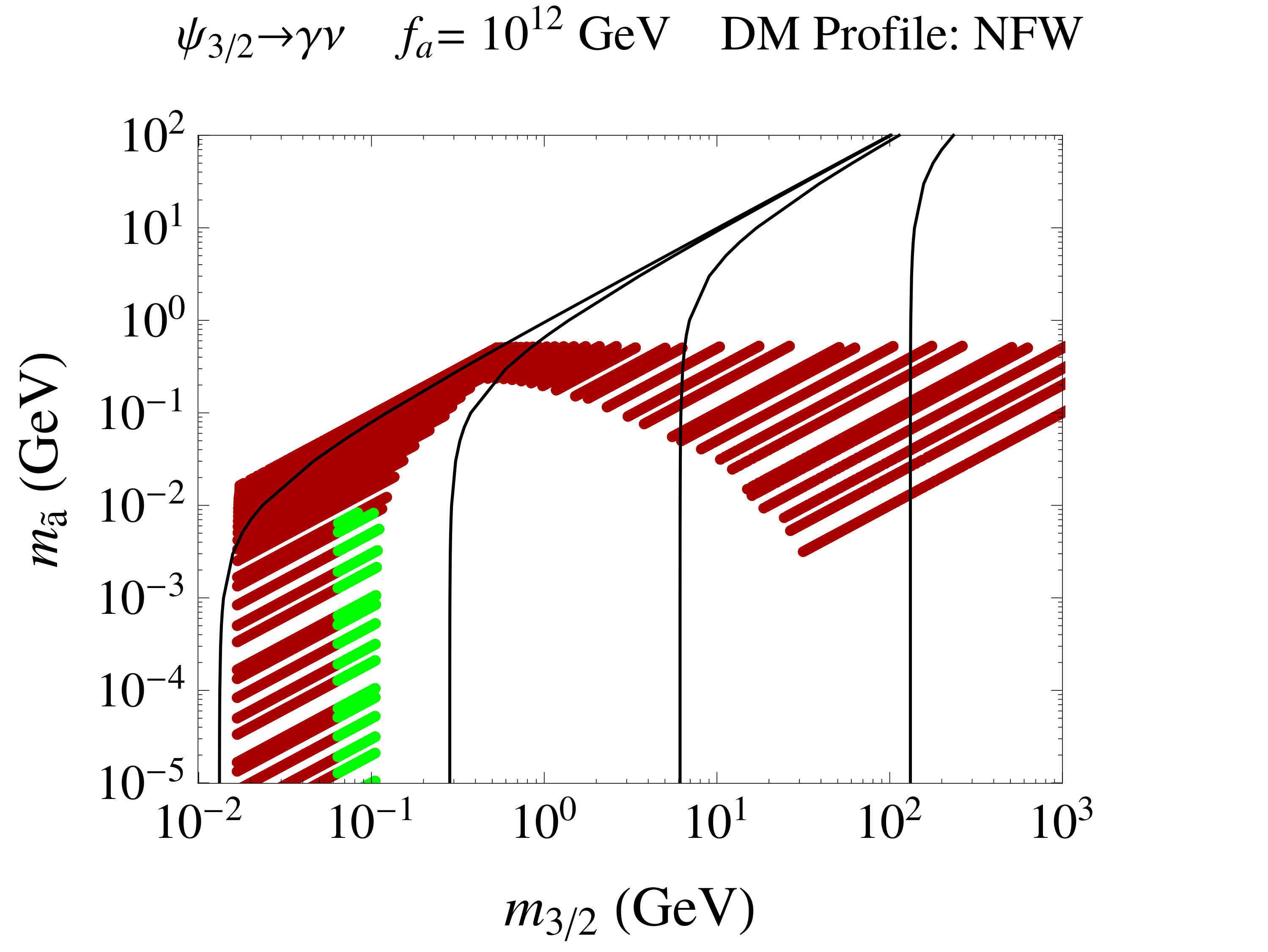}
\\
\hspace*{-0mm} 
\hspace*{-0mm}  \includegraphics[height=6cm]{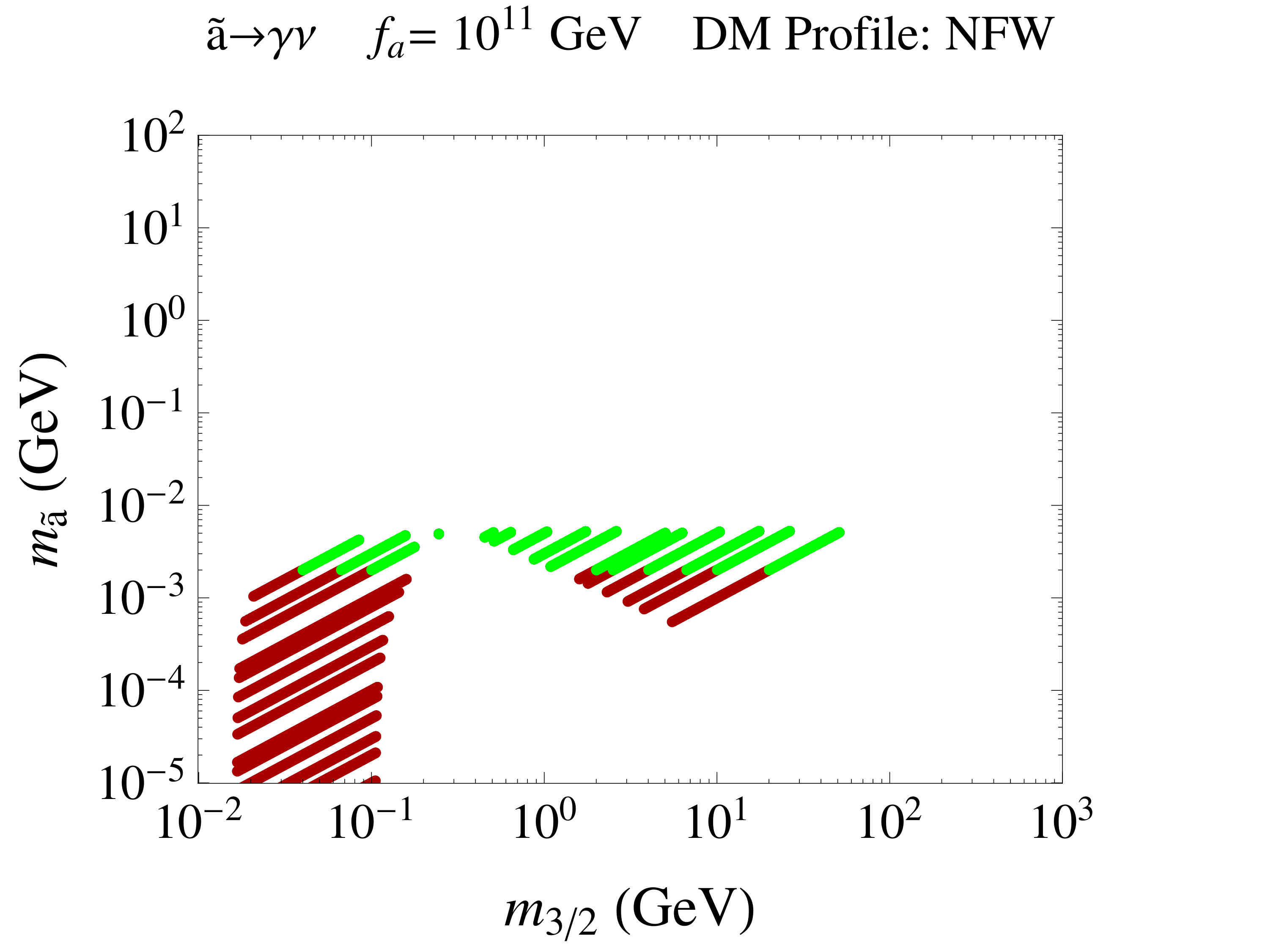} 
\hspace*{-0.5cm}\includegraphics[height=6cm]{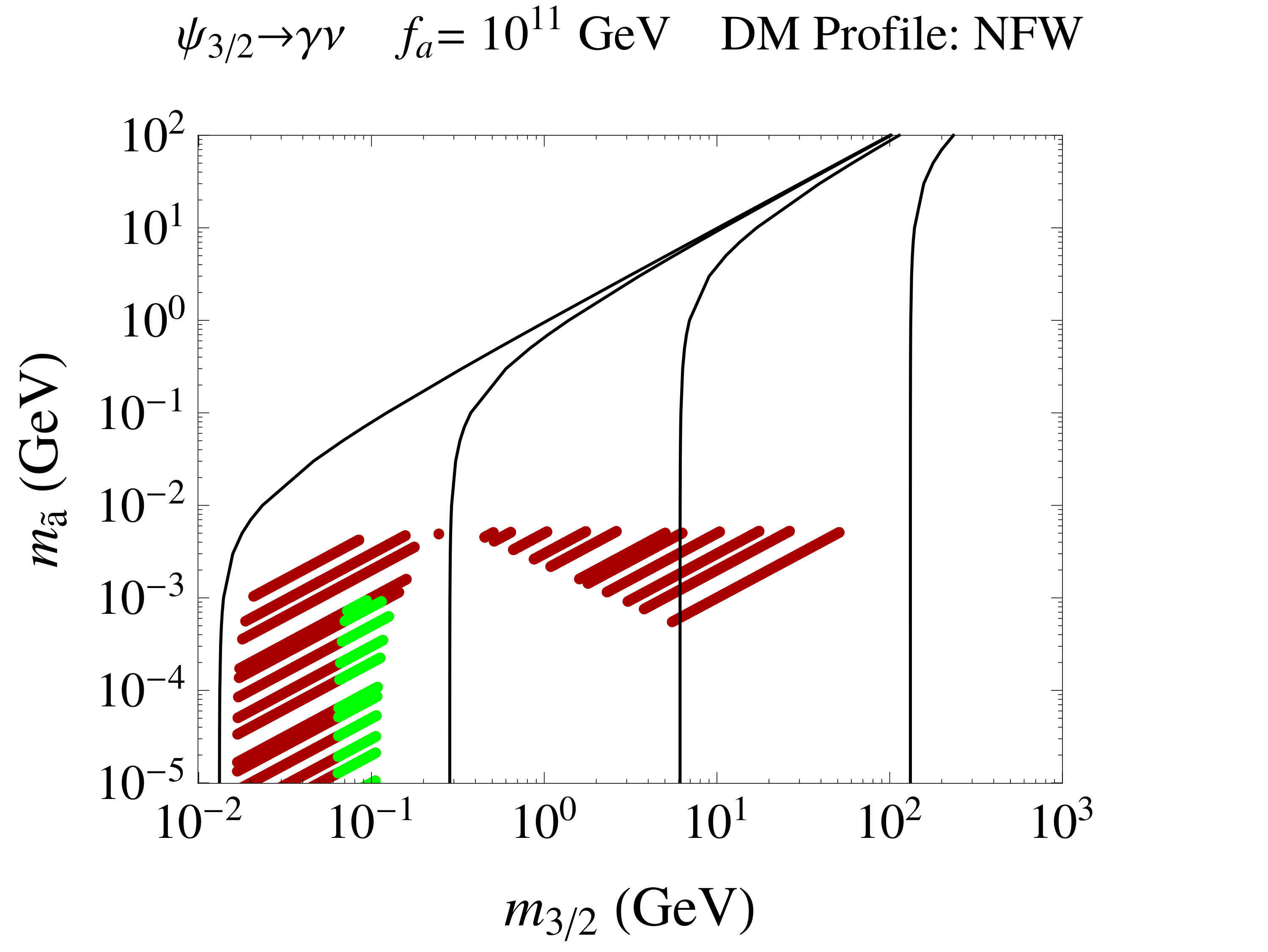}
\\ 
  \end{tabular}
  \caption{
  Mismas referencias que en la Figura~\ref{figconstrains1}, pero incrementando la sensibilidad de e-ASTROGAM un orden de magnitud con respecto a la propuesta actual.
}
    \label{figprofiles1}
\end{center}
\end{figure}

En cambio, las características de las distintas propuestas para la próxima generación de telescopios de rayos gamma constituye una fuente de incertezas más significativa.  En nuestro análisis nos basamos en la sensibilidad de la última propuesta presentada por la colaboración e-ASTROGAM. Sin embargo, si la sensibilidad de los instrumentos del futuros próximo se incrementan un orden de magnitud más (o se alcanza un efecto similar al mejorarse nuestra comprensión y modelaje del fondo de rayos gamma), se puede generar un importante impacto en la detectabilidad de la señal. Esto es factible considerando los asombrosos avances en las técnicas y tecnologías desarrolladas en los últimos años, y en la ardua competencia que atraviesan las colaboraciones experimentales en búsqueda de recursos renovando sus propuestas periódicamente. En este sentido, los resultados anteriores han sido conservativos. Por lo tanto, en la Figura~\ref{figprofiles1} asumiendo un perfil NFW de DM al igual que en la Figura~\ref{figconstrains1}, mostramos el espacio de parámetros detectable para un instrumento con la sensibilidad aumentada como hemos mencionado. Podemos ver que la región a explorar verde se ha expandido significativamente, en especial la zona donde esperaríamos detectar una doble línea. Para $f_a=10^{13}$~GeV, esta región se encuentra dentro de los rangos de masas $60\lesssim m_{3/2}\lesssim 200$ MeV y $4 \lesssim m_{\tilde a}\lesssim 100$ MeV. Aún para $f_a=10^{12}$~GeV donde no existía la posibilidad de detectar una doble línea en Figura~\ref{figconstrains1}, ahora es posible obtenerla dentro del siguiente rango de masas $60\lesssim m_{3/2}\lesssim 100$ MeV y $2 \lesssim m_{\tilde a}\lesssim 10$ MeV.


\newpage

\section{Caso 2: gravitino LSP - axino NLSP
}
\label{sec:gravitinoLSPaxinoNLSP}

Como es esperado, muchas de las definiciones para este escenario son similares a las estudiadas en el caso 1, Sección~\ref{sec:axinoLSPgravitinoNLSP}. Sin embargo, existen algunas diferencias que remarcaremos, las cuales conducen a predicciones donde las señales abarcan regiones distintas del espacio de parámetros permitido. En el marco de supergravedad existe un término de interacción que no viola paridad-R entre el gravitino, el axino y el axión. Para este escenario el axino es el NLSP y el gravitino el LSP, por lo tanto resulta que~\cite{Hamaguchi:2017}
\begin{equation}
\Gamma(\tilde{a} \rightarrow \psi_{3/2} \, a)\simeq\frac{m_{\tilde{a}}^5}{96\pi m_{3/2}^2 M_P^2}(1-r_{3/2})^2(1-r_{3/2}^{2})^3, 
\label{decayto32a}
\end{equation}
donde despreciamos la masa del axión y 
\bea
r_{3/2} \equiv \frac{m_{3/2}}{m_{\tilde{a}}}.
\eea
Este canal de decaimiento domina sobre el decaimiento a dos cuerpos mostrado en Ec.~(\ref{decay2bodyaxino}) el cual involucra parámetros que violan paridad-R. El tiempo de vida del axino estimado en Ec.~(\ref{axinolifetime}) suponiendo que era el LSP, resulta en este caso
\begin{equation}
\tau_{\tilde{a}}\simeq \Gamma^{-1} \left( \tilde{a} \rightarrow \psi_{3/2} \, a \right) \simeq 1.18 \times 10^{13} s \, \left( \frac{m_{3/2}}{0.1 \text{ GeV}} \right)^2 \, \left( \frac{1 \text{ GeV}}{m_{\tilde{a}}} \right)^5,
\end{equation}
donde en la segunda igualdad hemos despreciado la contribución de $r_{3/2}$ en Ec.~(\ref{decayto32a}), lo cual es válido solo si $m_{3/2}\ll m_{\tilde{a}}$.

Es importante notar que si bien ${\Gamma^{-1}}(\tilde{a}\rightarrow\gamma\nu_i) \gg t_{hoy}$, esto no sucede para ${\tau}_{\tilde{a}}$, indicando que este resultado afecta las ecuaciones de la densidad reliquia de ambos candidatos a materia oscura

\subsection{Abundancia de gravitino LSP y axino NLSP
}
\label{subsec: relic2}

A diferencia del gravitino, cuyo tiempo de vida es mucho más grande que la edad del universo, el axino NLSP tiene un tiempo de vida inferior, como vimos en Ec~(\ref{decayto32a}), por lo tanto su decaimiento afecta su densidad reliquia
\bea
\Omega_{\tilde{a}}h^2 = \Omega_{\tilde{a}}^{\text{TP}}h^2 e^{-(t_{hoy}-t_0)/ \tau_{\tilde{a}}},
\label{NLSPrelicgeneralgravLSP}
\eea
donde $\Omega_{\tilde{a}}^{\text{TP}}h^2$ corresponde a la densidad reliquia del axino NLSP si este fuese completamente estable y no decayese. Para el modelo de axiones KSVZ la densidad reliquia de origen termal fue dada en Ec.~(\ref{relicaxinos}).

Para calcular la densidad reliquia del gravitino LSP necesitamos considerar tanto la producción térmica, dada en la Ec.~(\ref{relicgravitinos}) como la producción no térmica generada por el decaimiento del axino NLSP. Entonces
\bea
\Omega_{3/2}h^2 = \Omega_{3/2}^{\text{TP}}h^2  + \Omega_{3/2}^{\text{NTP}}h^2,
\label{LSPrelicgeneralgravLSP}
\eea
donde
\bea
\Omega_{3/2}^{\text{NTP}}h^2 = r_{3/2} \, \Omega_{\tilde{a}}^{\text{TP}}h^2 \, (1-e^{-(t_{hoy}-t_0)/ \tau_{\tilde{a}}}).
\label{LSPrelicNTP}
\eea

Asumiendo que $T_R \gtrsim 10^4$ GeV, junto con valor medido de densidad de materia oscura fría por la colaboración Planck, encontramos los mismos límites que en caso anterior. Para cada valor de $f_a$, obtenemos un límite superior para $m_{\tilde a}$. Por ejemplo, hallamos que $m_{\tilde{a}} \lesssim 50, 0.5, 0.005$ GeV para $f_a=10^{13}, 10^{12}, 10^{11}$ GeV, respectivamente. Mientras que para el gravitino obtenemos un límite inferior para su masa, $m_{3/2} \gtrsim 0.017 \text{ GeV}$. Como en el escenario que estamos analizando el gravitino es el LSP, dicho límite es incompatible con el límite de la masa del axino NLSP para $f_a=10^{11}$ GeV, por lo tanto tomaremos con $f_a\geq 10^{12}$ GeV. Teniendo en cuenta la discusión para la producción de axiones mediante el mecanismo de \textit{misalignment} (ver Ec.~(\ref{relicaxions})) trabajaremos a continuación con el siguiente rango para la escala de PQ
\begin{equation}
10^{12} \leq f_a \leq 10^{13}\ \text{GeV}.
\label{pqscalegravLSP}
\end{equation}

Como esperamos, si $\tau_{\tilde{a}}\ll t_{hoy}$, recuperamos las relaciones usuales~\cite{Covi:1999ty,Choi:2011yf,Roszkowski:2014}
\bea
\Omega_{\tilde{a}}h^2 &\simeq& 0,\\
\Omega_{3/2}h^2 &\simeq& \Omega_{3/2}^{\text{TP}}h^2 + r_{3/2} \, \Omega_{\tilde{a}}^{\text{TP}}h^2.
\eea

En este caso, el decaimiento del axino NLSP a gravitinos LSP y axiones produce radiación oscura. Mientras que los axinos generados mediante este mecanismo son siempre ultrarelativistas, la condición de los gravitinos depende de la relación $r_{3/2}$. Definimos de manera análoga a Ec~(\ref{ddmfraction}) la fracción de axinos NLSP que decaen a radiación oscura
\bea
f_{ddm}^{\text{DR}} = f_{\tilde{a}} \left( 1 - r_{3/2} \right),
\label{ddmfractiongravLSP}
\eea
con
\bea
f_{\tilde{a}}=\frac{\Omega^{\text{TP}}_{\tilde{a}}}{\Omega_{cdm}^{\text{Planck}}}
\label{fractionaxino}
\eea
como la fracción de axino NLSP.

\subsection{Flujo de rayos gamma}
\label{sec: gammaflux}

Para calcular el flujo de rayos gamma proveniente del halo galáctico para un escenario con múltiples componentes de materia oscura que decaen podemos utilizar las ecuaciones descritas en la Sección~\ref{subsec: gamma-ray flux}, intercambiando los roles del axino y gravitino.

Asumiendo que no existen términos de interacción entre las distintas componentes de materia oscura, y que la distribución de cada una de ellas es homogénea en la distribución de materia oscura de la galaxia, para cada componente $i$ podemos escribir
\begin{equation}
\tau_{\text{DM}_i\text{-eff}}= f_{\text{DM}_i}^{-1} \, \tau_{\text{DM}_i}, \hspace{1cm} \text{with} \hspace{1cm} f_{\text{DM}_i} = \frac{\Omega_{\text{DM}_i}}{\Omega_{cdm}^{\text{Planck}}},
 \label{newtime1gravLSP}
\end{equation}
donde $f_{\text{DM}_i}$ es la fracción de la componente $i$ de la materia oscura, $\tau_{\text{DM}_i}$ es la inversa de la anchura de decaimiento a fotones, y $\tau_{\text{DM}_i\text{-eff}}$ el tiempo efectivo de vida que puede ser contrastado con los límites presentados por las colaboraciones experimentales.

Sin embargo, si existen interacciones entre los candidatos a materia oscura, como es nuestro caso debido al decaimiento del axino NLSP a gravitino LSP, sus fracciones en Ec.~(\ref{newtime1gravLSP}) tienen que reemplazarse por (ver Ecs.~(\ref{NLSPrelicgeneralgravLSP}) y~(\ref{LSPrelicgeneralgravLSP})):
\bea
 f_{\text{DM}_2} & \rightarrow & f_{\tilde{a}} \, e^{-(t_{hoy}-t_0)/\tau_{\tilde{a}}},
\label{nuevagravLSP}
\\
 f_{\text{DM}_1} & \rightarrow & f_{3/2} + r_{3/2} \, f_{\tilde{a}} \, \left(1 - e^{-(t_{hoy} - t_0)/\tau_{\tilde{a}}} \right).
\label{fractionstimegravLSP}
\eea
Por lo tanto, sus tiempos efectivos de vida resultan
\bea
 \tau_{\tilde{a}\text{-eff}} & = & \left( f_{\tilde{a}} \, e^{-(t_{hoy}-t_0)/\tau_{\tilde{a}}} \right)^{-1} \, \Gamma^{-1}\left( \tilde{a} \rightarrow \gamma \nu_i \right) ,
\label{newtime2gravLSP}
\\
 \tau_{3/2\text{-eff}} & = & \left[ f_{3/2} + r_{3/2} \, f_{\tilde{a}} \, \left(1 - e^{-(t_{hoy} - t_0)/\tau_{\tilde{a}}} \right) \right]^{-1} \, \Gamma^{-1}\left( \psi_{3/2} \rightarrow \gamma \nu_i \right).
\label{efflifetimesgravLSP}
\eea

\subsection{Resultados}
\label{sec: resultsgravLSP}

\subsubsection{Límites a partir de observaciones cosmológicas}
\label{sec: cosmoobstconstraints}

\begin{figure}[t!]
\begin{center}
 \begin{tabular}{cc}
 \hspace*{-4mm}
 \epsfig{file=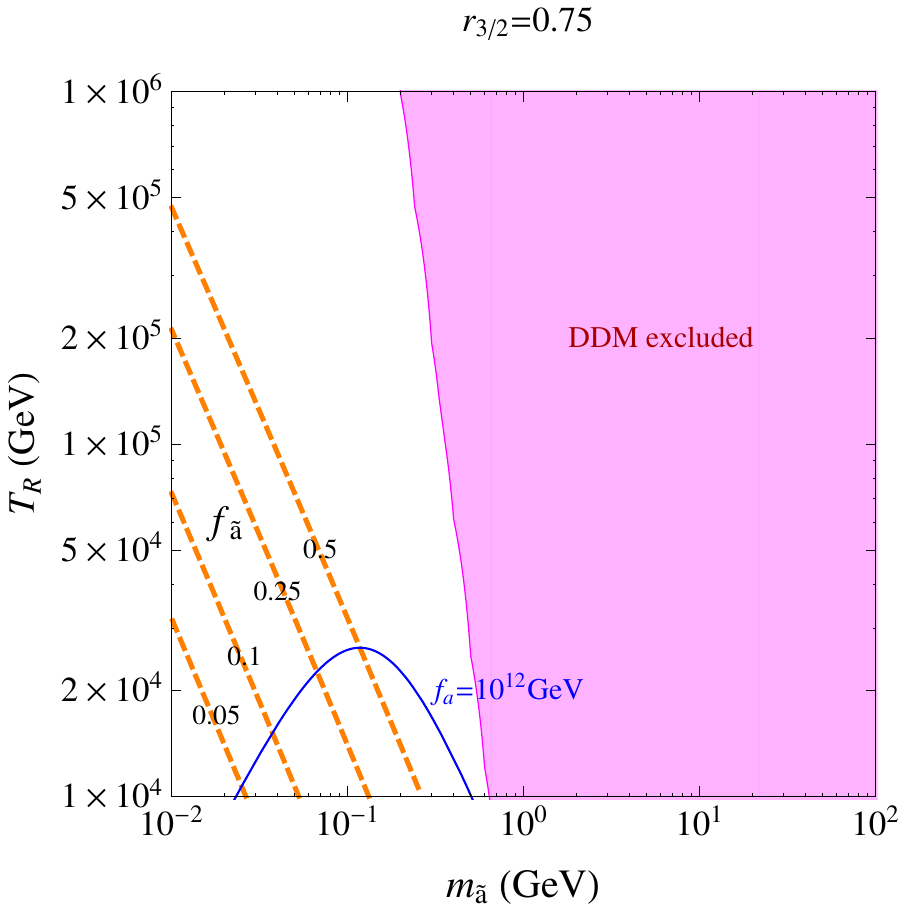,height=7cm} 
       \epsfig{file=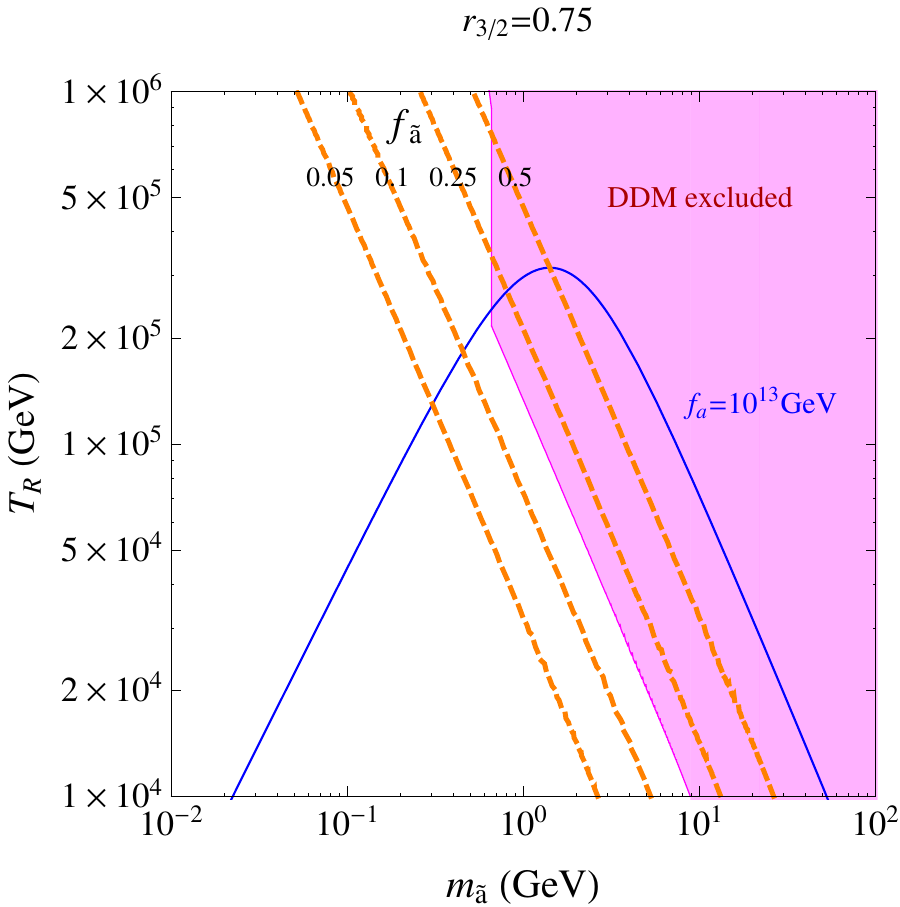,height=7cm}   
       \vspace*{-0.8cm}       
   \\ & 
    \end{tabular}
    \caption{Límites en la temperatura de \textit{reheating} en función de la masa del axino NLSP para es escenario con gravitino LSP y una relación entre sus masas $r_{3/2}=0.75$ y dos valores de la escala de PQ $f_a=10^{12}$ GeV (panel izquierdo) y $10^{13}$ GeV (panel derecho). 
    El resto de las referencias son iguales a las de la Figura~\ref{figconstrains1}, salvo que las líneas discontinuas naranjas corresponden a la fracción de axino NLSP. Los límites superiores $m_{\tilde{a}} \lesssim 0.5, 50$ GeV en los paneles izquierdos y derechos, respectivamente, fueron obtenidos con Ec.~(\ref{relicaxinos}) asumiendo $T_R \gtrsim 10^{4}$ GeV.
}
    \label{figgravitinoLSP1}
\end{center}
\end{figure}

Para ilustrar el espacio de parámetros disponible en este escenario, en la Figura~\ref{figgravitinoLSP1} mostramos  $T_R$ en función de $m_{\tilde a}$ para un valor fijo de $r_{3/2}=0.75$. El panel izquierdo corresponde a una escala de PQ igual a $f_a=10^{12}$ GeV, mientras que en el panel derecho $f_a=10^{13}$ GeV.

Las líneas azules satisfacen la producción de materia oscura medida por la colaboración Planck, por lo tanto, la región por encima se encuentra excluida por exceso de producción de materia oscura fría. Mientras que la región magenta se encuentra excluida por exceso de producción de radiación oscura. Los límites sobre $f_{ddm}^{\text{DM}}$ son convertidos a límites superiores de $f_{\tilde{a}}$, para valores de $r_{3/2}$ de acuerdo con Ec.~(\ref{ddmfractiongravLSP}).

Notamos algunas diferencias con respecto al caso 1 con axino LSP mostrado en la Figura~\ref{figaxinoLSP1}, donde podíamos mostrar para un dado $r_{\tilde a}$ varios contornos azules correspondientes a diferentes valores de $f_a$, y la región magenta de exclusión por radiación oscura. En el caso de gravitino LSP no podemos hacer lo mismo para un dado valor de $r_{3/2}$. La razón es que la densidad reliquia termal del axino depende de la escala de PQ, y por lo tanto las restricciones de DDM, es decir, de producción de radiación oscura, cambian cuando cambiamos la fracción de axinos $f_{\tilde{a}}$.

Otra diferencia importante es que en este escenario para masas cada vez más grandes del LSP, la densidad reliquia termal del NLSP incrementa. Sin embargo, en el caso anterior lo contrario sucedía. Esto modifica la forma de la exclusión de DDM, ya que el decaimiento del NLSP es la fuente de partículas ultrarelativistas. En el panel de la izquierda de la Figura~\ref{figgravitinoLSP1} se muestra una región con axinos NLSP con tiempos de vida largos (su decaimiento sucede luego de la era actual), mientras que en el panel de la derecha se muestra un axino con tiempo de vida intermedio (su decaimiento sucede entre recombinación y el presente).

\subsubsection{Límites por observaciones de rayos gamma y perspectivas de detección}
\label{detection}

\begin{figure}[t!]
\begin{center}
 \begin{tabular}{cc}
 \hspace*{-4mm}
 \epsfig{file=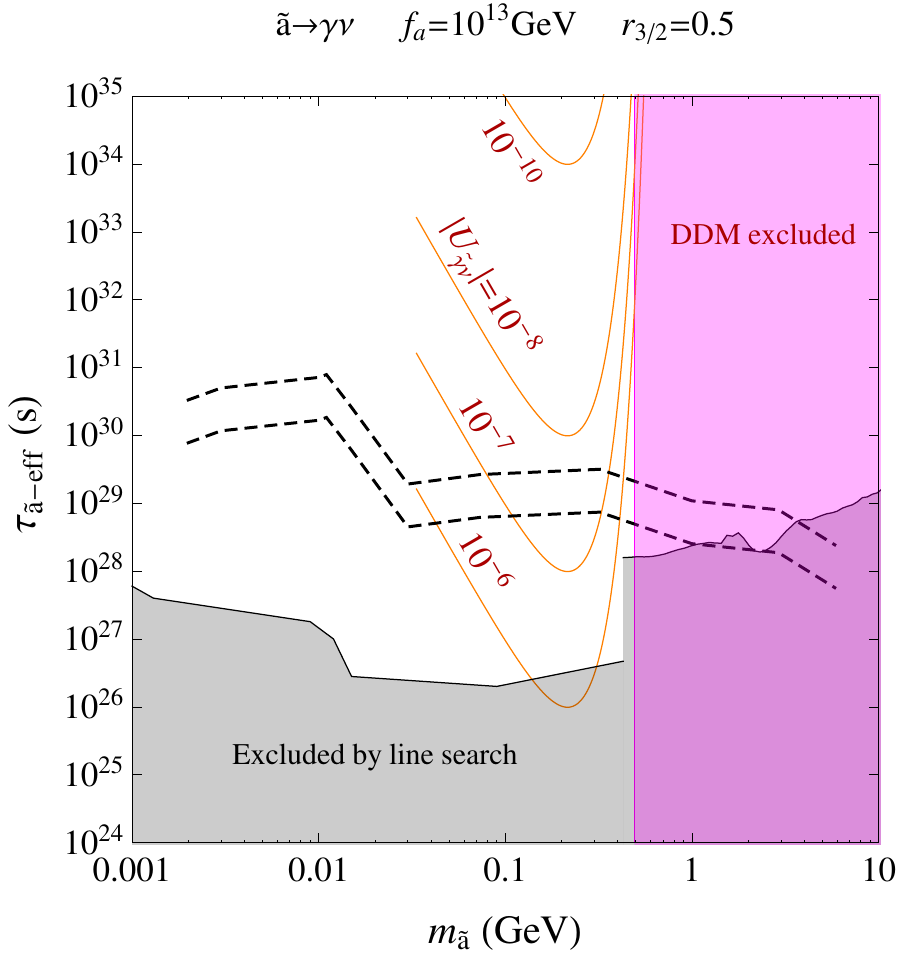,height=7cm}
       \epsfig{file=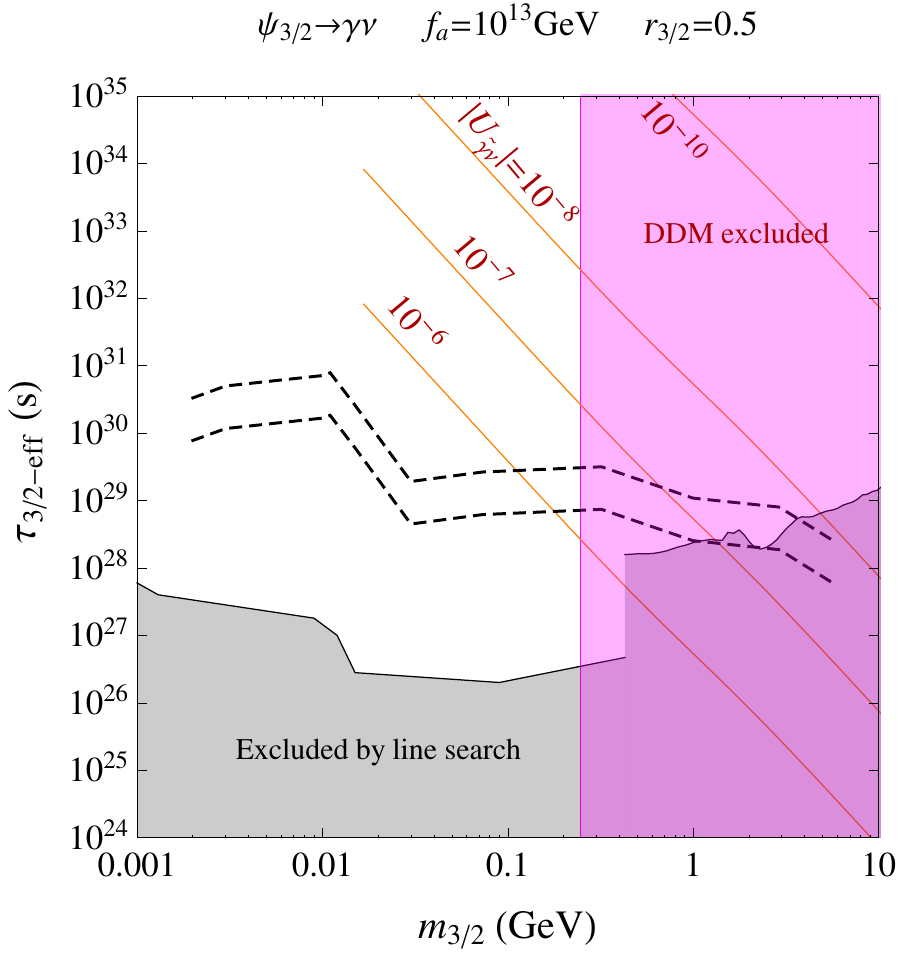,height=7cm}  
       \vspace*{-0.8cm}      
   \\ &
    \end{tabular}
    \caption{ Tiempo efectivo de vida en función de la masa del axino NLSP (panel izquierdo) y de la masa del gravitino LSP (panel derecho). El flujo de rayos gamma de cada candidato se analizó por separado. El resto de las referencias son iguales a las de la Figura~\ref{figaxinoLSP2}.
}
    \label{figgravitinoLSP2}
\end{center}
\end{figure}

Para analizar los efectos del decaimiento del axino NLSP sobre la búsqueda de líneas espectrales en la Figura~\ref{figgravitinoLSP2} mostramos el tiempo efectivo de vida de cada candidato en función de su respectivas masas, para un ejemplo con $f_a=10^{13}$ GeV y $r_{3/2}=0.5$, donde una señal doble podría ser detectada. El panel de la izquierda (derecha) corresponde a los límites del espacio de parámetros considerando la línea espectral producida por el axino NLSP (gravitino LSP) decayendo a $\gamma \nu$. Nuevamente, ambos paneles corresponden al mismo escenario, por lo tanto tienen que tenerse en cuenta las restricciones de ambos paneles para determinar la viabilidad de cada punto.

Las curvas naranjas muestran los parámetros predichos por el modelo $\mu\nu$SSM para varios valores representativos de $|U_{\tilde{\gamma} \nu}|$. En el panel de la izquierda, correspondiente a la emisión de rayos gamma provenientes de axinos, podemos ver los efectos de su reducción para $m_{\tilde{a}} \gtrsim 0.2$ GeV, de acuerdo con Ec.~(\ref{newtime2gravLSP}).
En el panel derecho se puede ver un tiempo de vida efectivo mayor al caso con solo gravitino DM debido a la contribución no termal.

Como podemos ver en la figura, regiones significativas por debajo de las líneas negras discontinuas pueden ser exploradas. En este caso, esto se debe principalmente a la emisión de líneas espectrales provenientes de axinos NSLP para  $0.03\lesssim m_{\tilde{a}} \lesssim 0.5$ GeV  y  $10^{-8} < |U_{\tilde{\gamma} \nu}| \leq 10^{-6}$ (ver panel izquierdo). Es más, en este ejemplo existe una pequeña región en el panel de la derecha correspondiente a una señal detectable proveniente de gravitino LSP, para $0.15 \lesssim m_{3/2} \lesssim 0.25$ GeV y $|U_{\tilde{\gamma} \nu}|\approx 10^{-6}$. 
Dada la relación entre masas utilizado en esta figura $r_{3/2}=0.5$, este rango de masas del gravitino corresponde a masas de axino  $0.3 \lesssim m_{\tilde{a}} \lesssim 0.5$ GeV, es decir dentro del rango donde se produce una línea detectable para el axino, por lo tanto esperamos una señal doble como característica distintiva de este escenario.

\begin{figure}[t!]
 \begin{center}
  \begin{tabular}{cc}
 \hspace*{-0mm}
 \includegraphics[height=6cm]{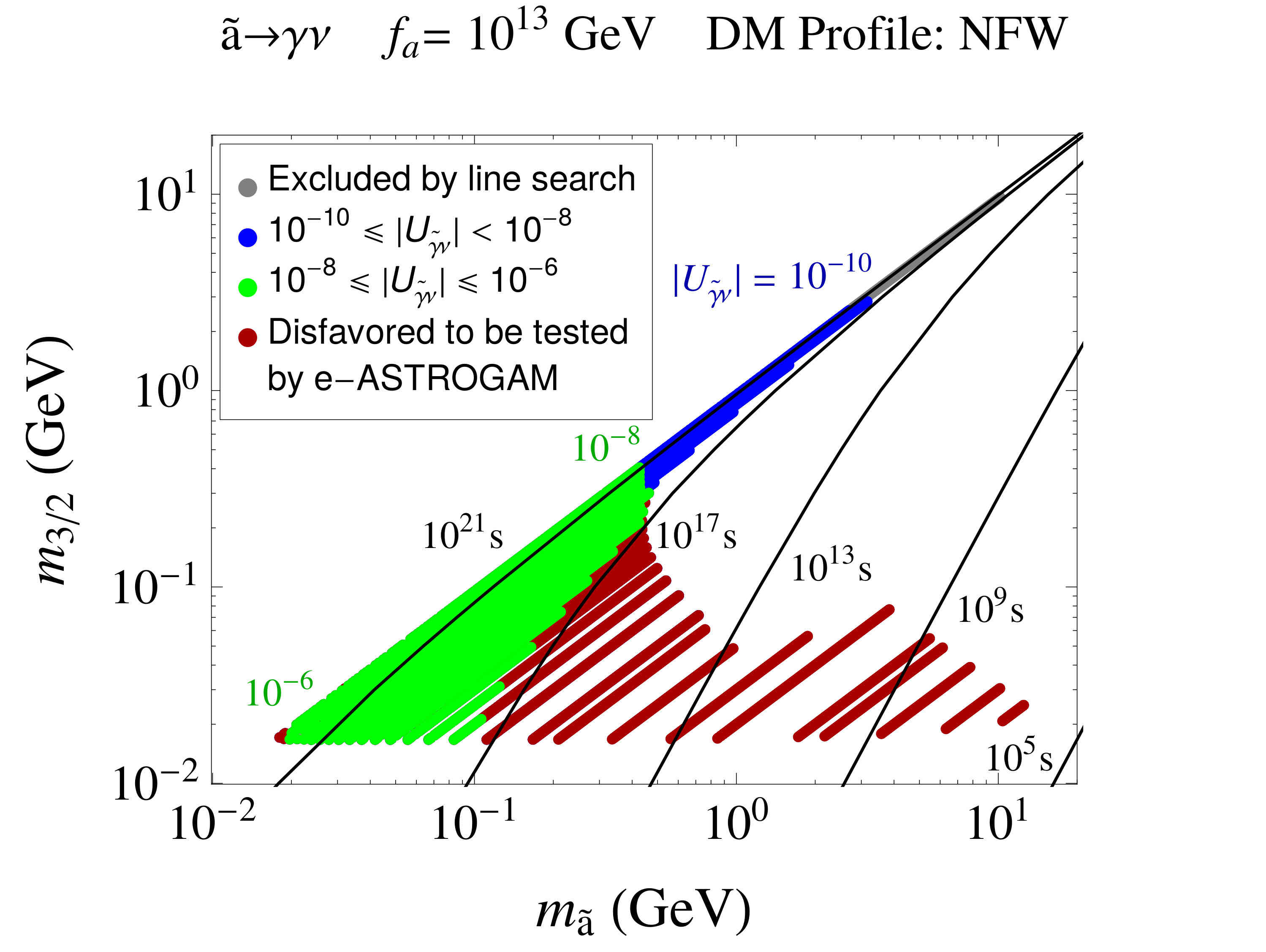} \hspace*{-0.5cm} \includegraphics[height=6cm]{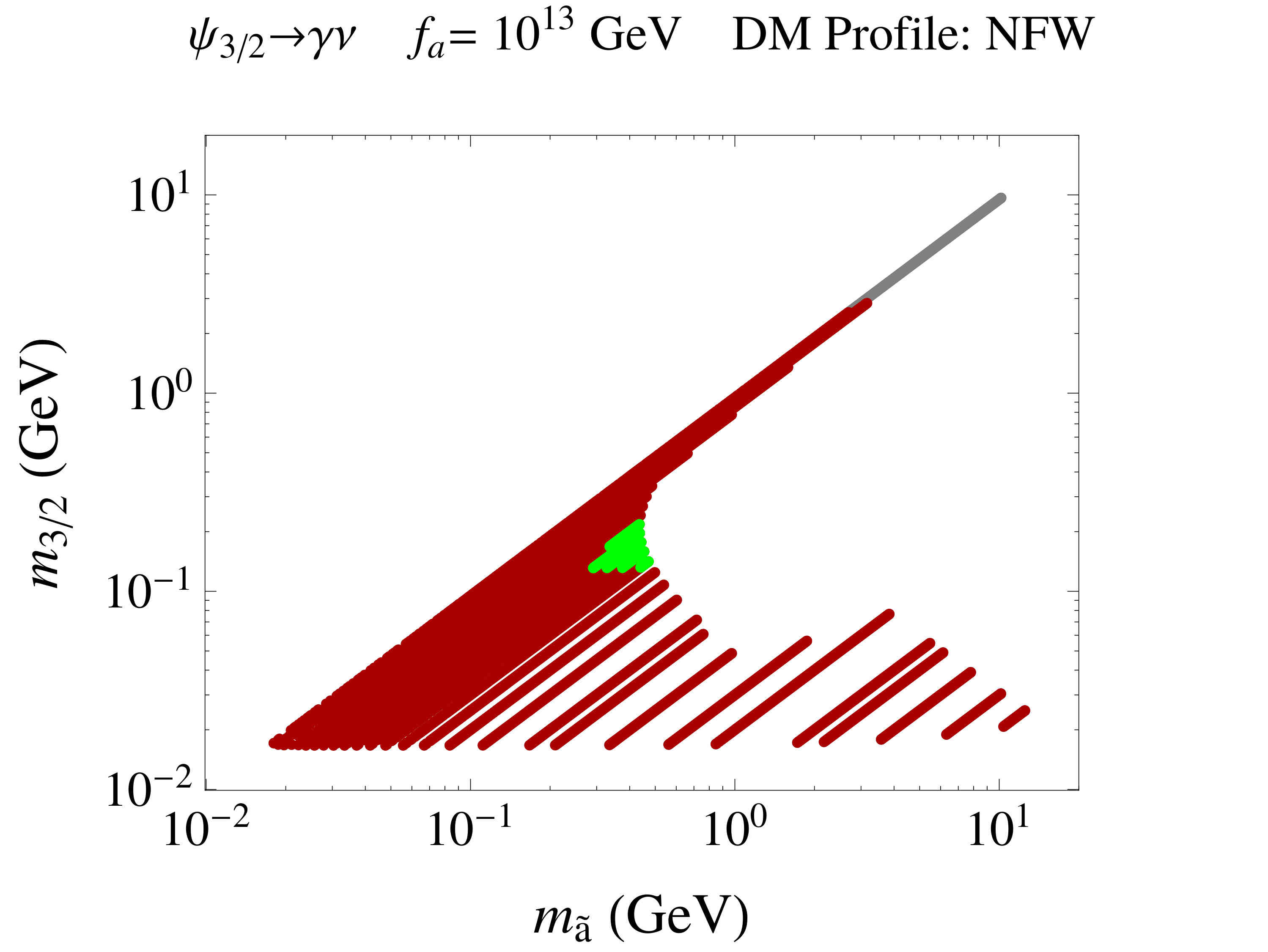}\\     
	\hspace*{-0mm} \includegraphics[height=6cm]{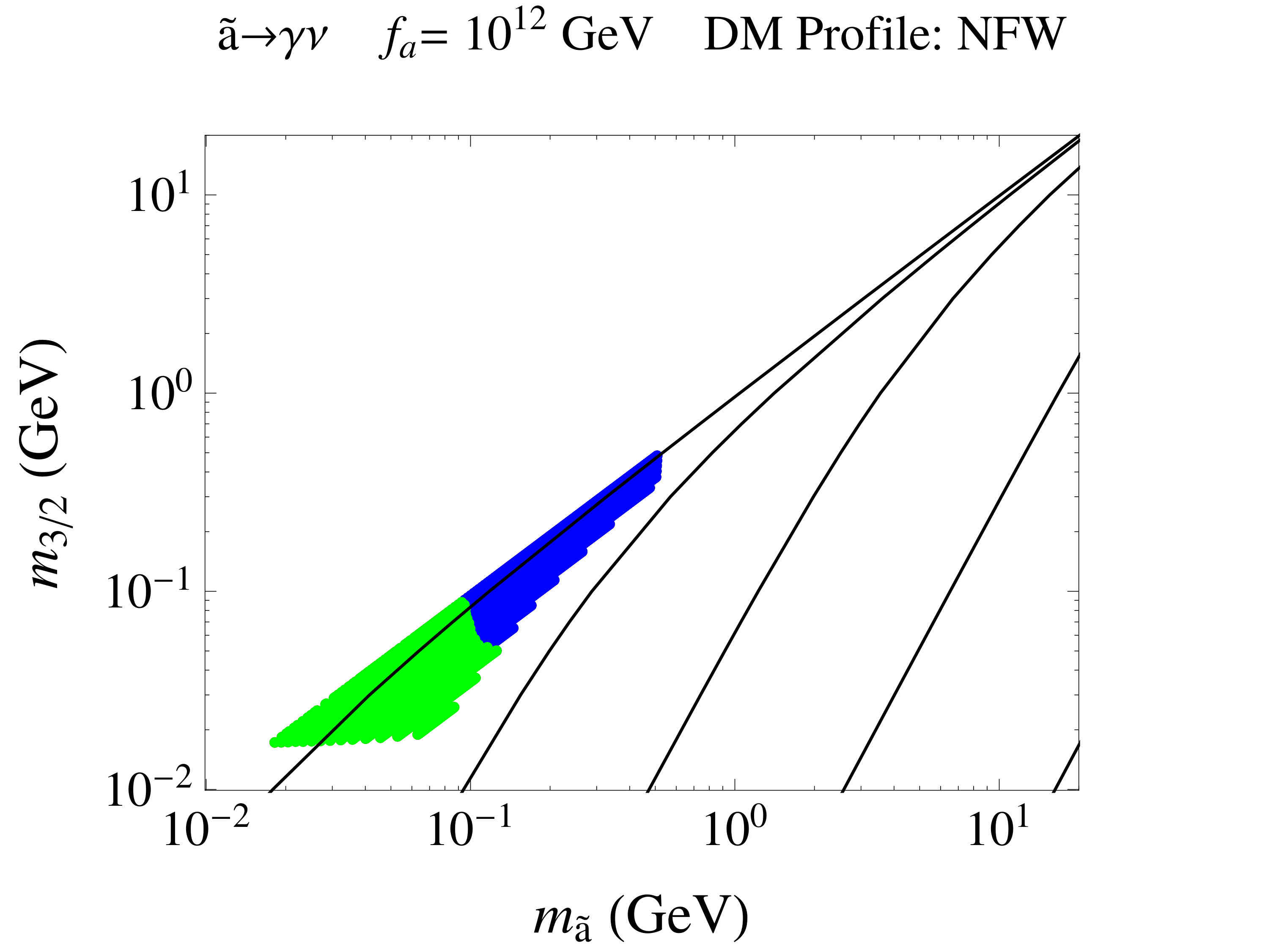} \hspace*{-0.5cm} \includegraphics[height=6cm]{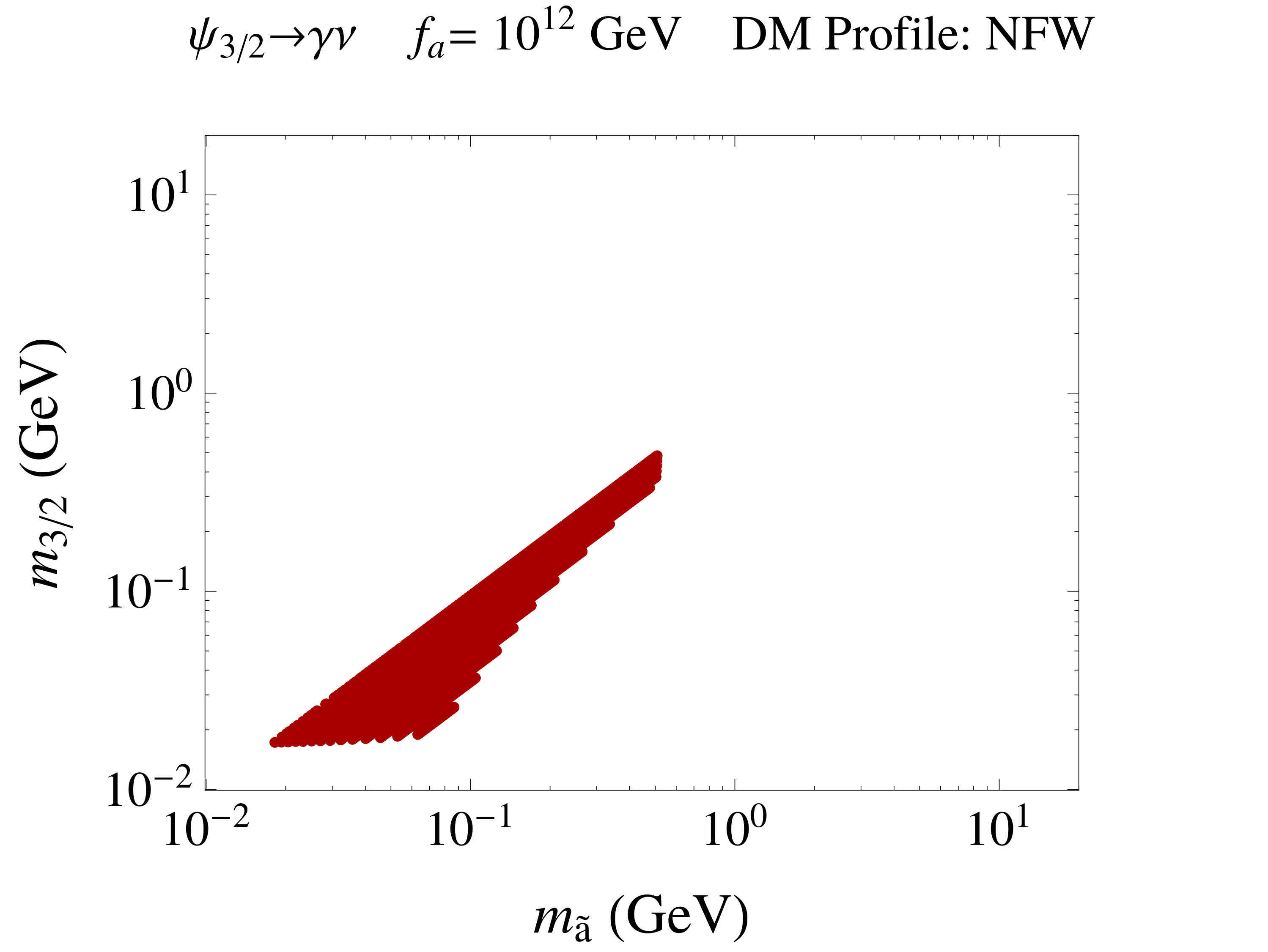}\\ 
  \end{tabular}
  \caption{Mismas referencias que en la Figura~\ref{figconstrains1}, pero para el caso gravitino LSP con axino NLSP.
}
    \label{figconstrains2}
\end{center}
\end{figure}

Para realizar un análisis completo del espacio de parámetros permitido, para  $m_{3/2} \gtrsim 0.017$ GeV hemos realizado un barrido numérico sobre el siguiente rango: $1.05\ m_{3/2} \lesssim m_{\tilde a} \lesssim 50, 0.5$ GeV para $f_a=10^{13}, 10^{12}$ GeV, respectivamente, teniendo en cuenta los límites discutidos en Ecs.~(\ref{relicaxinos}) y~(\ref{relicgravitinos}).
El resultado se muestra en la Figura~\ref{figconstrains2}, donde al igual que en la Figura~\ref{figconstrains1} se analizan las señales de rayos gamma de cada candidato por separado, axino NLSP en los paneles de la izquierda y gravitino LSP en la derecha. Las regiones verdes y azules corresponden a los puntos que pueden ser explorados por e-ASTROGAM asumiendo un perfil NFW de DM.

Para valores de $r_{3/2}$ cercanos a 1, es decir, la línea superior del borde, recuperamos el espacio de parámetros obtenido en el caso 1, donde el axino era el LSP y el gravitino el NLSP. Es decir, que no recuperamos en dicho límite el mismo espacio de parámetros que teníamos al considerar solo gravitino como DM, y por lo tanto el comportamiento del LSP difiere entre los casos 1 y 2 cuando $r_{LSP} \rightarrow 1$.

A pesar de que los límites de DDM para $f_{ddm}^{\text{DR}}$ se relajan ya que
$\tau_{\tilde a}^{-1}\simeq\Gamma(\tilde{a} \rightarrow \psi_{3/2} \, a) \rightarrow 0$ cuando $r_{3/2} \rightarrow 1$ (ver Ec.~(\ref{decayto32a})),  el efecto de las densidades reliquias iniciales de ambos candidatos no se puede ser ignorado (ver Ecs.~(\ref{nuevagravLSP}) y~(\ref{fractionstimegravLSP})). Para la región de masas permitidas, la densidad inicial de axino NLSP es relevante, y su decaimiento a fotón-neutrino puede causar señales que impongan límites al espacio de parámetros, que de otra forma no existirían considerando solamente las contribuciones del gravitino LSP. Es decir, en este caso por más que el gravitino sea el LSP, la exclusión y/o detección está dada principalmente por el axino NLSP. La causa reside en la diferencia entre las escalas de energías involucradas en el decaimiento a fotones de los candidatos a DM, $f_a \ll M_{Pl}$.

A partir de la  Figura~\ref{figconstrains2}, podemos concluir que una región significativa del espacio de parámetros podrá ser explorada por la próxima generación de telescopios de rayos gamma. Notemos que las líneas negras de los paneles izquierdos muestran el tiempo de vida del axino NLSP, y que la región donde esperamos una señal proveniente de este candidato corresponde a $\tau_{\tilde a}\simeq\Gamma^{-1} \left( \tilde{a} \rightarrow \psi_{3/2} \: a \right) > t_{hoy}$. Además, tenemos que  $\Gamma(\tilde{a}\rightarrow\gamma\nu_i) > \Gamma(\psi_{3/2}\rightarrow\gamma\nu_i)$ como fue discutido en la Sección~\ref{sec:DM multiplecomp}.
En particular, la región detectable se encuentra en el siguiente rango de masas: $20$ MeV $\lesssim m_{\tilde a} \lesssim 3$ GeV y $17$ MeV $\lesssim m_{3/2} \lesssim 3$ GeV.
De acuerdo con lo discutido arriba, también esperamos en el caso 2 una señal proveniente de gravitino LSP detectable en una región más pequeña. Dicha zona corresponde a los puntos verdes del panel superior derecho con $f_a=10^{13}$~GeV: $300\lesssim m_{\tilde a} \lesssim 500$ MeV y $150\lesssim m_{3/2} \lesssim 250$ MeV, con $0.2 \lesssim r_{3/2} \lesssim  0.55$. En dicha región esperamos que una doble línea espectral pueda ser medida por e-ASTROGAM. Notamos que esta región incluye a la discutida en el ejemplo de la Figura~\ref{figgravitinoLSP2}.

\begin{figure}[t!]
 \begin{center}
  \begin{tabular}{cc}
 \hspace*{-0mm}
 \includegraphics[height=6cm]{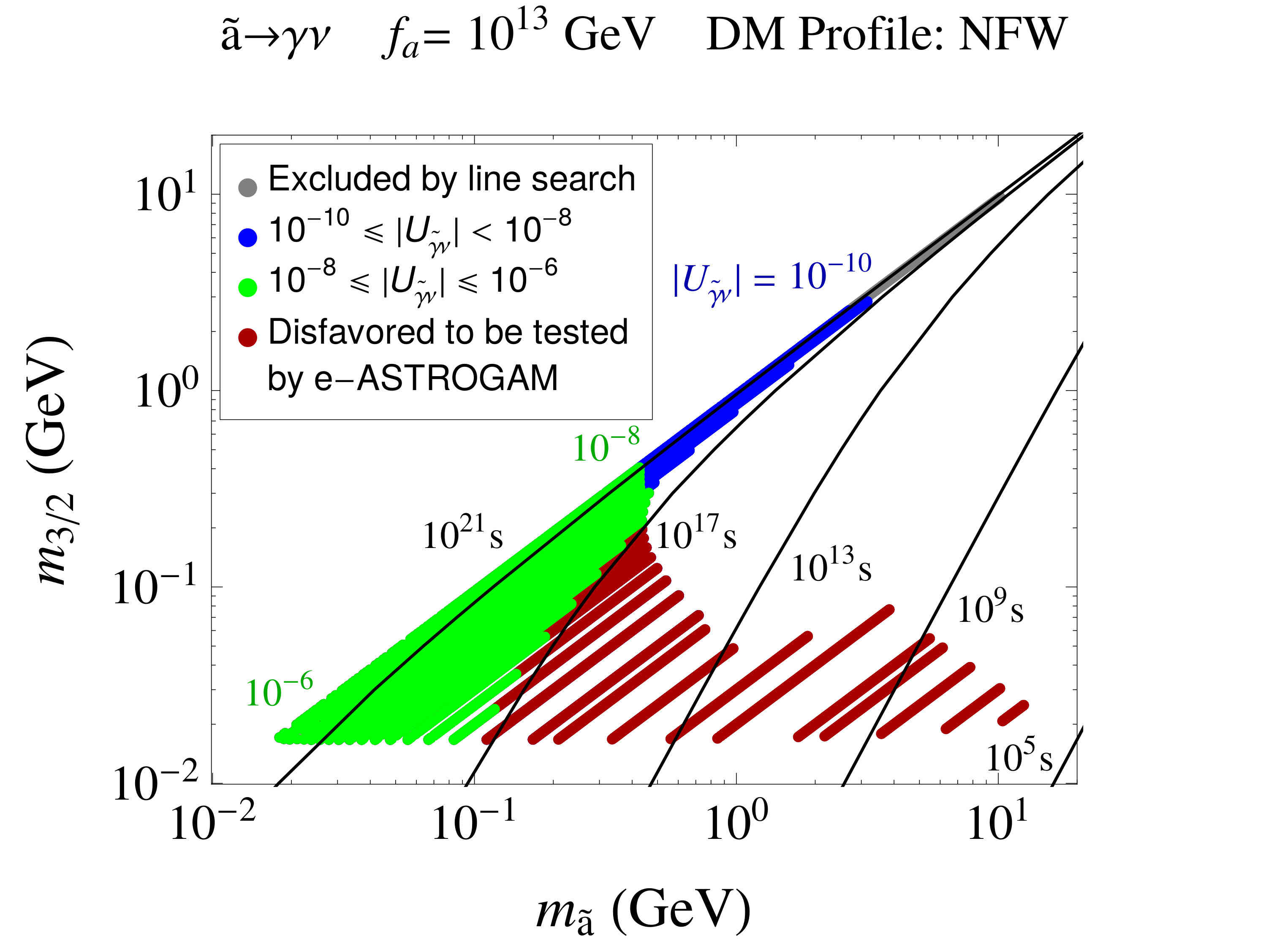} \hspace*{-0.5cm} \includegraphics[height=6cm]{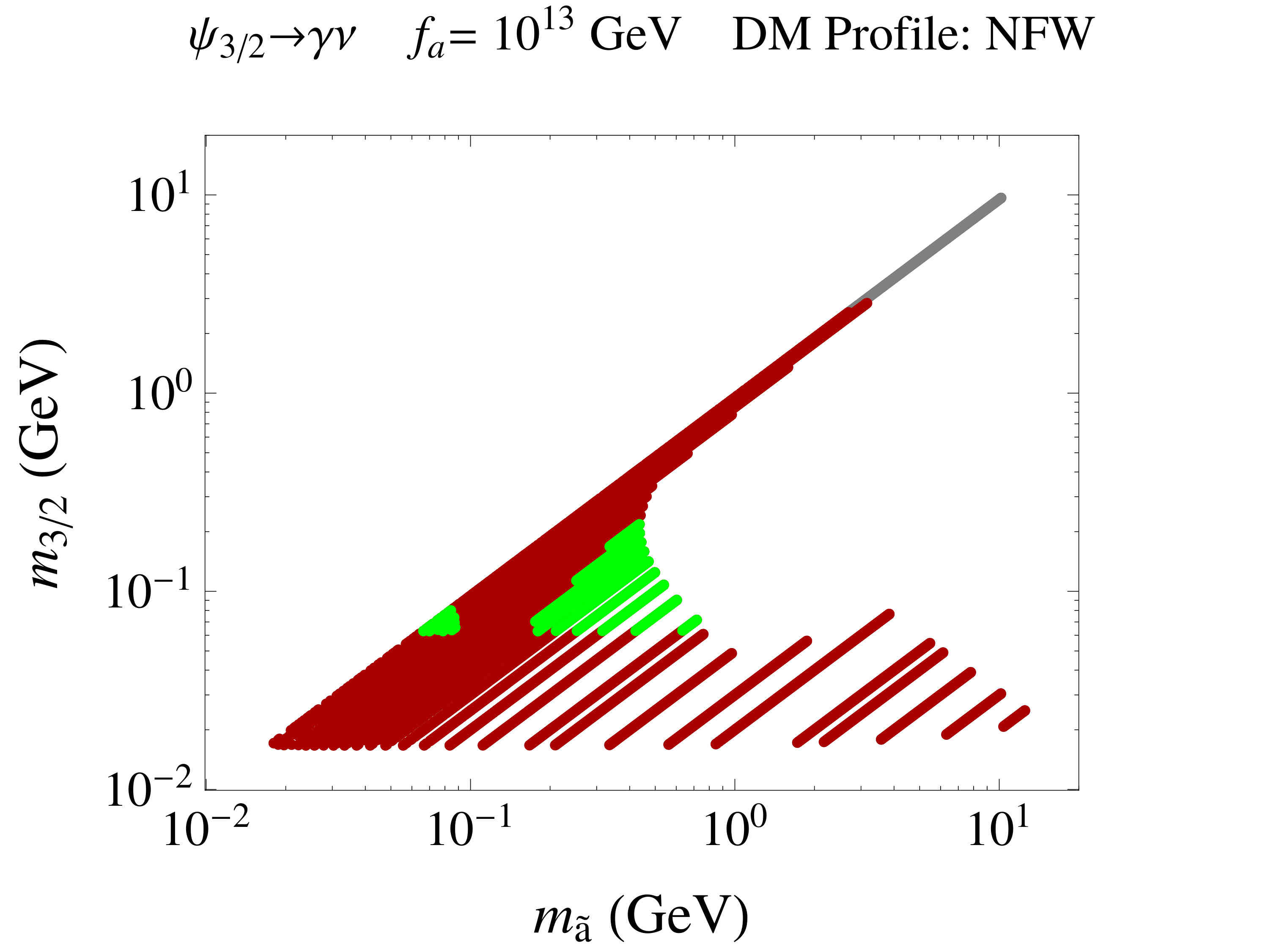}\\     
	\hspace*{-0mm} \includegraphics[height=6cm]{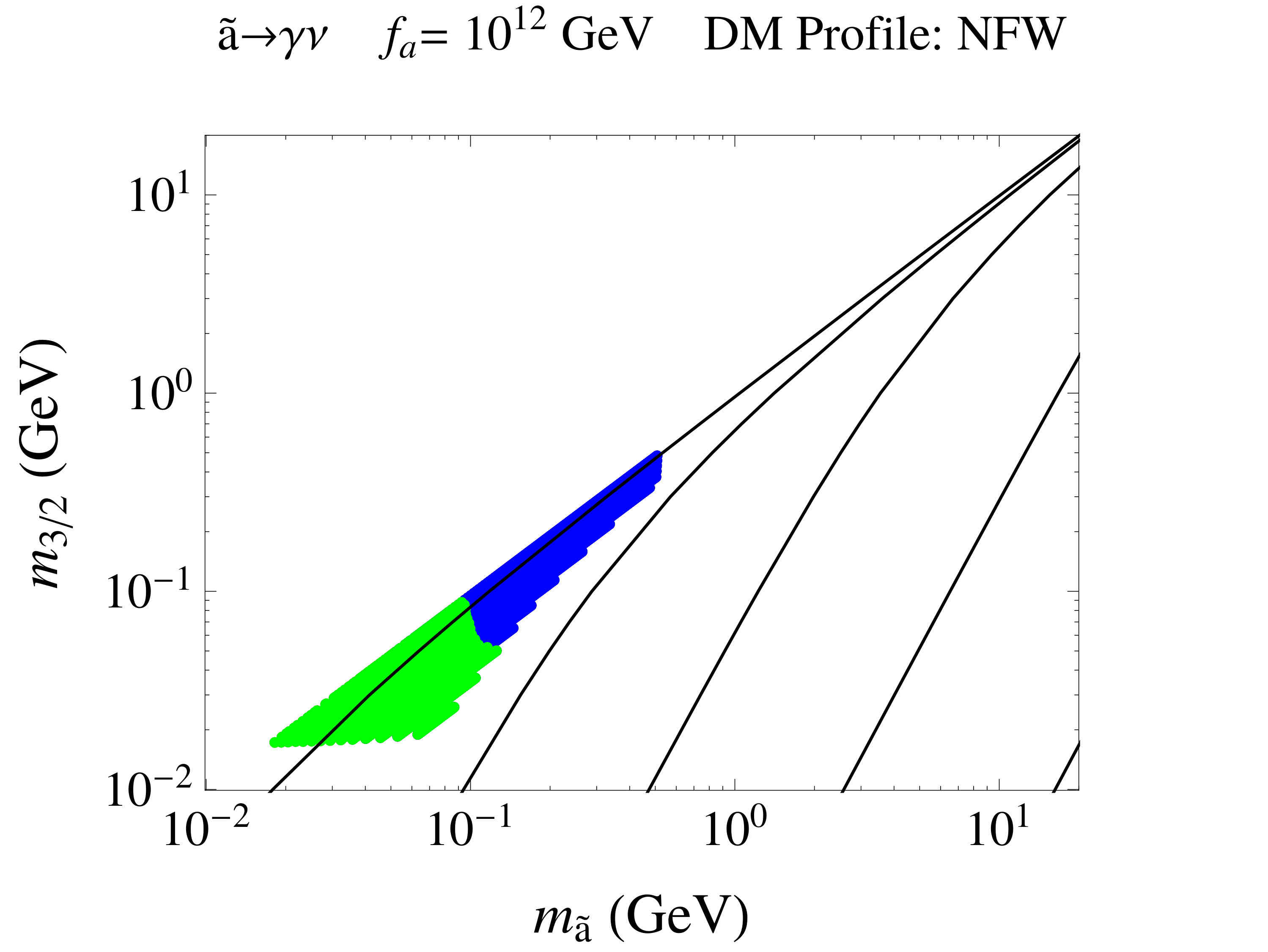} \hspace*{-0.5cm} \includegraphics[height=6cm]{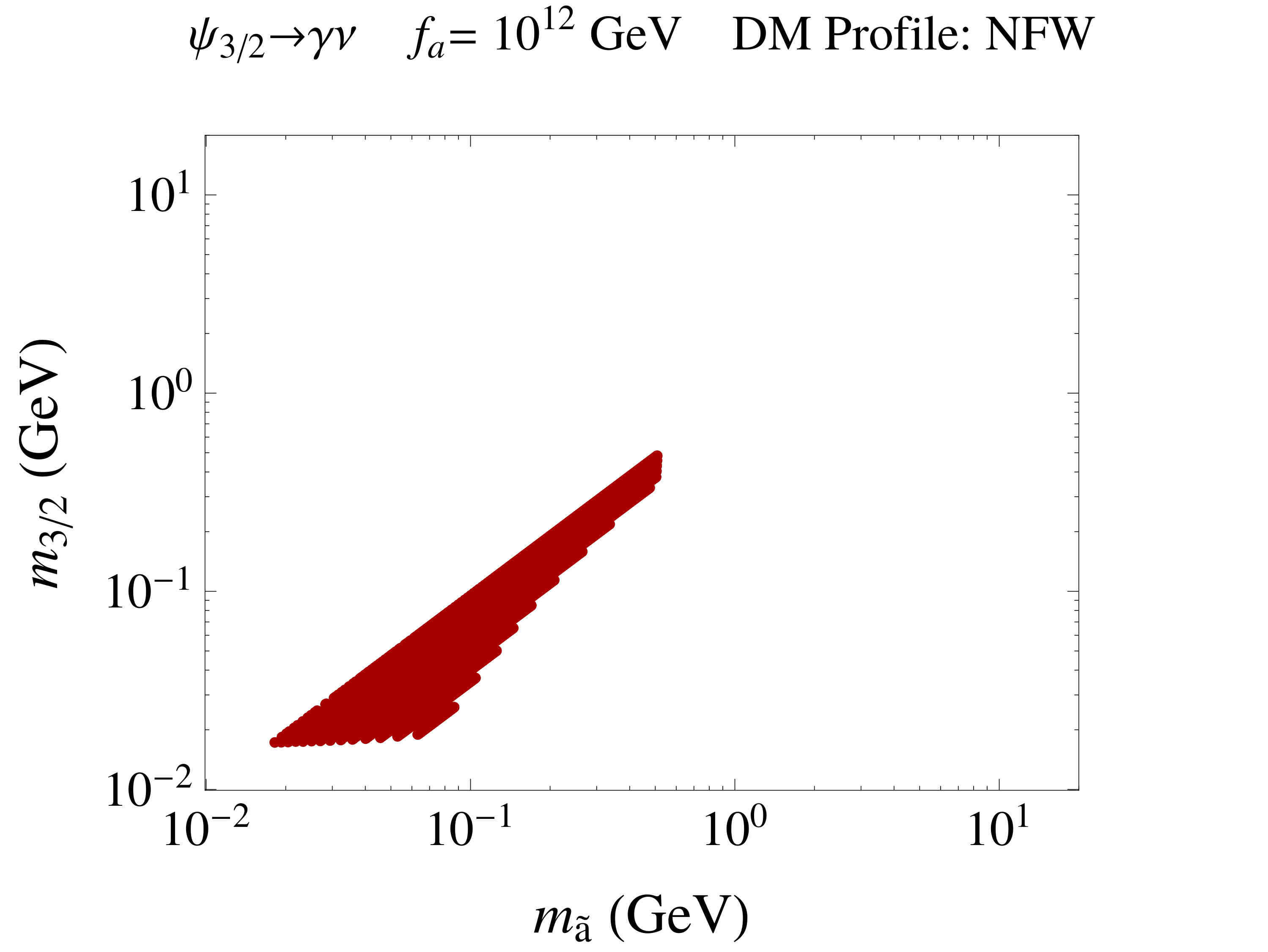}\\ 
  \end{tabular}
  \caption{Mismas referencias que en la Figura~\ref{figconstrains2}, pero incrementando la sensibilidad de e-ASTROGAM un orden de magnitud con respecto a la propuesta actual.}
    \label{figprofiles3}
\end{center}
\end{figure}

En la Figura~\ref{figconstrains2} utilizamos la sensibilidad de e-ASTROGAM asumiendo un perfil NFW de DM. En este caso, emplear distintos perfiles de densidades como Einasto B, no modifica significativamente el espacio de parámetros predicho.

Nuevamente la fuente de incertezas más importante corresponde a las características del instrumento. Si consideramos una sensibilidad incrementada un orden de magnitud con respecto a la última propuesta de e-ASTROGAM, se puede ver el impacto sobre la detección indirecta de materia oscura en la Figura~\ref{figprofiles3}. Podemos ver que la región explorable verde se extiende, y especialmente la región con doble línea se hace más importante. Para $f_a=10^{13}$~GeV, esta región es $60 \lesssim m_{\tilde a}\lesssim 700$ MeV y
$60\lesssim m_{3/2}\lesssim 250$ MeV.

\section{Conclusiones del capítulo}
\label{sec:conclusionsmultipleDM}

En este capítulo analizamos escenarios donde la materia oscura se encuentra compuesta por varias especies. En particular, estudiamos escenarios con los dos candidatos estudiados de forma individual en los capítulos anteriores: el gravitino y el axino. Si bien cada candidato puede decaer a fotón-neutrino y producir una señal, son buenos candidatos a materia oscura ya que sus decaimientos están suprimidos tanto por el parámetro de mezcla fotino-neutrino que viola paridad-R, como por la escala característica de cada partícula: $M_{Pl}$ y $f_a$.

Hemos visto que $\Gamma(\tilde{a}\rightarrow\gamma\nu_i)$ es típicamente más grande que 
$\Gamma(\psi_{3/2}\rightarrow\gamma\nu_i)$ (ver Ec.~(\ref{relation})), ya que $M_{Pl} \gg f_a$, a menos qué $r_{\tilde a}\equiv m_{\tilde{a}}/m_{3/2}$ sea muy chico. Por ejemplo, para $f_a=10^{13}, 10^{12}, 10^{11}$ GeV tiene que ser más chico que aproximadamente 0.02, 0.004, 0.001, respectivamente. Este hecho es muy importante pues nos indica que el flujo proveniente del decaimiento de axinos (ya sea como LSP o NLSP) será generalmente dominante en la búsqueda de líneas espectrales.

Posteriormente analizamos dos casos de DDM: axino-gravitino primero como LSP-NLSP y luego como NLSP-LSP, respectivamente. Para ambos escenarios existe un término de interacción entre gravitino-axino-axión que conserva paridad-R provocando el decaimiento del NLSP al LSP más un axión. En general, la anchura de decaimiento es más grande que las anchuras de decaimiento involucrando fotón-neutrino. Esto inyecta en el universo especies ultrarelativistas que pueden alterar la evolución del universo. Hemos discutido las tres regiones relevantes: la región de masas chicas o NLSP con tiempo de vida media grandes decayendo luego de la época actual, la región de masas intermedias donde el decaimiento del NLSP se produce entre recombinación y el presente, y la región de masas grandes o tiempo de vida corto donde el NLSP decae antes de recombinación.

Además, si tanto el axino como el gravitino pueden coexistir en el presente, ambos candidatos a materia oscura pueden ser fuentes de rayos gamma. Más aún, el decaimiento del NLSP al LSP puede tener un tiempo característico similar o incluso inferior a la edad del universo, por lo tanto modificamos el cálculo de las abundancias de cada candidato y por consiguiente las predicciones sobre el flujo de rayos gamma se ven afectadas.

Como primer caso en la Sección~\ref{sec:axinoLSPgravitinoNLSP} hemos estudiado el escenario con axino LPS y gravitino NLSP. Asumiendo $T_R \gtrsim 10^{4}$ GeV encontramos un límite inferior para la masa del gravitino de $m_{3/2} \gtrsim 0.017$ GeV. También encontramos las regiones del espacio de parámetros excluido por observaciones cosmológicas, que tienen en cuenta los importantes límites sobre la cantidad de gravitino NLSP que puede ser convertido en radiación oscura (ver Figura~\ref{figaxinoLSP1}).

Con respecto al flujo de rayos gamma predicho en este escenario DDM en el marco del $\mu\nu$SSM, sectores del espacio de parámetros han sido excluidos por búsquedas de líneas espectrales empleando instrumentos como \Fermi LAT, mientras que regiones significativas podrán ser explorada por la próxima generación de telescopios. En particular hemos considerado como referencia la sensibilidad propuesta para e-ASTROGAM y hallamos que explorará: $7 \: \text{MeV}  \lesssim m_{\tilde a} \lesssim 3$ GeV y $20 \: \text{MeV} \lesssim m_{3/2} \lesssim 1$ TeV. Como mencionamos anteriormente, la señal es dominada por el decaimiento del axino LSP (ver los paneles izquierdos de la Figura~\ref{figconstrains1}). 

Para $\tau_{3/2}\simeq {\Gamma}^{-1} (\psi_{3/2} \rightarrow a \, \tilde{a})~>~t_{hoy}$, una señal proveniente de gravitino NLSP puede ser medida en una pequeña región con $f_a=10^{13}$~GeV dentro del siguiente rango de masas $100\lesssim m_{3/2}\lesssim 200$ MeV y $10 \lesssim m_{\tilde a}\lesssim 60$ MeV (ver el panel superior derecho de la Figura~\ref{figconstrains1} y Figura~\ref{figprofiles11} para los perfiles NFW y Einasto B de materia oscura, respectivamente). En estos rangos, una doble señal proveniente del decaimiento axinos y gravitinos podría ser observada, generando un espectro que es muy difícil de replicar mediante fuentes astrofísicas o mediante otros modelos de materia oscura con un solo componente.

Como segundo caso en la Sección~\ref{sec:gravitinoLSPaxinoNLSP} hemos estudiado el escenario opuesto, con gravitino LPS y axino NLSP. En el contexto de DDM, encontramos que esta combinación de partículas puede reproducir la cantidad de DM medida y evitar el exceso de radiación oscura. Hallamos que el espacio de parámetros permitido es distinto, por ejemplo el límite inferior de la escala de PQ se incrementa de $f_a=10^{11}$ GeV en el primer caso, a $f_a=10^{12}$ GeV en el segundo (ver Ecs.~(\ref{pqscale}) y~(\ref{pqscalegravLSP})) para evitar exceso de producción de materia oscura fría asumiendo $T_R \gtrsim 10^{4}$ GeV (ver Figura~\ref{figgravitinoLSP1}).

En el caso anterior, donde el gravitino era el NLSP, al aumentar la masa se disminuía su producción termal y se relajaban los límites de DDM por emisión de radiación oscura. Para masas chicas de gravitino, el tiempo de vida del mismo supera la edad del universo, nuevamente relajando las restricciones.

Sin embargo, en el caso con axino NLSP la producción termal del NLSP además de depender de la escala de PQ, aumenta con su masa. Por lo que se modifica el espacio de parámetros excluido por exceso de especies ultrarelativistas, y se espera un espacio de parámetros disponible más reducido comparando el rango de masas del NLSP de ambos casos.

Estudiamos el flujo de rayos gamma predicho y presentamos las exclusiones realizadas con datos de \Fermi LAT así como las predicciones para la próxima generación de telescopios. En particular, esperamos que se pueda explorar el siguiente rango de masas $20 \: \text{MeV} \lesssim m_{\tilde a} \lesssim 3$ GeV y $17 \: \text{MeV} \lesssim m_{3/2} \lesssim 3$ GeV. La región más amplia de detección está determinada principalmente por el decaimiento axino, es decir, el NLSP y no por el LSP como se podría esperar ingenuamente. Como discutimos previamente, esto se debe a la diferencia entre las escalas de Planck y PQ, provocando que la anchura de decaimiento del axino sea más importante que la del gravitino. Por supuesto para tener señal de fotones del NLSP, este tiene que poder sobrevivir hasta la época actual, es decir $\tau_{\tilde a}\simeq\Gamma^{-1} \left( \tilde{a} \rightarrow \psi_{3/2} \: a \right) > t_{hoy}$, como se puede ver en los paneles izquierdos de la Figura~\ref{figconstrains2}.

Por otra parte, esperamos detectar una señal proveniente del decaimiento del gravitino LSP para $f_a=10^{13}$~GeV en el siguiente rango $300\lesssim m_{\tilde a} \lesssim 500$ MeV y $150\lesssim m_{3/2} \lesssim 250$ MeV, con $0.2 \lesssim r_{3/2} \lesssim  0.55$ (ver panel superior derecho de la Figura~\ref{figconstrains2}) y $r_{3/2} \equiv r_{\tilde{a}}^{-1}$. Para estos casos, una doble señal podría ser medida. Dicha región es complementaria a la hallada en el caso 1, por lo tanto es espacio de parámetros total donde esperamos obtener una doble línea es la unión de ambas regiones.

En cuanto a las incertezas que pueden ser un factor importante en la detección de líneas, el perfil de densidad de materia oscura solo afecta de manera significativa al primer caso. Por otro lado, aumentando la sensibilidad de la futura generación de telescopios de rayos gamma con respecto a e-ASTROGAM, se genera un cambio realmente notable en las perspectivas de detección para ambos casos. En particular, el espacio de parámetros donde esperaríamos encontrar la señal de doble línea característica de este tipo de modelos se expande significativamente, aumentando las posibilidades de medir una señal clara que podría ayudar a discriminar la naturaleza de la materia oscura.

\spacing{1}


\chapter{Conclusiones}
\label{conclusiones}

\spacing{1.5}

A pesar del enorme éxito del SM y del hito histórico que marcó el descubrimiento del bosón de Higgs, la única partícula escalar predicha por el modelo, existen motivos tanto teóricos como experimentales para creer que el SM debe ser extendido. Uno de ellos es la identidad de la materia oscura que, luego de varias décadas de trabajos teóricos y una gran cantidad de experimentos, permanece como uno de los grandes enigmas de la física contemporánea. En este contexto, los modelos supersimétricos resultan muy atractivos ya que proveen soluciones a dichos problemas y otorgan predicciones claras contrastables con experimentos realizables a corto plazo.

En esta tesis hemos tratado el problema de la naturaleza de la materia oscura desde un punto de vista fenomenológico trabajando en el $\mu\nu$SSM. Este modelo SUSY con contenido natural mínimo de partículas resulta libre del problema-$\mu$ y capaz de reproducir la física de neutrinos, gracias a la inclusión de neutrinos dextrógiros en el espectro. Para los candidatos considerados se estudiaron las señales características que pueden discriminar distintos escenarios, y restringir o verificar las predicciones con experimentos actuales o de próxima generación.

La simetría discreta llamada paridad-R ha guiado el enfoque y las búsquedas de muchos experimentos realizados en las últimas décadas con resultados nulos hasta ahora. En el $\mu\nu$SSM dicha simetría no se conserva causando que el LSP decaiga a partículas del SM. Por lo tanto, los candidatos a materia oscura típicos de estos modelos, usualmente el neutralino, no resultan viables. Sin embargo, existen otros candidatos SUSY naturales tales como el gravitino o el axino. El primero surge en construcciones generales de modelos SUSY como el compañero supersimétrico del gravitón dada la existencia de gravedad. El segundo surge como el compañero supersimétrico del axión, al aplicar en modelos SUSY la solución de Peccei-Quinn al problema de violación de CP en QCD. Si bien ambos pueden decaer a partículas del SM, sus interacciones se encuentran suprimidas por los pequeños parámetros de ruptura de paridad-R (relacionados con la física de neutrinos) y por las grandes escalas de Planck y de PQ. Esto causa que puedan tener un tiempo de vida mayor a la edad del universo. Debido a la escala de sus interacciones ambos son ejemplos de superWIMPs, en consecuencia son prácticamente imposibles de detectar en experimentos de detección directa de materia oscura. Por lo tanto, los experimentos de detección indirecta son ideales para estudiar este tipo de candidatos.

El gravitino como único candidato a materia oscura fue considerado en primer lugar. Hemos estudiado con detalle las consecuencias fenomenológicas de un gravitino inestable, así también como la importancia del parámetro de mezcla fotino-neutrino que viola paridad-R y está involucrado en el decaimiento del gravitino a fotón-neutrino. De esta manera, vimos la importante relación entre las cotas experimentales del sector de neutrinos y la posible detección indirecta de materia oscura. Realizamos un barrido numérico del espacio de parámetros del modelo y obtuvimos el rango más general del parámetro de mezcla que reproduce la física de neutrinos.

Con dichas herramientas analizamos el flujo de rayos gamma producido por el decaimiento del gravitino, no solo a fotón-neutrino sino a estados finales de tres cuerpos.  Estudiamos los distintos regímenes donde distintos canales dominan y extendimos de esta manera la predicción realizada en trabajos anteriores que asumen solo decaimientos a fotón-neutrino.

Generamos los flujos de rayos gamma esperados para estados finales con dos y tres cuerpos, los cuales tienen forma de línea espectral y de señal continua, respectivamente. La inclusión de canales con morfologías espectrales distintas implica contrastar los resultados con análisis específicos. Utilizamos los datos de búsqueda de líneas para el primer caso, y mediciones de EGB, es decir el fondo de rayos gamma, para la señal continua, ambos determinados por \Fermi LAT. Recuperamos las cotas establecidas por trabajos previos en los límites correspondientes y expandimos el espacio de parámetros permitido para gravitino LSP como DM cuando la contribución de los canales a tres cuerpos es importante.

Posteriormente, consideramos por primera vez en el $\mu\nu$SSM al axino como único candidato a materia oscura. Señalamos que su fenomenología resulta similar a la del gravitino, pero al mismo tiempo presenta características singulares, tendiendo a favorecer masas más chicas. Analizamos los dos modelos de axiones más populares, y en ambos la relación con el parámetro de mezcla fotino-neutrino que permite el decaimiento del axino a fotón-neutrino.

Utilizamos los límites actuales de las observaciones de \Fermi LAT y estudiamos las perspectivas de detección para la próxima generación de telescopios de rayos gamma. Distintas propuestas para explorar el cielo en el rango energético MeV-GeV han sido presentadas en los últimos años, las cuales cubrirán oportunamente el espacio de parámetros favorecido por el axino. Como referencia hemos considerado al proyecto e-ASTROGAM y las proyecciones para la búsqueda de líneas espectrales teniendo en cuenta su sensibilidad. Hemos estudiado los efectos de distintos perfiles de densidades y establecido las regiones de masas y tiempos de vida del axino LSP que podrán ser exploradas mediante la búsqueda de líneas.

Dada la naturalidad, las características, los parámetros que relacionan a ambos candidatos contemplados, y el interés que muestra la comunidad internacional en futuros experimentos de detección indirecta, hemos considerado un escenario con múltiples candidatos a materia oscura: axino y gravitinos coexistiendo.

Hemos analizado las características y diferencias de dos escenarios, axino LSP con gravitino NLSP y el caso opuesto. Debemos mencionar que además de los decaimientos a fotón-neutrino existe una interacción que causa el decaimiento del NLSP al LSP y un axión, lo que inyecta especies ultrarelativistas en el universo. Dependiendo del tiempo en que se produce dicho decaimiento, aplicamos cotas a la cantidad de radiación oscura y establecimos el espacio de parámetros compatible.

Dichos resultados fueron utilizados para analizar el flujo de rayos gamma de varios candidatos a materia oscura cuyas abundancias varían en el tiempo, y se concluyó que una importante región del espacio de parámetros podrá ser explorada por futuros experimentos. Como resultado se obtuvo que la señal se encuentra dominada por la emisión proveniente del axino, tanto para el caso que sea el LSP o el NLSP. Esto se explica por la diferencia entre las escalas $M_{Pl}$ y $f_a$ involucradas respectivamente en las interacciones de los gravitinos y los axinos. 
Sin embargo, en ambos casos encontramos que también es detectable una señal proveniente del gravitino en una región de masas más pequeña. Dicho rango está incluido en la región donde se predice una señal del axino, por lo que esperamos una señal en forma de dos líneas espectrales, situación difícil de reproducir mediante procesos astrofísicos.

Como último punto estudiamos las incertezas asociadas a la elección del perfil de densidad de materia oscura y a las especificaciones del instrumento. 
Así como la década anterior se caracterizó por un impresionante avance en la sensibilidad de los experimentos de detección directa, esperamos que la década que comienza presencie el mismo auge con los métodos de detección indirecta tal como parecen indicar los últimos años.

En resumen, hemos abordado el problema de la identidad de la materia oscura en el marco del $\mu\nu$SSM, el cual es un modelo SUSY activamente estudiado por la comunidad tanto desde el punto de vista teórico como fenomenológico. Estos análisis emplean enfoques complementarios a los aquí tratados como por ejemplo la búsqueda de vértices desplazados en el LHC. 

En esta tesis hemos estudiado dos candidatos naturales de los modelos SUSY más generales, el gravitino y el axino. Se comenzó con un análisis individual, para continuar con el análisis de escenarios donde ambos coexisten. 
Luego de aplicar los límites impuestos por observaciones cosmológicas y la relación de los candidatos con la física de neutrinos, determinamos el espacio de parámetros permitido y la perspectiva de detección mediante señales de rayos gamma. Para ello, se consideraron experimentos contemporáneos e instrumentos futuros actualmente propuestos por la comunidad.

Para finalizar, debemos decir que los análisis con múltiples candidatos a materia oscura, así como la posibilidad de hallar señales múltiples, son temas que se encuentran en pleno auge. 
Como ya hemos mencionado, en análisis cosmológicos se ha considerado recientemente sistemas con múltiples candidatos a materia oscura concluyendo que podrían aliviar distintos problemas, como por ejemplo la tensión en la medición del parámetro de Hubble. Aquí hemos mostrado un escenario concreto de física de partículas que podría enmarcarse en estas nuevas propuestas, lo que da lugar a posibles próximos trabajos.


\spacing{1}


\addcontentsline{toc}{chapter}{Apéndice}

\appendix

\chapter{Ecuaciones de minimización del potencial escalar neutro en el \texorpdfstring{$\mu\nu$}{munu}SSM} \label{minimizacionapendice}
\spacing{1.5}

Escribiremos las ocho condiciones de minimización del potencial escalar neutro, Ec.~(\ref{Velectro}), con respecto al módulo de $v_{d}$, $v_{u}$, $v_{\nu_i^c}$, $v_{\nu_i}$:

\begin{align}
	& \frac{1}{4}G^{2}(\sum_{i}v_{\nu_i}v_{\nu_i}+v_{d}^{2}-v_{u}^{2})v_{d} + m_{H_{d}}^{2}v_{d} + v_{d}v_{u}^{2}\sum_{i}\lambda_{i}^{2} - \sum_{i}(A^{\lambda}\lambda)_{i}v_{\nu_i^c}v_{u}  \cr
	& + \sum_{i,j}v_{d}\lambda_{i}\lambda_{j}v_{\nu_i^c}v_{\nu_j^c} - \sum_{i,j,k}\kappa_{ijk}\lambda_{k}v_{u}v_{\nu_i^c}v_{\nu_j^c} - \sum_{i,j,k}Y^{\nu}_{ij}\lambda_{k}v_{\nu_i}v_{\nu_j^c}v_{\nu_k^c} - \sum_{i,j}Y^{\nu}_{ij}\lambda_{j}v_{u}^{2}v_{\nu_i}=0,
\end{align}

\begin{align}
	& -\frac{1}{4}G^{2}(\sum_{i}v_{\nu_i}v_{\nu_i}+v_{d}^{2}-v_{u}^{2})v_{u} + m_{H_{u}}^{2}v_{u} + v_{u}v_{d}^{2}\sum_{i}\lambda_{i}^{2} - \sum_{i}(A^{\lambda}\lambda)_{i}v_{\nu_i^c}v_{d} \cr
	& + \sum_{i,j}(A^{\nu}Y^{\nu})_{ij}v_{\nu_i}v_{\nu_j^c} + \sum_{i,j}v_{u}\lambda_{i}\lambda_{j}v_{\nu_i^c}v_{\nu_j^c} - \sum_{i,j,k}\kappa_{ijk}\lambda_{k}v_{d}v_{\nu_i^c}v_{\nu_j^c} + \sum_{i,j,k}\sum_{l}Y^{\nu}_{jl}\kappa_{ilk}v_{\nu_j}v_{\nu_i^c}v_{\nu_k^c} \cr
	& - \sum_{i,j}2Y^{\nu}_{ij}\lambda_{j}v_{d}v_{u}v_{\nu_i} + \sum_{i,j,k}Y^{\nu}_{ik}Y^{\nu}_{jk}v_{u}v_{\nu_i}v_{\nu_j} + \sum_{i,j,k}Y^{\nu}_{ik}Y^{\nu}_{jk}v_{u}v_{\nu_i^c}v_{\nu_j^c}=0,
\end{align}

\begin{align}
	& \sum_{j}m_{\widetilde{\nu}_{ij}^{c}}^{2}v_{\nu_j^c} - (A^{\lambda}\lambda)_{i}v_{u}v_{d} + \sum_{j}(A^{\nu}Y^{\nu})_{ij}v_{\nu_j}v_{u} + \sum_{j,k}(A^{\kappa}\kappa)_{ijk}v_{\nu_j^c}v_{\nu_k^c} + \sum_{j}\lambda_{i}\lambda_{j}v_{d}^{2}v_{\nu_j^c} \cr
	& + \sum_{j}\lambda_{i}\lambda_{j}v_{u}^{2}v_{\nu_j^c} + \sum_{j,k,l}\sum_{m}2\kappa_{imk}\kappa_{lmj}v_{\nu_j^c}v_{\nu_k^c}v_{\nu_l^c} - \sum_{j,k}2\kappa_{ijk}\lambda_{k}v_{d}v_{u}v_{\nu_j^c} + \sum_{j,k,l}2Y^{\nu}_{jl}\kappa_{ikl}v_{u}v_{\nu_j}v_{\nu_k^c} \cr
	& - \sum_{j,k}Y^{\nu}_{ij}\lambda_{k}v_{d}v_{\nu_j}v_{\nu_k^c} - \sum_{j,k}Y^{\nu}_{jk}\lambda_{i}v_{d}v_{\nu_k}v_{\nu_j^c} + \sum_{j,k,l}Y^{\nu}_{ij}Y^{\nu}_{kl}v_{\nu_j}v_{\nu_l}v_{\nu_k^c} + \sum_{j,k}Y^{\nu}_{ik}Y^{\nu}_{jk}v_{u}^{2}v_{\nu_j^c}=0,
\end{align}

\begin{align}
	& \frac{1}{4}G^{2}(\sum_{j}v_{\nu_j}v_{\nu_j} + v_{d}^{2} - v_{u}^{2})v_{\nu_i} + \sum_{j}m_{\widetilde{L}_{ij}}^{2}v_{\nu_j} + \sum_{j}(A^{\nu}Y^{\nu})_{ij}v_{\nu_j^c}v_{u} + \sum_{j,k,l}Y^{\nu}_{il}\kappa_{jkl}v_{u}v_{\nu_j^c}v_{\nu_k^c} \cr
	& - \sum_{j,k}Y^{\nu}_{ij}\lambda_{k}v_{d}v_{\nu_j^c}v_{\nu_k^c} - \sum_{j}Y^{\nu}_{ij}\lambda_{j}v_{d}v_{u}^{2} + \sum_{j,k,l}Y^{\nu}_{ij}Y^{\nu}_{kl}v_{\nu_j^c}v_{\nu_k}v_{\nu_l^c} + \sum_{j,k}Y^{\nu}_{ik}Y^{\nu}_{jk}v_{u}^{2}v_{\nu_j}=0,
\label{nui}
\end{align}
con $G^{2}=g^{2}+g^{'2}$. En la derivación de las ecuaciones de arriba se ha asumido que $\kappa_{ijk}$, $(A_{\kappa} \kappa)_{ijk}$, $Y_{\nu_{ij}}$, $(A_{\nu} Y_{\nu})_{ij}$, $(m_{\widetilde{\nu}^c}^{2})_{ij}$ y $(m_{\widetilde{L}}^{2})_{ij}$ son simétricos en $i, j, k$.

Las masas de Dirac de los neutrinos está dada por $m_{D_{ij}}\equiv Y^{\nu}_{ij}v_{u}$. Experimentalmente se tiene que las masas de los neutrinos son muy pequeñas. Por lo tanto, los acoples de Yukawa de los neutrinos deben ser muy pequeños para obtener la escala de masas de neutrinos apropiada en el mecanismo de \textit{seesaw}. Si consideramos Ec.~(\ref{nui}) notamos el límite $Y^{\nu}_{ij}\rightarrow 0$ implica que $v_{\nu_i}\rightarrow 0$. 

En conclusión, para conseguir la correcta escala de masas de neutrinos tanto $Y^{\nu}_{ij}$ como  $v_{\nu_i}$ deben ser pequeños.

\spacing{1}


\chapter{Soluciones maximales de la física de neutrinos en el \texorpdfstring{$\mu\nu$}{munu}SSM}
\label{maximal}
\spacing{1.5}

En la Sección~\ref{seccionmeff} hemos descrito cómo funciona el mecanismo de \textit{seesaw} del $\mu \nu$SSM, en particular ilustrando distintos límites para la matriz efectiva de masas del sector de neutrinos, Ec.~(\ref{mefff}). Mostramos que si tomamos $v_{d}\rightarrow 0$ (ó $v_{\nu_i}>>\frac{Y^{\nu}_iv_{d}}{3\lambda}$, considerando $Y^{\nu}_{ii}=Y^{\nu}_i$), para los valores típicos de los parámetros, obtenemos [Ec.~(\ref{meffsimple})]:
\begin{equation}
	(m_{eff})_{ij}\approx \frac{v_{u}^{2}}{6\kappa v_{\nu^c}}Y^{\nu}_{i}Y^{\nu}_{j}(1-3\delta_{ij}) - \frac{v_{\nu_i}v_{\nu_j}}{2M},
	\label{meffap}
\end{equation}

A continuación veremos cómo encontrar soluciones, en el límite considerado, en un escenario maximal. En primera aproximación, los autoestados de masa poseen la misma composición de $\nu_{\mu}$ y $\nu_{\tau}$, como se ve en la Figura~\ref{figuraneutrinos}. Por lo tanto, vamos a considerar $Y^{\nu}_{2}=Y^{\nu}_{3}$ y $v_{\nu_2}=v_{\nu_3}$. Entonces la matriz de masas efectivas simplificada, Ec.~(\ref{meffap}), toma la siguiente forma:
\begin{equation}
m_{eff}=\left( {\begin{array}{ccc}
	d & c & c\\
	c & A & B\\
	c & B & A
	\end{array} } \right),
	\label{matrizap}
\end{equation}
donde
\begin{align}
	& d=-\frac{v_{u}^{2}}{3\kappa v_{\nu^c}}(Y^{\nu}_{1})^{2}-\frac{1}{2M}v_{\nu_1}^{2}, \cr
	& c=\frac{v_{u}^{2}}{6\kappa v_{\nu^c}}Y^{\nu}_{1}Y^{\nu}_{2}-\frac{1}{2M}v_{\nu_1}v_{\nu_2}, \cr
	& A=-\frac{v_{u}^{2}}{3\kappa v_{\nu^c}}(Y^{\nu}_{2})^{2}-\frac{1}{2M}v_{\nu_2}^{2}, \cr
	& B=-\frac{v_{u}^{2}}{6\kappa v_{\nu^c}}(Y^{\nu}_{2})^{2}-\frac{1}{2M}v_{\nu_1}^{2}.
\end{align}

Los autovalores de la matriz efectiva son:
\begin{align}
	& \frac{1}{2}(A+B-\sqrt{8c^{2}+(A+B-d)^{2}}+d), \cr
	& \frac{1}{2}(A+B+\sqrt{8c^{2}+(A+B-d)^{2}}+d), \cr
	& A-B.
\end{align}
y los autovectores correspondientes son:
\begin{align}
	& \left(-\frac{(A+B+\sqrt{8c^{2}+(A+B-d)^{2}}-d)}{2},c,c\right), \cr
	& \left(-\frac{(A+B-\sqrt{8c^{2}+(A+B-d)^{2}}-d)}{2},c,c\right), \cr
	& \left(0,-1,1\right),
\end{align}
los cuales no han sido normalizados por simplicidad. De manera explícita notamos que $\sin^{2}\theta_{13}=0$ y $\sin^{2}\theta_{23}=1/2$, como queremos en el régimen maximal. Por otra parte, tenemos libertad de elegir los parámetros libres para reproducir el valor experimental de $\sin^{2}\theta_{12}$.

Es importante remarcar que los valores de $\sin^{2}\theta_{13}$ y $\sin^{2}\theta_{23}$ hallados son consecuencia de haber elegido $Y^{\nu}_{2}=Y^{\nu}_{3}$ y $v_{\nu_2}=v_{\nu_3}$. Por lo tanto, dicho resultado es válido tanto en la fórmula simplificada de la matriz de masas, Ec.~(\ref{meffap}), como en la fórmula general, Ec.~(\ref{mefff}), pero con distintas expresiones de $A$, $B$, $c$ y $d$ en cada caso.
\bigskip

Si consideramos que $\sin^{2}\theta_{12}=0$, entonces podemos tomar $c=0$. Para ello, vamos a analizar el límite donde el neutrino electrónico se encuentra completamente desacoplado del resto, es decir $c\rightarrow 0$. En este escenario el segundo autovector no posee componente de $\nu_{e}$, por lo tanto $\sin^{2}\theta_{12}\rightarrow 0$, y su composición resulta mitad $\nu_{\mu}$ y mitad $\nu_{\tau}$. Si queremos analizar la situación fenomenológicamente viable donde $\sin^{2}\theta_{12}\neq 0$, haremos $c\neq 0$.

Los autovalores en el límite $c\rightarrow 0$ son:
\begin{align}
	& d, \cr
	& A+B, \cr
	& A-B,
\end{align}
donde
\begin{align}
	& \left|d\right|=\left|\frac{v_{u}^{2}}{3\kappa v_{\nu^c}}(Y^{\nu}_{1})^{2}+\frac{1}{2M}v_{\nu_1}^{2}\right|, \cr
	& \left|A+B\right|=\left|\frac{v_{u}^{2}}{6\kappa v_{\nu^c}}(Y^{\nu}_{2})^{2}+\frac{1}{M}v_{\nu_2}^{2}\right|, \cr
	& \left|A-B\right|=\frac{v_{u}^{2}}{2\kappa v_{\nu^c}}(Y^{\nu}_{2})^{2}.
\end{align}

Finalmente, hallamos que
\begin{align}
	& \Delta m^{2}_{atm}\approx\left|4AB\right|=\left|4\left(\frac{v_{u}^{4}(Y^{v}_{2})^{4}}{18\kappa^{2}v_{\nu^c}^2}-\frac{(v_{\nu_2})^{4}}{4M^{2}}-\frac{v_{u}^{2}(Y^{\nu}_{2})^{2}v_{\nu_2}^{2}}{12M\kappa v_{\nu^c}}\right)\right|, \cr
	& \Delta m^{2}_{sol}\approx\left(|(A-B)^{2}-d^{2}\right|=\left|\left(\frac{v_{u}^{2}(Y^{\nu}_{2})^{2}}{6\kappa v_{\nu^c}}+\frac{(v_{\nu_2})^{2}}{M}\right)^{2}-\left(\frac{v_{u}^{2}(Y^{\nu}_{1})^{2}}{3\kappa v_{\nu^c}}+\frac{(v_{\nu_1})^{2}}{2M}\right)^{2}\right|.
\end{align}

Es importante notar que necesitamos $\left|A-B\right|>\left|A+B\right|$ para obtener jerarquía normal, de lo contrario obtendríamos $\sin^{2}\theta_{12}=0$, aún cuando $c\neq 0$, pues el rol de $\theta_{12}$ y $\theta_{13}$ cambian. Para ello requerimos que $\left|M\right|>\left|2\kappa v_{\nu^c}\right|$, tomando $M<0$.

En contraparte, considerando jerarquía invertida, necesitamos que $\left|A-B\right|>\left|A+B\right|$ para obtener $\sin^{2}\theta_{12}\neq 0$ cuando $c\neq 0$ y tomaremos $M>0$.
\bigskip

El régimen trimaximal $\sin^{2}\theta_{13}=0$, $\sin^{2}\theta_{23}=1/2$ y $\sin^{2}\theta_{12}=1/3$ puede ser hallado considerando en (\ref{matrizap}) c=A+B-d.
En este caso, los autovalores son:
\begin{align}
	& -(A+B)+2d, \cr
	& 2(A+B)-d, \cr
	& A-B.
\end{align}

Remarcamos que el régimen maximal considerado está experimentalmente excluido. Mas es ilustrativo analizar dicho régimen por su simplicidad para estudiar el mecanismo de \textit{seesaw} del modelo. Para hallar soluciones más generales debemos romper la degeneración entre los $Y^{\nu}$ y $v_{\nu}$ de los $\nu_{\mu}$ y $\nu_{\tau}$, tal como hemos realizado en este trabajo.

\spacing{1}


\chapter{Anchura de decaimiento del gravitino a estados finales de tres cuerpos}

\label{3body}
\spacing{1.5}

El cálculo de los decaimientos del gravitino a nivel árbol mediante un fotón o un bosón $Z$, y mediante un bosón $W$, fueron realizados en Refs.~\cite{Choi:2010xn,Choi:2010jt,Grefe:2011dp,Diaz:2011pc}. Mostramos en este apéndice los resultados de sus anchuras de decaimiento con respecto a $s$~\cite{Grefe:2011dp}, donde $s$ es la masa invariante de los dos fermiones, $f$ y $\bar{f}$.
La anchura de decaimiento total puede ser obtenida integrando estos resultados en el rango de la masa invariante $0\leq s \leq m_{3/2}^2$. En el caso del intercambio de un fotón virtual, se debe integran sobre el rango  $4m_f^2\leq s \leq m_{3/2}^2$, para evitar un propagador divergente.

\noindent
\textbf{i. $\Psi_{3/2}\rightarrow\gamma^*/Z^* \, \nu_i\rightarrow f \,  \bar{f} \, \nu_i$} 
\bea
\begin{aligned}
 &\frac{d\Gamma(\Psi_{3/2}\rightarrow \gamma^*/Z^* \, \nu_i\rightarrow f \,  \bar{f} \, \nu_i)}{ds}\\ 
 & \approx \frac{m_{3/2}^3 \, \beta_s^2}{768 \, \pi^3 \, M_{Pl}^2}\left[ \frac{e^2 \, Q^2}{s} \, |U_{\tilde{\gamma} \nu_i}|^2 \, f_s \, \sqrt{1-4 \, \frac{m_f^2}{s}}\left( 1+2 \, \frac{m_f^2}{s}\right) \right.\\
 & + \frac{g_Z}{(s-m_Z^2)^2+m_Z^2 \, \Gamma_Z^2} \left\lbrace \vphantom{\frac12} g_Z \,  U_{\tilde{Z} \nu_i}^2 \, s \, (C_V^2+C_A^2) \, f_s \right. \\
 & - \frac{8}{3} \, \frac{m_Z}{m_{3/2}} \, g_Z \,  U_{\tilde{Z} \nu_i}  \, \left( \frac{v_{\nu_i}}{v}+ \sin\beta \, \text{Re}  \, U_{\tilde{H}_u^0 \nu_i} \, - \cos\beta \, \text{Re} \,  U_{\tilde{H}_d^0 \nu_i} \, \right)  \,  s \, (C_V^2+C_A^2) \, j_s \\
 & + \frac{1}{6} \, g_Z  \, m_Z^2  \, \left|\frac{v_{\nu_i}}{v}+ \sin\beta \,  U_{\tilde{H}_u^0 \nu_i} \, - \cos\beta \,  U_{\tilde{H}_d^0 \nu_i} \, \right|^2 \, (C_V^2+C_A^2) \, h_s \\
 & + e \, Q \left(  \,  \text{Re} \,  U_{\tilde{\gamma} \nu_i} \, (m_Z^2-s) \, + \, \text{Im} \,  U_{\tilde{\gamma} \nu_i} \,  m_z  \, \Gamma_Z \right)  \,  C_V \\
 & \left. \left. \times \left( \,  2  \, U_{\tilde{Z} \nu_i} \, f_s \, +\frac{8}{3} \, \frac{m_Z}{m_{3/2}} \left( \,  \frac{v_{\nu_i}}{v}+ \sin\beta \, \text{Re}  \, U_{\tilde{H}_u^0 \nu_i} \, -  \, \cos\beta \, \text{Re} \,  U_{\tilde{H}_d^0 \nu_i} \,  \right)   \, j_s \, \right) \vphantom{\frac12} \right\rbrace  \vphantom{\sqrt{1-4\frac{m_f^2}{s}}} \right]\ ,
\label{width1}
\end{aligned}
\eea
donde $g_Z=g_2 \, / \cos\theta_W$ es el acople de gauge del bosón $Z$ , $m_Z$ y $\Gamma_Z$ su masa y su anchura de decaimiento a dos fermiones, $v$ es el VEV del Higgs, $Q$ es la carga de los fermiones del estado final, y $C_V$ y $C_A$ son los coeficientes de la estructura $V-A$ del vértice del bosón $Z$ con dos fermiones
\bea
    C_V=\frac{1}{2}T^3-Q\sin^2\theta_W\ , \hspace{1cm} C_A=-\frac{1}{2}T^3\ . 
\label{}
\eea
Las funciones cinemáticas $\beta_s$, $f_s$, $j_s$ y $h_s$ están dadas por
\begin{equation*}
 \beta_s=1-\frac{s}{m_{3/2}^2}\ , \hspace{1cm} f_s=1+\frac{2}{3}\frac{s}{m_{3/2}^2} +\frac{1}{3}\frac{s^2}{m_{3/2}^4}\ ,
\end{equation*}

\bea
    j_s=1+\frac{1}{2}\frac{s}{m_{3/2}^2}\ , \hspace{1cm} h_s=1+10\frac{s}{m_{3/2}^2} +\frac{s^2}{m_{3/2}^4}\ . 
\label{kin}
\eea
Nótese que $U_{\tilde{\chi} \nu_i}$ denota la mezcla entre el neutralino $\tilde{\chi}$ y el neutrino $\nu_i$, obtenida a partir de la matriz de masas de los fermiones neutros (matriz de masas neutralino-neutrino).

\bigskip

\noindent \textbf{ii. $\Psi_{3/2}\rightarrow W^* \, l\rightarrow f  \, \bar{f}' \, l$} 
\bea
\begin{aligned}
 &\frac{d\Gamma(\Psi_{3/2}\rightarrow W^{+*} \, l^-_i\rightarrow f \,  \bar{f}' \, l^-_i)}{ds}\\ 
 & \approx \frac{g_2^2  \, m_{3/2}^3 \,  \beta_s^2}{1536 \, \pi^3 \, M_{Pl}^2 \, \left( (s-m_W^2)^2+m_W^2 \, \Gamma_W^2\right) } \left(\vphantom{\frac12}  \, s \,  U^2_{\tilde{W}^- l_i^-}  \, f_s \right. \\
 & \left. -\frac{8}{3} \, \frac{m_W}{m_{3/2}} \, s \, U_{\tilde{W}^- l_i^-} \,  
\left( \frac{v_{\nu_i}}{v}-\sqrt{2} \,  \cos \beta  \,  \text{Re} \,  U_{\tilde{H}_d^- l_i^-} \, \right)  \,  j_s \,  + \frac{1}{6} \,  m_W^2  \, \left|\frac{v_{\nu_i}}{v}-\sqrt{2} \,  \cos \beta  \, U_{\tilde{H}_d^- l_i^-} \, \right|^2 \,  h_s  \, \vphantom{\frac12}\right)\ ,
\label{width2}
\end{aligned}
\eea
donde $U_{\tilde{\chi}^- l_i^-}$ denota la mezcla entre el chargino $\tilde{\chi}^-$ y el leptón $l_i^-$, obtenida a partir de la matriz de masas de los fermiones cargados (matriz de masas chargino-leptón). Las funciones cinemáticas $\beta_s$, $f_s$, $j_s$ y $h_s$ son las mismas que se mostraron anteriormente.

\spacing{1}


\bibliography{munussm}
\bibliographystyle{JHEP}

\end{document}